\documentclass{aa}  

\usepackage{graphicx}
\usepackage{txfonts}
\usepackage{natbib}
\usepackage{arydshln}
\usepackage{color}
\usepackage{caption}
\usepackage{blindtext}
\usepackage[colorlinks, citecolor=blue, linkcolor=blue, urlcolor=blue]{hyperref}

\begin{document}

    \title{A high-precision abundance analysis of the nuclear benchmark star \object{HD~20}\thanks{$^{1)}$This  paper  includes  data  gathered  with  the  6.5 m \textit{Magellan}
          Telescopes located at Las Campanas Observatory, Chile. $^{2)}$Based in part on data products from observations made with ESO Telescopes under program IDs 090.B-0605(A) (PI: Chanam{\'e}) and 60.A-9036(A).}}
    \author{Michael Hanke\inst{1}
          \and
          Camilla Juul Hansen\inst{2}
          \and
          Hans-Günter Ludwig\inst{3}
          \and
          Sergio Cristallo\inst{4,5}
          \and
          Andrew McWilliam\inst{6}
          \and
          Eva K. Grebel\inst{1}
          \and
          Luciano Piersanti\inst{4,5}          
          }

   \institute{Astronomisches  Rechen-Institut, Zentrum  für  Astronomie  der  Universität  Heidelberg, Mönchhofstr. 12-14,  D-69120  Heidelberg, Germany\\
              e-mail: \href{mailto:mhanke@ari.uni-heidelberg.de}{mhanke@ari.uni-heidelberg.de}
              \and
              Max Planck Institute for Astronomy, Königstuhl 17, D-69117 Heidelberg, Germany
              \and
              Landessternwarte, Zentrum für Astronomie der Universität Heidelberg, Königstuhl 12, D-69117 Heidelberg, Germany
              \and
              INAF -- Osservatorio Astronomico d’Abruzzo, via M. Maggini snc, Teramo, Italy
              \and
              INFN -- Sezione di Perugia, Via A. Pascoli snc, Perugia, Italy
              \and
              Carnegie Observatories, 813 Santa Barbara St., Pasadena CA 91101, USA
              }

   \date{\today}
 
\abstract{Metal-poor stars with available detailed information about their chemical inventory pose powerful empirical benchmarks for nuclear astrophysics. Here we present our spectroscopic chemical abundance investigation of the metal-poor ($\mathrm{[Fe/H]}=-1.60\pm0.03$~dex), $r$-process-enriched ($\mathrm{[Eu/Fe]}=0.73\pm0.10$~dex) halo star HD~20 using novel and archival high-resolution data at outstanding signal-to-noise ratios (up to $\sim1000$~{\AA}$^{-1}$). By combining one of the first asteroseismic gravity measurements in the metal-poor regime from a TESS light curve with the spectroscopic analysis of iron lines under non-local thermodynamic equilibrium conditions, we derive a set of highly accurate and precise stellar parameters. These allow us to delineate a reliable chemical pattern that is comprised of solid detections of 48 elements, including 28 neutron-capture elements. Hence, we establish HD~20 among the few benchmark stars that have almost complete patterns and possess low systematic dependencies on the stellar parameters. Our light-element ($Z\leq30$) abundances are representative of other, similarly metal-poor stars in the Galactic halo with contributions from core-collapse supernovae of type II. In the realm of the neutron-capture elements, our comparison to the scaled solar $r$-pattern shows that the lighter neutron-capture elements ($Z\lesssim60$) are poorly matched. In particular, we find imprints of the weak $r$-process acting at low metallicities. Nonetheless, by comparing our detailed abundances to the observed metal-poor star BD~+17~3248, we find a persistent residual pattern involving mainly the elements Sr, Y, Zr, Ba, and La. These are indicative of enrichment contributions from the $s$-process and we show that mixing with material from predicted yields of massive, rotating AGB stars at low metallicity considerably improves the fit. Based on a solar ratio of heavy- to light-$s$ elements -- at odds with model predictions for the $i$-process -- and a missing clear residual pattern with respect to other stars with claimed contributions from this process, we refute (strong) contributions from such astrophysical sites providing intermediate neutron densities. Finally, nuclear cosmochronology is used to tie our detection of the radioactive element Th to an age estimate for HD~20 of $11.0\pm3.8$~Gyr.}
\keywords{Stars: abundances -- Stars: chemically peculiar -- Stars: individual: HD 20 -- Stars: evolution -- Nuclear reactions, nucleosynthesis, abundances -- Galaxy: halo} 
\titlerunning{Detailed abundance analysis of HD~20}
\maketitle

\section{Introduction}\label{Sec: Introduction}
Among the cornerstones of Galactic archeology are studies of metal-poor stars as bearers of fossil records of Galactic evolution. In this respect, revealing the kinematics and chemistry of this relatively rare subclass of stars provides vital insights into the build-up of galaxy components like the Galactic halo and the origin of the chemical elements.

Nucleosynthesis of iron-peak elements, from Si to approximately Zn (atomic numbers $14\leq Z \leq30$), is thought to be dominated by explosive nucleosynthesis, from both thermonuclear supernovae (Type~Ia exploding white dwarfs), and core-collapse supernovae (CCSNe, massive stars), whereas major production of elements from Li up to and including Si is thought to be dominated by hydrostatic burning processes \citep{Woosley95, Nomoto06, Kobayashi19}.

Beyond the iron peak, electrostatic Coulomb repulsion ensures that charged-particle reactions play a minuscule role in element synthesis (with the possible exception of proton-rich nuclei). Temperatures high enough for charged particles to overcome the Coulomb barrier photo-dissociate the larger nuclei. Thus, most of the elements heavier than the iron peak result from neutron captures, which are divided into the slow ($s$) and rapid ($r$) processes by their capture rates with respect to the $\beta$-decay timescale \citep{Burbidge57, Cameron57}. The involved neutron densities differ by many orders of magnitude and are thought to be $n<10^{8}$~cm$^{-3}$ and $n\gtrsim10^{20}$~cm$^{-3}$ for the $s$- and $r$-process, respectively \citep{Busso01, Meyer94}. In recent years, an additional, so-called intermediate ($i$) process -- representing neutron densities in between typical $r$- and $s$-values -- is gaining attention as models are capable of reproducing some peculiar chemical patterns typically found in C-rich metal-poor stars \citep[e.g.,][]{Roederer16, Hampel16, Koch19, Hampel19}. 

The main $s$-process is believed to be active during the thermally pulsing phases of asymptotic giant branch (AGB) stars, which provide the required low neutron fluxes \citep[e.g.,][]{Gallino98, Straniero06, Lugaro12, Karakas14}, whereas several sites have been proposed to generate neutron-rich environments for the $r$-process to occur. Viable candidates are neutrino-driven winds in CCSNe \citep{Arcones07, Wanajo13}, jets in magneto-rotational supernovae \citep[MR SNe,][]{Cameron03, Moesta18}, and neutron star mergers \citep[NSMs, e.g., ][]{Lattimer74, Chornock17}. The latter site lately gained a lot of attention since, for example, \citet{Pian17} found indications for short-lived $r$-process isotopes in the spectrum of the electromagnetic afterglow of the gravitational wave event GW170817 that was detected and confirmed as an NSM by the LIGO experiment \citep{Abbott17}. The authors, however, could not single out individual elements. Only later, direct spectroscopic investigations revealed the newly produced neutron-capture element Sr in this NSM \citep{Watson19}. Nonetheless, as stressed by  -- for example -- \citet{Cote19} and \citet{Ji19}, other sites like MR SNe may still be needed to explain the full budget of $r$-process elements observed in the Galaxy. Nuclear benchmark stars allow for detailed studies of each of the neutron-capture processes.

From an observational point of view, there have been a number of spectroscopic campaigns that specifically targeted metal-poor stars to constrain the nucleosynthesis of heavy elements in the early Milky Way, among which are -- to name a few -- \citet{Beers05}, \citet{Hansen12,Hansen14a}, and the works by the $r$-process alliance \citep[e.g.,][and follow-up investigations]{Hansen18a, Sakari18}. Following \citet{Beers05}, the rare class of $r$-process-rich stars is commonly subdivided by a somewhat arbitrary cut into groups of moderately enhanced $r$-I ($0.3\leq[$Eu/Fe]\footnote{Throughout this paper, we employ the standard bracket notation $[X/Y] = \left(\log{\epsilon(X)} - \log{\epsilon(Y)}\right) - \left(\log{\epsilon(X)} - \log{\epsilon(Y)}\right)_\odot$, with $\log{\epsilon(X)}=\log{(n_X/n_\mathrm{H})} + 12$ being the abundance of the chemical element $X$.}$\leq +1.0$~dex; $\mathrm{[Ba/Eu]<0}$~dex) and strongly enhanced $r$-II ($\mathrm{[Eu/Fe]}> +1.0$~dex; $\mathrm{[Ba/Eu]<0}$~dex) stars. In the context of this classification, our benchmark star HD~20 falls in the $r$-I category. Recently, \citet{Gull18} reported on the first finding of an $r$-I star with a combined ``$r$+$s$'' pattern, which was explained by postulating mass transfer from a companion that evolved through the AGB phase.

\begin{table}
\caption{Comparison of abundances for HD~20 in common between \citet{Burris00} and \citet{Barklem05}. Typical errors are 0.20 to 0.25 dex.}
\label{Table: Previous studies}
\centering
\resizebox{0.85\columnwidth}{!}{%
\begin{tabular}{lccc}
\hline\hline
\\[-5pt]
$X$                 & \multicolumn{2}{c}{$\log{\epsilon(X)}$ [dex]} & $\Delta$ [dex]\\
  & \citet{Burris00} & \citet{Barklem05}\\
\hline\\[-5pt]
Fe  & $\phantom{-}6.28$ &  $\phantom{-}5.92$ & 0.36\\
Sr  & $\phantom{-}1.56$ &  $\phantom{-}1.51$ & 0.05\\
Y   & $\phantom{-}0.80$ &  $\phantom{-}0.62$ & 0.18\\
Zr  & $\phantom{-}1.67$ &  $\phantom{-}1.40$ & 0.27\\
Ba  & $\phantom{-}1.32$ &  $\phantom{-}0.86$ & 0.46\\
La  & $\phantom{-}0.22$ &  $-0.08$           & 0.30\\
Nd  & $\phantom{-}0.69$ &  $\phantom{-}0.26$ & 0.43\\
Eu  & $-0.11$           &  $-0.27$           & 0.16\\
\hline
\end{tabular}}
\end{table}
Here, we present a comprehensive spectroscopic abundance analysis of HD~20, an $r$-process-rich star at the peak of the halo metallicity distribution function ([Fe/H$]=-1.60$~dex) with a heavy-element pattern that suggests pollution with $s$-process material. 

Based on the full 6D phase-space information from the second data release (DR2) of the \textit{Gaia} mission \citep{Gaia18}, \citet{Roederer18a} concluded that HD~20 may be chemodynamically associated with two other metal-poor halo stars with observed $r$-process excess. Based on its kinematics -- characterized by a highly eccentric orbit ($e=0.975^{+0.002}_{-0.004}$) and a close pericentric passage ($r_\mathrm{peri}=0.19^{+0.04}_{-0.02}$~kpc) -- and its low metallicity, the authors speculate that HD~20 and its associates may have been accreted from a disrupted satellite.

Among others, HD~20 has been a subject of two previous abundance studies by \citet{Burris00} and \citet{Barklem05} who reported eight and ten abundances for elements with $Z\geq30$, respectively. Both groups employed medium-resolution ($R=\lambda/FWHM\sim20\,000$) spectra at signal-to-noise ratios (S/N) slightly above 100~pixel$^{-1}$. Table \ref{Table: Previous studies} lists the findings for the eight elements that are in common between both works and we note systematic disagreements -- in a sense that the abundances by \citet{Burris00} generally are above \citet{Barklem05} -- exceeding even the considerable quoted errors of about 0.2~dex. The authors adopted very similar effective temperatures ($T_\mathrm{eff}$) for their analyses (5475~K versus 5445~K), while the employed stellar surface gravities ($\log{g}$) and microturbulent velocities ($v_\mathrm{mic}$) differ strongly by +0.41~dex and $-0.30$~km~s$^{-1}$. Inconsistencies between the studies are likely to be tied to these discrepancies as already recognized by \citet[][see also Appendix \ref{Sec: Abundance systematics due to atmosphere} for a detailed discussion of the impact of model parameters on individual stellar abundances]{Barklem05}.

Our work aims at painting a complete picture of the chemical pattern in HD~20 consisting of 58 species from the primordial light element Li to the heavy $r$-process element U. To this end, a compilation of high-quality, newly obtained and archival spectra was used, allowing for many elemental detections with high internal precisions. Furthermore, specific attention was devoted to the determination of accurate stellar parameters in order to mitigate the effect of systematic error contributions to the robustness of the deduced pattern. In this respect, an essential building block of our analysis is a highly accurate and precise stellar surface gravity from an asteroseismic analysis of the light curve that was obtained by NASA's \textit{Transiting Exoplanet Survey Satellite} \citep[TESS,][]{Ricker15}. Hence, we establish HD~20 as a new metal-poor benchmark star -- both in terms of fundamental properties as well as complete abundance patterns -- which, in light of its bright nature ($V\approx9$~mag), provides an ideal calibrator for future spectroscopic surveys. 

This paper is organized as follows: In Sect. \ref{Sec: Observations} we introduce the spectroscopic, photometric, and astrometric data sets employed throughout the analyses. Sect. \ref{Sec: Stellar parameters} is dedicated to the detailed discussion of our derived stellar parameters, followed by Sect. \ref{Sec: Abundance analysis}, which presents a description of the adopted procedures for the abundance analysis. Our results for HD~20 and constraints drawn from its abundance pattern can be found in Sect. \ref{Sec: Results and Discussion}. Finally, in Sect. \ref{Sec: Conclusion}, we summarize our findings and provide an outlook for further studies.

\section{Observations and data reduction}\label{Sec: Observations}
\subsection{Spectroscopic observations}\label{Subsec: Spectroscopic data}
We obtained a spectrum of HD~20 in the night of August 15, 2013 using both arms of the \textit{Magellan} Inamori Kyocera Echelle (MIKE) spectrograph \citep{Bernstein03}. An exposure of 1093~s integration time was taken using a slit width of 0.5\arcsec and a 2x1 on-chip-binning readout mode. This setup allowed for a full wavelength coverage from 3325 to 9160~{\AA} at a resolution of $R\approx45\,000$.

\begin{figure}
    \centering
    \resizebox{\hsize}{!}{\includegraphics{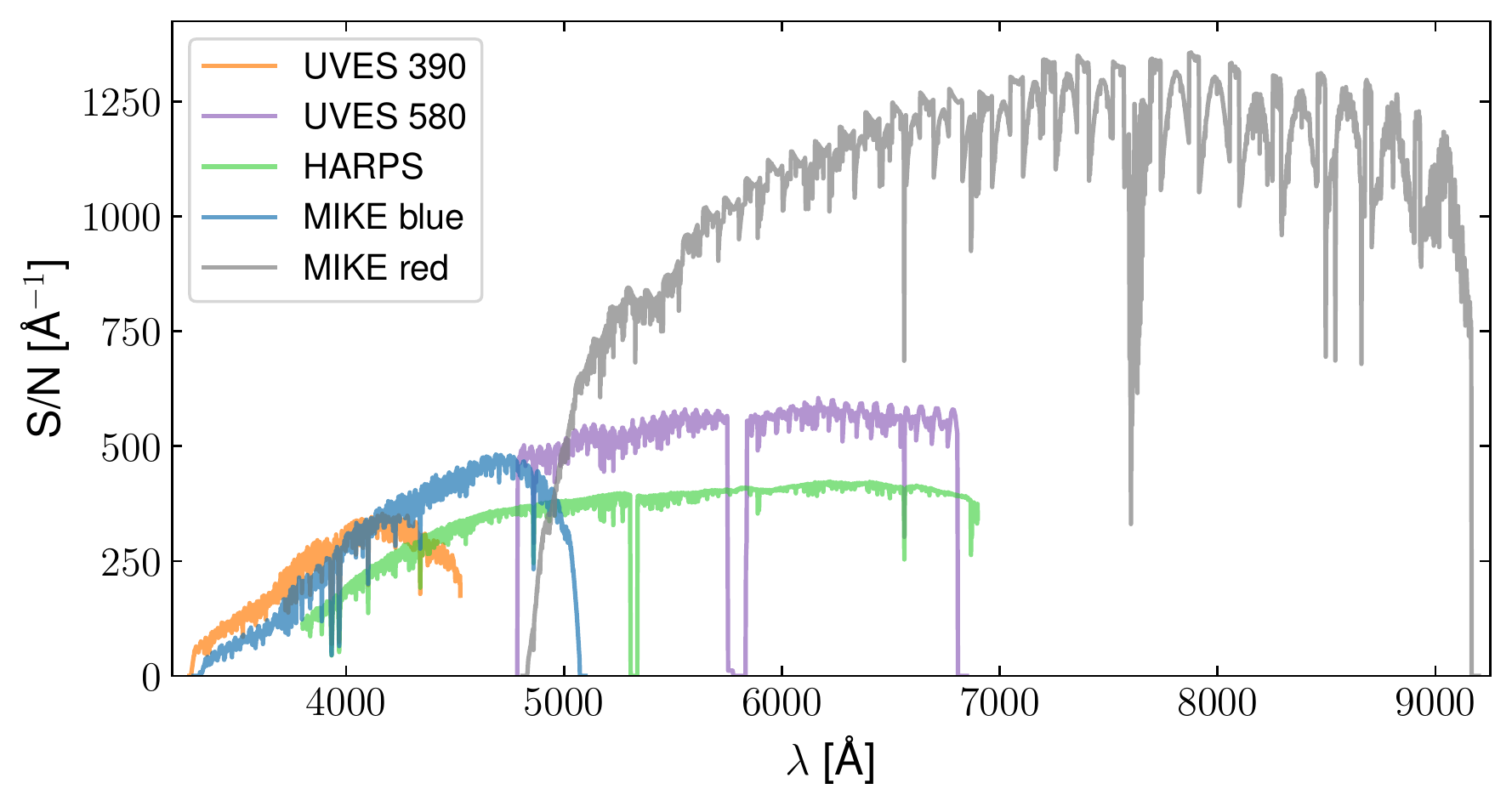}}
      \caption{S/N as a function of wavelength for the employed spectra of HD~20 from all three high-resolution spectrographs. Since the dispersion spacing between adjacent pixels varies among the instruments, we present the S/N per 1~{\AA}.
              }
      \label{Fig:SN_allspec}
\end{figure}
The raw science frame was reduced by means of the pipeline reduction package by \citet{Kelson03}, which performs flat-field division, sky modeling and subtraction, order tracing, optimal extraction, and wavelength calibration based on frames obtained with the built-in ThAr lamp. For the MIKE red spectrum, the reduction routine combined 26 ten-second "milky flat" exposures, taken using a quartz lamp and diffuser, resulting in a S/N of approximately 100 per 2x1 binned CCD pixel near the middle of the array, per exposure. This gave a total S/N of about 500 pixel$^{-1}$ in the combined flat. Due to lower flux in the blue quartz lamp, the milky flat exposure time was set to 20~s per frame. In addition, the 26 blue-side milky flat exposures were supplemented with seven ten-second exposures of a hot star, HR~7790, taken with the diffuser. The median seeing of 0.72{\arcsec}, corresponding to 5.4 CCD pixels FWHM, indicates that the flux for each wavelength point was taken from approximately 2 FWHM, or about 11 pixels. At the H$\alpha$ wavelength the pixels are 0.047~{\AA} wide, indicating roughly 21 pixels per {\AA}. These details suggest that the S/N of the final, extracted, flat field flux is 5000 to 7000~{\AA}$^{-1}$, significantly greater than the S/N of the stellar spectrum. The resulting S/N of the extracted object spectrum ranges from about 40~{\AA}$^{-1}$ at the blue-most edge to more than 1200~{\AA}$^{-1}$ redward of 7000~{\AA}. We present the detailed distribution of S/N with wavelength in Fig. \ref{Fig:SN_allspec}.

Our MIKE observation was complemented by data retrieved through the ESO Advanced Data Products (ADP) query form, with two additional, reduced high-resolution spectra for this star: The first is a 119~s reduced exposure (ID 090.B-0605(A)) from the night of October 13, 2012 using the UVES spectrograph with a dichroic \citep{Dekker00} at the ESO/VLT Paranal Observatory. For the blue arm, a setup with an effective resolution of $R\sim58\,600$ centered at a wavelength of 390~nm (UVES 390) was chosen, whereas the red arm was operated at $R\sim66\,300$ with a central wavelength of 580~nm (UVES 580). Especially the UVES 390 exposure poses an additional asset, since it supersedes our MIKE spectrum in the UV at higher S/N and -- more importantly -- bluer wavelength coverage and considerably higher resolution. 

The second ESO spectrum was taken on December 29, 2006 (ID 60.A-9036(A)) employing the HARPS spectrograph \citep{Mayor03} at the 3.6 m Telescope at the ESO La Silla Observatory. With a similar wavelength coverage and at substantially lower S/N than the UVES spectra, this observation adds a very high resolution of 115\,000 that was used to corroborate our findings for the intrinsic line broadening (Sect. \ref{Subsec: Rotational velocity}). The S/N values reached with both ESO spectrographs are shown in Fig. \ref{Fig:SN_allspec} alongside the distribution for MIKE.

\subsection{Radial velocities and binarity}\label{Subsec: Radial velocities and binarity}
\begin{figure}
    \centering
    \resizebox{\hsize}{!}{\includegraphics{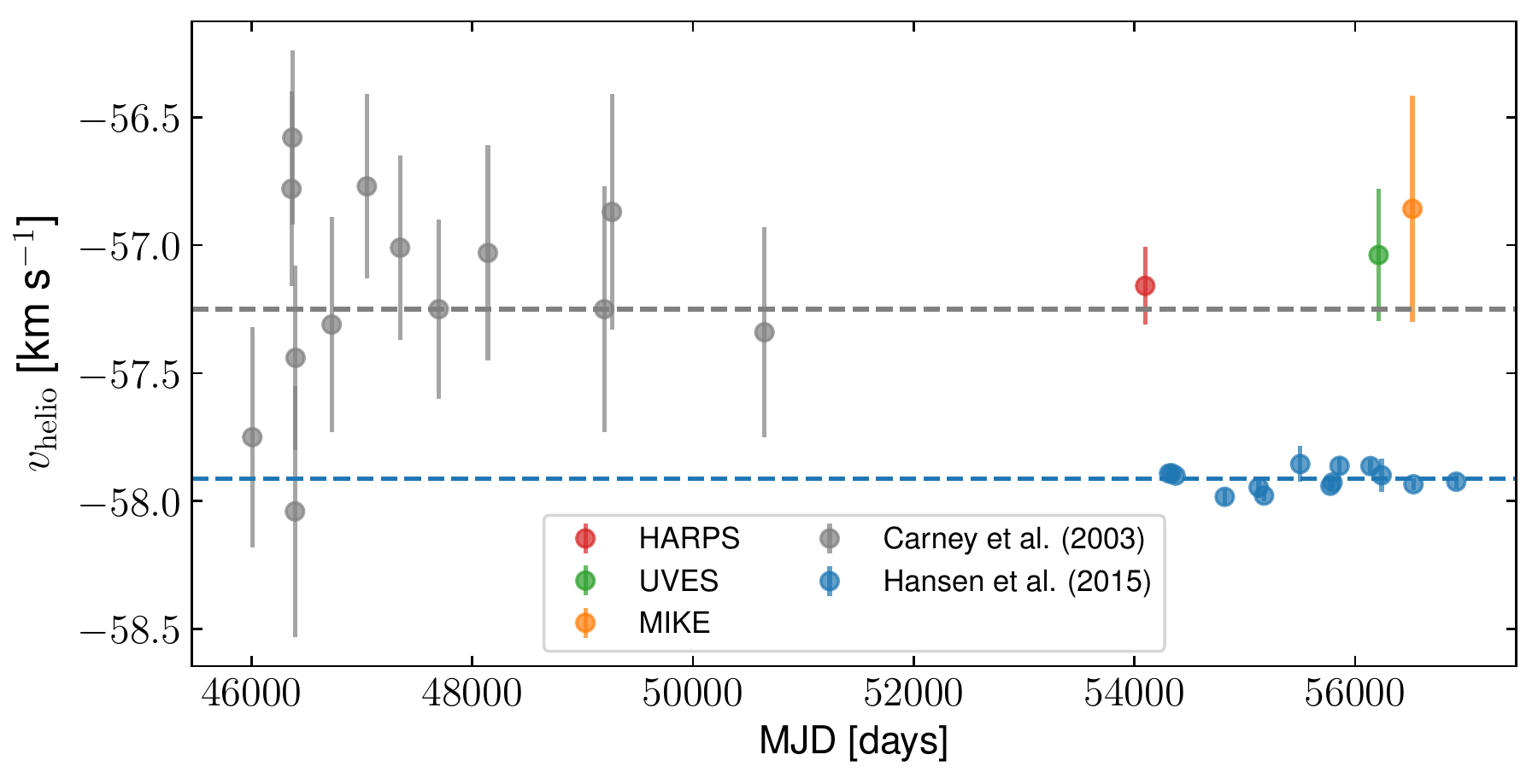}}
      \caption{Comparison of literature time series for $v_\mathrm{helio}$ by \citet[][gray filled circles]{Carney03} and \citet[][blue filled circles]{Hansen15} to measurements from the three spectra employed throughout this study (see legend). The gray and blue dashed lines resemble the median values for the two reference samples.
              }
      \label{Fig:RV_comparisons}
\end{figure}
All spectra were shifted to the stellar rest frame after determining radial velocities through cross correlation with a synthetic template spectrum of parameters that are representative for HD~20 (see Sect. \ref{Sec: Stellar parameters} and Table \ref{Table: Target information}). For the HARPS and UVES spectra, we established the radial velocity zero point using standard stars that were observed in the same nights \citep[HD~69830 and HD~7041, respectively, with reference values from][]{Soubiran18}, whereas we used the telluric O$_2$ B-band at $\sim6900$~{\AA} to calibrate the MIKE spectrum. This way, we found $v_\mathrm{helio}=-57.16\pm0.15$, $-57.04\pm0.26$, and $-56.86\pm0.44$~km~s$^{-1}$ from the HARPS, UVES, and MIKE spectra of HD~20. These findings are consistent with the mean value $-57.18\pm0.11$~km~s$^{-1}$ from the radial-velocity monitoring program by \citet{Carney03} and considerably above the reported value by \citet{Hansen15} of $-57.914\pm0.041$~km~s$^{-1}$. A graphical juxtaposition is shown in Fig. \ref{Fig:RV_comparisons}. We note that -- owing to the usage of different spectrographs and resolutions -- our radial velocity analysis is by no means homogeneous and slight discrepancies are therefore to be expected. Nevertheless, the observed offset with respect to \citet{Hansen15} is significant. The anomaly with respect to \citet{Carney03} has already been noted by \citet{Hansen15} and was linked to a difference in the applied scales. Apart from this systematic bias, over a time span of $10\,011$ days, there is no indication of real radial velocity variations. As a consequence, a binary nature of HD~20 can be ruled out with high confidence.

\subsection{Photometry and astrometric information}\label{Subsec: Photometric and astrometric data}
\begin{table}
\caption{Fundamental properties and stellar parameters entering this work.}
\label{Table: Target information}
\centering
\resizebox{\columnwidth}{!}{%
\begin{tabular}{lcccl}
\hline\hline
Quantity         & Value             & Unit & \multicolumn{1}{c}{Source$^a$} & Note\\
\\[-5pt]
\multicolumn{5}{c}{(Astro-) physical constants}\\
\hline\\[-5pt]  
$L_0$                 & $3.0128\cdot10^{28}$          &  W   & 1 & \\
$L_\sun$              & $3.828\cdot10^{26}$           &  W   & 1 & \\
$M_\mathrm{bol,\sun}$ & 4.74                          &  mag & 1 & \\
$R_\sun$              & $6.9577\pm0.0014\cdot10^{8}$  &  m   & 2 & \\
$T_\mathrm{eff,\sun}$ & 5771                          &  K   & 2 & \\
$\log{g}_\sun$        & 4.438                         &  dex & 3 & \\
\hline\\[-5pt]
\multicolumn{5}{c}{Observables}\\
\hline\\[-5pt]                               
$B$                       & $9.65\pm0.02$   & mag  & 4 & \\
$V$                       & $9.059\pm0.013$   & mag  & 5 & \\
$J_\mathrm{2MASS}$        & $7.704\pm0.030$   & mag  & 6 & \\
$H_\mathrm{2MASS}$        & $7.348\pm0.029$   & mag  & 6 & \\
$K_\mathrm{s,2MASS}$      & $7.249\pm0.031$   & mag  & 6 & \\
$b-y$                     & $0.434\pm0.003$   & mag  & 7 & \\
$E(B-V)$                  & $0.0149\pm0.0005$ & mag  & 8 & \\
$G$                       & 8.849             & mag  & 9 & \\ 
$G_\mathrm{BP}-G_\mathrm{RP}$ & 0.886         & mag  & 9 & \\
$\varpi$                     & $1.945\pm0.053$   & mas  & 9 & \\
$\mu_\alpha \cos{\delta}$ & $132.434\pm0.066$ & mas yr$^{-1}$  & 9 & \\
$\mu_\delta$              & $-39.917\pm0.058$ & mas yr$^{-1}$  & 9 & \\
$v_\mathrm{helio}$        & $-57.914\pm0.041$   & km s$^{-1}$    & 10 & \\[5pt] 
$f_\mathrm{max}$        & $27.19^{+1.34}_{-1.17}$ & $\mu$Hz & 11 & Sect. \ref{Subsec: TESS asteroseismology}\\[5pt] 
\hline\\[-5pt]
\multicolumn{5}{c}{Deduced quantities}\\
\hline\\[-5pt]
$d$                  & $507\pm13$        & pc                 & 11 & Sect. \ref{Subsec: Photometric and astrometric data}\\[5pt]
$\log{g}$            & $2.366^{+0.020}_{-0.021}$        & dex    & 11 & Sect. \ref{Subsec: TESS asteroseismology}\\[5pt]
$T_\mathrm{eff}$     & $5246^{+76}_{-50}$       & K                  & 11 & Sect. \ref{Subsubsec: Bayesian inference}\\[5pt]
$v_\mathrm{mic}$     & $1.95^{+0.09}_{-0.06}$     & km s$^{-1}$        & 11 & Sect. \ref{Subsubsec: Bayesian inference}\\[5pt]
$[$M/H]              & $-1.60\pm0.03$    & dex                & 11 & Sect. \ref{Subsubsec: Bayesian inference}\\[5pt]
$[$Fe/H]             & $-1.60\pm0.03$    & dex                & 11 & Sect. \ref{Subsec: Model atmosphere parameters}\\[5pt]
$v_\mathrm{mac}$           & $5.82\pm0.03$     & km s$^{-1}$        & 11 & Sect. \ref{Subsec: Rotational velocity}\\[5pt]
$L/L_\odot$          & $60.9^{+4.6}_{-4.3}$ &  & 11 & Sect. \ref{Subsec: Other structural parameters}\\[5pt]
$R/R_\odot$          & $9.44^{+0.46}_{-0.43}$  &  & 11 & Sect. \ref{Subsec: Other structural parameters}\\[5pt]
$m/M_\odot$          & $0.76\pm0.08$ &         & 11 & Sect. \ref{Subsec: Other structural parameters}\\[5pt]
$^{12}\mathrm{C}/^{13}\mathrm{C}$ & $3.92^{+1.68}_{-0.98}$ &     & 11 & Sect. \ref{Subsubsec: C, N, and O}\\[5pt]
$[\alpha$/Fe]$^b$    & $0.45$           & dex                & 11 & Sect. \ref{Subsec: up to the iron peak}\\[5pt]
age                  & $11.0\pm3.8$      & Gyr                & 11 & Sect. \ref{Subsec:Cosmochronological age}\\[5pt]
\hline
\end{tabular}}
\tablefoot{
\tablefoottext{a}{References: (1): \citet{Mamajek15}; (2): \citet{Heiter15} and references therein; (3): \citet{Prsa16}; (4): \citet{Hoeg00}; (5): \citet{AnthonyTwarog94}; (6): \citet{Skrutskie06}; (7): \citet{Hauck98}; (8): \citet{Schlafly11}; (9): \citet{Gaia18}; (10): \citet{Hansen15}; (11): This study.}
\tablefoottext{b}{$[\alpha$/Fe] = $\frac{1}{5} \mathrm{[(Mg + Si + S + Ca + Ti)/Fe]}$.}
}
\end{table}
Visual to near-infrared broadband photometric information for HD~20 was compiled from the literature and is listed in Table \ref{Table: Target information} together with the respective errors and sources.

$BVR_\mathrm{C}I_\mathrm{C}$ photometry was presented in \citet{Beers07} in a program that was targeting specific stars such as HD~20. Their results were also employed by \citet{Barklem05} and follow-up works by relying on the deduced parameters. The authors report $V=9.236\pm0.001$~mag, which is in strong disagreement to other findings in the literature. For example, the Hipparcos catalog \citep{HIPPARCOS97} lists $V=9.04$~mag \citep[used for temperature estimates in the spectroscopic studies of][]{Gratton00,Fulbright03}, while \citet{AnthonyTwarog94} provide a consistent value of $V=9.059\pm0.013$~mag \citep[used, e.g., by][]{Carney03}. Furthermore, we estimate $V\approx9.00\pm0.05$~mag from \textit{Gaia} photometry and the analytical relation for $(G-V)$ as a function of $G_\mathrm{BP}$ and $G_\mathrm{RP}$\footnote{Sect. 5.3.7 of the \textit{Gaia} Data Release 2 Documentation release 1.2: \url{https://gea.esac.esa.int/archive/documentation/GDR2/}}. For completeness, we mention here the finding of $V=9.40$~mag by \citet{Ducati02}, which again poses a strong deviation. We point out that HD~20 does not expose any signs of photometric variability as revealed by time-resolved photometry over 6.6~yr from DR9 of the All-Sky Automated Survey for Supernovae \citep[ASAS-SN,][]{Jayasinghe19} showing -- again in agreement with most of the literature -- $V=9.01\pm0.08$~mag. 

Despite the rather low quoted internal uncertainties, we hence discard the photometry by \citet{Beers07} and \citet{Ducati02} from consideration as we suspect inaccuracies in the calibration procedures. A disruptive factor might be a blend contribution by a star about 14{\arcsec} to the southeast, though we deem this an unlikely option since \textit{Gaia} DR2 reports it to be much fainter ($G=8.849$~mag versus 14.675~mag). Consequently, we resorted to magnitudes for the $B$-band from the Tycho-2 catalog \citep{Hoeg00} and for $V$ by \citet{AnthonyTwarog94}. For the near-infrared $JHK_\mathrm{s}$ photometry we queried the 2MASS catalog \citep{Skrutskie06} and the Strömgren color $b-y$ is taken from \citet{Hauck98}.

In terms of reddening we applied $E(B-V)=0.0149\pm0.0005$~mag, which was extracted from the reddening maps by \citet{Schlafly11}. Whenever dereddened colors or extinction-corrected magnitudes were employed, we adopted the optical extinction ratio $R_V=A(V)/E(B-V)=3.1$ attributed to the low-density interstellar medium (ISM) together with the reddening ratios $E(\mathrm{color})/E(B-V)$ compiled in Table 1 of \citet{Ramirez05}. Considering the overall very low reddening of HD~20, uncertainties in the latter ratios ought to have negligible impact on the quantities deduced from photometry.

A parallax of $\varpi=1.945\pm0.053$~mas was retrieved from \textit{Gaia} DR2 from which we computed a geometric distance to HD~20 of $d=507\pm13$~pc\footnote{Though mathematically incorrect, the error on the inverse parallax can be considered symmetric in light of the small relative parallax error.}. Here, we accounted for the quasar-based parallax zero point for \textit{Gaia} DR2 of $-0.029$~mas \citep{Lindegren18}. Our finding is fully in line with the distance $507^{+14}_{-13}$~pc derived in the Bayesian framework of \citet{BailerJones18}.

\section{Stellar parameters}\label{Sec: Stellar parameters}
A crucial part of any spectroscopic analysis aiming at high-accuracy chemical abundances is the careful determination of the stellar parameters entering the model atmospheres needed when solving for the radiative transfer equations. Here, we outline the inference method applied for determining the parameters; effective temperature, surface gravity, microturbulence, metallicity, and line broadening. 

Our adopted stellar parameters (Table \ref{Table: Target information}) are based on a spectroscopic analysis of Fe lines that were corrected for departures from the assumption of local thermodynamic equilibrium (LTE) together with asteroseismic information from the TESS mission, whereas several other techniques -- both spectroscopic and photometric -- including their caveats are discussed in Appendix \ref{Sec: Alternative methods for determining stellar parameters}.

\subsection{Surface gravity from TESS asteroseismology}\label{Subsec: TESS asteroseismology}
 \begin{figure}
    \centering
    \resizebox{\hsize}{!}{\includegraphics{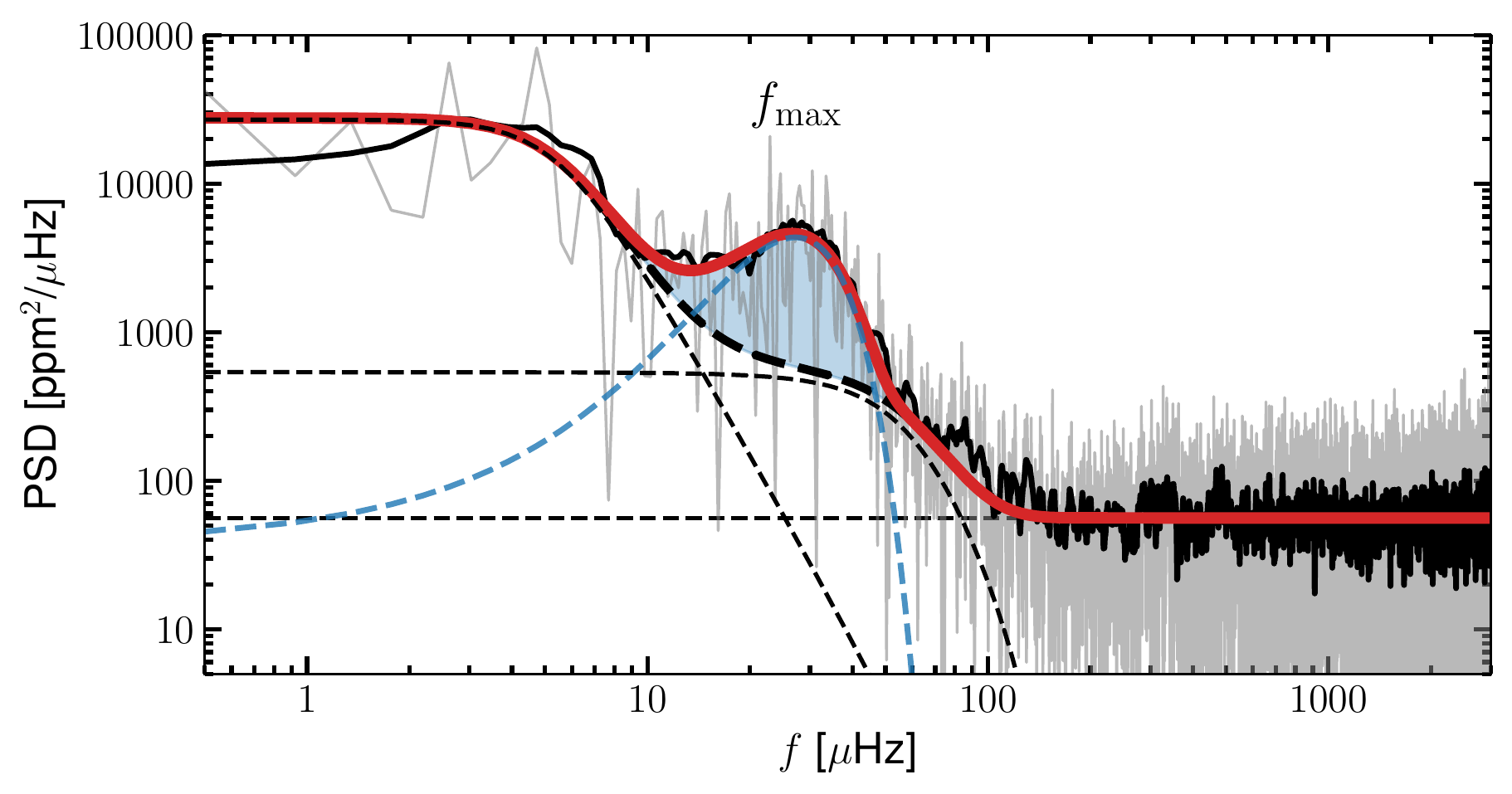}}
      \caption{Power spectral density (PSD) for HD~20 based on TESS light-curve data. The thick black line depicts a smoothed version of the PSD (thin gray line) and the best-fit model is shown in red. The blue shaded area indicates the power excess, whereas individual model components are represented by thin blue and black dashed lines. 
              }
      \label{Fig:asterosmeismic_TESS}
\end{figure}
Recently, \citet{Creevey19} showed in their time-resolved radial velocity analysis of the benchmark star HD~122563 that the asteroseismic scaling relation
\begin{equation}\label{Eq: astero scaling relation}
 \log{g}_\mathrm{seis.} = \log{g}_\odot + \log{\left(\frac{f_\mathrm{max}}{f_{\mathrm{max},\odot}} \sqrt{\frac{T_\mathrm{eff}}{T_{\mathrm{eff},\odot}}}\right)}
\end{equation}
based on the frequency $f_\mathrm{max}$ of maximum power of solar-like oscillations holds even in the regime of metal-poor and evolved stars. This motivated the exploration of the feasibility of an asteroseismic gravity determination for HD~20. 

Fortunately, TESS measured a 27.4 days light curve with a two-minute cadence for this star during Sector~2. We employed the \textit{lightkurve} Python package \citep{lightkurve} to retrieve and reduce the data in order to calculate the power spectrum seen in Fig. \ref{Fig:asterosmeismic_TESS}. A power excess is identifiable around the frequency $f_\mathrm{max}\approx27$~$\mu$Hz, which we attribute to solar-like oscillations. 

We performed a fit to the obtained power spectrum following the prescriptions by \citet{Campante19}. Therefore, we assumed a multi-component background model consisting of super-Lorentzian profiles that account for various granulation effects \citep[see, e.g.,][for details]{Corsaro17} as well as a constant noise component. The decision on the number of super-Lorentzian components for the background was made based on Bayesian model comparison using Bayes factors from evidences that were estimated with the \textit{Background}\footnote{\url{https://github.com/EnricoCorsaro/Background}} extension to the high-DImensional And multi-MOdal NesteD Sampling \citep[DIAMONDS\footnote{\url{https://github.com/EnricoCorsaro/DIAMONDS}},][]{Corsaro14} algorithm. We found that a model with three super-Lorentzian components has an insignificantly stronger support compared to a two-component one. The latter observation indicates that -- given the data -- the meso-granulation around frequencies of $f_\mathrm{max}/3\approx9$~$\mu$Hz is indistinguishable from the component due to super-granulation and/or other low-frequency signals since they occupy a similar frequency range in HD~20. Thus, we adopted only two super-Lorentzians for the background fit. Finally, a Gaussian profile was used to represent the power excess on top of the background model. 

In order to sample and optimize the high-dimensional parameter space of all involved model coefficients, we again made use of DIAMONDS. The resulting best-fit model, as well as its individual components, are depicted in Fig. \ref{Fig:asterosmeismic_TESS}. We estimated $f_\mathrm{max}=27.19^{+1.34}_{-1.17}$~$\mu$Hz which translates into $\log{g}=2.368^{+0.021}_{-0.019}$~dex from Eq. \ref{Eq: astero scaling relation} using $f_{\mathrm{max},\odot}=3050$~$\mu$HZ \citep{Kjeldsen95} and our adopted $T_\mathrm{eff}$. Owing to a weak coupling of the asteroseismic gravity to the temperature, we do not consider it in isolation, but refer the reader to Sect. \ref{Subsubsec: Bayesian inference}, where we outline the procedure to reach simultaneous parameter convergence.

\subsection{Iron lines}
A list of suitable \ion{Fe}{i} and \ion{Fe}{ii} lines for the purpose of deriving accurate stellar parameters was compiled using the Atomic Spectra Database \citep{NIST_ASD} of the National Institute of Standards and Technology (NIST). To this end, in order to mitigate biases by uncertain oscillator strengths ($\log{gf}$), only those lines were considered that are reported to have measured $\log{gf}$ values with accuracy levels $\leq10\%$ (grade B or better in the NIST evaluation scheme) for \ion{Fe}{i} and $\leq25\%$ (grade C or better) for \ion{Fe}{ii} lines. The lines retrieved this way were checked to be isolated by means of spectrum synthesis (see Sect. \ref{Subsec: Line list}) and their EW was measured by EWCODE (Sect. \ref{Subsec: Equivalent widths}). From these, we added the ones that were measured with more than $5\sigma$ significance to the final list. Laboratory line strengths for the resulting 133 \ion{Fe}{i} transitions were measured and reported by \citet{Fuhr88}, \citet{BWL}, \citet{BKK}, and \citet{BK}. For the 13 \ion{Fe}{ii} lines that survived the cleaning procedure, the data are from \citet{Schnabel04}.

\subsection{Spectroscopic model atmosphere parameters}\label{Subsec: Model atmosphere parameters}
Throughout our analyses we employed the LTE radiation transfer code MOOG \citep[][July 2017 release]{Sneden73} including an additional scattering term in the source function as described by \citet{Sobeck11}\footnote{\url{https://github.com/alexji/moog17scat} as of November 2018}. Our atmosphere models are based on the grid of 1D, static, and plane parallel ATLAS9 atmospheres by \citet{Castelli03} with opacity distribution functions that account for $\alpha$-enhancements ([$\alpha$/Fe]=+0.4, Sect. \ref{Subsec: up to the iron peak}). Models for parameters between the grid points were constructed via interpolation in the grid. Here, we used the iron abundance [Fe/H] as proxy for the models' overall metallicities [M/H], since we assume that all elements other than the $\alpha$-elements follow the solar elemental distribution scaled by [Fe/H]. We note that the fact that HD~20 shows enhancements in the neutron-capture elements (Sect. \ref{Subsec.Neutron-capture elements}) does not prevent this assumption, as the elements in question are only detectable in trace amounts with negligible impact on atmospheric properties such as temperature, density, gas-, or electron pressure.

\begin{figure}
    \centering
    \resizebox{\hsize}{!}{\includegraphics{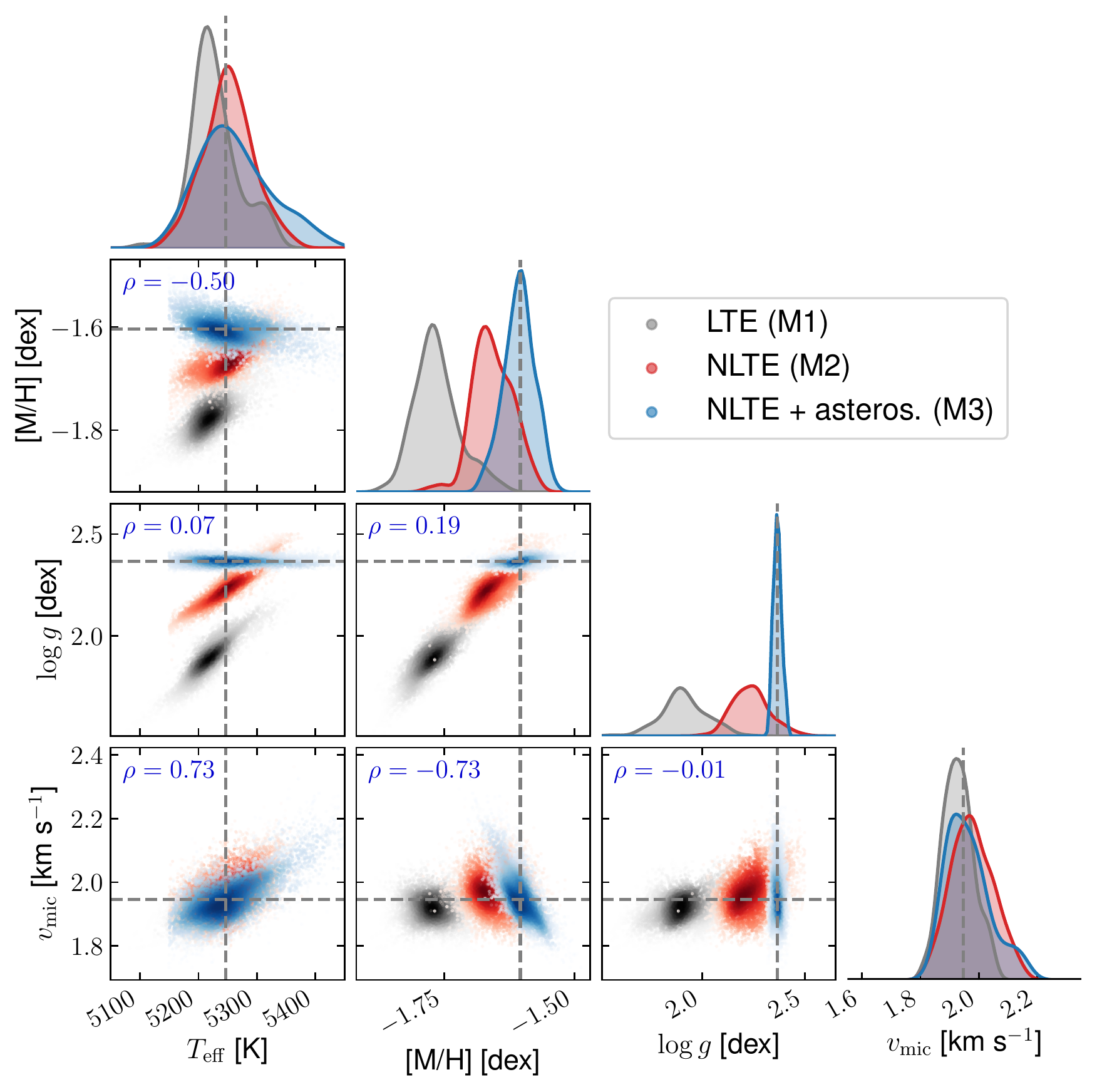}}
      \caption{Samples drawn from the posterior distribution of the stellar parameters (Eq. \ref{Eq: posterior}). Shown are the three different approaches M1 (gray), M2 (red), and M3 (blue) with the darkness of the colors illustrating the local density as estimated from a Gaussian kernel density estimate. The sample sizes are $2\cdot10^4$ and the adopted stellar parameters from method M3 (Tables \ref{Table: Target information} and \ref{Table: Posterior statistics}) based on the median of the distributions are indicated using horizontal and vertical dashed lines. The correlation coefficients for pairs of two parameters in M3 are presented in the \textit{top left corner} of each panel. The marginalized, one-dimensional distributions for the individual parameters are depicted by smoothed histograms at the top of each column.  
              }
      \label{Fig:MCMC_stellpar}
\end{figure}

Our $T_\mathrm{eff}$ estimate is based on the spectroscopic excitation balance of \ion{Fe}{i} lines. This technique relies on tuning the model temperature such that lines at different lower excitation potential ($\chi_\mathrm{ex}$) yield the same abundance -- in other words a zero-slope of $\log{\epsilon(\ion{Fe}{i})}$ with $\chi_\mathrm{ex}$ is enforced. In this respect it is important to account for the circumstance that \ion{Fe}{i} transitions are prone to substantial non-LTE (NLTE) effects in metal-poor stars, in a sense that not only the overall abundance is shifted toward higher values, but the magnitude of the effect varies with $\chi_\mathrm{ex}$, too. Hence, as pointed out by \citet{Lind12}, the $T_\mathrm{eff}$ for which the excitation trend is leveled is shifted to systematically offset temperatures from the LTE case (see Fig. \ref{Fig:MCMC_stellpar}). To overcome this problem, we computed NLTE abundance departures by interpolation in a close-meshed, precomputed grid of corrections that was created specifically for this project and parameter space \citep[priv. comm.: M. Bergemann and M. Kovalev, see][for details]{Bergemann12a, Lind12}.

The microturbulence parameter $v_\mathrm{mic}$ is an ad-hoc parameter that approximatively accounts for the effects of otherwise neglected turbulent motions in the atmosphere, which mainly affect the theoretical line strength of strong lines. Here, we tuned $v_\mathrm{mic}$ in order to erase trends of the inferred, NLTE-corrected abundances for \ion{Fe}{i} features with the reduced line strength, $\mathrm{RW} = \log{\left(\mathrm{EW}/\lambda\right)}$.

\begin{figure}
    \centering
    \resizebox{0.9\hsize}{!}{\includegraphics{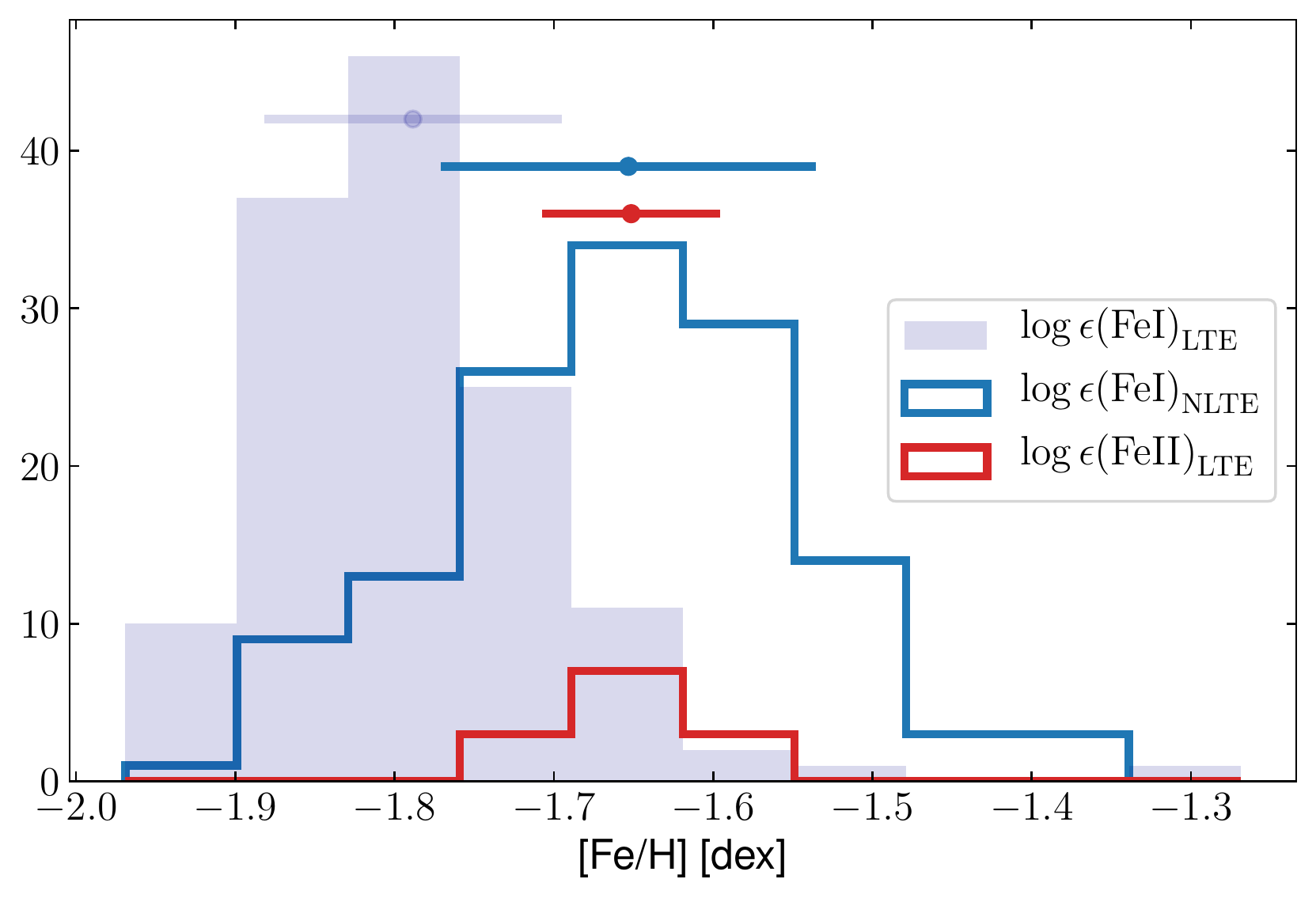}}
      \caption{Diagnostic plot for spectroscopic ionization balance. Shown are the histograms of the Fe abundance distributions ([Fe/H$]=\log{\epsilon(Fe)}-7.50$~dex) at the adopted gravity ($\log{g}=2.24$~dex from method M2, see Sect. \ref{Subsubsec: Bayesian inference}) both in LTE (gray filled) and NLTE (blue) for \ion{Fe}{i} and in LTE for \ion{Fe}{ii} (red). NLTE corrections for \ion{Fe}{ii} remain well below 0.01~dex and are therefore neglected here. Points with error bars and arbitrary ordinate offsets at the top of the panel denote the means and standard deviations for each of the distributions of the same color.
              }
      \label{Fig:ionization_balance}
\end{figure}
Even though we prefer our highly accurate asteroseismic measurement over requiring spectroscopic ionization balance for determining $\log{g}$, we discuss this method here to compare our findings to more classical spectroscopic parameter estimation methods that are widely used throughout the literature. The procedure is based on balancing abundances of neutral lines and singly ionized lines that are sensitive to changes in gravity (see also Appendix \ref{Sec: Abundance systematics due to atmosphere}). Hence, by tuning the model gravity to erase discrepancies between the abundances deduced from both ionization states of the same element, $\log{g}$ can be inferred. Commonly, especially for FGK stars, the high number of available Fe lines in both ionization stages qualifies this species as an ideal indicator. While the modeling of \ion{Fe}{ii} line strengths is insensitive to departures from LTE (<0.01~dex), trustworthy gravities from the ionization balance can only be obtained once departures from LTE are removed from the \ion{Fe}{i} abundances \citep[e.g.,][]{Lind12}. In particular, by neglecting NLTE influences, one would considerably underestimate $\log{\epsilon(\ion{Fe}{i})}$ and consequently $\log{g}$. This can be seen in Fig. \ref{Fig:ionization_balance}, where we compare \ion{Fe}{i} under the LTE assumption to NLTE-corrected \ion{Fe}{i}. Illustrated is the best abundance agreement -- that is a perfect overlap of both the $\log{\epsilon(\ion{Fe}{i})}_\mathrm{NLTE}$ and $\log{\epsilon(\ion{Fe}{Ii})}_\mathrm{LTE}$\footnote{Since we find corrections for \ion{Fe}{ii} that amount to less than 0.01~dex, we can assume $\log{\epsilon(\ion{Fe}{Ii})}_\mathrm{LTE}=\log{\epsilon(\ion{Fe}{Ii})}_\mathrm{NLTE}$.} abundance distributions -- obtained for $\log{g}=2.24$~dex and [M/H$] = -1.65$~dex.

When assessing the error budget on [Fe/H], we caution that in this study's realm of very high S/N spectra, random noise is not the prevailing origin for the line-by-line scatter of 0.10~dex and 0.03~dex for $\log{\epsilon(\ion{Fe}{i})}_\mathrm{NLTE}$ and $\log{\epsilon(\ion{Fe}{ii})}_\mathrm{LTE}$, respectively. In fact, looking at the abundance errors for individual lines from EW errors only (Table \ref{Table: Line-by-line abundances}), the random component remains well below 0.03~dex in the majority of cases. We conclude that the scatter is mostly of non-stochastic nature -- for example due to uncertain oscillator strengths and flaws in the 1D assumption -- and hence a division of the rms scatter by the square root of the number of lines is not a statistically meaningful quantifier of the metallicity error (see Appendix \ref{Subsec: Spectrograph comparison} for more detailed discussions).

\subsubsection{Bayesian inference}\label{Subsubsec: Bayesian inference}
We emphasize that spectroscopic stellar parameters are strongly interdependent, that is, uncertainties and systematic errors of one quantity should not be considered in isolation. The usage of asteroseismic information mitigates this circumstance only to some degree as we show below. Hence, all model parameters need to be iterated until simultaneous convergence is reached. For this purpose, we used the \textit{emcee} \citep{ForemanMackey13} Python implementation of a Markov chain Monte Carlo (MCMC) sampler in order to draw samples from the posterior probability distribution $P$ of the four model parameters $T_\mathrm{eff}$, [M/H], $\log{g}$, and $v_\mathrm{mic}$,
\begin{equation}\label{Eq: posterior}
 P(\vec{x}|\vec{y}) \propto \mathcal{L}(\vec{y}|\vec{x})\cdot p(\vec{x}),
\end{equation}
with $\mathcal{L}$ being the likelihood function and $p$ the prior. Here, $\vec{y}$ denotes the measured EWs and $\vec{x}$ represents the set of model atmosphere parameters. A flat prior of unity was assumed within the parameter space covered by our grid of NLTE corrections, and zero otherwise. We explored three different likelihoods representing the purely spectroscopic LTE (M1) and NLTE (M2) methods, as well as a mixed ``NLTE + asteroseismology'' (M3) approach. The likelihoods take the form
\begin{equation}\label{Eq: likelihood}
 \mathcal{L} = \exp{\left(-\frac{a^2_{\chi_\mathrm{ex}}}{2\sigma_a^2} - \frac{b^2_{\mathrm{RW}}}{2\sigma_b^2} - \frac{\Delta^2_\mathrm{[M/H],\ion{Fe}{ii}}}{2\sigma^2_\mathrm{\ion{Fe}{ii}}} - \Gamma_i\right)},
\end{equation}
where $a_{\chi_\mathrm{ex}}$ and $b_{\mathrm{RW}}$ are the slopes of the deduced LTE (M1) or NLTE (M2 and M3) abundances, $\log{\epsilon_\mathrm{\ion{Fe}{i}}(\vec{y}, \vec{x}, \chi_\mathrm{ex})}$, with $\chi_\mathrm{ex}$ and RW for any given set of parameters $\vec{x}$. The variances of the latter slopes were determined from repeated linear fits to bootstrapped samples by means of robust least squares involving a smooth L1 loss function. We prefer this non-parametric approach over ordinary least squares because of the systematically underestimated abundance errors from EW uncertainties alone (see previous Sect.). The third term in Eq. \ref{Eq: likelihood} represents the difference between the model metallicity and \ion{Fe}{ii} abundance, whereas $\Gamma_i$ introduces the gravity sensitivity. For approaches M1 and M2 it represents the ionization (im-)balance
\begin{equation}
 \Gamma_\mathrm{MI/MII} = \frac{\Delta^2_{\ion{Fe}{i},\ion{Fe}{ii}}}{2\sigma^2_\Delta}
\end{equation}
in the LTE and NLTE case, while we do not enforce ionization balance for M3, but use the asteroseismic information through
\begin{equation}
 \Gamma_\mathrm{M3} = \frac{\log{g}-\log{g}_\mathrm{seis.}(T_\mathrm{eff}, f_\mathrm{max})}{2\sigma^2_{\log{g}_\mathrm{seis.}}}.
\end{equation}
In this expression $\log{g}_\mathrm{seis.}$ is calculated from Eq. \ref{Eq: astero scaling relation}. We emphasize that -- while being clearly subject to biases in LTE -- a perfect ionization balance may not be desirable even in the 1D NLTE case (M2), because it still lacks proper descriptions of hydrodynamical and 3D conditions. These might pose other sources for differences between abundances from \ion{Fe}{i} and \ion{Fe}{ii} at the true $\log{g}$. In fact, there is a remaining marginal ionization imbalance $\log{\epsilon{\ion{Fe}{ii}}}-\log{\epsilon{\ion{Fe}{i}}}=0.08\pm0.10$ when adopting approach M3.

\begin{table}
\caption{Median values and 68.2\% confidence intervals for the stellar parameters from the posterior distributions for the three different likelihood functions (see main text for details). The method adopted throughout this work is M3.}
\label{Table: Posterior statistics}
\centering
\resizebox{\columnwidth}{!}{%
\begin{tabular}{lcccc}
\hline\hline
\\[-5pt]
Method & $T_\mathrm{eff}$ & [M/H] & $\log{g}$ & $v_\mathrm{mic}$\\
       & [K]              & [dex] & [dex]     & km~s$^{-1}$\\
\hline\\[-5pt]
LTE (M1) & $5220^{+35}_{-32}$ & $-1.78^{+0.04}_{-0.04}$ & $1.894^{+0.115}_{-0.115}$ & $1.92^{+0.05}_{-0.05}$\\[5pt]
NLTE (M2) & $5250^{+48}_{-43}$ & $-1.65^{+0.04}_{-0.04}$ & $2.244^{+0.102}_{-0.095}$ & $1.97^{+0.07}_{-0.06}$\\[5pt]
NLTE + asteros. (M3) & $5246^{+76}_{-50}$ & $-1.60^{+0.03}_{-0.03}$ & $2.366^{+0.020}_{-0.021}$ & $1.95^{+0.09}_{-0.06}$\\[5pt]

\hline
\end{tabular}}
\end{table}
Figure \ref{Fig:MCMC_stellpar} shows various representations of the multidimensional posterior distributions for M1, M2, and M3. As expected, we found strong correlations between $T_\mathrm{eff}$, [M/H], and $\log{g}$ in the purely spectroscopically informed methods M1 and M2. Using approach M3, we can effectively lift the degeneracies with $\log{g}$ as quantified by insignificant Pearson correlation coefficients (Fig. \ref{Fig:MCMC_stellpar}). For each approach, we deduced the optimal parameters and error margins from the median, 15.9$^\mathrm{th}$, and 84.1$^\mathrm{th}$ percentiles, respectively. These are listed in Table \ref{Table: Posterior statistics}. It is evident that M1 significantly underestimates both [M/H] and $\log{g}$ due to deducing lower \ion{Fe}{i} abundances that have a direct impact on the ionization balance and therefore the inferred gravity. M2 and M3, however, yield results that are in good agreement with the strongest deviation amounting to just $1.2\sigma$ in $\log{g}$. This highlights the importance of considering NLTE effects already at the stage of stellar parameter inference and shows that 1D NLTE ionization balance is capable of producing gravities that are as accurate as the highly trustworthy asteroseismic scaling relations. Since the precision of the latter is better by about a factor of five, we adopt the parameters inferred from M3 throughout this work. We corroborated this set of fundamental stellar parameters using several independent techniques, including temperatures from the shapes of the Balmer lines in HD~20's spectrum. The reader is referred to Appendix~\ref{Sec: Alternative methods for determining stellar parameters} for a detailed outline and comparison.

\subsection{Line broadening}\label{Subsec: Rotational velocity}
\begin{figure*}
    \centering
    \resizebox{0.9\hsize}{!}{\includegraphics{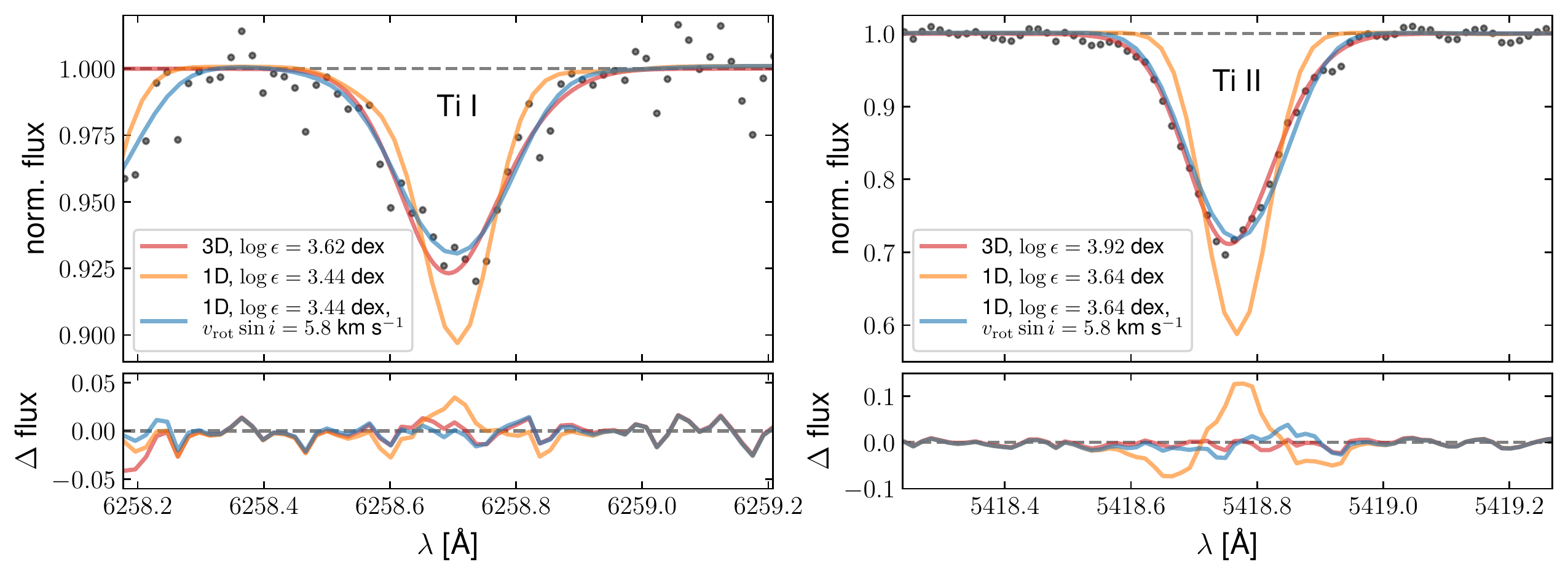}}
      \caption{Comparison of synthetic line shapes against the observed profiles in the UVES 580 spectrum for two representative Ti lines. Red spectra resemble 3D syntheses, while blue and orange colors indicate 1D syntheses with and without additional broadening. The instrumental profile ($R=66\,230$) was mimicked by convolution with a Gaussian kernel for all three types of synthesis. No rotational broadening was applied to the 3D syntheses.
              }
      \label{Fig:3D_Ti}
\end{figure*}
\citet{Carney03} reported a rotational velocity of $v_\mathrm{rot}\sin{i} = 5.9$~km s$^{-1}$ for HD~20, which is unexpectedly high given the evolutionary state of this star where any initial rotation is expected to be eliminated. The authors caution, however, that the face value just below their instrumental resolution of 8.5~km~s$^{-1}$ might be biased due to a number of systematic influences on their method, amongst which is turbulent broadening \citep[see also, e.g., ][]{Preston19}. Turbulent and rotational broadening have almost identical impacts on the line shape, a degeneracy that can only be broken using spectra of very high resolution and S/N \citep{Carney08}. Hence -- despite the name -- we rather consider $v_\mathrm{rot}\sin{i}$ a general broadening parameter.

Given that rotation or any other line broadening mechanism are key quantities that critically affect the precision and accuracy of abundances from spectrum synthesis (Sect. \ref{Sec: Abundance analysis}), we tackled this property from a theoretical point of view. To this end, a collection of isolated \ion{Ti}{i} and \ion{Ti}{ii} features were simulated using LTE radiative transfer in a CO$^5$BOLD model atmosphere \citep{Freytag12}, which realistically models the microphysics of stellar atmospheres under 3D, hydrodynamical conditions. We note that the chosen atmospheric parameters ($T_\mathrm{eff}=5500$~K, $\log{g}=2.5$~dex, [M/H$]=-2.0$~dex) only roughly match our findings -- hence deviations in the abundance scales can be expected. The overall line-shape, however, is expected to be reasonably accurately reproduced. Our synthetic profiles were compared to their observed counterparts in the UVES 580 spectrum, which offers the best trade-off between resolution and S/N in the considered wavelength regimes. The nominal velocity resolution is 4.5~km~s$^{-1}$. Comparisons for two representative lines are presented in Fig. \ref{Fig:3D_Ti}. The 3D profiles are shown next to rotationally broadened, 1D versions and we find that no additional rotational broadening is required in the 3D case as the line shape can be fully recovered by properly accounting for microphysics together with the instrumental resolution. Thus, we conclude that -- if at all -- HD~20 is rotating only slowly (i.e., $v\sin{i}\lesssim1$~km~s$^{-1}$). On top of the overall line broadening, slight profile asymmetries are correctly reproduced by the 3D models.   

In order to improve our 1D spectrum syntheses beyond broadening by the instrumental line spread function, we analyzed the deviation of individual, isolated Fe features from their 1D LTE line shape. The comparison was performed against the UVES 580 and the HARPS spectrum. Based on 171 lines in common for both spectra, we found that a broadening velocity of $v_\mathrm{mac} = 5.82 \pm 0.03$~km~s$^{-1}$ can successfully mimic the line shape from both spectrographs. The latter value is in good agreement with the value 5.9~km~s$^{-1}$ found by \citet{Carney03}, who do not list an error specific to HD~20 but quote general standard errors between 0.5 and 3~km~s$^{-1}$ for their entire sample of stars.

\subsection{Other structural parameters}\label{Subsec: Other structural parameters}
Given our spectroscopic temperature and metallicity, we can deduce HD~20's luminosity through
\begin{equation}\label{Eq: luminosity}
 \frac{L}{L_\odot} = \left(\frac{d}{10\mathrm{ pc}}\right)^2 \frac{L_0}{L_\odot}\cdot 10^{-0.4\left(V - A(V) + BC_V(T_\mathrm{eff}, \mathrm{[Fe/H]})\right)}
\end{equation}
with the zero-point luminosity $L_0$ (see Table \ref{Table: Target information}) and the bolometric correction $BC_V$ from the calibration relation by \citet[henceforth AAM99]{Alonso99}, which itself depends on $T_\mathrm{eff}$ and [Fe/H]. We find $L/L_\odot=60.9^{+4.6}_{-4.3}$, in line with the value $58.6\pm2.2$ reported in \textit{Gaia} DR2. The error on $L$ was computed through a Monte Carlo error propagation assuming Gaussian error distributions for the input variables and an additional uncertainty for $BC_V$ of 0.05~mag. The asymmetric error limits stem from the 15.9 and 84.1 percentiles of the final parameter distributions, respectively.

We can furthermore infer the stellar radius using 
\begin{equation}
 \frac{R}{R_\odot}=\sqrt{\frac{L}{L_\odot}}\left(\frac{T_\mathrm{eff}}{T_{\mathrm{eff},\odot}}\right)^{-2}
\end{equation}
resulting in $9.44^{+0.46}_{-0.43}$. This compares to $8.69^{+0.19}_{-0.80}$ from \textit{Gaia} DR2, where the slight discrepancy can be explained by a higher temperature estimate from \textit{Gaia} (see discussion in Sect. \ref{Subsubsec: Color - [Fe/H] - Teff relations}).

Finally, it is possible to deduce a mass estimate using the basic stellar structure equation
\begin{equation}\label{Eq: Stellar structure}
\log{\frac{m}{m_\sun}} = \log{\frac{g}{g_\odot}} - 4\log\frac{T_\mathrm{eff}}{T_\mathrm{eff,\sun}}+0.4(M_\mathrm{bol,\sun}-M_\mathrm{bol}).
\end{equation}
The involved solar reference values can be found in Table \ref{Table: Target information}. As for Eq. \ref{Eq: luminosity}, the bolometric magnitude $M_\mathrm{bol}$ can be computed from the $V$-band photometry and the $BC_V$ relation by AAM99. We find a mass of $(0.76\pm0.08)M_\odot$.

\section{Abundance analysis}\label{Sec: Abundance analysis}
\begin{table}
\caption{Final adopted abundances.}
\label{Table: Final abundances}
\centering
\resizebox{\columnwidth}{!}{%
\addtolength{\tabcolsep}{-3pt}
\begin{tabular}{@{\extracolsep{6pt}}l@{}r@{}r@{}rc@{}r@{}r@{}r}
\hline\hline\\[-7pt]
 & \multicolumn{3}{c}{LTE} & \multicolumn{3}{c}{NLTE} & \\[2pt]
\cline{2-4} \cline{5-7}\\[-5pt]
$X$ & \multicolumn{1}{c}{$\langle\log{\epsilon(X)}\rangle^a$} & \multicolumn{1}{c}{[$X$/Fe]$^b$} & $n$ & $\langle\log{\epsilon(X)}\rangle^a$ & \multicolumn{1}{c}{[$X$/Fe]} & $n$ & $\log{\epsilon(X)}_\sun^c$\\[3pt]
 & \multicolumn{1}{c}{[dex]} & \multicolumn{1}{c}{[dex]} & & \multicolumn{1}{c}{[dex]} & \multicolumn{1}{c}{[dex]} & & \multicolumn{1}{c}{[dex]}\\
\hline\\[-5pt]
\ion{Li}{i } & $<-0.34$ & $< 0.41$ &   1 & \dots & \multicolumn{1}{c}{\dots} & \dots &  1.05\\[2pt]
C (CH) & $  6.25\pm0.05$ & $ -0.38\pm0.07$ &   1 & \dots & \multicolumn{1}{c}{\dots} & \dots &  8.43\\[2pt]
N (NH) & $  6.21\pm0.10$ & $  0.18\pm0.11$ &   1 & \dots & \multicolumn{1}{c}{\dots} & \dots &  7.83\\[2pt]
\ion{O }{i } & $  7.79\pm0.18$ & $  0.70\pm0.18$ &   2 & $ 7.79\pm0.18$ & $ 0.70\pm0.18$ &   2 &  8.69\\[2pt]
\ion{Na}{i } & $  4.50\pm0.10$ & $  0.06\pm0.11$ &   2 & $ 4.42\pm0.10$ & $-0.14\pm0.12$ &   2 &  6.24\\[2pt]
\ion{Mg}{i } & $  6.25\pm0.10$ & $  0.45\pm0.11$ &   3 & $ 6.24\pm0.10$ & $ 0.32\pm0.12$ &   3 &  7.60\\[2pt]
\ion{Al}{i } & $  4.54\pm0.15$ & $ -0.11\pm0.16$ &   5 & $ 4.68\pm0.06$ & $-0.09\pm0.09$ &   5 &  6.45\\[2pt]
\ion{Si}{i } & $  6.21\pm0.13$ & $  0.50\pm0.14$ &  16 & $ 6.23\pm0.03$ & $ 0.40\pm0.08$ &   5 &  7.51\\[2pt]
\ion{Si}{ii} & $  6.49\pm0.10$ & $  0.58\pm0.10$ &   2 & \dots & \multicolumn{1}{c}{\dots} & \dots &  7.51\\[2pt]
\ion{S }{i } & $  6.03\pm0.04$ & $  0.71\pm0.06$ &   4 & \dots & \multicolumn{1}{c}{\dots} & \dots &  7.12\\[2pt]
\ion{K }{i } & $  4.15\pm0.10$ & $  0.92\pm0.11$ &   2 & $ 3.60\pm0.10$ & $ 0.25\pm0.12$ &   2 &  5.03\\[2pt]
\ion{Ca}{i } & $  4.92\pm0.06$ & $  0.38\pm0.08$ &  25 & \dots & \multicolumn{1}{c}{\dots} & \dots &  6.34\\[2pt]
\ion{Sc}{ii} & $  1.66\pm0.06$ & $  0.11\pm0.06$ &  11 & \dots & \multicolumn{1}{c}{\dots} & \dots &  3.15\\[2pt]
\ion{Ti}{i } & $  3.34\pm0.10$ & $  0.19\pm0.11$ &  20 & $ 3.81\pm0.10$ & $ 0.54\pm0.13$ &  15 &  4.95\\[2pt]
\ion{Ti}{ii} & $  3.67\pm0.10$ & $  0.31\pm0.10$ &  28 & \dots & \multicolumn{1}{c}{\dots} & \dots &  4.95\\[2pt]
\ion{V }{i } & $  2.07\pm0.12$ & $ -0.06\pm0.13$ &   4 & \dots & \multicolumn{1}{c}{\dots} & \dots &  3.93\\[2pt]
\ion{V }{ii} & $  2.50\pm0.10$ & $  0.16\pm0.10$ &   2 & \dots & \multicolumn{1}{c}{\dots} & \dots &  3.93\\[2pt]
\ion{Cr}{i } & $  3.67\pm0.12$ & $ -0.17\pm0.13$ &  12 & \dots & \multicolumn{1}{c}{\dots} & \dots &  5.64\\[2pt]
\ion{Cr}{ii} & $  4.25\pm0.15$ & $  0.21\pm0.15$ &  15 & \dots & \multicolumn{1}{c}{\dots} & \dots &  5.64\\[2pt]
\ion{Mn}{i } & $  3.33\pm0.08$ & $ -0.30\pm0.10$ &   4 & $ 3.73\pm0.08$ & $-0.02\pm0.11$ &   4 &  5.43\\[2pt]
\ion{Fe}{i } & $  5.70\pm0.07$ & $  0.00\pm0.09$ & 133 & $ 5.82\pm0.10$ & $ 0.00\pm0.13$ & 133 &  7.50\\[2pt]
\ion{Fe}{ii} & $  5.90\pm0.03$ & $  0.00\pm0.04$ &  13 & $ 5.90\pm0.03$ & $ 0.00\pm0.04$ &  13 &  7.50\\[2pt]
\ion{Co}{i } & $  3.25\pm0.11$ & $  0.06\pm0.12$ &   6 & $ 3.61\pm0.10$ & $ 0.30\pm0.12$ &   3 &  4.99\\[2pt]
\ion{Ni}{i } & $  4.43\pm0.07$ & $  0.01\pm0.09$ &  32 & \dots & \multicolumn{1}{c}{\dots} & \dots &  6.22\\[2pt]
\ion{Cu}{i } & $  1.76\pm0.10$ & $ -0.63\pm0.11$ &   3 & \dots & \multicolumn{1}{c}{\dots} & \dots &  4.19\\[2pt]
\ion{Zn}{i } & $  2.88\pm0.10$ & $  0.12\pm0.11$ &   2 & \dots & \multicolumn{1}{c}{\dots} & \dots &  4.56\\[2pt]
\ion{Ga}{i } & $  1.03\pm0.20$ & $ -0.21\pm0.21$ &   1 & \dots & \multicolumn{1}{c}{\dots} & \dots &  3.04\\[2pt]
\ion{Rb}{i } & $< 1.52$ & $< 0.45$ &   1 & \dots & \multicolumn{1}{c}{\dots} & \dots &  2.87\\[2pt]
\ion{Sr}{i } & $  1.00\pm0.10$ & $ -0.07\pm0.11$ &   1 & $ 1.40\pm0.10$ & $ 0.21\pm0.12$ &   1 &  2.87\\[2pt]
\ion{Sr}{ii} & $  1.50\pm0.26$ & $  0.23\pm0.26$ &   1 & \dots & \multicolumn{1}{c}{\dots} & \dots &  2.87\\[2pt]
\ion{Y }{ii} & $  0.54\pm0.09$ & $ -0.07\pm0.09$ &   7 & \dots & \multicolumn{1}{c}{\dots} & \dots &  2.21\\[2pt]
\ion{Zr}{ii} & $  1.26\pm0.09$ & $  0.28\pm0.09$ &   5 & $ 1.41\pm0.10$ & $ 0.42\pm0.10$ &   2 &  2.58\\[2pt]
\ion{Mo}{i } & $  0.48\pm0.10$ & $  0.40\pm0.11$ &   1 & \dots & \multicolumn{1}{c}{\dots} & \dots &  1.88\\[2pt]
\ion{Ru}{i } & $  0.55\pm0.10$ & $  0.60\pm0.11$ &   1 & \dots & \multicolumn{1}{c}{\dots} & \dots &  1.75\\[2pt]
\ion{Rh}{i } & $ -0.19\pm0.40$ & $  0.70\pm0.40$ &   1 & \dots & \multicolumn{1}{c}{\dots} & \dots &  0.91\\[2pt]
\ion{Pd}{i } & $ -0.12\pm0.20$ & $  0.11\pm0.21$ &   1 & \dots & \multicolumn{1}{c}{\dots} & \dots &  1.57\\[2pt]
\ion{Ag}{i } & $ -0.29\pm0.20$ & $  0.57\pm0.21$ &   1 & \dots & \multicolumn{1}{c}{\dots} & \dots &  0.94\\[2pt]
\ion{Ba}{ii} & $  0.93\pm0.10$ & $  0.35\pm0.10$ &   2 & $ 0.74\pm0.10$ & $ 0.16\pm0.10$ &   2 &  2.18\\[2pt]
\ion{La}{ii} & $ -0.09\pm0.06$ & $  0.41\pm0.06$ &   6 & \dots & \multicolumn{1}{c}{\dots} & \dots &  1.10\\[2pt]
\ion{Ce}{ii} & $  0.19\pm0.03$ & $  0.21\pm0.04$ &  11 & \dots & \multicolumn{1}{c}{\dots} & \dots &  1.58\\[2pt]
\ion{Pr}{ii} & $ -0.35\pm0.10$ & $  0.53\pm0.10$ &   3 & \dots & \multicolumn{1}{c}{\dots} & \dots &  0.72\\[2pt]
\ion{Nd}{ii} & $  0.21\pm0.06$ & $  0.39\pm0.06$ &  22 & \dots & \multicolumn{1}{c}{\dots} & \dots &  1.42\\[2pt]
\ion{Sm}{ii} & $ -0.06\pm0.04$ & $  0.58\pm0.05$ &  12 & \dots & \multicolumn{1}{c}{\dots} & \dots &  0.96\\[2pt]
\ion{Eu}{ii} & $ -0.35\pm0.10$ & $  0.73\pm0.10$ &   3 & \dots & \multicolumn{1}{c}{\dots} & \dots &  0.52\\[2pt]
\ion{Gd}{ii} & $  0.04\pm0.15$ & $  0.56\pm0.15$ &   5 & \dots & \multicolumn{1}{c}{\dots} & \dots &  1.07\\[2pt]
\ion{Tb}{ii} & $ -0.74\pm0.10$ & $  0.55\pm0.10$ &   2 & \dots & \multicolumn{1}{c}{\dots} & \dots &  0.30\\[2pt]
\ion{Dy}{ii} & $  0.21\pm0.07$ & $  0.70\pm0.07$ &   4 & \dots & \multicolumn{1}{c}{\dots} & \dots &  1.10\\[2pt]
\ion{Ho}{ii} & $ -0.49\pm0.10$ & $  0.63\pm0.10$ &   2 & \dots & \multicolumn{1}{c}{\dots} & \dots &  0.48\\[2pt]
\ion{Er}{ii} & $ -0.04\pm0.09$ & $  0.64\pm0.09$ &   4 & \dots & \multicolumn{1}{c}{\dots} & \dots &  0.92\\[2pt]
\ion{Tm}{ii} & $ -0.87\pm0.00$ & $  0.63\pm0.02$ &   4 & \dots & \multicolumn{1}{c}{\dots} & \dots &  0.10\\[2pt]
\ion{Yb}{ii} & $ -0.06\pm0.20$ & $  0.70\pm0.20$ &   1 & \dots & \multicolumn{1}{c}{\dots} & \dots &  0.84\\[2pt]
\ion{Lu}{ii} & $ -0.57\pm0.10$ & $  0.93\pm0.10$ &   1 & \dots & \multicolumn{1}{c}{\dots} & \dots &  0.10\\[2pt]
\ion{Hf}{ii} & $ -0.23\pm0.10$ & $  0.52\pm0.10$ &   3 & \dots & \multicolumn{1}{c}{\dots} & \dots &  0.85\\[2pt]
\ion{Os}{i } & $  0.40\pm0.10$ & $  0.80\pm0.11$ &   2 & \dots & \multicolumn{1}{c}{\dots} & \dots &  1.40\\[2pt]
\ion{Ir}{i } & $  0.42\pm0.10$ & $  0.84\pm0.11$ &   1 & \dots & \multicolumn{1}{c}{\dots} & \dots &  1.38\\[2pt]
\ion{Pb}{i } & $< 0.37$ & $< 0.42$ &   1 & \dots & \multicolumn{1}{c}{\dots} & \dots &  1.75$^d$\\[2pt]
\ion{Th}{ii} & $ -0.85\pm0.10$ & $  0.73\pm0.10$ &   1 & \dots & \multicolumn{1}{c}{\dots} & \dots &  0.02\\[2pt]
\ion{U }{ii} & $<-1.21$ & $< 0.93$ &   1 & \dots & \multicolumn{1}{c}{\dots} & \dots & -0.54\\[2pt]

\hline
\end{tabular}
\addtolength{\tabcolsep}{3pt}}
\tablefoot{
\tablefoottext{a}{For $n\geq4$, the error is considered as the mad of the line-by-line abundance distribution scaled by the factor 1.48 to be concordant with a normal distribution. Otherwise, a floor error of 0.10~dex is assumed (see main text for details).}
\tablefoottext{b}{With the exception of \ion{O}{i}, [$X$/Fe$]_\mathrm{LTE}$ is given relative to the LTE abundance of the Fe species at the same ionization stage.}
\tablefoottext{c}{The solar reference abundances are from \citet{Asplund09}.}
\tablefoottext{d}{The Pb abundance was taken from meteoroids.}
}
\end{table}
The abundances presented here were computed using either EWs (Sect. \ref{Subsec: Equivalent widths}) or spectrum synthesis for such cases where blending was found to be substantial. For this purpose we employed the spectra providing the highest S/N at any given wavelength, that is, UVES~390 blueward of $\sim4300$~{\AA}, MIKE blue for $4300\lesssim \lambda \lesssim5000$~{\AA}, and MIKE red in the regime $5000\lesssim\lambda\lesssim8000$~{\AA} (cf. Fig. \ref{Fig:SN_allspec}). Despite the circumstance that MIKE reaches substantially more redward, we do not consider it there because of considerable fringing. The radiation transfer was solved using MOOG and an ATLAS9 model for our exact specifications (previous Sects. and Table \ref{Table: Target information}) that was constructed via interpolation. Our computations involved molecular equilibrium computations involving a network consisting of the species H$_2$, CH, NH, OH, C$_2$, CN, CO, N$_2$, NO, O$_2$, TiO, H$_2$O, and CO$_2$. Individual, line-by-line abundances can be found in Table \ref{Table: Line-by-line abundances}, while we summarize the adopted final abundances and their associated errors in Table \ref{Table: Final abundances}. In order to reduce the impact of outliers, abundances were averaged using the median. For ensembles of four and more lines, we computed the corresponding errors via the median absolute deviation (mad) which is scaled by the factor 1.48 in order to be conform with Gaussian standard errors. As noted already in Sect. \ref{Subsec: Model atmosphere parameters}, for the vast majority of species, the magnitude of the line-by-line scatter is inconsistent with merely the propagation of random spectrum noise, but accounts for additional -- possibly systematic -- sources of error further down in the abundance analysis. Consequently, we set a floor uncertainty of 0.10~dex for those species with less than four available lines, where the mad would not be a robust estimator for the scatter. For a discussion of this as well as of influences from uncertain stellar parameters, we refer the reader to Appendices \ref{Subsec: Spectrograph comparison} and \ref{Sec: Abundance systematics due to atmosphere}. For elements with only one line measured with the line abundance uncertainty alone exceeding the floor error, we adopted the error on the line abundance instead. 

\subsection{Line list}\label{Subsec: Line list}
Suitable lines for an abundance analysis of HD~20 were compiled and identified using literature atomic data. We retrieved all line data that are available through the Vienna Atomic Line Database \citep[VALD, ][]{Piskunov95, Ryabchikova15} in the wavelength range from 3280~{\AA} to 8000~{\AA}, representing the combined wavelength coverage of the spectra at hand. In a first run, we synthesized a spectrum from this line list and discarded all profiles that did not exceed a line depth of $0.1\%$ of the continuum level. The remaining features were visually checked for their degree of isolation and usability by comparing the observed spectra with syntheses with varying elemental abundances. The resulting list with the adopted line parameters and original sources thereof can be found in Table \ref{Table: Line-by-line abundances}. Additional hyperfine structure (HFS) line lists were considered for the elements Li \citep{Hobbs99}, Sc \citep{Kurucz95}, V \citep{Lawler14}, Mn \citep{DenHartog11}, Co \citep{Kurucz95}, Cu \citep{Kurucz95}, Ag \citep{Hansen12}, Ba \citep{McWilliam98}, La \citep{Lawler01a}, Pr \citep{Sneden09}, Eu \citep{Lawler01b}, Tb \citep{Lawler01c}, Ho \citep{Lawler04}, Yb \citep{Sneden09}, and Lu \citep{LSCI}.

\subsection{Equivalent widths}\label{Subsec: Equivalent widths}
The majority of the spectral features identified to be suitable for our analysis are sufficiently isolated so that an EW analysis could be pursued. We measured EWs from the spectra of all three spectrographs using our own semi-automated Python tool EWCODE \citep{Hanke17}. In brief, EWCODE places a local, linear continuum estimate that is based on the neighboring wavelength ranges next to the profile of interest and fits Gaussian profiles. The user is prompted with the fit and can interactively improve the fit by, for example, introducing additional blends or refining the widths of the continuum ranges. Our measurements for individual lines along with EWCODE's error estimates are listed in Table \ref{Table: Line-by-line abundances}.

\subsection{Notes on individual elements}\label{Subsec: Notes on individual elements}
In the following, we will comment in detail on the analysis of abundances from several features that needed special attention exceeding the standard EW or spectrum synthesis analysis. Furthermore, whenever available, we comment on NLTE corrections that were applied to the LTE abundances. 

\subsubsection{Lithium (Z = 3)}\label{Subsubsec: Li}
The expected strongest feature of \ion{Li}{i} is the resonance transition at 6707.8~{\AA}. Despite our high-quality data, within the noise boundaries, the spectrum of HD~20 appears perfectly flat with no feature identifiable whatsoever. For the region in question we estimate from our MIKE spectrum $S/N\approx1050$~pixel$^{-1}$, which would allow for $3\sigma$ detections of Gaussian-like features with EWs of at least 0.3~m{\AA} as deduced from the formalism provided in \citet{Battaglia08}. The latter EW translates into an upper limit $\log{\epsilon(\mathrm{Li})}<-0.34$~dex. 

\subsubsection{Carbon, nitrogen, and oxygen (Z = 6, 7, and 8)}\label{Subsubsec: C, N, and O}
\begin{figure*}
    \centering
    \resizebox{0.87\hsize}{!}{\includegraphics{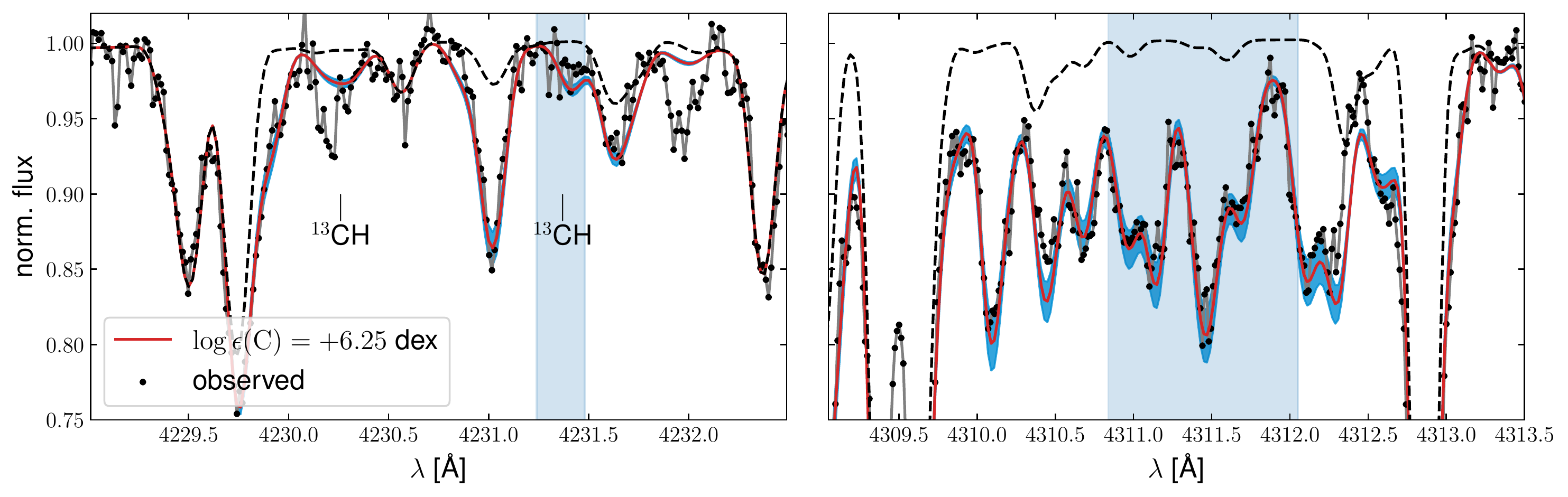}}
      \caption{C abundance and $^{12}$C/$^{13}$C from the CH $G$-band in the UVES~390 spectrum. \textit{Left panel}: Region around the two features that are dominated by $^{13}$CH, one of which is used to pinpoint $^{12}$C/$^{13}$C (blue rectangle). The bluer feature at $\sim4230$~{\AA} was not considered due to an unidentified blend (see main text). The observed spectrum is represented by black dots connected by gray lines and the best-fit synthesis (red) and its abundance error margin of 0.05~dex are depicted in blue, respectively. The dashed spectrum shows a synthesis without any C. \textit{Right panel}: Same as \textit{left panel} but in the range used to constrain the C (CH) abundance.
              }
      \label{Fig:4300_CH_0}
\end{figure*}
\begin{figure}
    \centering
    \resizebox{\hsize}{!}{\includegraphics{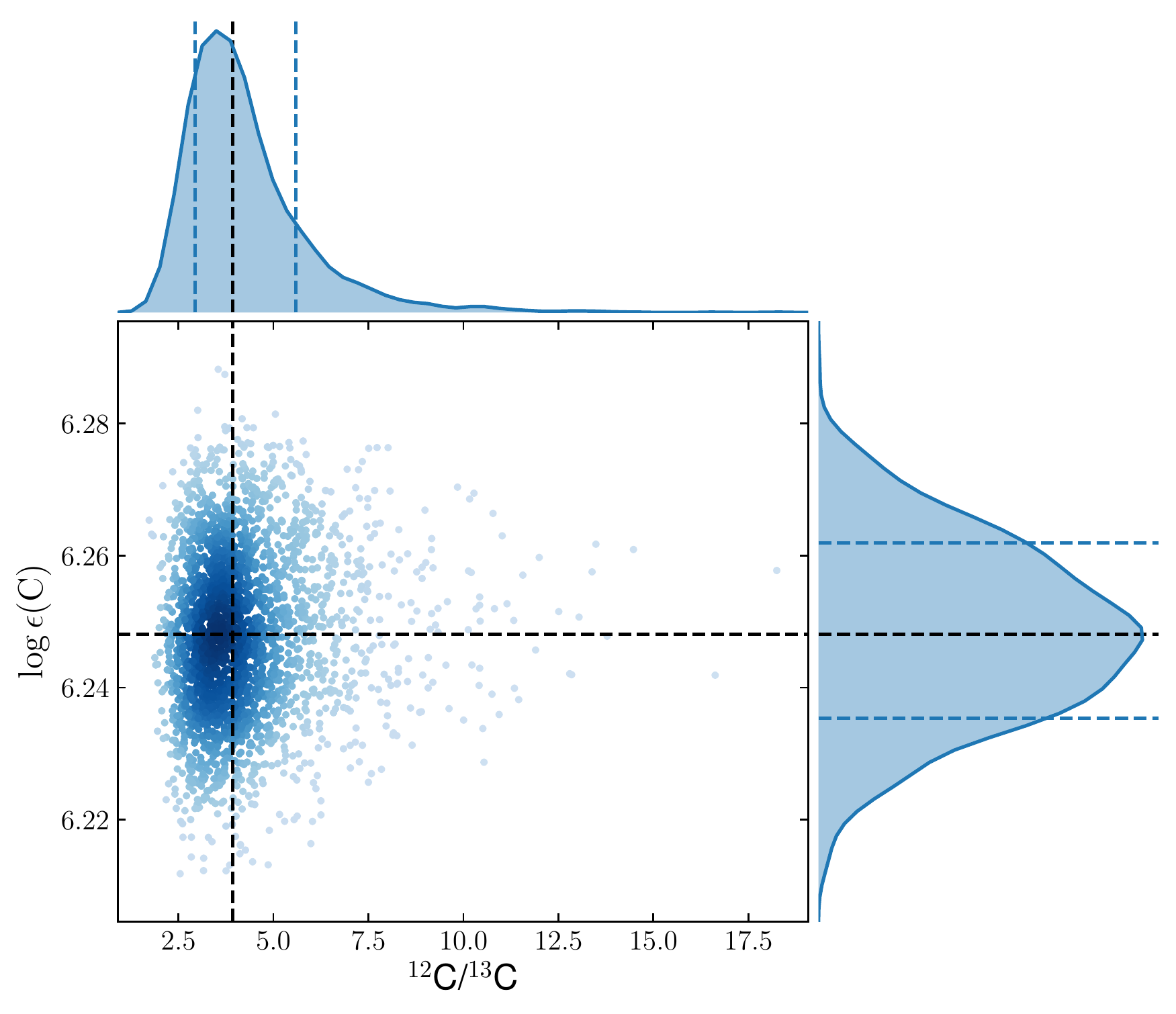}}
      \caption{Two-dimensional representation of the MCMC sample used to fit $\log{\epsilon{(\mathrm{C})}}$ and $^{12}$C/$^{13}$C simultaneously including the marginal distributions. Median values and asymmetric limits are displayed by dashed lines.
              }
      \label{Fig:4300_CH_1}
\end{figure}
Our C abundances are based on synthesis of the region around the CH $G$-band at $\sim4300$~{\AA} with molecular line data for $^{12}$CH and $^{13}$CH from \citet{Masseron14}. We identified a range between 4310.8~{\AA} and 4312.1~{\AA} that in HD~20 is almost devoid of atomic absorption and hence is ideal for CH synthesis irrespective of other elemental abundances. We show this range in Fig. \ref{Fig:4300_CH_0}. Only very substantial changes in the model isotopic ratio $^{12}\mathrm{C}/^{13}\mathrm{C}$ have a notable effect on this region, manifesting mostly in an effective blue- or redshift of the molecular features. In contrast, the two $^{13}$CH profiles near $\sim4230$~{\AA} (left panel of Fig. \ref{Fig:4300_CH_0}) are rather sensitive to the isotopic ratio. As cautioned by \citet{Spite06}, the blueward profile has a dominant blend they attribute to an unidentified transition from an $r$-process element. Given the $r$-process-rich nature of our star, we do not consider this feature here. Employing both ranges, one for the C abundance and one for $^{12}\mathrm{C}/^{13}\mathrm{C}$, the two measures can be effectively decoupled as can be seen in Fig. \ref{Fig:4300_CH_1}, where we present the results of an MCMC sampling run used to draw from the posterior distribution of the fitted parameters in the regions indicated in Fig. \ref{Fig:4300_CH_0}. From this distribution we determine $^{12}\mathrm{C}/^{13}\mathrm{C}=3.92^{+1.68}_{-0.98}$. Though nominally less, an error of 0.05~dex was adopted for $\log{\epsilon}(\mathrm{C})=6.25$~dex in order to account for the circumstance that the continuum level in the right-hand spectrum had to be established from a region more than one {\AA} away on either side, thus introducing a slight normalization uncertainty.   

\begin{figure*}
    \centering
    \resizebox{0.9\hsize}{!}{\includegraphics{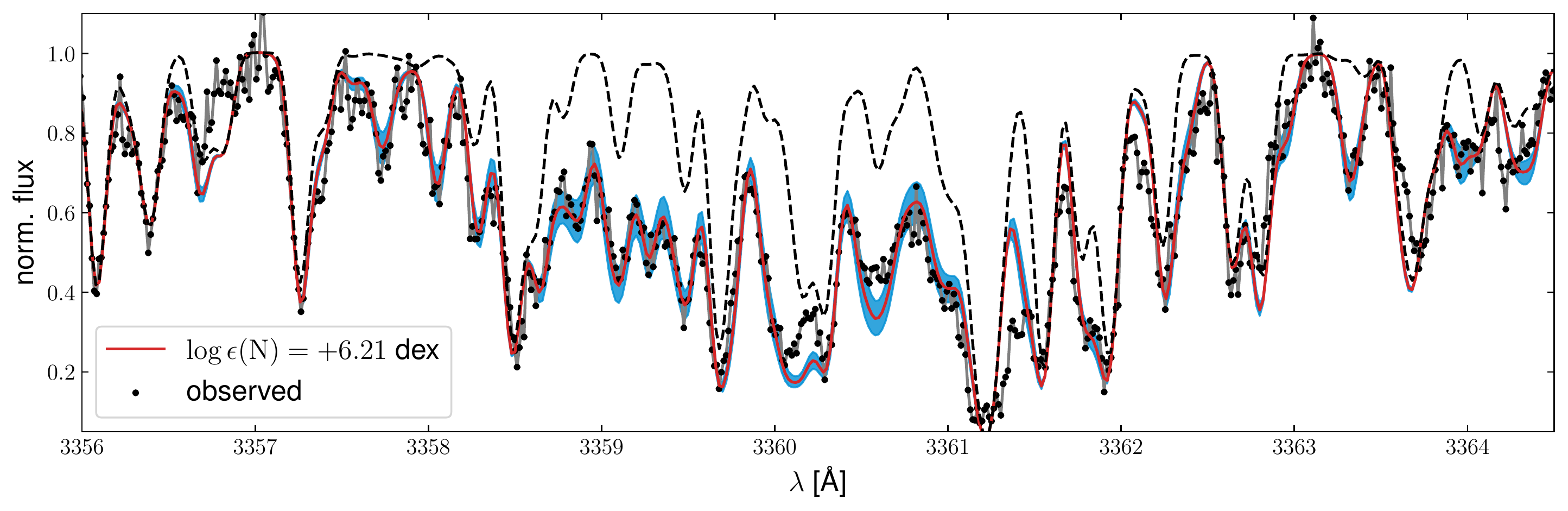}}
      \caption{Same as \textit{right panel} of Fig. \ref{Fig:4300_CH_0}, but for a synthesis of the NH-band at $\sim3360$~{\AA}. A synthesis without any N is indicated by the black dashed curve. The blue error range corresponds to an abundance variation of $\pm0.10$~dex. 
              }
      \label{Fig:3360_NH}
\end{figure*}
We determined the N abundance in a similar fashion employing the NH-band at $\sim3360$~{\AA} (see Fig. \ref{Fig:3360_NH}). From our synthesis we inferred $\log{\epsilon}(\mathrm{N})=6.21\pm0.10$~dex. The present data do not permit the determination of the isotopic ratio $^{14}\mathrm{N}/^{15}\mathrm{N}$.

Unfortunately, the frequently used [\ion{O}{i}] line at 6300.3~{\AA} is strongly blended with telluric absorption features in all available spectra and hence rendered useless for precise abundance studies. Nonetheless, the high S/N of the MIKE spectra allowed for the measurement of the much weaker [\ion{O}{i}] transitions at 5577.3~{\AA} and 6363.8~{\AA}, from which we deduced a mean abundance of $\log{\epsilon}(\mathrm{\ion{O}{i}})_\mathrm{LTE}=7.79\pm0.18$~dex, or $\mathrm{[O/Fe]}=0.70$~dex. The forbidden lines ought to have negligible LTE corrections, because they have metastable upper levels. Hence, the collisional rate is higher than the radiative rate and LTE is obtained, in other words $\log{\epsilon}(\mathrm{\ion{O}{i}})_\mathrm{NLTE}=\log{\epsilon}(\mathrm{\ion{O}{i}})_\mathrm{LTE}$. Severe changes in the O abundance result in non-negligible effects on the molecular equilibrium, in particular through their impact on the formation of CO. For this reason, the overabundance found here was considered in all syntheses, including the ones for CH and NH outlined above.

We note here that abundances from the O triplet at $\sim7773$~{\AA} could be firmly detected and are listed in Table \ref{Table: Line-by-line abundances}. However, we discard them ($\log{\epsilon}(\mathrm{\ion{O}{i}})_\mathrm{LTE}\approx8.22$~dex) from consideration in this work, since they are in strong disagreement to the abundances from the forbidden lines. The formation of the lines in question is subject to considerable NLTE effects as shown by, for example, \citet{Sitnova13}. Using the MPIA NLTE spectrum tools\footnote{\url{http://nlte.mpia.de/gui-siuAC_secE.php}} to retrieve corrections for individual line abundances, we found an average 1D NLTE bias of $-0.14$~dex, which is not enough to erase the discrepancy. We therefore suspect much stronger effects when considering line formation in NLTE using 3D dynamical models \citep[e.g.,][]{Amarsi19}.

\subsubsection{Sodium (Z = 11)}\label{Subsubsec: Na}
Equivalent widths from the two weak Na lines at 5682~{\AA} and 5688~{\AA} were employed to compute an abundance of $\log{\epsilon}(\mathrm{Na})_\mathrm{LTE}=4.50\pm0.10$~dex. We emphasize the artificial increase of the latter uncertainty to 0.10~dex as discussed earlier. According to the INSPECT database\footnote{\url{www.inspect-stars.com}} \citep{Lind11}, for these lines and HD~20's parameters a mean NLTE correction of $-0.08$~dex should be applied, leading to $\log{\epsilon}(\mathrm{Na})_\mathrm{NLTE}=4.42$~dex and consequently [Na/Fe$]=-0.14$~dex. The frequently used \ion{Na}{i} transitions at 6154~{\AA} and 6160~{\AA} could not be firmly detected in any of our spectra owing to HD~20's rather high temperature, which strongly reduces the strength of these lines.  

\subsubsection{Magnesium (Z = 12)}\label{Subsubsec: Mg}
The three \ion{Mg}{i} lines employed for abundance determinations in this work were corrected for departures from the LTE assumptions by means of the MPIA NLTE spectrum tools, which is based on \citet{Bergemann17a,Bergemann17b}. The mean correction is only $+0.04$~dex, indicating that the effects are not severe for the selected lines.

\subsubsection{Aluminum (Z = 13)}\label{Subsubsec: Al}
Our Al abundance for HD~20 is based on five neutral transitions. While spectrum syntheses revealed the 3944~{\AA} profile to be severely blended, the other strong UV resonance feature at 3961~{\AA} was found to be sufficiently isolated for getting a robust abundance. In addition, the high S/N of our MIKE spectrum allowed for the detection of two pairs of weak, high-excitation lines at $\sim6697$~{\AA} and $\sim7835$~{\AA}, respectively. In LTE, there is a considerable difference of almost 1~dex between the abundances from the resonance line ($\log{\epsilon}(\mathrm{Al})_\mathrm{LTE}=3.58$~dex), and the four weak lines ($\log{\epsilon}(\mathrm{Al})_\mathrm{LTE}=4.54$~dex). As shown by \citet{Nordlander17}, this can be explained by substantial NLTE effects on Al line formation in metal-poor giants like HD~20. Indeed, by interpolation in their pre-computed grid, we found corrections of 1.02~dex for the strong line and 0.14 to 0.20~dex for the weak lines, which alleviates the observed discrepancy. We emphasize that [Al/Fe] (Table \ref{Table: Final abundances}) remains unaltered by going from LTE to NLTE, because both the \ion{Fe}{i} transitions and the majority of our \ion{Al}{i} lines experience the same direction and magnitude of corrections. We note here that \citet{Barklem05} report on a strong depletion in LTE of [Al/Fe$]=-0.80$~dex \citep[on the scale of][]{Asplund09} based on the UV resonance line, only. Hence, that finding at face value should be treated with caution since severe NLTE biases can be expected.

\subsubsection{Silicon (Z = 14)}\label{Subsubsec: Si}
Five of our 16 \ion{Si}{i} lines with measured EWs have a correspondence in the MPIA NLTE database \citep[][]{Bergemann13}. The deduced corrections for HD~20's stellar parameters are marginal at a level of $-0.01$ to $-0.04$~dex. As a consequence, the ionization imbalance of $-0.28$~dex between \ion{Si}{i} and \ion{Si}{ii} that prevails in LTE cannot be compensated this way. Lacking NLTE corrections for our two \ion{Si}{ii} transitions, however, we cannot draw definite conclusions at this point.

\subsubsection{Sulfur (Z = 16)}\label{Subsubsec: S}
We detected in total four S features that are spread over two wavelength windows at $\sim4695$~{\AA} and $\sim6757$~{\AA}, corresponding to the second and eighth \ion{S}{i} multiplet. Using spectrum synthesis, we found a mean abundance $\log{\epsilon}(\ion{S}{i})_\mathrm{LTE}=6.03\pm0.04$~dex that is mainly driven by the strongest profile at 4694.1~{\AA}. Concerning influences of NLTE on \ion{S}{i}, in the literature there is no study dealing with the second multiplet. For the eighth multiplet, however, \citet{Korotin08,Korotin09} and \citet{Korotin17} showed that the expected corrections for HD~20 are minor and remain well below $0.10$~dex. Since we detected no considerable difference in our LTE analysis between the eighth and second multiplet, we conclude that the correction  -- if any -- for the second multiplet is probably small, too.

\subsubsection{Potassium (Z = 19)}\label{Subsubsec: K}
The K abundance presented here is based on the EWs of two red resonance lines at 7665~{\AA} and 7699~{\AA}, respectively. These lines are expected to be subject to severe departures from LTE. \citet{Mucciarelli17} showed for giants in four globular clusters that the magnitude of the NLTE correction strongly increases with increasing $T_\mathrm{eff}$, $\log{g}$, and $\log{\epsilon(\mathrm{K})_\mathrm{LTE}}$. One of their clusters, NGC~6752, exhibits a similar metallicity ($-1.55$~dex) as HD~20 and we estimate from their Fig. 3 a correction of our LTE abundance of at least $-0.5$ to $-0.6$~dex. For our adopted NLTE abundance (Table \ref{Table: Final abundances}) we assume a shift by $-0.55$~dex.

\subsubsection{Titanium (Z = 22)}\label{Subsubsec; Ti}
Our LTE analysis of Ti lines shows an ionization imbalance of $(\log{\epsilon(\ion{Ti}{i})}-\log{\epsilon(\ion{Ti}{ii})})_\mathrm{LTE}=-0.33$~dex. We have determined line-by-line NLTE corrections for our \ion{Ti}{i} abundances from the grid by \citet{Bergemann11} amounting to values ranging from $+0.4$ to $+0.6$~dex. It is noteworthy that corrections to \ion{Ti}{ii} are insignificant in the present regime of stellar parameters \citep[cf.,][]{Bergemann11}. The newly derived NLTE abundances switch the sign of the ionization imbalance with a reduced amplitude ($(\log{\epsilon(\ion{Ti}{i})}-\log{\epsilon(\ion{Ti}{ii})})_\mathrm{NLTE}=+0.14$~dex). Inconsistencies in other metal-poor stars manifesting themselves in ionization imbalances even in NLTE have already been noted by \citet{Bergemann11} and were explained by inaccurate or missing atomic data. More recently, \citet{Sitnova16} found lower NLTE corrections and therefore weaker -- but still non-zero -- ionization imbalances for stars in common with \citet{Bergemann11}, which they mainly attributed to the inclusion of high-excitation levels of \ion{Ti}{i} in their model atom. In light of prevailing uncertainties of \ion{Ti}{i} NLTE calculations, we do not believe that the ionization imbalance of Ti contradicts our results from Sect. \ref{Subsec: Model atmosphere parameters}. 

\subsubsection{Manganese (Z = 25)}\label{Subsubsec: Mn}
Following \citet{Bergemann08}, our four abundances from \ion{Mn}{i} lines should experience a considerable mean NLTE adjustment of $+0.40$~dex and thus are consistent with a solar [Mn/Fe]. More recently, \citet{Mishenina15} casted some doubt on the robustness of the aforementioned NLTE calculations by showing the absence of systematic discrepancies in LTE between multiplets that according to \citet{Bergemann08} ought to have different NLTE corrections. Nonetheless, \citet{Bergemann19} corroborated the strong NLTE corrections found in the earlier study. Moreover, the authors remark that \ion{Mn}{i} transitions at a lower excitation potential of more than 2~eV are not strongly affected by convection -- that is 3D effects -- and are recommended as 1D NLTE estimator. Since the latter is satisfied for all of our four used Mn lines, our 1D NLTE abundance ought to be an accurate estimate. 

\subsection{Cobalt (Z = 27)}\label{Subsubsec: Co}
The Co NLTE corrections were obtained from \citet{Bergemann10}. For three out of the six measured lines corrections are available and amount to $+0.46$~dex on average.

\subsubsection{Copper (Z = 29)}\label{Subsubsec: Cu}
We measured three profiles of \ion{Cu}{i} in our spectra, two of which originate from low-excitation ($\sim1.5$~eV) states. Albeit for dwarfs, at [Fe/H$]\sim-1.5$~dex, \citet{Yan15} predicted for these two transitions at 5105.5~{\AA} and 5782.2~{\AA} stronger NLTE corrections compared to the ones for our high-excitation ($\sim3.8$~eV) line. This is somewhat reflected in our LTE abundances where the lower-excitation lines yield a lower value by about $0.3$~dex. Lacking a published pre-computed grid, it is hard to predict the exact amount of NLTE departures for our giant star and its temperature. Yet, \citet{Shi18} and  \citet{Korotin18} showed that the corrections correlate much stronger with [Fe/H] than they do with $\log{g}$ or $T_\mathrm{eff}$. We make no attempt to rectify our Cu abundances at this point, but judging from the literature we note that the corrections are probably on the order of $+0.2$~dex for the low-excitation- and $+0.1$~dex for the high-excitation lines.

\subsubsection{Strontium (Z = 38)}\label{Subsubsec: Sr}
In principle, our spectra cover the UV resonance lines of \ion{Sr}{ii} at 4077~{\AA} and 4215~{\AA}, though we found those to be strongly saturated and we could not reproduce the line shape through LTE synthesis. Furthermore, the lines in question are subject to a substantial degree of blending by several atomic and molecular transitions \citep[see also][]{Andrievsky11}. Fortunately, it was possible to measure EWs of the much weaker lines at 4607~{\AA} (\ion{Sr}{i}) and at 4161~{\AA} (\ion{Sr}{ii}). For these we deduced abundances of 1.00~dex and 1.50~dex, respectively, which indicates a substantial discrepancy between the two ionization stages. The latter can be attributed to considerable NLTE departures for the neutral transition. \citet{Bergemann12}\footnote{\citet{Bergemann12} mention a \ion{Sr}{ii} line at 4167.8~{\AA} in their Table 1. However the line parameters provided are for the line at 4161.8~{\AA}. NLTE corrections are not provided for this transition.} and \citet{Hansen13} performed extensive NLTE calculations for this line from which we extract a correction of +0.4~dex for HD~20's stellar parameters. Thus, the observed difference is effectively erased, although we emphasize the lack of published \ion{Sr}{ii} corrections for the line and stellar parameters in question, which -- in turn -- may re-introduce a slight disagreement.

\subsubsection{Zirconium (Z = 40)}\label{Subsubsec: Zr}
Two out of our five measured \ion{Zr}{ii} lines were investigated for NLTE effects by \citet{Velichko10}. The authors note that departures mainly depend on metallicity and gravity, whereas there is only a weak coupling to $T_\mathrm{eff}$. From their published grid of corrections we extrapolate corrections of 0.15~dex and 0.18~dex for our abundances from the lines at 4209.0~{\AA} and 5112.3~{\AA}, respectively.   

\subsubsection{Barium (Z = 56)}\label{Subsubsec: Ba}
In HD~20, the \ion{Ba}{ii} profile at 4554~{\AA} is strongly saturated and thus largely insensitive to abundance. We further excluded the 6141~{\AA} line because of blending by an Fe feature. Our abundance hence is based on synthesis of the two clean and only moderately strong transitions at 5853~{\AA} and 6496~{\AA}, yielding $\log{\epsilon(\ion{Ba}{ii})})_\mathrm{LTE}=0.77$~dex and 1.09~dex, respectively. In light of the recent work on NLTE line formation by \citet{Mashonkina19}, the presented disagreement can be expected in LTE, as in our parameter regime NLTE corrections for the two lines differ. Indeed, interpolation in their published grid\footnote{The grid does not reach down to $\log{g}=2.37$~dex, but instead ends at $\log{g}=3.0$~dex. Consequently, a linear extrapolation was performed. We note, however, that this seems uncritical since gravity is not a governing parameter in the considered regime.} resulted in corrections of $-0.10$~dex and $-0.27$~dex, hence reducing the gap to 0.15~dex, which can be explained by the combined statistical uncertainties. 

\subsubsection{Lutetium (Z = 71)}\label{Subsubsec: Lu}
\begin{figure}
    \centering
    \resizebox{\hsize}{!}{\includegraphics{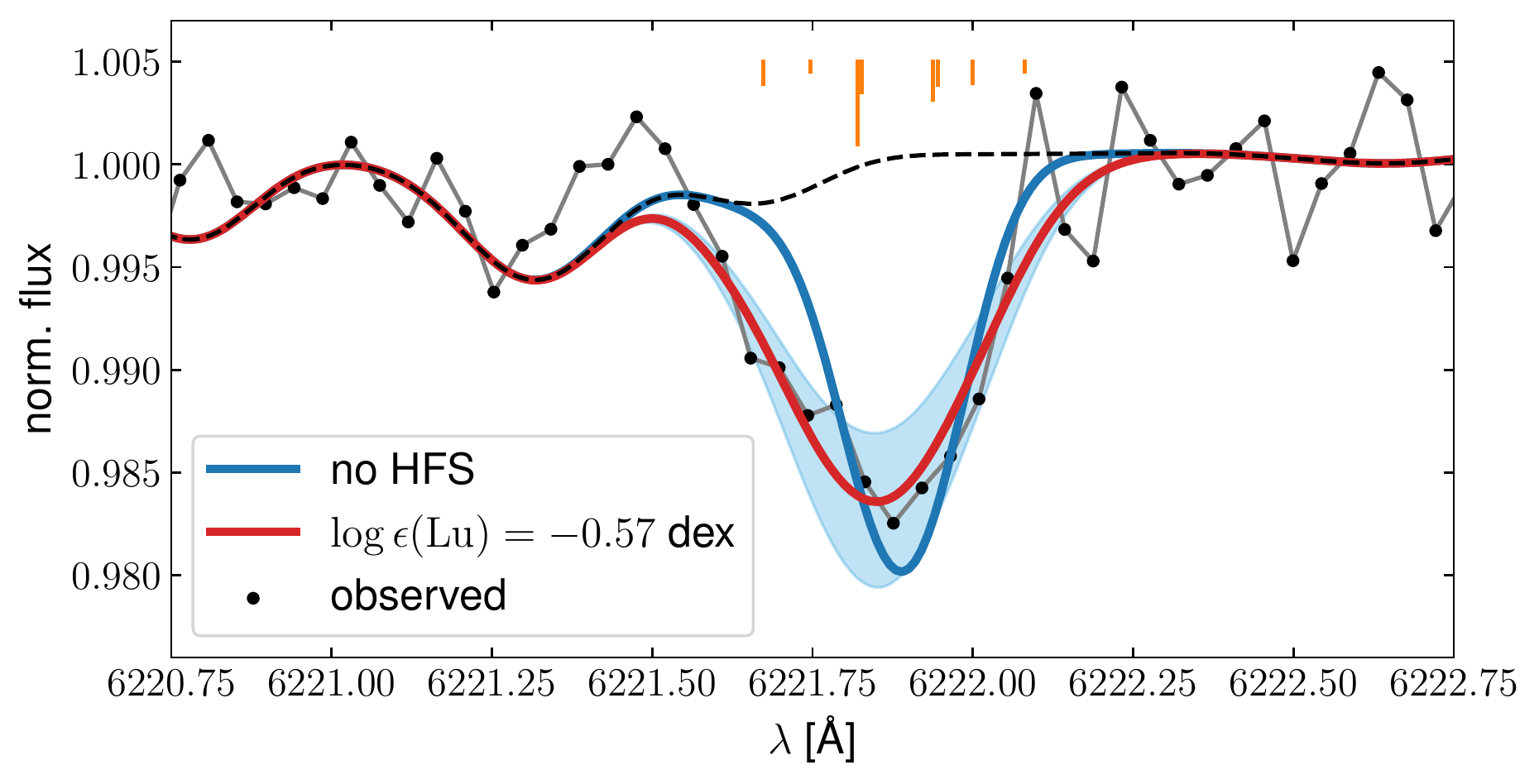}}
      \caption{Synthesis of the \ion{Lu}{ii} line at 6221.9~{\AA}. The red line represents the best abundance match with an error of 0.1~dex (blue shaded region). The broad range of HFS components for $^{175}$Lu from \citet{LSCI} are indicated by vertical orange lines at the top and have been taken into account for this synthesis. The impact of the negligence of HFS on the line shape is indicated by the blue line.
              }
      \label{Fig:Lu_6221}
\end{figure}
The very high S/N of about 1000~pixel$^{-1}$ in the MIKE spectrum in concert with an overall high Lu abundance ([Lu/Fe$]=0.93$~dex) allowed for a solid detection (4.7~m{\AA}) of the otherwise very weak \ion{Lu}{ii} profile at 6221.9~{\AA}. We mention the line here explicitly, because it was found to have an exceptionally pronounced HFS structure as we show in Fig. \ref{Fig:Lu_6221} where two syntheses are compared; one including HFS and one neglecting it. The line components were taken from \citet{LSCI}. We note that we consider only the $^{175}$Lu isotope here, because the only other stable isotope, $^{176}$Lu, is expected to be a minority component judging from its solar fractional abundance \citep[2.59\%,][]{LSCI}. Despite the considerable additional line broadening due to atmospheric effects (Sect. \ref{Subsec: Rotational velocity}), hyperfine splitting is still the dominant source of broadening, thus highlighting the importance of including it in our analysis.

\subsubsection{Upper limits on rubidium, lead, and uranium (Z = 37, 82, and 92)}\label{Subsubsec: Upper limits on Lead and Uranium}
\begin{figure}
    \centering
    \resizebox{\hsize}{!}{\includegraphics{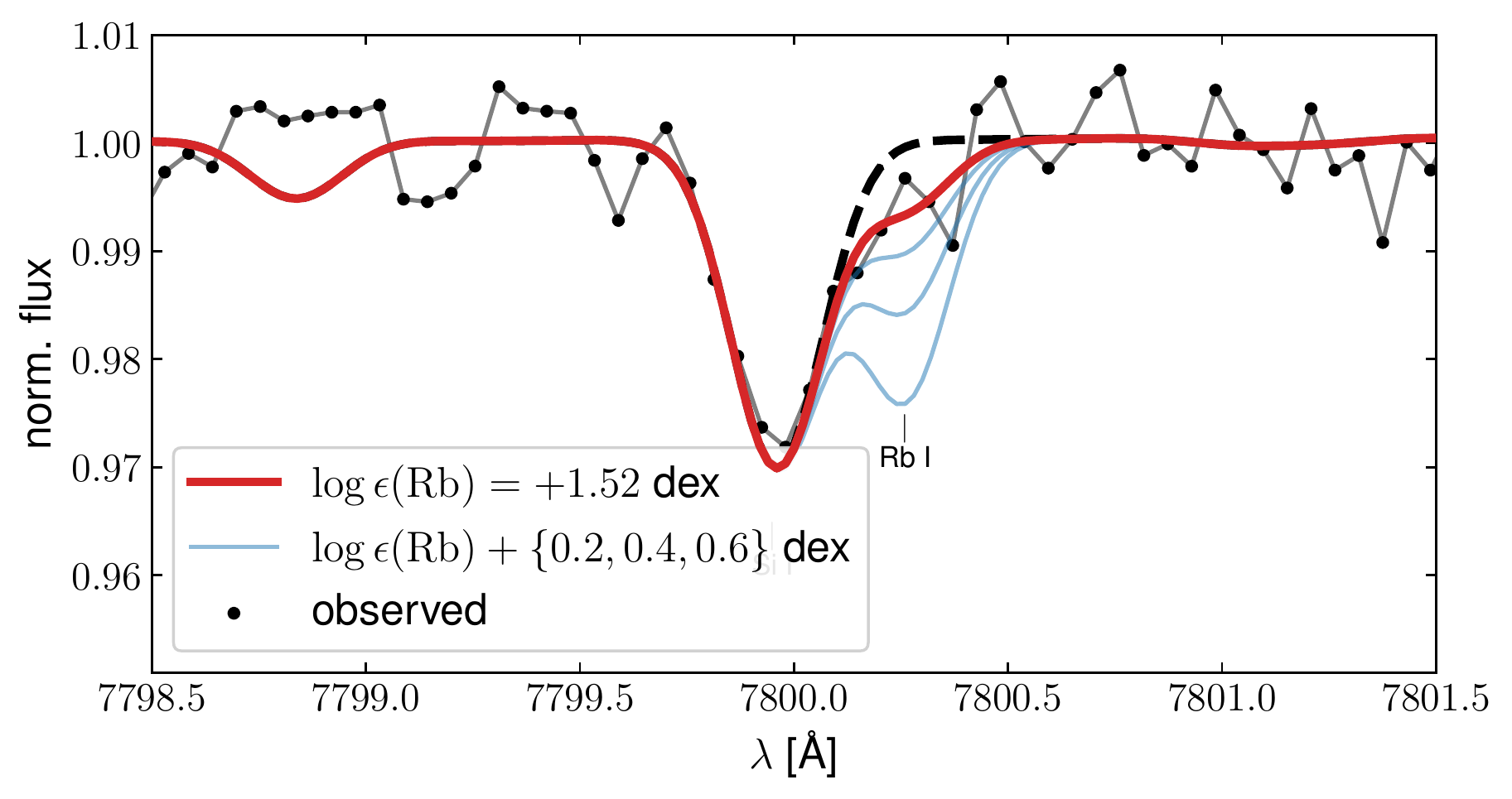}}
      \caption{Upper limit on Rb from the \ion{Rb}{i} line at 7800.3~{\AA}. The red model denotes the adopted upper limit of $+1.52$~dex, whereas blue lines are syntheses with Rb abundances successively increased by 0.2~dex.
              }
      \label{Fig:Rb_7800}
\end{figure}
\begin{figure}
    \centering
    \resizebox{\hsize}{!}{\includegraphics{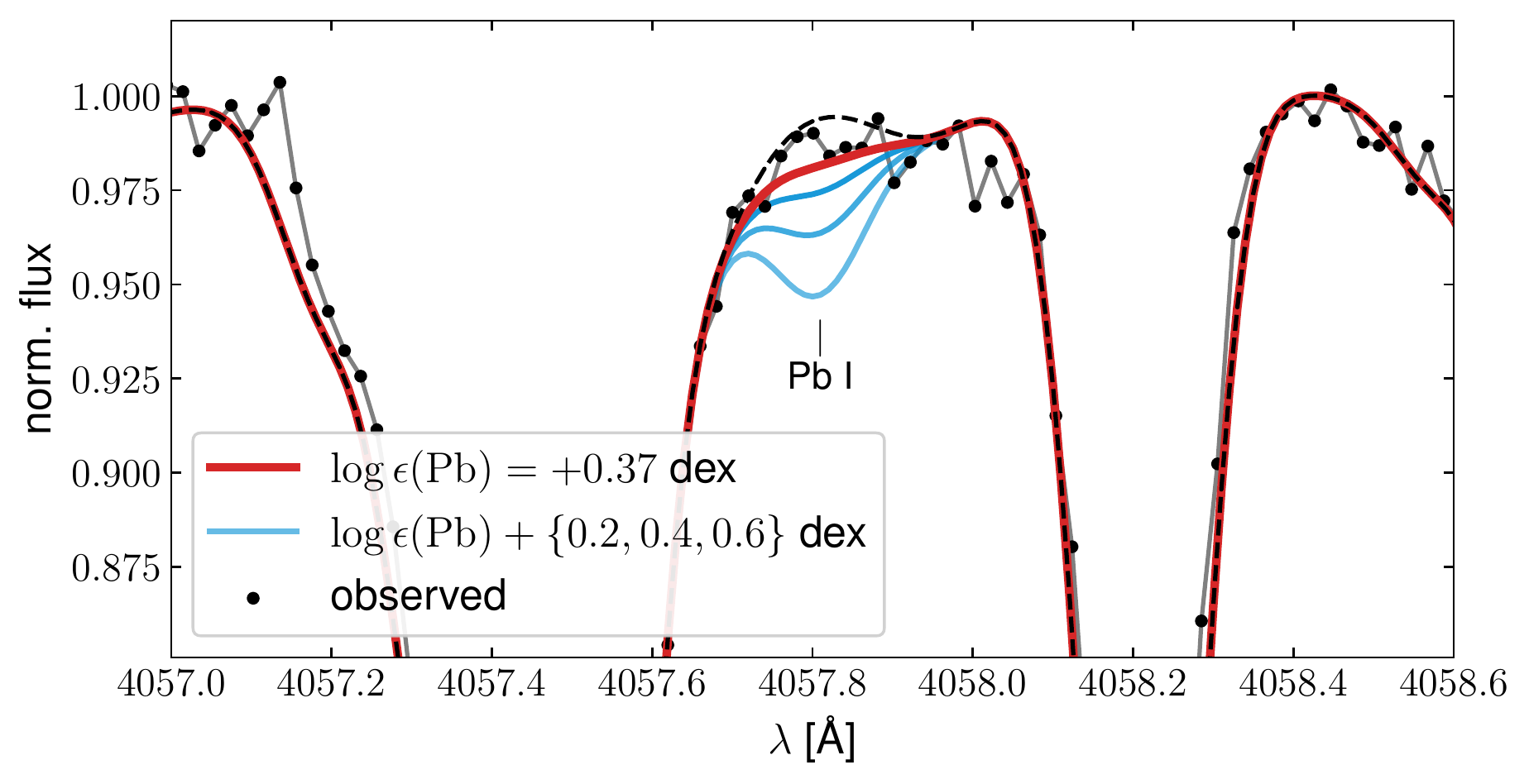}}
      \caption{Same as Fig. \ref{Fig:Rb_7800} but for the \ion{Pb}{i} transition at 4057.8~{\AA} and an upper limit of $+0.37$~dex. 
              }
      \label{Fig:Pb_4057}
\end{figure}
\begin{figure}
    \centering
    \resizebox{\hsize}{!}{\includegraphics{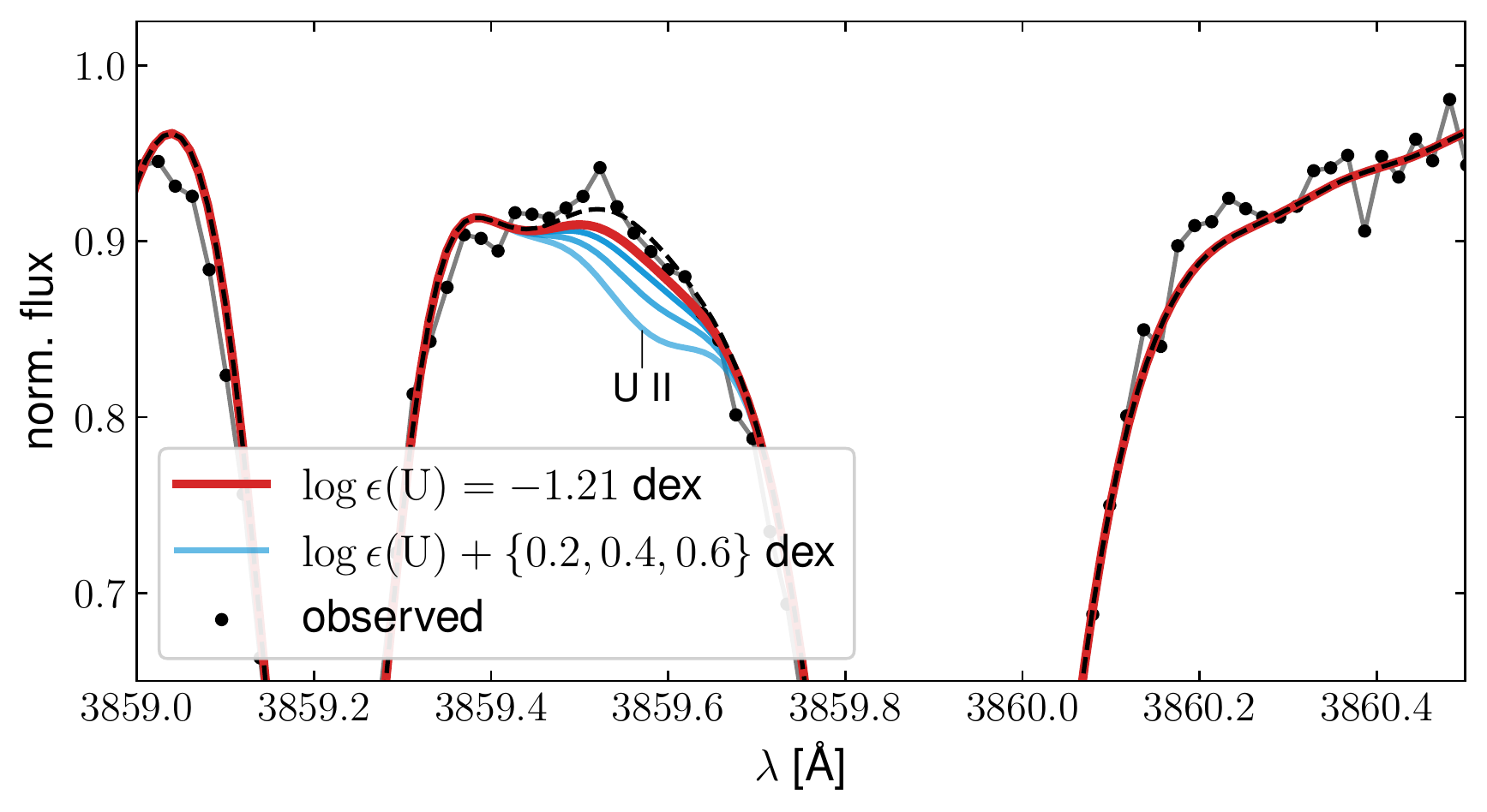}}
      \caption{Same as Fig. \ref{Fig:Rb_7800} but for the \ion{U}{ii} feature at 3859.6~{\AA} and an upper limit of $-1.21$~dex.
              }
      \label{Fig:U_3859}
\end{figure}
For Rb, Pb, and U it was not possible to obtain solid detections despite the high-quality spectra at hand. Nonetheless, we could estimate reasonable upper limits based on the lines at 7800.3~{\AA} (\ion{Rb}{i}), 4057.8~{\AA} (\ion{Pb}{i}), and 3859.6~{\AA} (\ion{U}{ii}). Since there is a considerable amount of blending by a variety of species involved in shaping the spectrum in the three wavelength regimes, we cannot estimate the upper limit in the same way as for Li (Sect. \ref{Subsubsec: Li}). Thus, we used synthesis at varying abundances of the target elements in order to establish the highest abundance that is still consistent with the noise level present in the spectral regions (Figs. \ref{Fig:Rb_7800}, \ref{Fig:Pb_4057}, and \ref{Fig:U_3859}). This way, we found $\log{\epsilon(\mathrm{Rb})}<1.52$~dex, $\log{\epsilon(\mathrm{Pb})}<0.37$~dex, and $\log{\epsilon(\mathrm{U})}<-1.21$~dex, respectively.

\section{Results and Discussion}\label{Sec: Results and Discussion}
\subsection{Light elements (Z $\leq$ 8)}\label{Subsec: Light elements}
\begin{figure}
    \centering
    \resizebox{0.95\hsize}{!}{\includegraphics{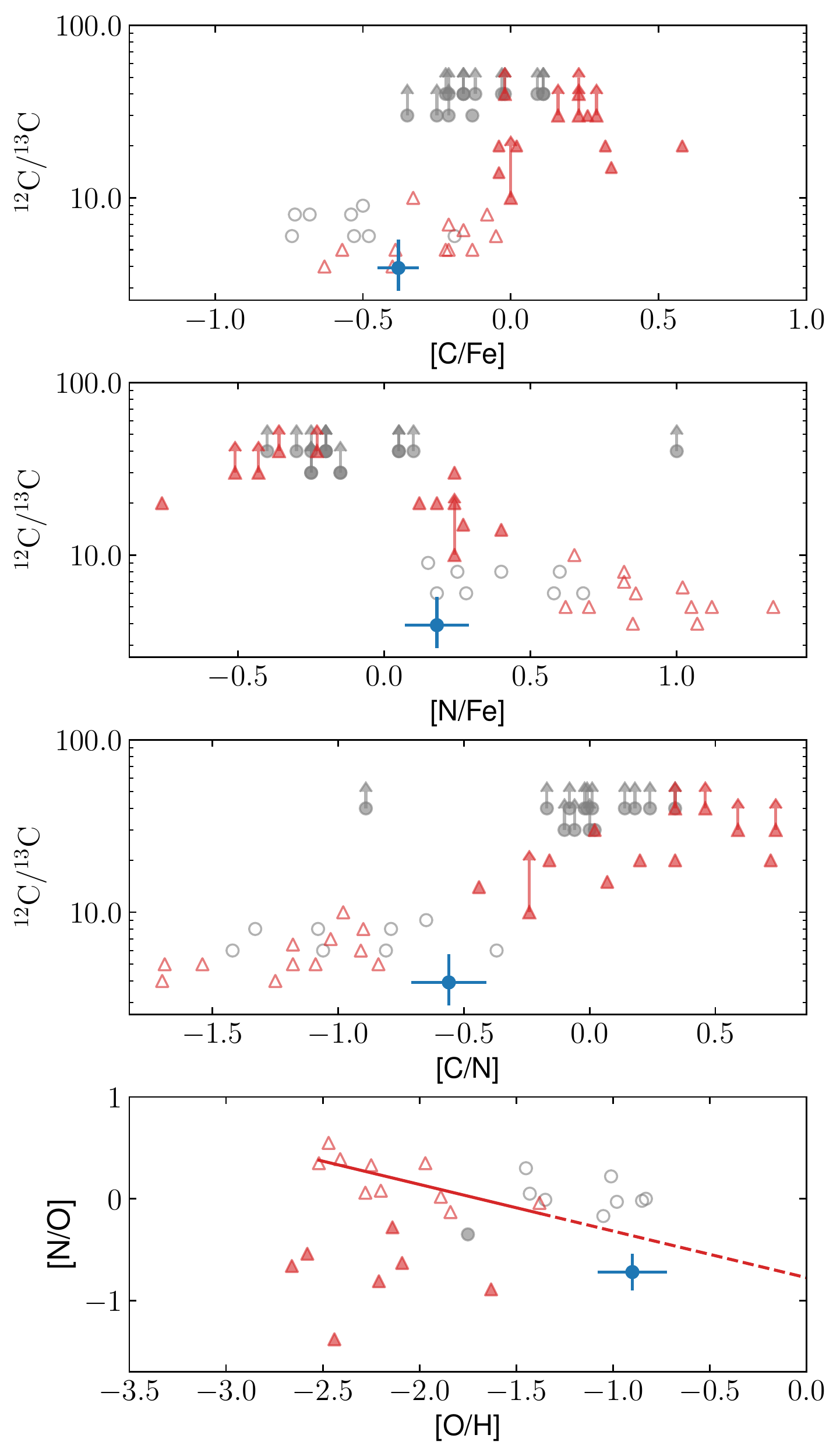}}
      \caption{Comparison of CNO elemental abundances of mixed and unmixed stars with HD~20 shown in blue for comparison. Gray circles resemble the study by \citet{Gratton00} while red triangles indicate the stars published in \citet{Spite05, Spite06}. Two C-rich stars were excluded from the latter sample. Lower limits on $^{12}$C/$^{13}$C are indicated by upward pointing arrows and the classification into mixed and unmixed stars according to the authors are represented by open and filled symbols, respectively. The red line in the \textit{lower panel} mimics the linear relation between [N/O] and [O/H] for mixed stars as reported by \citet{Spite05}, whereas the dashed line extrapolates the same relation to higher values of [O/H].
              }
      \label{Fig:CNO_mixing}
\end{figure}
Our Li, C, and N abundances show imprints of a pattern that is commonly attributed to internal mixing occuring when a star reaches the RGB bump where processed material from the H-burning shell gets dredged up to the convective layer. Observationally, the effect can be seen in the stellar surface abundances of bright giants \citep[brighter than the RGB-bump at $\log{L/L_\odot}\sim1.8$, e.g.,][]{Gratton00} and horizontal branch stars that show non-detections of Li and depletions of [C/Fe] in lockstep with low $^{12}$C/$^{13}$C ratios and enhancements in [N/Fe]. Indeed, for HD~20 we could not detect Li and found [C/Fe$]=-0.38$~dex, a value that is representative for the samples of mixed stars by \citet{Gratton00} and \citet{Spite06}. On the other hand, as can be seen in Fig. \ref{Fig:CNO_mixing}, the marginal enhancement in [N/Fe] ($0.18\pm0.11$~dex) and as a consequence the comparatively high [C/N] ($-0.56$~dex) render HD~20 at rather extreme positions among the mixed populations. A further puzzling observation is the strong O overabundance of [O/Fe$]=0.70$~dex that places HD~20 slightly below the general trend of [N/O] with [O/H] by \citet{Spite05} that appears generic for mixed stars (lower panel of Fig. \ref{Fig:CNO_mixing}). We lack a suitable explanation for a mechanism that could produce such large O excesses. Deep mixing with O-N cycle material can be ruled out as origin, as the O-N cycle would produce N at the expense of O and therefore show depletions -- which is exactly the opposite of the observed O enhancement.

\subsection{HD 20's evolutionary state}\label{Subsec: HD 20s evolutionary state}
\begin{figure}
    \centering
    \resizebox{0.951\hsize}{!}{\includegraphics{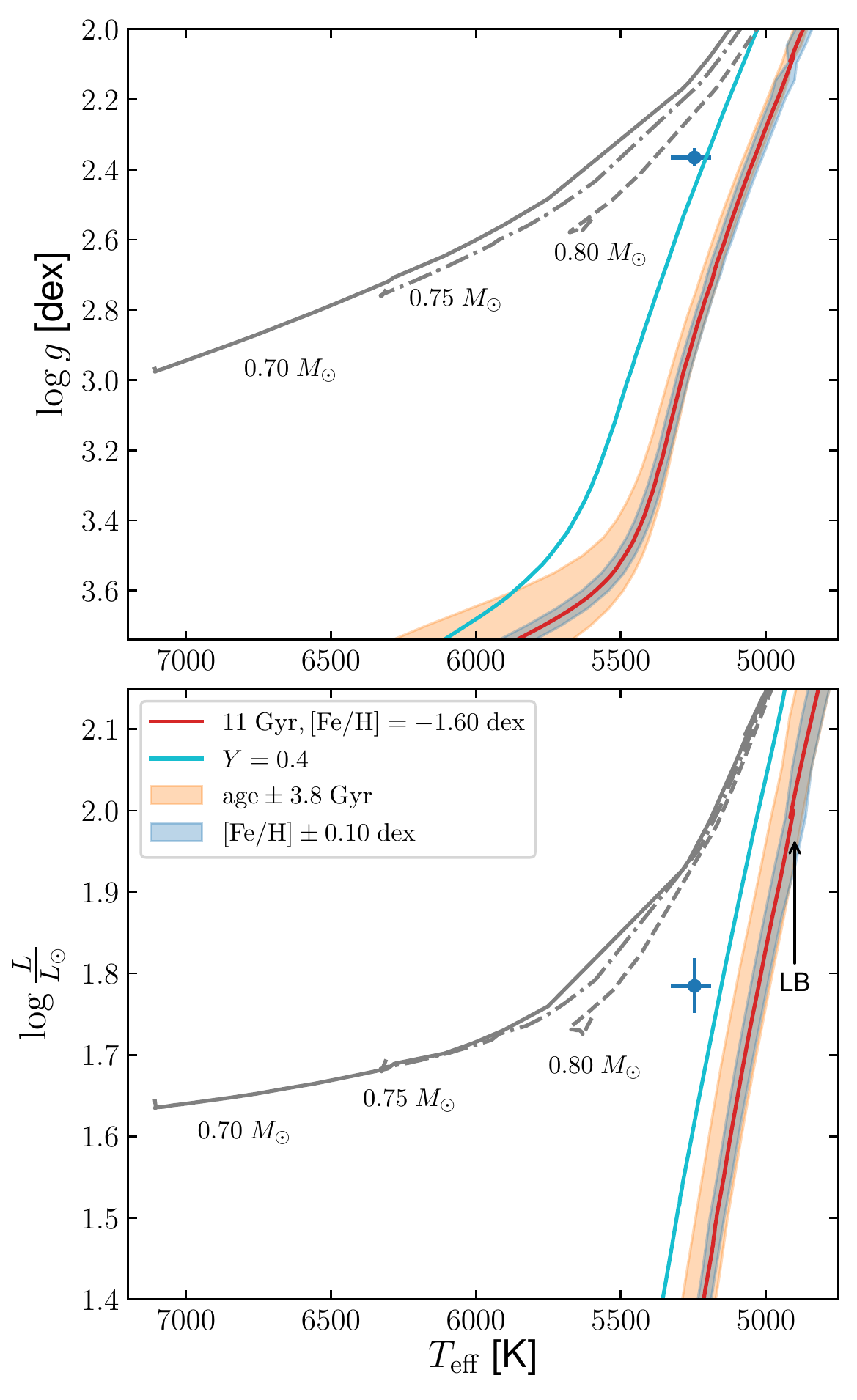}}
      \caption{Kiel diagram (\textit{upper panel}) and Hertzsprung-Russell diagram (\textit{lower panel}) with isochrones and helium burning tracks. HD~20's position is depicted by a blue filled circle with error bars. In the \textit{upper panel} the error on the gravity is smaller than the circle size. The red line represents a He-normal 11~Gyr isochrone at [Fe/H$]=-1.60$~dex, and [$\alpha$/Fe$]=+0.4$~dex with age and metallicity error margins shown by orange and blue ranges. The RGB luminosity bump for this particular model at $\log{L/L_\odot}\sim2.0$ is highlighted in the \textit{lower panel} by an arrow and the label ``LB''. The light blue curve is a model with the same parameters except for $Y=0.4$. He-burning tracks for three different masses are shown by gray lines of different line styles with the stellar masses being indicated next to the respective tracks.
      }
      \label{Fig:HRD_tracks}
\end{figure}
Earlier works on HD~20 assumed it to be a red horizontal branch star \citep[e.g.,][]{Gratton00, Carney03}. Given our newly derived set of fundamental parameters, we can neither reject nor confirm this hypothesis. In Fig. \ref{Fig:HRD_tracks}, we illustrate HD~20's position in the space of the structural parameters $T_\mathrm{eff}$, $\log{L/L_\odot}$, and $\log{g}$ together with an isochrone from the Dartmouth Stellar Evolution Database \citep{Dotter08}. The model parameters were selected to resemble the findings in the present work, that is, an age of 11~Gyr (Sect. \ref{Subsec:Cosmochronological age}), [Fe/H$]=-1.60$~dex, as well as [$\alpha$/Fe$]=0.40$~dex (Sect. \ref{Subsec: up to the iron peak}). The impacts from uncertainties in the two input parameters that affect the isochrone most -- the stellar age and [Fe/H] -- are indicated by representative error margins. While we adopted a standard scaling for the He mass fraction ($Y=0.245+1.5\cdot Z$) for the latter model, we furthermore show the case of an extreme He enhancement of $Y=0.4$. In addition, a set of He-burning tracks for three different stellar masses (0.70, 0.85, and $0.9M_\sun$) from the Dartmouth database are depicted in the same plot. 

Given its luminosity and/or gravity ($\log{L/L_\odot}=1.78$ and $\log{g}=2.366$), HD~20 appears too warm for a $\sim11$~Gyr old classical red giant, though the implied mass from the isochrone of $0.84M_\sun$ resides within one standard deviation of our mass estimate ($0.76\pm0.08M_\sun$). On the other hand, taking our asteroseismic mass and $L$ for granted, HD~20 would be between 250~K and 350~K too cool to be consistent with the models for the horizontal branch, depending on whether a one-sigma or spot-on agreement is desired. This appears infeasible even for slightly warmer photometric temperature scales (Appendix \ref{Subsubsec: Color - [Fe/H] - Teff relations}). Still, the circumstance that our star is significantly fainter than the luminosity bump of the presented isochrone at $\log{L/L_\odot}\sim2.0$ while nonetheless exhibiting mixing signatures (see previous Sect.) points towards a scenario where HD~20 has already evolved all the way through the red giant phase and is in fact now a horizontal branch star.

An alternative hypothesis for explaining HD~20's position in the Hertzsprung-Russell diagram would be a non-standard He content as the model with strongly increased $Y$ poses a considerably better fit to the observations. Such extreme levels of He have been found for second-generation stars in the most massive globular clusters \citep{Milone18, Zennaro19}. Nevertheless, characteristic chemical signatures of these peculiar stars are strong enhancements in light elements such as N, Na, and Al in lockstep with depletions of O and Mg \citep[e.g.,][]{Bastian18}; none of which were found here (see Sects. \ref{Subsec: Light elements} and \ref{Subsec: up to the iron peak}). As a consequence, it is unlikely that HD~20 is a classical red giant star with high $Y$.

Unfortunately, our TESS light curve of HD~20 cannot be used to analyze the period spacing of the $l=1$ mixed gravity and pressure modes to distinguish between helium-burning and non-helium-burning evolutionary stages as described by, for instance, \citet{Bedding11} and \citet{Mosser12b}. For achieving this, a much longer time baseline than the available 27 days would be required in order to allow for a finer scanning of the frequencies around $f_\mathrm{max}$ and the identification of subordinate peaks in the power spectrum.

\subsection{Abundances up to Zn (11 $\leq$ Z $\leq$ 30)}\label{Subsec: up to the iron peak}
\begin{figure*}
    \centering
    \resizebox{\hsize}{!}{\includegraphics{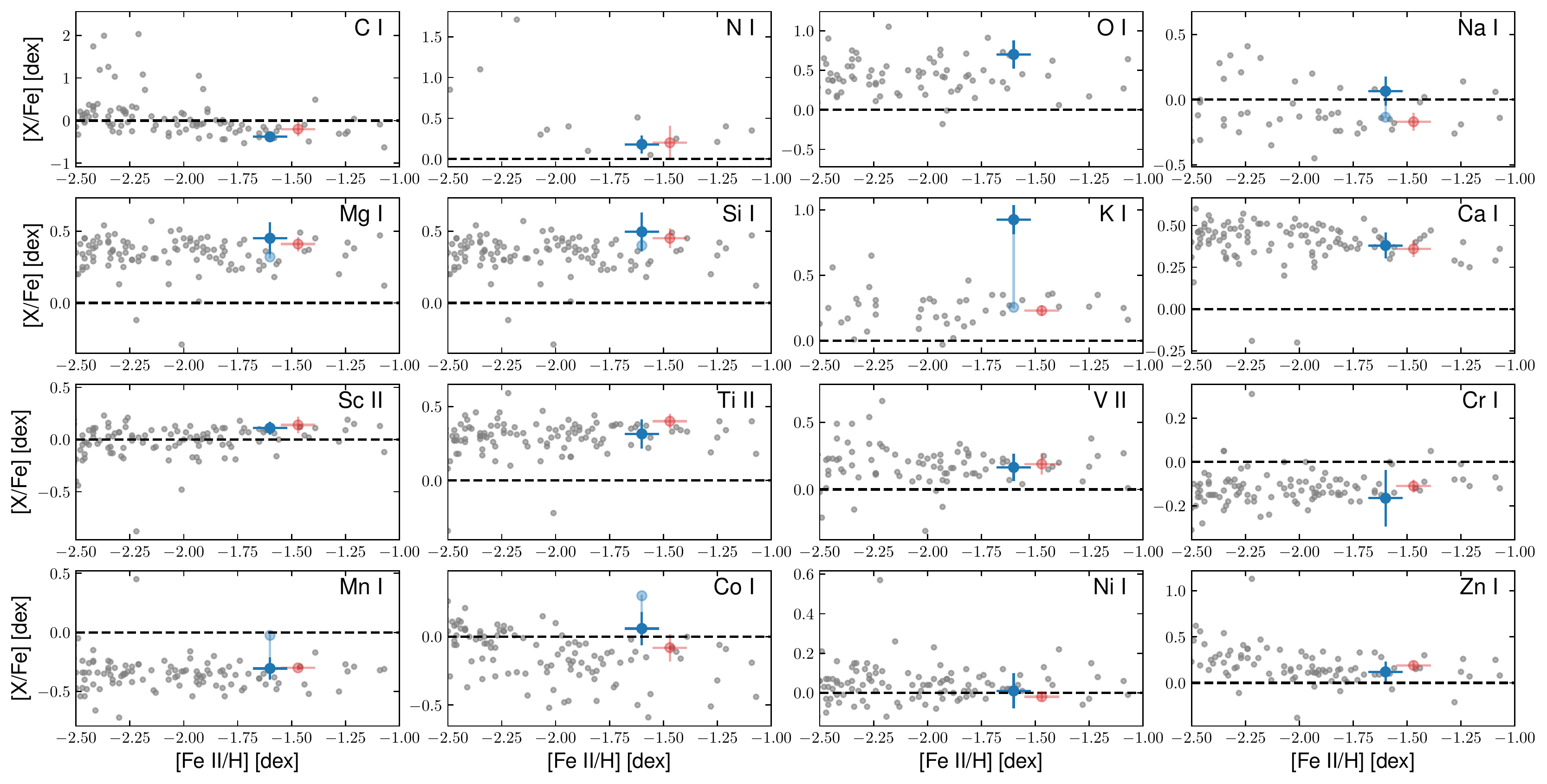}}
      \caption{Comparison of HD~20 (blue circle) to the metal-poor field star compilation (gray dots) by \citet{Roederer14} and the red horizontal branch star HD~222925 \citep[][red circle]{Roederer18}. Dark blue circles and error bars indicate the result in LTE while the light blue circles indicate the NLTE-corrected ones. In the reference samples, corrections have been applied to \ion{O}{i}, \ion{Na}{i}, and \ion{K}{i}. On the abscissa we show abundances from \ion{Fe}{ii} since these are less prone to departures from the LTE assumption (Sect. \ref{Subsec: Model atmosphere parameters}).
              }
      \label{Fig:halo_comparison}
\end{figure*}
\begin{figure}
    \centering
    \resizebox{\hsize}{!}{\includegraphics{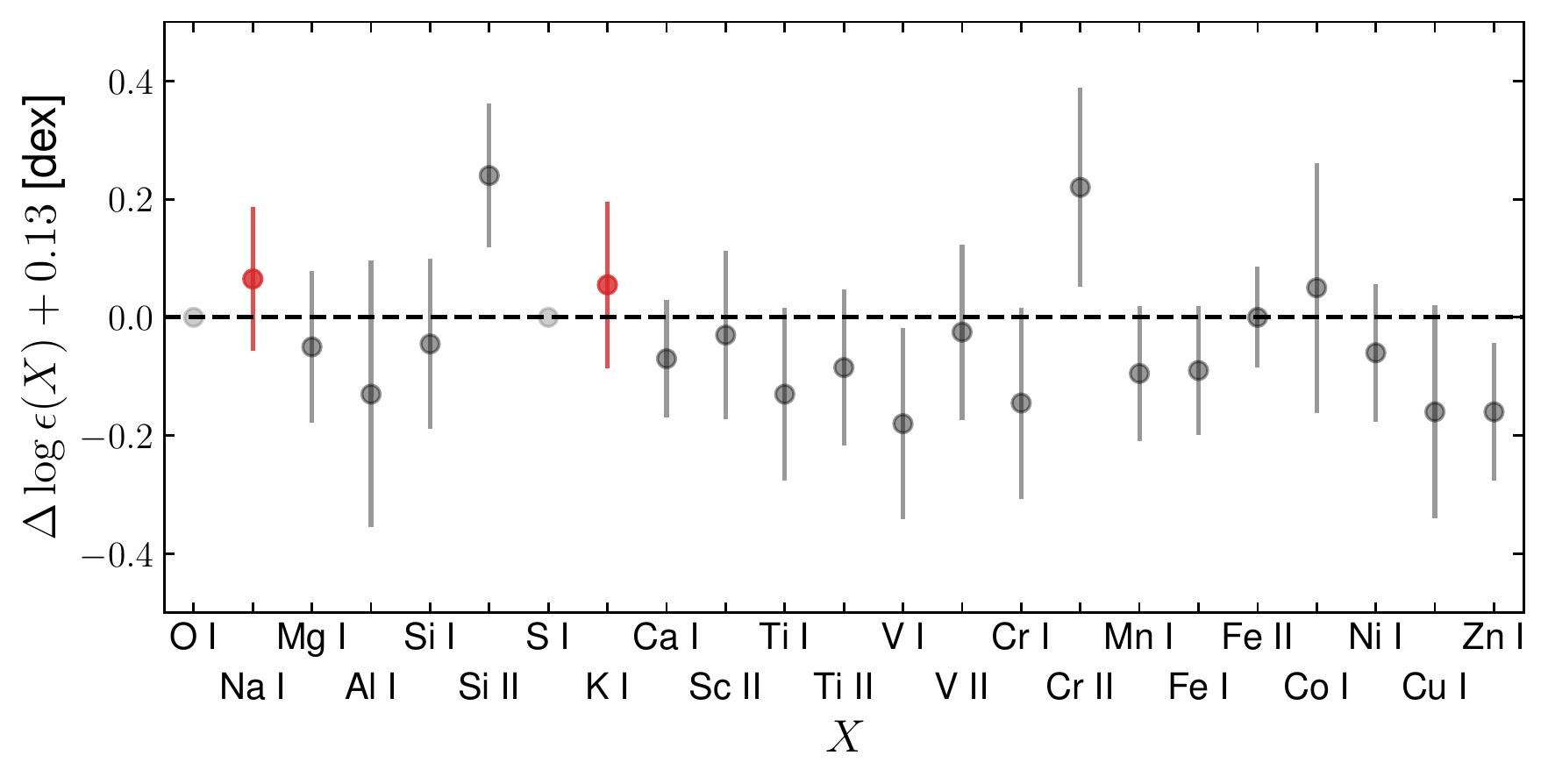}}
      \caption{Residual abundance pattern from O to Zn between HD~20 and HD~222925 after scaling by the difference in $\log{\epsilon(\ion{Fe}{ii})}$ of 0.13~dex. NLTE abundances were used for both stars for \ion{Na}{i} and \ion{K}{i} (red filled circles).
              }
      \label{Fig:light_element_pattern}
\end{figure}
We could deduce abundances for 22 species of 17 chemical elements in the range $8 \leq Z \leq 30$. For the $\alpha$-elements Mg, Si, S, Ca, and Ti we report a mean enhancement of [$\alpha$/Fe$]=0.45$~dex in LTE, which is in disagreement with the finding by \citet{Barklem05} where a conversion to the \citet{Asplund09} scale yields $\frac{1}{3}[(\mathrm{Mg}\mathrm{+Ti}+\mathrm{Ca})/\mathrm{Fe}]\approx0.23$~dex. The discrepancy is alleviated when using the same elements for comparison, that is, $\frac{1}{3}[(\mathrm{Mg}+\ion{Ti}{i}+\mathrm{Ca})/\mathrm{Fe}]=0.34\pm0.13$~dex or $\frac{1}{3}[(\mathrm{Mg}+\ion{Ti}{ii}+\mathrm{Ca})/\mathrm{Fe}]=0.38\pm0.07$~dex. In light of Appendix \ref{Sec: Abundance systematics due to atmosphere}, the origin for the observed difference is likely to be tied to their substantially hotter $T_\mathrm{eff}$ (see discussion in Sect. \ref{Subsubsec: Color - [Fe/H] - Teff relations}). Our value is typical for MW field stars at this [Fe/H] where nucleosynthetic processes in massive stars have played a dominant role in the enrichment of the ISM and supernovae of type Ia (mostly Fe-peak yields) have not yet started to contribute \citep[e.g.,][]{McWilliam97}. A minimum $\chi^2$ fit to the SN yields from \citet{Heger10} using StarFit\footnote{\url{http://starfit.org/}} \citep[see][for detailed discussions]{Placco16, Chan17, Fraser17} shows that the lighter elements of HD~20 -- in NLTE -- can be well reproduced by a $\sim11.6M_\sun$ faint CCSN with an explosion energy of $0.6\cdot10^{51}$~erg. We stress that at HD~20's metallicity we are likely not dealing with a single SN enrichment. Nevertheless, we are looking for a dominant contribution, which might survive even if it is highly integrated over time.

Overall, we find an excellent agreement of the deduced abundances with the field population at similar metallicities as demonstrated in Fig. \ref{Fig:halo_comparison}, where our findings are overlayed on top of the sample of metal-poor stars by \citet{Roederer14}. For elements with two available species we only present one representative. There are only two departures from the general trends: O and Co, which both are enhanced in comparison. However, as already noted in \citet{Roederer14}, the reference sample shows trends with stellar parameters -- most notably $T_\mathrm{eff}$ -- and thus evolutionary state. For elements heavier than N, mixing (Sect. \ref{Subsec: Light elements}) cannot be responsible for these trends, hence indicating contributions from systematic error sources in the abundance analyses. We therefore compare HD~20 to \object{HD~222925}, a star that was recently studied in great detail by \citet{Roederer18} and found to occupy a similar parameter space ($T_\mathrm{eff}=5636$~K, $\log{g}=2.54$~dex, and [Fe/H$]=-1.47$~dex). Its light-element abundances are also indicated in Fig. \ref{Fig:halo_comparison} and we present a differential comparison in Fig. \ref{Fig:light_element_pattern}. After correcting for the difference in metallicity (0.13~dex), we find a remarkable match between the two stars in the considered range (reduced $\chi^2$ of 0.49). Similarities between the two stars have already been reported in the literature from a kinematical point of view \citep{Roederer18a} and based on their metallicity \citep{Barklem05, Roederer18}. We emphasize, however, that the similarities do not extend to the neutron-capture regime, since HD~222925 is an $r$-II and HD~20 an $r$-I star with possible $s$-process contamination, as outlined in the following section.

\subsection{Neutron-capture elements (Z > 30)}\label{Subsec.Neutron-capture elements}
\begin{figure*}
    \centering
    \resizebox{0.8\hsize}{!}{\includegraphics{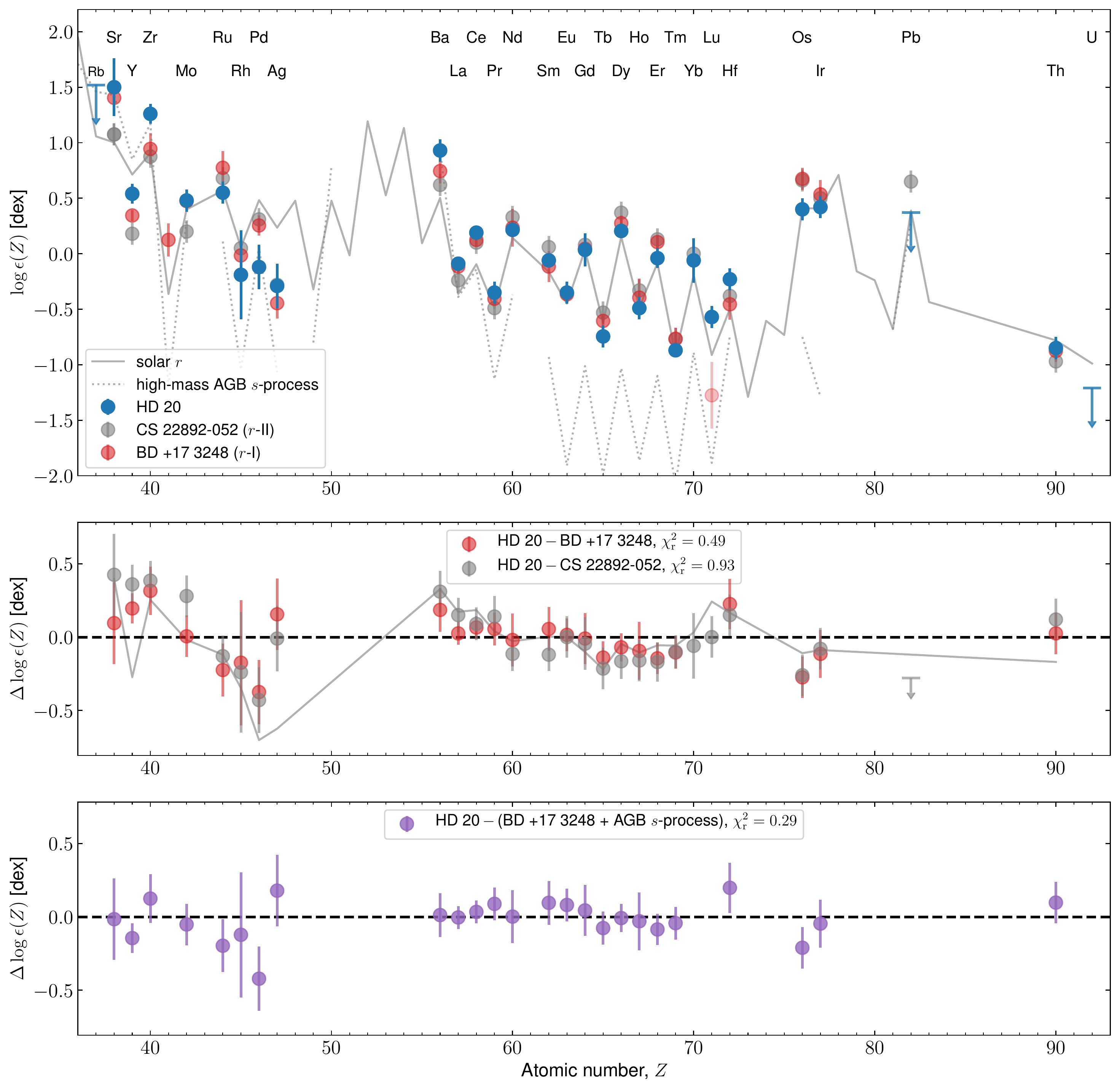}}
      \caption{Neutron-capture abundance pattern in LTE. \textit{Upper panel}: HD~20's heavy element abundances are indicated in blue. Shown in gray and red are abundances of the $r$-II star CS~22892-052 by \citet{Sneden03} and the $r$-I star BD~+17~3248 by \citet{Cowan02} with updates from \citet{Cowan05} and \citet{Roederer10}. The omitted Lu abundance for BD~+17~3248 (see main text) is depicted in light red. Both patterns were scaled to achieve the overall best match to HD~20 in the entire considered range. The gray solid line denotes the solar-scaled $r$ pattern from \citet{Sneden08} and the best-fit AGB model (see text) is represented by dotted lines. \textit{Middle panel}: Residual pattern between HD~20 and the solar $r$ pattern (gray line), CS~22892-052 (gray), and BD~+17~3248 (red). \textit{Lower panel}: Residual pattern after mixing a contribution from BD~+17~3248 with $s$-process material from the AGB yield model. 
              }
      \label{Fig:s_and_r}
\end{figure*}
In order to delineate the nucleosynthetic processes that contributed to the observed abundances of heavy elements ($Z>30$) in HD~20, we compare to a set of observed and predicted patterns. Following the classification scheme by \citet{Beers05}, our findings of [Eu/Fe$]=+0.73$~dex and [Ba/Eu$]=-0.38$~dex place HD~20 in the regime of a typical $r$-I star. As indicated by the comparison in the top and middle panels of Fig. \ref{Fig:s_and_r}, HD~20's heavy-element pattern from Nd to Ir ($60\leq Z \leq77$) is consistent with the scaled solar $r$-process by \citet{Sneden08} when considering observational errors. In the light neutron-capture regime from Sr to Ag ($38\leq Z \leq47$), however, the agreement is poor. This behavior is archetypal for $r$-process rich stars \citep[e.g.,][]{Roederer18} and led to the postulation of the existence of an additional, low-metallicity primary production channel of yet to be identified origin \citep[the so-called weak $r$ or lighter element primary process,][]{McWilliam98, Travaglio04, Hansen12, Hansen14a}. 

In Fig. \ref{Fig:s_and_r}, we further compare to the well-studied benchmark $r$-II and $r$-I stars \object{CS~22892-052} \citep{Sneden03} and \object{BD~+17~3248} \citep{Cowan02, Cowan05, Roederer10}. The latter is a red horizontal branch star that is reasonably close to HD~20 in stellar parameter space ($T_\mathrm{eff}=5200$~K, $\log{g}=1.80$~dex, [M/H$]=-2.0$~dex, $v_\mathrm{mic}=1.9$~km~s$^{-1}$) -- a circumstance that effectively reduces the impact of systematics (e.g., due to NLTE effects, see also Appendix \ref{Sec: Abundance systematics due to atmosphere}) on differential comparisons. In our analyses, we omitted the Lu abundance for BD~+17~3248 from the UV \ion{Lu}{ii} line reported by \citet{Roederer10}, because -- regardless of the substantial quoted error of 0.3~dex -- it appears to represent a strong, likely unphysical outlier. We stress that neither of the abundance patterns attributed to the two stars is necessarily a tracer of a pure nuclear process. In contrast, they are likely to represent integrated signatures with different contributions from both the main and weak primary $r$-components \citep[cf.,][]{Li13, Hansen14a}. The abundances in the range $38 \leq Z\leq90$ for the two reference stars were scaled such that the reduced sum of the normalized quadratic deviations, $\chi^2_\mathrm{r}$, was minimized (see middle panel of Fig. \ref{Fig:s_and_r}). Both patterns reproduce the depression of Y between Sr and Zr ([Y/<(Sr,Zr)>$]=-0.33$~dex) and the deviation of Ag from the solar $r$-process. This points towards an enrichment contribution to HD~20 by the weak $r$-process as postulated earlier.

Nevertheless, the overall residual abundances from Sr to Zr as well as from Ba to Pr appear enhanced with respect to the heavy $r$-nuclei ($Z\geq60$). Another particularly outstanding residual feature is a statistically significant downward trend from Ba to Yb, which seems slightly less pronounced in the comparison involving BD~+17~3248. In solar system material, the lighter elements in question have dominant contributions from the $s$-process \citep{Bisterzo14,Prantzos19}, leading to the intriguing conclusion that -- despite its moderately low metallicity -- HD~20's natal cloud might have been polluted with $s$-process material. In order to test this hypothesis, we mixed the pattern of BD~+17~3248 as proxy for an integrated $r$-process pattern with main $s$-process yield models for thermally pulsing AGB stars with a standard $^{13}$C pocket from the FUll-Network Repository of Updated Isotopic Tables \& Yields \citep[F.R.U.I.T.Y.][]{Cristallo11}. An upper metallicity limit was placed at [Fe/H$]=-1.6$ ($Z=0.0003$), since it is infeasible for AGB polluters to have had higher [Fe/H] than HD~20 itself. We retrieved models for all remaining metallicities, stellar masses, and rotational velocities available through F.R.U.I.T.Y.. In addition, a set of newly computed models with initial rotational speeds of 30 and 60~km~s$^{-1}$ for stellar masses of 2 and $5M_\odot$ was included. 

In the past, rotation has been considered as a potential process able to reproduce the observed spread in $s$-process elements at various metallicities \citep[see][and references therein]{Piersanti13}. However, depending on the adopted physical prescriptions, different results have been obtained \citep[see, e.g.,][]{Langer99, Siess04, denHartogh19}. Moreover, it has to be taken into account that recent asteroseismic measurements of low-mass stars in the Galactic disk \citep[see, e.g.,][]{Mosser12} demonstrated that stars belonging to the red clump region are characterized by slowly rotating cores. This latter feature tends to exclude the possibility to have fast-rotating cores for low-mass AGB stars in the solar neighborhood (which is an essential condition in order to have sizeable effects lead by rotation-induced mixing). However, the same has not yet been confirmed for stars with larger masses ($M>3M_\sun$) and/or at low metallicities ([Fe/H$]<-1$~dex).

The optimal mixture of integrated $r$- and main $s$-contributions to the overall neutron-capture budget of HD~20 was obtained by minimizing the expression
\begin{equation}\label{Eq:mixed_r_s}
 \chi^2=\sum_i (\log{(a\cdot\epsilon_{r,i} + b\cdot\epsilon_{s,i})} - \log{\epsilon_{\mathrm{HD 20, i}}})^2/\sigma_i^2,
\end{equation}
with $a$ and $b$ being the weight coefficients for the two $r$ and $s$ template patterns and the index $i$ denoting those individual elements in the range $38\leq Z \leq90$ with available entries for the HD~20 pattern, the BD~+17~3248 pattern, and the AGB yield tables. 

\begin{figure}
    \centering
    \resizebox{0.95\hsize}{!}{\includegraphics{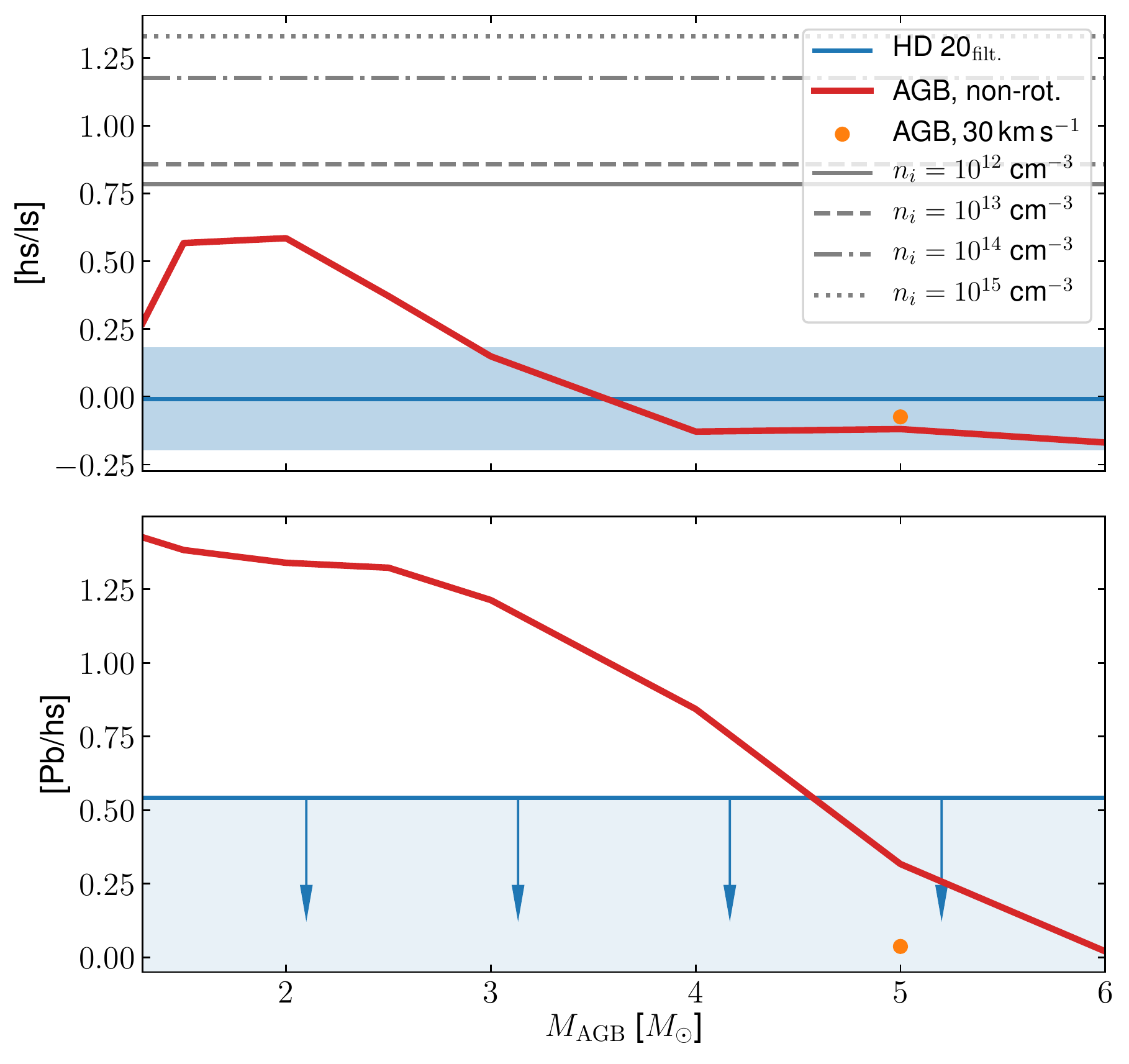}}
      \caption{Comparison of [hs/ls] and upper limit on [Pb/hs] for HD~20 against AGB $s$-process models of different initial masses. The value determined for HD~20 is indicated by blue horizontal lines and error margins, while models of $Z=0.0001$ without rotation are shown in red. The adopted best-fit model with a rotation of 30~km~s$^{-1}$ is depicted in orange. For juxtaposition, we show $i$-process predictions for [hs/ls] from \citet{Hampel16} for four different neutron densities, $n_i$, in the \textit{upper panel} using black dotted, dash-dotted, dashed, and solid lines (see legend).
              }
      \label{Fig:hsls_AGB_vs_iproc}
\end{figure}
A decisive observational quantity for pinpointing the AGB model mass is the ratio [hs/ls] of mean abundances for the heavy-$s$ (hs, represented by Ba, La, and Ce) and light-$s$ (ls, represented by Sr, Y, and Zr) elements. The models predict supersolar [hs/ls] at low masses ($\lesssim3M_\odot$) with a decreasing trend with increasing model mass. Close-to solar ratios are found in the region between $3M_\sun$ and $5M_\sun$. This behavior is demonstrated in Fig. \ref{Fig:hsls_AGB_vs_iproc}, where we also indicate the solar [hs/ls] measured for HD~20 (0.00~dex\footnote{Here we mention a ratio that was filtered for the $r$-process contribution (see later in this Sect.) as compared to the unfiltered value of 0.18~dex.}). We conclude that main $s$-process contributions are likely to originate from high-mass ($>3M_\odot$) AGB stars. This is bolstered by only being able to deduce an upper limit for HD~20's Pb abundance -- an element that is predicted to have strong contributions from models with masses $<5M_\odot$ \citep[e.g.,][]{Bisterzo12, Cristallo15}. The large contribution to Pb comes from the radiative burning of the $^{13}$C($\alpha$,$n$)$^{16}$O reaction \citep[see][]{Straniero95}, which is the dominant source in low-mass AGB stars. On the other hand, in more massive AGBs major neutron bursts come from the $^{22}$Ne($\alpha$,$n$)$^{25}$Mg reaction, which is efficiently activated at the bottom of the convective shells during thermal pulses. These episodes commonly lead to minor Pb production\footnote{Telling the whole truth, also massive AGBs can produce large amounts of Pb, but this occurs at very low metallicities only (i.e., [Fe/H$]<-2$~dex).}. At the same time, it is expected to find large Rb excesses from these massive AGB stars and their $^{22}$Ne($\alpha$,$n$)$^{25}$Mg neutron source, manifesting in, for example, supersolar [Rb/Zr] \citep{Garcia-Hernandez09, Perez-Mesa17}. For HD~20, we found $\mathrm{[Rb/Zr]}<0.55$~dex from the upper limit on the Rb abundance and after filtering our Zr finding from its dominant $r$-process contribution (see later in this Sect.). This upper limit is $\sim0.2$~dex higher than the largest predictions from our employed, massive (i.e., 4-5$M_\sun$) AGB models. A robust measurement of Rb could be used to place further constraints on the exact initial mass of the polluting AGB star. In order to achieve this, spectra with an even higher S/N in the region around 7800~{\AA} are required.  

\begin{figure*}
    \centering
    \resizebox{\hsize}{!}{\includegraphics{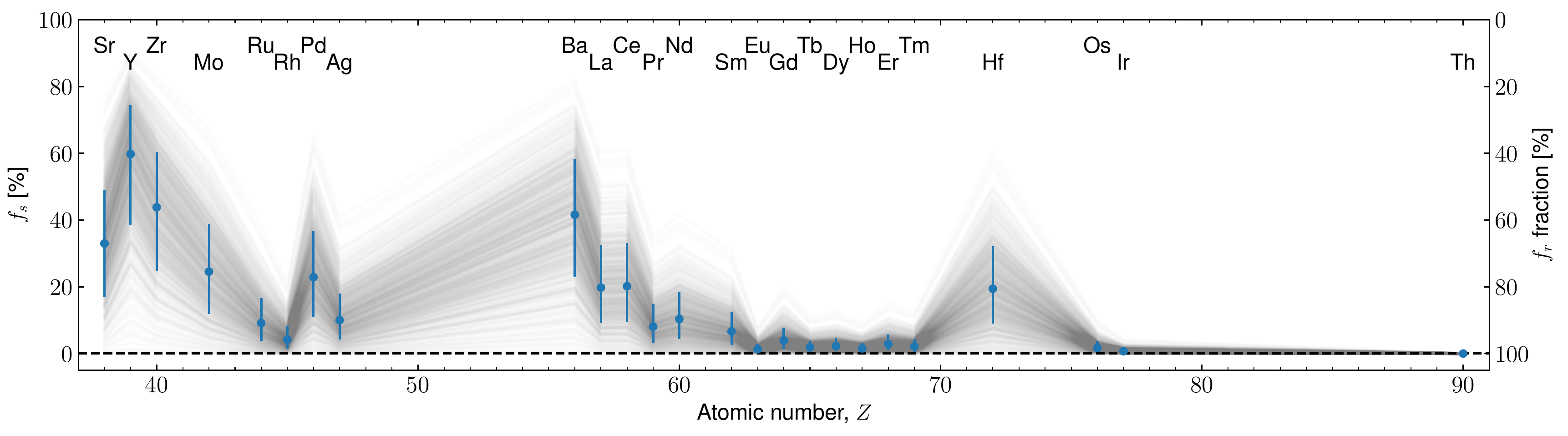}}
      \caption{Estimated $r$- (right-hand scale) and $s$-fractions (left-hand scale) in HD~20 based on Eq. \ref{Eq:r and s fraction} with BD~+17~3248 as proxy for an $r$ pattern and the best-fit AGB model representing the $s$-enrichment site. Shown are only those elements that have a measured abundance in BD~+17~3248. 
              }
      \label{Fig:r_and_s_fractions}
\end{figure*}
By minimizing Eq. \ref{Eq:mixed_r_s} we found the best-fit ($\chi^2_\mathrm{r}=0.29$) AGB model to be the one with $5M_\odot$, $Z=0.0001$ ([Fe/H$]\approx-2.15$~dex), and a rotational velocity of 30~km~s$^{-1}$. Here, the model with non-zero angular momentum poses a slightly better fit than its non-rotating counterpart with all other parameters kept fixed (see also Fig. \ref{Fig:hsls_AGB_vs_iproc} top panel). The adopted mixture can successfully reproduce the entire neutron-capture pattern in HD~20. This includes the observations for the commonly employed tracers [hs/ls] and [Ba/Eu], as well as the downward trend from Ba to Yb that persists when assuming an $r$-only enrichment.

Using yields from the aforementioned main $s$-model and the BD~+17~3248 pattern together with the best-fit model parameters for Eq. \ref{Eq:mixed_r_s}, we can estimate the fractional (integrated) $r$- and (main) $s$-process contributions to individual elements in HD~20 through
\begin{equation}\label{Eq:r and s fraction}
f_{r,i} = \frac{a\cdot\epsilon_{r,i}}{a\cdot\epsilon_{r,i} + b\cdot\epsilon_{s,i}}; \ \ f_{s,i} = 1-f_{r,i}.
\end{equation}
In order to properly account for fit uncertainties, we sampled the posterior distribution of the parameters $a$ and $b$ with \textit{emcee} using the abundance errors. In Fig. \ref{Fig:r_and_s_fractions} we show 800 individual realizations of the samples. From these, the fractions and asymmetric limits were estimated from the median, the 15.9$^\mathrm{th}$, and the 84.1$^\mathrm{th}$ percentiles, respectively. These are listed in Table \ref{Table: s and r fractions}.
\begin{table}
\caption{Estimated fractional contributions from the $r$- and $s$-process for elements with $Z\geq38$ in HD~20.}
\label{Table: s and r fractions}
\centering
\resizebox{0.5\columnwidth}{!}{%
\addtolength{\tabcolsep}{-3pt}
\begin{tabular}{ccrr}
\hline\hline\\[-7pt]
$Z$ & Element & \multicolumn{1}{c}{$f_r$} & \multicolumn{1}{c}{$f_s$}\\
 & & \multicolumn{1}{c}{[\%]} & \multicolumn{1}{c}{[\%]}\\
\hline\\[-5pt]
38 & Sr & $67.0^{+15.6}_{-15.7}$ & $33.0^{+15.7}_{-15.6}$\\[4pt]
39 & Y & $40.2^{+21.0}_{-14.4}$ & $59.8^{+14.4}_{-21.0}$\\[4pt]
40 & Zr & $56.2^{+18.9}_{-16.3}$ & $43.8^{+16.3}_{-18.9}$\\[4pt]
42 & Mo & $75.5^{+12.4}_{-14.0}$ & $24.5^{+14.0}_{-12.4}$\\[4pt]
44 & Ru & $90.9^{+5.0}_{-7.1}$ & $9.1^{+7.1}_{-5.0}$\\[4pt]
45 & Rh & $95.8^{+2.3}_{-3.6}$ & $4.2^{+3.6}_{-2.3}$\\[4pt]
46 & Pd & $77.1^{+11.6}_{-13.5}$ & $22.9^{+13.5}_{-11.6}$\\[4pt]
47 & Ag & $90.0^{+5.5}_{-7.6}$ & $10.0^{+7.6}_{-5.5}$\\[4pt]
56 & Ba & $58.4^{+18.3}_{-16.3}$ & $41.6^{+16.3}_{-18.3}$\\[4pt]
57 & La & $80.2^{+10.3}_{-12.5}$ & $19.8^{+12.5}_{-10.3}$\\[4pt]
58 & Ce & $79.8^{+10.4}_{-12.6}$ & $20.2^{+12.6}_{-10.4}$\\[4pt]
59 & Pr & $92.0^{+4.4}_{-6.4}$ & $8.0^{+6.4}_{-4.4}$\\[4pt]
60 & Nd & $89.7^{+5.7}_{-7.9}$ & $10.3^{+7.9}_{-5.7}$\\[4pt]
62 & Sm & $93.4^{+3.7}_{-5.4}$ & $6.6^{+5.4}_{-3.7}$\\[4pt]
63 & Eu & $98.7^{+0.7}_{-1.2}$ & $1.3^{+1.2}_{-0.7}$\\[4pt]
64 & Gd & $96.1^{+2.2}_{-3.4}$ & $3.9^{+3.4}_{-2.2}$\\[4pt]
65 & Tb & $98.1^{+1.1}_{-1.7}$ & $1.9^{+1.7}_{-1.1}$\\[4pt]
66 & Dy & $97.8^{+1.3}_{-2.0}$ & $2.2^{+2.0}_{-1.3}$\\[4pt]
67 & Ho & $98.5^{+0.9}_{-1.4}$ & $1.5^{+1.4}_{-0.9}$\\[4pt]
68 & Er & $97.2^{+1.6}_{-2.5}$ & $2.8^{+2.5}_{-1.6}$\\[4pt]
69 & Tm & $97.8^{+1.2}_{-1.9}$ & $2.2^{+1.9}_{-1.2}$\\[4pt]
72 & Hf & $80.6^{+10.1}_{-12.3}$ & $19.4^{+12.3}_{-10.1}$\\[4pt]
76 & Os & $98.3^{+1.0}_{-1.5}$ & $1.7^{+1.5}_{-1.0}$\\[4pt]
77 & Ir & $99.3^{+0.4}_{-0.6}$ & $0.7^{+0.6}_{-0.4}$\\[4pt]
90 & Th & $100.0^{+0.0}_{-0.0}$ & $0.0^{+0.0}_{-0.0}$\\[4pt]

\hline
\end{tabular}
\addtolength{\tabcolsep}{3pt}}
\end{table}
We find significant $s$-process fractions above 30\% for the elements Sr, Y, Zr, and Ba, whereas only Y might have had a dominant ($f_s>50\%$) enrichment contribution from the $s$-process. This could be corroborated by measuring isotopic fractions for selected elements from spectra at very high resolution \citep[e.g.,][]{Mashonkina19}, though we note that HD~20 shows considerable intrinsic line broadening signatures (Sect. \ref{Subsec: Rotational velocity}) that may exceed the hyperfine splitting effect.

An important question to answer with respect to our proposed $s$-process imprint is whether the finding is caused by mixing in the ISM prior to the formation of HD~20, or as a result of surface pollution via mass transfer in a binary system \citep[e.g.,][]{Gull18}. The latter option was ruled out with high confidence in Sect. \ref{Subsec: Radial velocities and binarity}, where we showed a lack of radial velocity variation. Therefore, a binary signal could only be hidden if the orbit would be seen almost perfectly face-on. Consequently, we strongly prefer the scenario where HD~20 had its chemical pattern composition mixed in the ISM.

\subsection{i-process considerations}\label{Subsex: i-process considerations}
Another metal-poor star with signatures of simultaneous overabundances in both $s$- and $r$-process material is HD~94028. Among others, this star has been studied spectroscopically by \citet{Roederer12} and \citet{Roederer16}, who complemented the abundance pattern from elements typically found in the optical with more exotic species (e.g., Ge, As, Se) that are only measurable in ultraviolet spectra gathered with the \textit{Hubble Space Telescope} (HST). The authors concluded that several abundance ratios -- most notably supersolar [As/Ge], [Mo/Fe], and [Ru/Fe] -- are poorly described by combinations of $s$- and $r$-process patterns and therefore suggested an additional contribution by the $i$-process. However, more recently, \citet{Han18} indicated that both [As/Ge] and [Mo/Ru] may be well explained by weak $r$-nucleosynthesis without the need for an additional $i$-process. The $i$-process was also proposed by \citet{Koch19} as a candidate to reproduce their observed pattern for a metal-poor bulge star \citep[labeled $\#$10464, following the naming convention in][]{Koch16}. The authors find that either a mixture of an $i$-pattern with a main $s$-pattern or an $i$-process with two proton ingestion events reproduces their observations best.

\begin{figure}
    \centering
    \resizebox{\hsize}{!}{\includegraphics{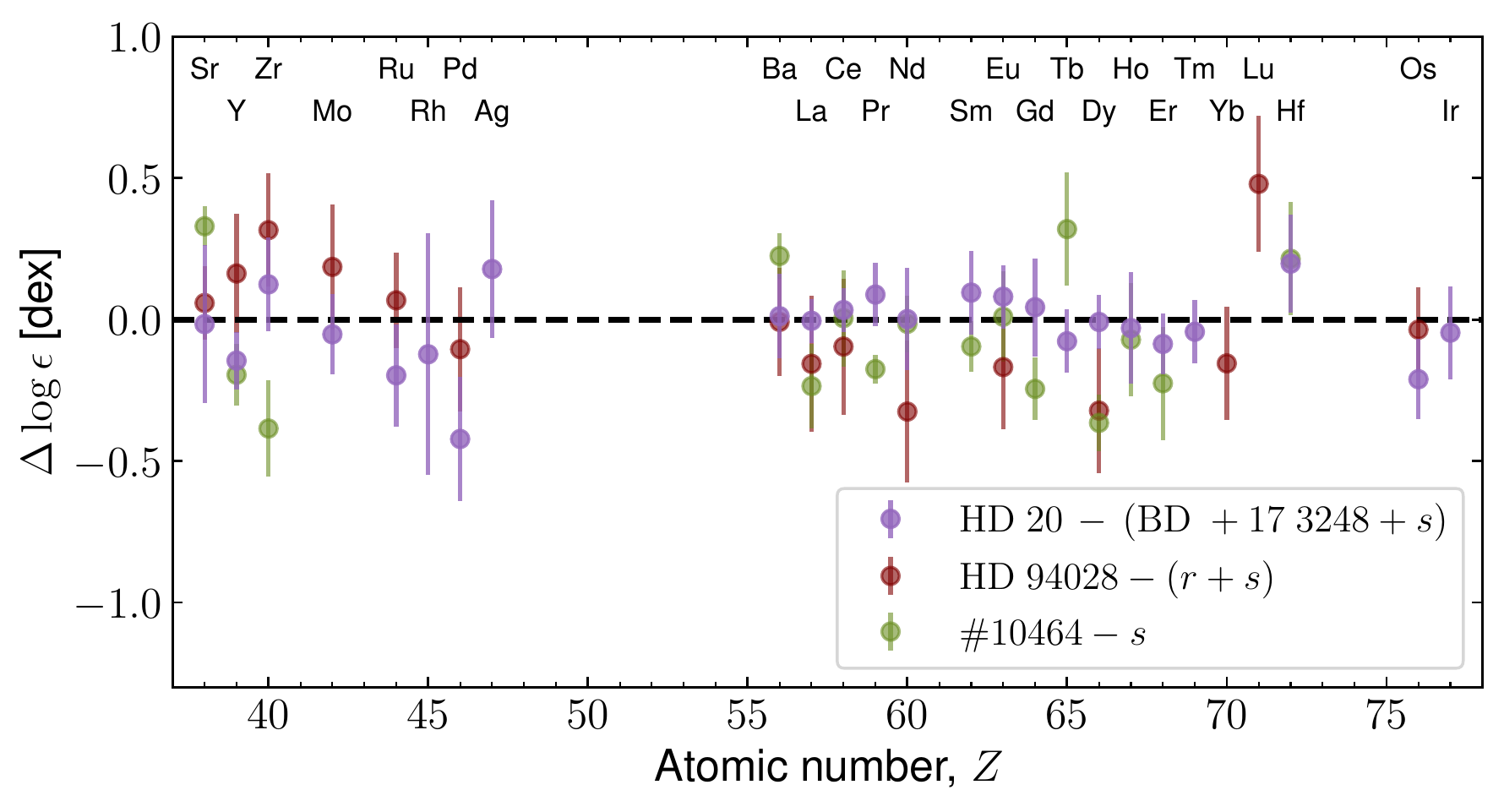}}
      \caption{Comparison of the residual HD~20 pattern (purple, same as \textit{lower panel} of Fig. \ref{Fig:s_and_r}) to the patterns of HD~94028 (red) and $\#$10464 (green) after subtracting $r+s$ and $s$ contributions, respectively. The residual pattern for HD~94028 was determined following the procedure outlined in Sect. 5.2. in \citet{Roederer16}, whereas a $Z=0.0001$, $2M_\sun$ AGB model was assumed for the $s$-enrichment in $\#$10464.  
              }
      \label{Fig:iproc_HD20_vs_Roed_and_Koch}
\end{figure}
Based on Fig. \ref{Fig:hsls_AGB_vs_iproc}, the residual [hs/ls] of HD~20 is seen to be well described by a $5M_\sun$ rotating AGB star, while the $i$-process of intermediate neutron densities predicts much too high [hs/ls] ratios. In any case, we compare HD~20 to the two supposedly $i$-enriched stars to search for $i$-process indications in the patterns. By comparing to the filtered patterns of $\mathrm{HD~94028}-(r+s)$ and $\mathrm{\#10464}-s$ (Fig. \ref{Fig:iproc_HD20_vs_Roed_and_Koch}), no clear $i$-process features stand out, and we cannot claim any $i$-process contribution in HD~20. However, some weak $r$-enrichment might have taken place. Until further $i$-process indications, such as elemental ratios [As/Ge] or strong pattern trends can robustly be associated with the $i$-process, it is hard to observationally investigate such contaminations. In order to test [As/Ge] we would need HST data.

\subsection{Cosmochronological age}\label{Subsec:Cosmochronological age}
\begin{table}
\caption{Age estimates from different radioactive chronometers.}
\label{Table: ages}
\centering
\resizebox{0.7\columnwidth}{!}{%
\addtolength{\tabcolsep}{-3pt}
\begin{tabular}{lccc}
\hline\hline\\[-7pt]
ratio & $\log{\epsilon(\mathrm{Th/r})}_0^a$ & $\log{\epsilon(\mathrm{Th/r})}$ & age\\
 & [dex] & [dex] & [Gyr]\\
\hline\\[-5pt]
Th/Eu & $-0.276$ & $-0.50\pm0.14$ & $10.0\pm6.5$\\
Th/Hf & $-0.063$ & $-0.62\pm0.14$ & $26.0\pm6.5$\\
Th/Os & $-1.009$ & $-1.25\pm0.14$ & $11.3\pm6.5$\\
Th/Ir & $-1.022$ & $-1.27\pm0.14$ & $11.6\pm6.5$\\
Th/U  & $\phantom{-}0.192$ & $>0.36$ & $>7.8^b$\\
\hline
\end{tabular}
\addtolength{\tabcolsep}{3pt}}
\tablefoot{
\tablefoottext{a}{Production ratios from method \textit{``fit1''} in Table 2 of \citet{Kratz07}.}
\tablefoottext{b}{Calculated using Eq. (2) in \citet{Cayrel01}.}
}
\end{table}
Having measured a reliable abundance for the radioactive element Th enables an estimate of HD~20's age from nuclear cosmochronology. The only isotope of Th with a lifetime that is relevant on cosmological timescales is $^{232}$Th ($\tau_{1/2}=14.05$~Gyr). The currently observed ratio $\log{\epsilon(\mathrm{Th/r})}$ of Th and other, stable $r$-elements can be related to a decay time using a theoretical initial production ratio, $\log{\epsilon(\mathrm{Th/r})}_0$, together with the age relation
\begin{equation}
 \Delta t = 46.7\mathrm{~Gyr}\cdot\left(\log{\epsilon(\mathrm{Th/r})}_0 - \log{\epsilon(\mathrm{Th/r})}\right)
\end{equation}
as outlined by \citet{Cayrel01}. For Table \ref{Table: ages} we considered the reference elements Eu and Hf as well as the third-peak elements Os and Ir. Moreover, we obtained a lower-limit age of 7.8~Gyr from our upper limit on the U abundance. Despite considerable ambiguities in theoretical production ratios \citep[e.g.,][]{Schatz02, Cowan99}, the dominant source of error for the inferred ages is the combined uncertainty of the abundances for each pair, which amounts to $\sqrt{2}\cdot0.1\mathrm{~dex}=0.14$~dex. The latter uncertainty linearly propagates into an age error of 6.5~Gyr \citep[see also][for a detailed discussion of other error sources]{Ludwig10}. According to \citet{Cayrel01}, the observational and theoretical uncertainties are minimized by using Os and Ir as baseline for the chronometers, since they are closest to Th in atomic number. However, we note that both Os and Ir were determined from the neutral species while our Th abundance was deduced from the singly ionized state, which potentially introduces biases due to NLTE effects\footnote{Furthermore, as demonstrated in Appendix \ref{Sec: Abundance systematics due to atmosphere}, among all relevant elements the two referred ones are most sensitive to uncertainties in the model temperature.}. As indicated by \citet{Hansen18}, NLTE effects on \ion{Th}{ii} abundances may be alleviated by introducing a full, 3D NLTE treatment. Hence the obtained abundance would be close to our 1D LTE estimate. 

The age of 26.0~Gyr from Th/Hf appears unreasonably high and we note that \citet{Roederer09} reported on a similar behavior for this chronometer. We thus suspect that the initial production rates are overestimated, which might be connected to a breakdown of the robustness of the heavy $r$-pattern in the region around Hf (M. Eichler priv. comm.). Removing our estimated high $s$-process contribution (19.4\%) for Hf only slightly decreases the deduced age by about 4~Gyr. In any case, we exclude the corresponding age from consideration and calculate a mean age of $11.0\pm3.8$~Gyr from the remaining three actual detections (10.0, 11.3, and 11.6~Gyr, thereby excluding the lower limit involving U). 

\section{Summary and Conclusions}\label{Sec: Conclusion}
We present a detailed investigation of the chemical composition of the metal-poor ([Fe/H$]=-1.60$~dex), $r$-process enhanced ($r$-I) Galactic halo star HD~20. Using newly obtained and archival very high signal-to-noise and high-resolution spectra in concert with extensive photometry and astrometry from the \textit{Gaia} and TESS missions, we carefully investigate the key fundamental stellar parameters, which are independently confirmed by a number of alternative approaches. These allow for a high-precision spectroscopic chemical analysis, yielding abundances for 25 species of 20 elements with $Z\leq30$, as well as for 29 species of 28 neutron-capture elements. Hence, we report on abundances for in total 48 elements, thereby adding 26 elements to the largest existing study of this star by \citet{Barklem05}. Moreover, we deduce meaningful upper limits for Li, Rb, Pb, and U. This renders our presented abundance pattern one of the most complete available to date and therefore adds HD~20 to the short list of benchmark stars for nuclear astrophysics involving traces of only $r$+$s$ processes. 

Regarding the light elements up to Zn we find a behavior typical for the Galactic halo at comparable metallicities indicative of an enrichment history dominated by CCSNe prior to the onset of contributions by supernovae of type Ia. Using yield models, we could show that faint CCSNe of progenitor masses around $\sim11.6M_\sun$ and explosion energies $\sim0.6\cdot10^{51}$~erg can explain the light-element pattern in HD~20. While the heavy neutron-capture elements are found to closely follow the solar $r$-process distribution, strong deviations are found with respect to the first-peak elements, primarily due to depletions in Y and Ag. We attribute this observation to the additional primary (weak) $r$-process acting at low metallicity that was postulated based on observations of other metal-poor stars \citep[e.g.,][]{Hansen12}. This emphasizes that the solar-scaled $r$-pattern cannot pose as a universal proxy for the $r$-process, particularly in the lighter neutron-capture regime.

In comparing our observed neutron-capture abundances to the benchmark $r$-I star BD~+17~3248 -- which was chosen in order to lessen the gravity of systematic abundance errors -- we find that several elements (Sr, Y, Zr, Ba, La) that are commonly associated with the $s$-process appear to be enhanced in HD~20 with respect to a pure $r$-process pattern. We obtain a considerably better fit of the overall distribution by introducing a dilution with material from main $s$-process yield predictions of a low-metallicity, massive, and rotating AGB star. Based on this model, we estimate a dominant $s$-process fraction for Y ($59.8^{+14.4}_{-21.0}$\%), whereas several other elements may still have a significant contribution from this production channel (Table \ref{Table: s and r fractions}, Fig. \ref{Fig:r_and_s_fractions}). Given the here presented abundance pattern for HD~20, we prefer an $r$+$s$ mixing scenario and refute $i$-process contributions until more robust abundance ratios or patterns will be put forward. Based on the lacking evidence of HD~20 being part of a binary system, we propose that the mixing happened in the ISM prior to the star's formation as opposed to surface pollution due to mass transfer from a companion. 

HD~20's age is estimated at $11.0\pm3.8$~Gyr based on nuclear cosmochronology from abundance ratios involving the radioactive element Th. We caution, however, that there are statistical and systematic error sources of both observational and theoretical nature that may bias this measure. Nonetheless, it appears safe to assume that the star is a representative of the old Galactic halo.

A future perspective for work on HD~20 is to complement our abundance pattern with UV spectra from HST. Deriving abundances from UV lines is extremely important in order to obtain more complete patterns. Key elements like As and Au carry important information on the neutron-capture environment and can only be assessed in HST data. Arsenic could contain crucial clues on the $i$-process, which we cannot explore in the ground-based, spectroscopically derived abundances, and Au is a good $r$-process indicator. An additional element that is more easily measured in the UV is Pb, which is an important $s$-process tracer for which we could only deduce an upper limit abundance in this study. Furthermore, understanding how and if the neutron-capture processes are formed and incorporated into later generations of stars is crucial to understand the need for an $i$-process versus efficient and fast mixing of $r$+$s$-process material in the ISM. Here, HD~20 offers promising insights into the neutron-capture processes as it is slightly enhanced and we detect clear traces of both $r$ and $s$. It poses a powerful benchmark and it is far less polluted than the sun.  

\begin{acknowledgements}
M.H., H.-G.L., and E.K.G. gratefully acknowledge support by the Deutsche Forschungsgemeinschaft (DFG, German Research Foundation) -- Project-ID 138713538 -- SFB 881 (``The Milky Way System'', subprojects A03, A04, and A08). C.J.H. acknowledges support from the Max Planck Society and the ‘ChETEC’ COST Action (CA16117), supported by COST (European Cooperation in Science and Technology). The authors are grateful to M. Catelan, A. Koch, Z. Prudil, and A. Gallagher for fruitful discussions. Furthermore, the help by M. Kovalev and M. Bergemann in setting up a grid of NLTE corrections for Fe lines is highly appreciated. A. Heger and C. Chan are acknowledged for their detailed support for the application of StarFit. We thank T. Nordlander and K. Lind for providing access to their grid of NLTE corrections for Al. We value the swift and constructive report issued by the anonymous referee. This work has made use of the VALD database, operated at Uppsala University, the Institute of Astronomy RAS in Moscow, and the University of Vienna. This paper presents results from the European Space Agency (ESA) space mission \textit{Gaia}. \textit{Gaia} data are being processed by the \textit{Gaia} Data Processing and Analysis Consortium (DPAC). Funding for the DPAC is provided by national institutions, in particular the institutions participating in the \textit{Gaia} MultiLateral Agreement (MLA). The \textit{Gaia} mission website is \url{https://www.cosmos.esa.int/gaia}. This paper includes data collected by the TESS mission. Funding for the TESS mission is provided by the NASA Explorer Program. This research made use of Lightkurve, a Python package for Kepler and TESS data analysis \citep{lightkurve}.
\end{acknowledgements}

\bibliographystyle{aa.bst}
\bibliography{sources.bib}

\begin{thebibliography}{238}
\expandafter\ifx\csname natexlab\endcsname\relax\def\natexlab#1{#1}\fi

\bibitem[{{Abbott} {et~al.}(2017){Abbott}, {Abbott}, {Abbott}, {Acernese},
  {Ackley}, {Adams}, {Adams}, {Addesso}, {Adhikari}, {Adya}, {Affeldt},
  {Afrough}, {Agarwal}, {Agathos}, {Agatsuma}, {Aggarwal}, {Aguiar}, {Aiello},
  {Ain}, {Ajith}, {Allen}, {Allen}, {Allocca}, {Altin}, {Amato}, {Ananyeva},
  {Anderson}, {Anderson}, {Angelova}, {Antier}, {Appert}, {Arai}, {Araya},
  {Areeda}, {Arnaud}, {Arun}, {Ascenzi}, {Ashton}, {Ast}, {Aston}, {Astone},
  {Atallah}, {Aufmuth}, {Aulbert}, {AultONeal}, {Austin}, {Avila-Alvarez},
  {Babak}, {Bacon}, {Bader}, {Bae}, {Bailes}, {Baker}, {Baldaccini},
  {Ballardin}, {Ballmer}, {Banagiri}, {Barayoga}, {Barclay}, {Barish},
  {Barker}, {Barkett}, {Barone}, {Barr}, {Barsotti}, {Barsuglia}, {Barta},
  {Barthelmy}, {Bartlett}, {Bartos}, {LIGO Scientific Collaboration}, \& {Virgo
  Collaboration}}]{Abbott17}
{Abbott}, B.~P., {Abbott}, R., {Abbott}, T.~D., {et~al.} 2017, \prl, 119,
  161101

\bibitem[{{Alonso} {et~al.}(1994){Alonso}, {Arribas}, \&
  {Martinez-Roger}}]{Alonso94}
{Alonso}, A., {Arribas}, S., \& {Martinez-Roger}, C. 1994, \aaps, 107, 365

\bibitem[{{Alonso} {et~al.}(1999{\natexlab{a}}){Alonso}, {Arribas}, \&
  {Mart{\'\i}nez-Roger}}]{Alonso99a}
{Alonso}, A., {Arribas}, S., \& {Mart{\'\i}nez-Roger}, C. 1999{\natexlab{a}},
  \aaps, 139, 335

\bibitem[{{Alonso} {et~al.}(1999{\natexlab{b}}){Alonso}, {Arribas}, \&
  {Mart{\'{\i}}nez-Roger}}]{Alonso99}
{Alonso}, A., {Arribas}, S., \& {Mart{\'{\i}}nez-Roger}, C. 1999{\natexlab{b}},
  \aaps, 140, 261 (AAM99)

\bibitem[{{Alonso} {et~al.}(2001){Alonso}, {Arribas}, \&
  {Mart{\'{\i}}nez-Roger}}]{Alonso01}
{Alonso}, A., {Arribas}, S., \& {Mart{\'{\i}}nez-Roger}, C. 2001, \aap, 376,
  1039

\bibitem[{{Amarsi} {et~al.}(2019){Amarsi}, {Nissen}, \&
  {Sk{\'u}lad{\'o}ttir}}]{Amarsi19}
{Amarsi}, A.~M., {Nissen}, P.~E., \& {Sk{\'u}lad{\'o}ttir}, {\'A}. 2019, \aap,
  630, A104

\bibitem[{{Amarsi} {et~al.}(2018){Amarsi}, {Nordlander}, {Barklem}, {Asplund},
  {Collet}, \& {Lind}}]{Amarsi18}
{Amarsi}, A.~M., {Nordlander}, T., {Barklem}, P.~S., {et~al.} 2018, \aap, 615,
  A139

\bibitem[{{Andrae} {et~al.}(2018){Andrae}, {Fouesneau}, {Creevey}, {Ordenovic},
  {Mary}, {Burlacu}, {Chaoul}, {Jean-Antoine-Piccolo}, {Kordopatis}, {Korn},
  {Lebreton}, {Panem}, {Pichon}, {Th{\'e}venin}, {Walmsley}, \&
  {Bailer-Jones}}]{Andrae18}
{Andrae}, R., {Fouesneau}, M., {Creevey}, O., {et~al.} 2018, \aap, 616, A8

\bibitem[{{Andrievsky} {et~al.}(2011){Andrievsky}, {Spite}, {Korotin},
  {Fran{\c{c}}ois}, {Spite}, {Bonifacio}, {Cayrel}, \& {Hill}}]{Andrievsky11}
{Andrievsky}, S.~M., {Spite}, F., {Korotin}, S.~A., {et~al.} 2011, \aap, 530,
  A105

\bibitem[{{Anthony-Twarog} \& {Twarog}(1994)}]{AnthonyTwarog94}
{Anthony-Twarog}, B.~J. \& {Twarog}, B.~A. 1994, \aj, 107, 1577

\bibitem[{{Arcones} {et~al.}(2007){Arcones}, {Janka}, \& {Scheck}}]{Arcones07}
{Arcones}, A., {Janka}, H.~T., \& {Scheck}, L. 2007, \aap, 467, 1227

\bibitem[{{Asplund} {et~al.}(2009){Asplund}, {Grevesse}, {Sauval}, \&
  {Scott}}]{Asplund09}
{Asplund}, M., {Grevesse}, N., {Sauval}, A.~J., \& {Scott}, P. 2009, \araa, 47,
  481

\bibitem[{{Bailer-Jones} {et~al.}(2018){Bailer-Jones}, {Rybizki}, {Fouesneau},
  {Mantelet}, \& {Andrae}}]{BailerJones18}
{Bailer-Jones}, C.~A.~L., {Rybizki}, J., {Fouesneau}, M., {Mantelet}, G., \&
  {Andrae}, R. 2018, \aj, 156, 58

\bibitem[{{Bard} {et~al.}(1991){Bard}, {Kock}, \& {Kock}}]{BKK}
{Bard}, A., {Kock}, A., \& {Kock}, M. 1991, \aap, 248, 315

\bibitem[{{Bard} \& {Kock}(1994)}]{BK}
{Bard}, A. \& {Kock}, M. 1994, \aap, 282, 1014

\bibitem[{{Barklem} {et~al.}(2005){Barklem}, {Christlieb}, {Beers}, {Hill},
  {Bessell}, {Holmberg}, {Marsteller}, {Rossi}, {Zickgraf}, \&
  {Reimers}}]{Barklem05}
{Barklem}, P.~S., {Christlieb}, N., {Beers}, T.~C., {et~al.} 2005, \aap, 439,
  129

\bibitem[{{Barklem} {et~al.}(2002){Barklem}, {Stempels}, {Allende Prieto},
  {Kochukhov}, {Piskunov}, \& {O'Mara}}]{Barklem02}
{Barklem}, P.~S., {Stempels}, H.~C., {Allende Prieto}, C., {et~al.} 2002, \aap,
  385, 951

\bibitem[{{Bastian} \& {Lardo}(2018)}]{Bastian18}
{Bastian}, N. \& {Lardo}, C. 2018, \araa, 56, 83

\bibitem[{{Battaglia} {et~al.}(2008){Battaglia}, {Irwin}, {Tolstoy}, {Hill},
  {Helmi}, {Letarte}, \& {Jablonka}}]{Battaglia08}
{Battaglia}, G., {Irwin}, M., {Tolstoy}, E., {et~al.} 2008, \mnras, 383, 183

\bibitem[{{Bedding} {et~al.}(2011){Bedding}, {Mosser}, {Huber},
  {Montalb{\'a}n}, {Beck}, {Christensen-Dalsgaard}, {Elsworth}, {Garc{\'\i}a},
  {Miglio}, {Stello}, {White}, {De Ridder}, {Hekker}, {Aerts}, {Barban},
  {Belkacem}, {Broomhall}, {Brown}, {Buzasi}, {Carrier}, {Chaplin}, {di Mauro},
  {Dupret}, {Frandsen}, {Gilliland }, {Goupil}, {Jenkins}, {Kallinger},
  {Kawaler}, {Kjeldsen}, {Mathur}, {Noels}, {Silva Aguirre}, \&
  {Ventura}}]{Bedding11}
{Bedding}, T.~R., {Mosser}, B., {Huber}, D., {et~al.} 2011, \nat, 471, 608

\bibitem[{{Beers} \& {Christlieb}(2005)}]{Beers05}
{Beers}, T.~C. \& {Christlieb}, N. 2005, \araa, 43, 531

\bibitem[{{Beers} {et~al.}(2007){Beers}, {Flynn}, {Rossi}, {Sommer-Larsen},
  {Wilhelm}, {Marsteller}, {Lee}, {De Lee}, {Krugler}, {Deliyannis}, {Simmons},
  {Mills}, {Zickgraf}, {Holmberg}, {{\"O}nehag}, {Eriksson}, {Terndrup},
  {Salim}, {Andersen}, {Nordstr{\"o}m}, {Christlieb}, {Frebel}, \&
  {Rhee}}]{Beers07}
{Beers}, T.~C., {Flynn}, C., {Rossi}, S., {et~al.} 2007, \apjs, 168, 128

\bibitem[{{Bergemann}(2011)}]{Bergemann11}
{Bergemann}, M. 2011, \mnras, 413, 2184

\bibitem[{{Bergemann} {et~al.}(2017{\natexlab{a}}){Bergemann}, {Collet},
  {Amarsi}, {Kovalev}, {Ruchti}, \& {Magic}}]{Bergemann17a}
{Bergemann}, M., {Collet}, R., {Amarsi}, A.~M., {et~al.} 2017{\natexlab{a}},
  \apj, 847, 15

\bibitem[{{Bergemann} {et~al.}(2017{\natexlab{b}}){Bergemann}, {Collet},
  {Sch{\"o}nrich}, {Andrae}, {Kovalev}, {Ruchti}, {Hansen}, \&
  {Magic}}]{Bergemann17b}
{Bergemann}, M., {Collet}, R., {Sch{\"o}nrich}, R., {et~al.}
  2017{\natexlab{b}}, \apj, 847, 16

\bibitem[{{Bergemann} {et~al.}(2019){Bergemann}, {Gallagher}, {Eitner},
  {Bautista}, {Collet}, {Yakovleva}, {Mayriedl}, {Plez}, {Carlsson},
  {Leenaarts}, {Belyaev}, \& {Hansen}}]{Bergemann19}
{Bergemann}, M., {Gallagher}, A.~J., {Eitner}, P., {et~al.} 2019, \aap, 631,
  A80

\bibitem[{{Bergemann} \& {Gehren}(2008)}]{Bergemann08}
{Bergemann}, M. \& {Gehren}, T. 2008, \aap, 492, 823

\bibitem[{{Bergemann} {et~al.}(2012{\natexlab{a}}){Bergemann}, {Hansen},
  {Bautista}, \& {Ruchti}}]{Bergemann12}
{Bergemann}, M., {Hansen}, C.~J., {Bautista}, M., \& {Ruchti}, G.
  2012{\natexlab{a}}, \aap, 546, A90

\bibitem[{{Bergemann} {et~al.}(2013){Bergemann}, {Kudritzki}, {W{\"u}rl},
  {Plez}, {Davies}, \& {Gazak}}]{Bergemann13}
{Bergemann}, M., {Kudritzki}, R.-P., {W{\"u}rl}, M., {et~al.} 2013, \apj, 764,
  115

\bibitem[{{Bergemann} {et~al.}(2012{\natexlab{b}}){Bergemann}, {Lind},
  {Collet}, {Magic}, \& {Asplund}}]{Bergemann12a}
{Bergemann}, M., {Lind}, K., {Collet}, R., {Magic}, Z., \& {Asplund}, M.
  2012{\natexlab{b}}, \mnras, 427, 27

\bibitem[{{Bergemann} {et~al.}(2010){Bergemann}, {Pickering}, \&
  {Gehren}}]{Bergemann10}
{Bergemann}, M., {Pickering}, J.~C., \& {Gehren}, T. 2010, \mnras, 401, 1334

\bibitem[{{Bergemann} {et~al.}(2016){Bergemann}, {Serenelli}, {Sch{\"o}nrich},
  {Ruchti}, {Korn}, {Hekker}, {Kovalev}, {Mashonkina}, {Gilmore}, {Randich},
  {Asplund}, {Rix}, {Casey}, {Jofre}, {Pancino}, {Recio-Blanco}, {de Laverny},
  {Smiljanic}, {Tautvaisiene}, {Bayo}, {Lewis}, {Koposov}, {Hourihane},
  {Worley}, {Morbidelli}, {Franciosini}, {Sacco}, {Magrini}, {Damiani}, \&
  {Bestenlehner}}]{Bergemann16}
{Bergemann}, M., {Serenelli}, A., {Sch{\"o}nrich}, R., {et~al.} 2016, \aap,
  594, A120

\bibitem[{{Bernstein} {et~al.}(2003){Bernstein}, {Shectman}, {Gunnels},
  {Mochnacki}, \& {Athey}}]{Bernstein03}
{Bernstein}, R., {Shectman}, S.~A., {Gunnels}, S.~M., {Mochnacki}, S., \&
  {Athey}, A.~E. 2003, in Society of Photo-Optical Instrumentation Engineers
  (SPIE) Conference Series, Vol. 4841, \procspie, ed. M.~{Iye} \& A.~F.~M.
  {Moorwood}, 1694--1704

\bibitem[{{Bessell}(1983)}]{Bessell83}
{Bessell}, M.~S. 1983, \pasp, 95, 480

\bibitem[{{Bi{\'e}mont} {et~al.}(2000){Bi{\'e}mont}, {Garnir}, {Palmeri}, {Li},
  \& {Svanberg}}]{Biemont00}
{Bi{\'e}mont}, E., {Garnir}, H.~P., {Palmeri}, P., {Li}, Z.~S., \& {Svanberg},
  S. 2000, \mnras, 312, 116

\bibitem[{{Biemont} {et~al.}(1984){Biemont}, {Grevesse}, {Kwiatkovski}, \&
  {Zimmermann}}]{BGKZ}
{Biemont}, E., {Grevesse}, N., {Kwiatkovski}, M., \& {Zimmermann}, P. 1984,
  \aap, 131, 364

\bibitem[{{Biemont} {et~al.}(1993){Biemont}, {Quinet}, \& {Zeippen}}]{BQZ}
{Biemont}, E., {Quinet}, P., \& {Zeippen}, C.~J. 1993, \aaps, 102, 435

\bibitem[{{Bisterzo} {et~al.}(2012){Bisterzo}, {Gallino}, {Straniero},
  {Cristallo}, \& {K{\"a}ppeler}}]{Bisterzo12}
{Bisterzo}, S., {Gallino}, R., {Straniero}, O., {Cristallo}, S., \&
  {K{\"a}ppeler}, F. 2012, \mnras, 422, 849

\bibitem[{{Bisterzo} {et~al.}(2014){Bisterzo}, {Travaglio}, {Gallino},
  {Wiescher}, \& {K{\"a}ppeler}}]{Bisterzo14}
{Bisterzo}, S., {Travaglio}, C., {Gallino}, R., {Wiescher}, M., \&
  {K{\"a}ppeler}, F. 2014, \apj, 787, 10

\bibitem[{{Bizzarri} {et~al.}(1993){Bizzarri}, {Huber}, {Noels}, {Grevesse},
  {Bergeson}, {Tsekeris}, \& {Lawler}}]{BHN}
{Bizzarri}, A., {Huber}, M.~C.~E., {Noels}, A., {et~al.} 1993, \aap, 273, 707

\bibitem[{{Blackwell} \& {Shallis}(1977)}]{Blackwell77}
{Blackwell}, D.~E. \& {Shallis}, M.~J. 1977, \mnras, 180, 177

\bibitem[{{Burbidge} {et~al.}(1957){Burbidge}, {Burbidge}, {Fowler}, \&
  {Hoyle}}]{Burbidge57}
{Burbidge}, E.~M., {Burbidge}, G.~R., {Fowler}, W.~A., \& {Hoyle}, F. 1957,
  Reviews of Modern Physics, 29, 547

\bibitem[{{Burris} {et~al.}(2000){Burris}, {Pilachowski}, {Armandroff},
  {Sneden}, {Cowan}, \& {Roe}}]{Burris00}
{Burris}, D.~L., {Pilachowski}, C.~A., {Armandroff}, T.~E., {et~al.} 2000,
  \apj, 544, 302

\bibitem[{{Busso} {et~al.}(2001){Busso}, {Gallino}, {Lambert}, {Travaglio}, \&
  {Smith}}]{Busso01}
{Busso}, M., {Gallino}, R., {Lambert}, D.~L., {Travaglio}, C., \& {Smith},
  V.~V. 2001, \apj, 557, 802

\bibitem[{{Cameron}(1957)}]{Cameron57}
{Cameron}, A.~G.~W. 1957, \pasp, 69, 201

\bibitem[{{Cameron}(2003)}]{Cameron03}
{Cameron}, A.~G.~W. 2003, \apj, 587, 327

\bibitem[{{Campante} {et~al.}(2019){Campante}, {Corsaro}, {Lund}, {Mosser},
  {Serenelli}, {Veras}, {Adibekyan}, {Antia}, {Ball}, {Basu}, {Bedding},
  {Bossini}, {Davies}, {Delgado Mena}, {Garc{\'\i}a}, {Handberg}, {Hon},
  {Kane}, {Kawaler}, {Kuszlewicz}, {Lucas}, {Mathur}, {Nardetto}, {Nielsen},
  {Pinsonneault}, {Reffert}, {Silva Aguirre}, {Stassun}, {Stello}, {Stock},
  {Vrard}, {Y{\i}ld{\i}z}, {Chaplin}, {Huber}, {Bean}, {{\c{C}}elik Orhan},
  {Cunha}, {Christensen-Dalsgaard}, {Kjeldsen}, {Metcalfe}, {Miglio},
  {Monteiro}, {Nsamba}, {{\"O}rtel}, {Pereira}, {Sousa}, {Tsantaki}, \&
  {Turnbull}}]{Campante19}
{Campante}, T.~L., {Corsaro}, E., {Lund}, M.~N., {et~al.} 2019, \apj, 885, 31

\bibitem[{{Carney} {et~al.}(2008){Carney}, {Gray}, {Yong}, {Latham}, {Manset},
  {Zelman}, \& {Laird}}]{Carney08}
{Carney}, B.~W., {Gray}, D.~F., {Yong}, D., {et~al.} 2008, \aj, 135, 892

\bibitem[{{Carney} {et~al.}(2003){Carney}, {Latham}, {Stefanik}, {Laird}, \&
  {Morse}}]{Carney03}
{Carney}, B.~W., {Latham}, D.~W., {Stefanik}, R.~P., {Laird}, J.~B., \&
  {Morse}, J.~A. 2003, \aj, 125, 293

\bibitem[{{Castelli} \& {Kurucz}(2003)}]{Castelli03}
{Castelli}, F. \& {Kurucz}, R.~L. 2003, in IAU Symposium, Vol. 210, Modelling
  of Stellar Atmospheres, ed. N.~{Piskunov}, W.~W. {Weiss}, \& D.~F. {Gray},
  A20

\bibitem[{{Cayrel} {et~al.}(2001){Cayrel}, {Hill}, {Beers}, {Barbuy}, {Spite},
  {Spite}, {Plez}, {Andersen}, {Bonifacio}, {Fran{\c{c}}ois}, {Molaro},
  {Nordstr{\"o}m}, \& {Primas}}]{Cayrel01}
{Cayrel}, R., {Hill}, V., {Beers}, T.~C., {et~al.} 2001, \nat, 409, 691

\bibitem[{{Chan} \& {Heger}(2017)}]{Chan17}
{Chan}, C. \& {Heger}, A. 2017, in 14th International Symposium on Nuclei in
  the Cosmos (NIC2016), 020209

\bibitem[{{Chornock} {et~al.}(2017){Chornock}, {Berger}, {Kasen},
  {Cowperthwaite}, {Nicholl}, {Villar}, {Alexand er}, {Blanchard}, {Eftekhari},
  {Fong}, {Margutti}, {Williams}, {Annis}, {Brout}, {Brown}, {Chen}, {Drout},
  {Farr}, {Foley}, {Frieman}, {Fryer}, {Herner}, {Holz}, {Kessler}, {Matheson},
  {Metzger}, {Quataert}, {Rest}, {Sako}, {Scolnic}, {Smith}, \&
  {Soares-Santos}}]{Chornock17}
{Chornock}, R., {Berger}, E., {Kasen}, D., {et~al.} 2017, \apjl, 848, L19

\bibitem[{{Corliss} \& {Bozman}(1962)}]{CB}
{Corliss}, C.~H. \& {Bozman}, W.~R. 1962, {Experimental transition
  probabilities for spectral lines of seventy elements; derived from the NBS
  Tables of spectral-line intensities}

\bibitem[{{Corsaro} \& {De Ridder}(2014)}]{Corsaro14}
{Corsaro}, E. \& {De Ridder}, J. 2014, \aap, 571, A71

\bibitem[{{Corsaro} {et~al.}(2017){Corsaro}, {Mathur}, {Garc{\'\i}a}, {Gaulme},
  {Pinsonneault}, {Stassun}, {Stello}, {Tayar}, {Trampedach}, {Jiang},
  {Nitschelm}, \& {Salabert}}]{Corsaro17}
{Corsaro}, E., {Mathur}, S., {Garc{\'\i}a}, R.~A., {et~al.} 2017, \aap, 605, A3

\bibitem[{{C{\^o}t{\'e}} {et~al.}(2019){C{\^o}t{\'e}}, {Eichler}, {Arcones},
  {Hansen}, {Simonetti}, {Frebel}, {Fryer}, {Pignatari}, {Reichert},
  {Belczynski}, \& {Matteucci}}]{Cote19}
{C{\^o}t{\'e}}, B., {Eichler}, M., {Arcones}, A., {et~al.} 2019, \apj, 875, 106

\bibitem[{{Cowan} {et~al.}(1999){Cowan}, {Pfeiffer}, {Kratz}, {Thielemann},
  {Sneden}, {Burles}, {Tytler}, \& {Beers}}]{Cowan99}
{Cowan}, J.~J., {Pfeiffer}, B., {Kratz}, K.~L., {et~al.} 1999, \apj, 521, 194

\bibitem[{{Cowan} {et~al.}(2005){Cowan}, {Sneden}, {Beers}, {Lawler},
  {Simmerer}, {Truran}, {Primas}, {Collier}, \& {Burles}}]{Cowan05}
{Cowan}, J.~J., {Sneden}, C., {Beers}, T.~C., {et~al.} 2005, \apj, 627, 238

\bibitem[{{Cowan} {et~al.}(2002){Cowan}, {Sneden}, {Burles}, {Ivans}, {Beers},
  {Truran}, {Lawler}, {Primas}, {Fuller}, {Pfeiffer}, \& {Kratz}}]{Cowan02}
{Cowan}, J.~J., {Sneden}, C., {Burles}, S., {et~al.} 2002, \apj, 572, 861

\bibitem[{{Cowley} \& {Corliss}(1983)}]{CC}
{Cowley}, C.~R. \& {Corliss}, C.~H. 1983, \mnras, 203, 651

\bibitem[{{Creevey} {et~al.}(2019){Creevey}, {Grundahl}, {Th{\'e}venin},
  {Corsaro}, {Pall{\'e}}, {Salabert}, {Pichon}, {Collet}, {Bigot}, {Antoci}, \&
  {Andersen}}]{Creevey19}
{Creevey}, O., {Grundahl}, F., {Th{\'e}venin}, F., {et~al.} 2019, \aap, 625,
  A33

\bibitem[{{Cristallo} {et~al.}(2011){Cristallo}, {Piersanti}, {Straniero},
  {Gallino}, {Dom{\'\i}nguez}, {Abia}, {Di Rico}, {Quintini}, \&
  {Bisterzo}}]{Cristallo11}
{Cristallo}, S., {Piersanti}, L., {Straniero}, O., {et~al.} 2011, \apjs, 197,
  17

\bibitem[{{Cristallo} {et~al.}(2015){Cristallo}, {Straniero}, {Piersanti}, \&
  {Gobrecht}}]{Cristallo15}
{Cristallo}, S., {Straniero}, O., {Piersanti}, L., \& {Gobrecht}, D. 2015,
  \apjs, 219, 40

\bibitem[{{Dekker} {et~al.}(2000){Dekker}, {D'Odorico}, {Kaufer}, {Delabre}, \&
  {Kotzlowski}}]{Dekker00}
{Dekker}, H., {D'Odorico}, S., {Kaufer}, A., {Delabre}, B., \& {Kotzlowski}, H.
  2000, in \procspie, Vol. 4008, Optical and IR Telescope Instrumentation and
  Detectors, ed. M.~{Iye} \& A.~F. {Moorwood}, 534--545

\bibitem[{{Den Hartog} {et~al.}(2003){Den Hartog}, {Lawler}, {Sneden}, \&
  {Cowan}}]{HLSC}
{Den Hartog}, E.~A., {Lawler}, J.~E., {Sneden}, C., \& {Cowan}, J.~J. 2003,
  \apjs, 148, 543

\bibitem[{{Den Hartog} {et~al.}(2006){Den Hartog}, {Lawler}, {Sneden}, \&
  {Cowan}}]{DLSC}
{Den Hartog}, E.~A., {Lawler}, J.~E., {Sneden}, C., \& {Cowan}, J.~J. 2006,
  \apjs, 167, 292

\bibitem[{{Den Hartog} {et~al.}(2011){Den Hartog}, {Lawler}, {Sobeck},
  {Sneden}, \& {Cowan}}]{DenHartog11}
{Den Hartog}, E.~A., {Lawler}, J.~E., {Sobeck}, J.~S., {Sneden}, C., \&
  {Cowan}, J.~J. 2011, \apjs, 194, 35

\bibitem[{{den Hartogh} {et~al.}(2019){den Hartogh}, {Hirschi}, {Lugaro},
  {Doherty}, {Battino}, {Herwig}, {Pignatari}, \& {Eggenberger}}]{denHartogh19}
{den Hartogh}, J.~W., {Hirschi}, R., {Lugaro}, M., {et~al.} 2019, \aap, 629,
  A123

\bibitem[{{Dotter} {et~al.}(2008){Dotter}, {Chaboyer}, {Jevremovi{\'c}},
  {Kostov}, {Baron}, \& {Ferguson}}]{Dotter08}
{Dotter}, A., {Chaboyer}, B., {Jevremovi{\'c}}, D., {et~al.} 2008, \apjs, 178,
  89

\bibitem[{{Ducati}(2002)}]{Ducati02}
{Ducati}, J.~R. 2002, VizieR Online Data Catalog, 2237

\bibitem[{{ESA}(1997)}]{HIPPARCOS97}
{ESA}, ed. 1997, ESA Special Publication, Vol. 1200, {The HIPPARCOS and TYCHO
  catalogues. Astrometric and photometric star catalogues derived from the ESA
  HIPPARCOS Space Astrometry Mission}

\bibitem[{{Foreman-Mackey} {et~al.}(2013){Foreman-Mackey}, {Hogg}, {Lang}, \&
  {Goodman}}]{ForemanMackey13}
{Foreman-Mackey}, D., {Hogg}, D.~W., {Lang}, D., \& {Goodman}, J. 2013, \pasp,
  125, 306

\bibitem[{{Fraser} {et~al.}(2017){Fraser}, {Casey}, {Gilmore}, {Heger}, \&
  {Chan}}]{Fraser17}
{Fraser}, M., {Casey}, A.~R., {Gilmore}, G., {Heger}, A., \& {Chan}, C. 2017,
  \mnras, 468, 418

\bibitem[{{Freytag} {et~al.}(2012){Freytag}, {Steffen}, {Ludwig},
  {Wedemeyer-B{\"o}hm}, {Schaffenberger}, \& {Steiner}}]{Freytag12}
{Freytag}, B., {Steffen}, M., {Ludwig}, H.~G., {et~al.} 2012, Journal of
  Computational Physics, 231, 919

\bibitem[{{Fuhr} {et~al.}(1988){Fuhr}, {Martin}, \& {Wiese}}]{Fuhr88}
{Fuhr}, J.~R., {Martin}, G.~A., \& {Wiese}, W.~L. 1988, Journal of Physical and
  Chemical Reference Data, 17

\bibitem[{{Fulbright} \& {Johnson}(2003)}]{Fulbright03}
{Fulbright}, J.~P. \& {Johnson}, J.~A. 2003, \apj, 595, 1154

\bibitem[{{Gaia Collaboration} {et~al.}(2018){Gaia Collaboration}, {Brown},
  {Vallenari}, {Prusti}, {de Bruijne}, {Babusiaux}, {Bailer-Jones}, {Biermann},
  {Evans}, {Eyer}, \& et~al.}]{Gaia18}
{Gaia Collaboration}, {Brown}, A.~G.~A., {Vallenari}, A., {et~al.} 2018, \aap,
  616, A1

\bibitem[{{Gallino} {et~al.}(1998){Gallino}, {Arlandini}, {Busso}, {Lugaro},
  {Travaglio}, {Straniero}, {Chieffi}, \& {Limongi}}]{Gallino98}
{Gallino}, R., {Arlandini}, C., {Busso}, M., {et~al.} 1998, \apj, 497, 388

\bibitem[{{Garc{\'\i}a-Hern{\'a}ndez}
  {et~al.}(2009){Garc{\'\i}a-Hern{\'a}ndez}, {Manchado}, {Lambert}, {Plez},
  {Garc{\'\i}a-Lario}, {D'Antona}, {Lugaro}, {Karakas}, \& {van
  Raai}}]{Garcia-Hernandez09}
{Garc{\'\i}a-Hern{\'a}ndez}, D.~A., {Manchado}, A., {Lambert}, D.~L., {et~al.}
  2009, \apjl, 705, L31

\bibitem[{{Garz}(1973)}]{GARZ}
{Garz}, T. 1973, \aap, 26, 471

\bibitem[{{Gonz{\'a}lez Hern{\'a}ndez} \& {Bonifacio}(2009)}]{Gonzalez09}
{Gonz{\'a}lez Hern{\'a}ndez}, J.~I. \& {Bonifacio}, P. 2009, \aap, 497, 497

\bibitem[{{Gratton} {et~al.}(2000){Gratton}, {Sneden}, {Carretta}, \&
  {Bragaglia}}]{Gratton00}
{Gratton}, R.~G., {Sneden}, C., {Carretta}, E., \& {Bragaglia}, A. 2000, \aap,
  354, 169

\bibitem[{{Gull} {et~al.}(2018){Gull}, {Frebel}, {Cain}, {Placco}, {Ji},
  {Abate}, {Ezzeddine}, {Karakas}, {Hansen}, {Sakari}, {Holmbeck}, {Santucci},
  {Casey}, \& {Beers}}]{Gull18}
{Gull}, M., {Frebel}, A., {Cain}, M.~G., {et~al.} 2018, \apj, 862, 174

\bibitem[{{Hampel} {et~al.}(2019){Hampel}, {Karakas}, {Stancliffe}, {Meyer}, \&
  {Lugaro}}]{Hampel19}
{Hampel}, M., {Karakas}, A.~I., {Stancliffe}, R.~J., {Meyer}, B.~S., \&
  {Lugaro}, M. 2019, \apj, 887, 11

\bibitem[{{Hampel} {et~al.}(2016){Hampel}, {Stancliffe}, {Lugaro}, \&
  {Meyer}}]{Hampel16}
{Hampel}, M., {Stancliffe}, R.~J., {Lugaro}, M., \& {Meyer}, B.~S. 2016, \apj,
  831, 171

\bibitem[{{Han} {et~al.}(2018){Han}, {Zhang}, {Yang}, {Niu}, \&
  {Zhang}}]{Han18}
{Han}, W., {Zhang}, L., {Yang}, G., {Niu}, P., \& {Zhang}, B. 2018, \apj, 856,
  58

\bibitem[{{Hanke} {et~al.}(2018){Hanke}, {Hansen}, {Koch}, \&
  {Grebel}}]{Hanke18}
{Hanke}, M., {Hansen}, C.~J., {Koch}, A., \& {Grebel}, E.~K. 2018, \aap, 619,
  A134

\bibitem[{{Hanke} {et~al.}(2017){Hanke}, {Koch}, {Hansen}, \&
  {McWilliam}}]{Hanke17}
{Hanke}, M., {Koch}, A., {Hansen}, C.~J., \& {McWilliam}, A. 2017, \aap, 599,
  A97

\bibitem[{{Hannaford} {et~al.}(1982){Hannaford}, {Lowe}, {Grevesse}, {Biemont},
  \& {Whaling}}]{HLGBW}
{Hannaford}, P., {Lowe}, R.~M., {Grevesse}, N., {Biemont}, E., \& {Whaling}, W.
  1982, \apj, 261, 736

\bibitem[{{Hansen} {et~al.}(2013){Hansen}, {Bergemann}, {Cescutti},
  {Fran{\c{c}}ois}, {Arcones}, {Karakas}, {Lind}, \& {Chiappini}}]{Hansen13}
{Hansen}, C.~J., {Bergemann}, M., {Cescutti}, G., {et~al.} 2013, \aap, 551, A57

\bibitem[{{Hansen} {et~al.}(2018{\natexlab{a}}){Hansen}, {El-Souri}, {Monaco},
  {Villanova}, {Bonifacio}, {Caffau}, \& {Sbordone}}]{Hansen18}
{Hansen}, C.~J., {El-Souri}, M., {Monaco}, L., {et~al.} 2018{\natexlab{a}},
  \apj, 855, 83

\bibitem[{{Hansen} {et~al.}(2014){Hansen}, {Montes}, \& {Arcones}}]{Hansen14a}
{Hansen}, C.~J., {Montes}, F., \& {Arcones}, A. 2014, \apj, 797, 123

\bibitem[{{Hansen} {et~al.}(2012){Hansen}, {Primas}, {Hartman}, {Kratz},
  {Wanajo}, {Leibundgut}, {Farouqi}, {Hallmann}, {Christlieb}, \&
  {Nilsson}}]{Hansen12}
{Hansen}, C.~J., {Primas}, F., {Hartman}, H., {et~al.} 2012, \aap, 545, A31

\bibitem[{{Hansen} {et~al.}(2015){Hansen}, {Andersen}, {Nordstr{\"o}m},
  {Beers}, {Yoon}, \& {Buchhave}}]{Hansen15}
{Hansen}, T.~T., {Andersen}, J., {Nordstr{\"o}m}, B., {et~al.} 2015, \aap, 583,
  A49

\bibitem[{{Hansen} {et~al.}(2018{\natexlab{b}}){Hansen}, {Holmbeck}, {Beers},
  {Placco}, {Roederer}, {Frebel}, {Sakari}, {Simon}, \& {Thompson}}]{Hansen18a}
{Hansen}, T.~T., {Holmbeck}, E.~M., {Beers}, T.~C., {et~al.}
  2018{\natexlab{b}}, \apj, 858, 92

\bibitem[{{Hauck} \& {Mermilliod}(1998)}]{Hauck98}
{Hauck}, B. \& {Mermilliod}, M. 1998, \aaps, 129, 431

\bibitem[{{Hawkins} {et~al.}(2016){Hawkins}, {Jofr{\'e}}, {Heiter}, {Soubiran},
  {Blanco-Cuaresma}, {Casagrande}, {Gilmore}, {Lind}, {Magrini}, {Masseron},
  {Pancino}, {Randich}, \& {Worley}}]{Hawkins16}
{Hawkins}, K., {Jofr{\'e}}, P., {Heiter}, U., {et~al.} 2016, \aap, 592, A70

\bibitem[{{Heger} \& {Woosley}(2010)}]{Heger10}
{Heger}, A. \& {Woosley}, S.~E. 2010, \apj, 724, 341

\bibitem[{{Heiter} {et~al.}(2015){Heiter}, {Jofr{\'e}}, {Gustafsson}, {Korn},
  {Soubiran}, \& {Th{\'e}venin}}]{Heiter15}
{Heiter}, U., {Jofr{\'e}}, P., {Gustafsson}, B., {et~al.} 2015, \aap, 582, A49

\bibitem[{{Hobbs} {et~al.}(1999){Hobbs}, {Thorburn}, \& {Rebull}}]{Hobbs99}
{Hobbs}, L.~M., {Thorburn}, J.~A., \& {Rebull}, L.~M. 1999, \apj, 523, 797

\bibitem[{{H{\o}g} {et~al.}(2000){H{\o}g}, {Fabricius}, {Makarov}, {Urban},
  {Corbin}, {Wycoff}, {Bastian}, {Schwekendiek}, \& {Wicenec}}]{Hoeg00}
{H{\o}g}, E., {Fabricius}, C., {Makarov}, V.~V., {et~al.} 2000, \aap, 355, L27

\bibitem[{{Ivarsson} {et~al.}(2003){Ivarsson}, {Andersen}, {Nordstr{\"o}m},
  {Dai}, {Johansson}, {Lundberg}, {Nilsson}, {Hill}, {Lundqvist}, \&
  {Wyart}}]{IAN}
{Ivarsson}, S., {Andersen}, J., {Nordstr{\"o}m}, B., {et~al.} 2003, \aap, 409,
  1141

\bibitem[{{Jayasinghe} {et~al.}(2019){Jayasinghe}, {Stanek}, {Kochanek},
  {Shappee}, {Holoien}, {Thompson}, {Prieto}, {Dong}, {Pawlak}, {Pejcha},
  {Shields}, {Pojmanski}, {Otero}, {Hurst}, {Britt}, \& {Will}}]{Jayasinghe19}
{Jayasinghe}, T., {Stanek}, K.~Z., {Kochanek}, C.~S., {et~al.} 2019, \mnras,
  485, 961

\bibitem[{{Ji} {et~al.}(2019){Ji}, {Drout}, \& {Hansen}}]{Ji19}
{Ji}, A.~P., {Drout}, M.~R., \& {Hansen}, T.~T. 2019, \apj, 882, 40

\bibitem[{{Jofr{\'e}} {et~al.}(2014){Jofr{\'e}}, {Heiter}, {Soubiran},
  {Blanco-Cuaresma}, {Worley}, {Pancino}, {Cantat-Gaudin}, {Magrini},
  {Bergemann}, {Gonz{\'a}lez Hern{\'a}ndez}, {Hill}, {Lardo}, {de Laverny},
  {Lind}, {Masseron}, {Montes}, {Mucciarelli}, {Nordlander}, {Recio Blanco},
  {Sobeck}, {Sordo}, {Sousa}, {Tabernero}, {Vallenari}, \& {Van Eck}}]{Jofre14}
{Jofr{\'e}}, P., {Heiter}, U., {Soubiran}, C., {et~al.} 2014, \aap, 564, A133

\bibitem[{{Jones} {et~al.}(2013){Jones}, {Noll}, {Kausch}, {Szyszka}, \&
  {Kimeswenger}}]{Jones13}
{Jones}, A., {Noll}, S., {Kausch}, W., {Szyszka}, C., \& {Kimeswenger}, S.
  2013, \aap, 560, A91

\bibitem[{{Karakas} \& {Lattanzio}(2014)}]{Karakas14}
{Karakas}, A.~I. \& {Lattanzio}, J.~C. 2014, \pasa, 31, e030

\bibitem[{{Kelson}(2003)}]{Kelson03}
{Kelson}, D.~D. 2003, \pasp, 115, 688

\bibitem[{{Kjeldsen} \& {Bedding}(1995)}]{Kjeldsen95}
{Kjeldsen}, H. \& {Bedding}, T.~R. 1995, \aap, 293, 87

\bibitem[{{Kobayashi} {et~al.}(2019){Kobayashi}, {Leung}, \&
  {Nomoto}}]{Kobayashi19}
{Kobayashi}, C., {Leung}, S.-C., \& {Nomoto}, K. 2019, arXiv e-prints,
  arXiv:1906.09980

\bibitem[{{Koch} {et~al.}(2016){Koch}, {McWilliam}, {Preston}, \&
  {Thompson}}]{Koch16}
{Koch}, A., {McWilliam}, A., {Preston}, G.~W., \& {Thompson}, I.~B. 2016, \aap,
  587, A124

\bibitem[{{Koch} {et~al.}(2019){Koch}, {Reichert}, {Hansen}, {Hampel},
  {Stancliffe}, {Karakas}, \& {Arcones}}]{Koch19}
{Koch}, A., {Reichert}, M., {Hansen}, C.~J., {et~al.} 2019, \aap, 622, A159

\bibitem[{{Korotin} {et~al.}(2017){Korotin}, {Andrievsky}, {Caffau}, \&
  {Bonifacio}}]{Korotin17}
{Korotin}, S., {Andrievsky}, S., {Caffau}, E., \& {Bonifacio}, P. 2017, in
  Astronomical Society of the Pacific Conference Series, Vol. 510, Stars: From
  Collapse to Collapse, ed. Y.~Y. {Balega}, D.~O. {Kudryavtsev}, I.~I.
  {Romanyuk}, \& I.~A. {Yakunin}, 141

\bibitem[{{Korotin}(2008)}]{Korotin08}
{Korotin}, S.~A. 2008, Odessa Astronomical Publications, 21, 42

\bibitem[{{Korotin}(2009)}]{Korotin09}
{Korotin}, S.~A. 2009, Astronomy Reports, 53, 651

\bibitem[{{Korotin} {et~al.}(2018){Korotin}, {Andrievsky}, \&
  {Zhukova}}]{Korotin18}
{Korotin}, S.~A., {Andrievsky}, S.~M., \& {Zhukova}, A.~V. 2018, \mnras, 480,
  965

\bibitem[{Kramida {et~al.}(2018)Kramida, {Yu.~Ralchenko}, Reader, \& {and NIST
  ASD Team}}]{NIST_ASD}
Kramida, A., {Yu.~Ralchenko}, Reader, J., \& {and NIST ASD Team}. 2018, {NIST
  Atomic Spectra Database (ver. 5.6.1), [Online]. Available:
  {\tt{https://physics.nist.gov/asd}} [2019, August 19]. National Institute of
  Standards and Technology, Gaithersburg, MD.}

\bibitem[{{Kratz} {et~al.}(2007){Kratz}, {Farouqi}, {Pfeiffer}, {Truran},
  {Sneden}, \& {Cowan}}]{Kratz07}
{Kratz}, K.-L., {Farouqi}, K., {Pfeiffer}, B., {et~al.} 2007, \apj, 662, 39

\bibitem[{{Kurucz} \& {Bell}(1995)}]{Kurucz95}
{Kurucz}, R.~L. \& {Bell}, B. 1995, {Atomic line list}

\bibitem[{{Kwiatkowski} {et~al.}(1982){Kwiatkowski}, {Zimmermann}, {Biemont},
  \& {Grevesse}}]{KZBa}
{Kwiatkowski}, M., {Zimmermann}, P., {Biemont}, E., \& {Grevesse}, N. 1982,
  \aap, 112, 337

\bibitem[{{Langer} {et~al.}(1999){Langer}, {Heger}, {Wellstein}, \&
  {Herwig}}]{Langer99}
{Langer}, N., {Heger}, A., {Wellstein}, S., \& {Herwig}, F. 1999, \aap, 346,
  L37

\bibitem[{{Lattimer} \& {Schramm}(1974)}]{Lattimer74}
{Lattimer}, J.~M. \& {Schramm}, D.~N. 1974, \apjl, 192, L145

\bibitem[{{Lawler} {et~al.}(2001{\natexlab{a}}){Lawler}, {Bonvallet}, \&
  {Sneden}}]{Lawler01a}
{Lawler}, J.~E., {Bonvallet}, G., \& {Sneden}, C. 2001{\natexlab{a}}, \apj,
  556, 452

\bibitem[{{Lawler} \& {Dakin}(1989)}]{LD}
{Lawler}, J.~E. \& {Dakin}, J.~T. 1989, Journal of the Optical Society of
  America B Optical Physics, 6, 1457

\bibitem[{{Lawler} {et~al.}(2007){Lawler}, {den Hartog}, {Labby}, {Sneden},
  {Cowan}, \& {Ivans}}]{LDLS}
{Lawler}, J.~E., {den Hartog}, E.~A., {Labby}, Z.~E., {et~al.} 2007, \apjs,
  169, 120

\bibitem[{{Lawler} {et~al.}(2006){Lawler}, {Den Hartog}, {Sneden}, \&
  {Cowan}}]{LD-HS}
{Lawler}, J.~E., {Den Hartog}, E.~A., {Sneden}, C., \& {Cowan}, J.~J. 2006,
  \apjs, 162, 227

\bibitem[{{Lawler} {et~al.}(2013){Lawler}, {Guzman}, {Wood}, {Sneden}, \&
  {Cowan}}]{LGWSC}
{Lawler}, J.~E., {Guzman}, A., {Wood}, M.~P., {Sneden}, C., \& {Cowan}, J.~J.
  2013, \apjs, 205, 11

\bibitem[{{Lawler} {et~al.}(2004){Lawler}, {Sneden}, \& {Cowan}}]{Lawler04}
{Lawler}, J.~E., {Sneden}, C., \& {Cowan}, J.~J. 2004, \apj, 604, 850

\bibitem[{{Lawler} {et~al.}(2009){Lawler}, {Sneden}, {Cowan}, {Ivans}, \& {Den
  Hartog}}]{LSCI}
{Lawler}, J.~E., {Sneden}, C., {Cowan}, J.~J., {Ivans}, I.~I., \& {Den Hartog},
  E.~A. 2009, \apjs, 182, 51

\bibitem[{{Lawler} {et~al.}(2008){Lawler}, {Sneden}, {Cowan}, {Wyart}, {Ivans},
  {Sobeck}, {Stockett}, \& {Den Hartog}}]{LSCW}
{Lawler}, J.~E., {Sneden}, C., {Cowan}, J.~J., {et~al.} 2008, \apjs, 178, 71

\bibitem[{{Lawler} {et~al.}(2001{\natexlab{b}}){Lawler}, {Wickliffe}, {Cowley},
  \& {Sneden}}]{LWCS}
{Lawler}, J.~E., {Wickliffe}, M.~E., {Cowley}, C.~R., \& {Sneden}, C.
  2001{\natexlab{b}}, \apjs, 137, 341

\bibitem[{{Lawler} {et~al.}(2001{\natexlab{c}}){Lawler}, {Wickliffe}, {den
  Hartog}, \& {Sneden}}]{Lawler01b}
{Lawler}, J.~E., {Wickliffe}, M.~E., {den Hartog}, E.~A., \& {Sneden}, C.
  2001{\natexlab{c}}, \apj, 563, 1075

\bibitem[{{Lawler} {et~al.}(2014){Lawler}, {Wood}, {Den Hartog}, {Feigenson},
  {Sneden}, \& {Cowan}}]{Lawler14}
{Lawler}, J.~E., {Wood}, M.~P., {Den Hartog}, E.~A., {et~al.} 2014, \apjs, 215,
  20

\bibitem[{{Lawler} {et~al.}(2001{\natexlab{d}}){Lawler}, {Wyart}, \&
  {Blaise}}]{Lawler01c}
{Lawler}, J.~E., {Wyart}, J.-F., \& {Blaise}, J. 2001{\natexlab{d}}, \apjs,
  137, 351

\bibitem[{{Li} {et~al.}(2013){Li}, {Shen}, {Liang}, {Cui}, \& {Zhang}}]{Li13}
{Li}, H., {Shen}, X., {Liang}, S., {Cui}, W., \& {Zhang}, B. 2013, \pasp, 125,
  143

\bibitem[{{Lightkurve Collaboration} {et~al.}(2018){Lightkurve Collaboration},
  {Cardoso}, {Hedges}, {Gully-Santiago}, {Saunders}, {Cody}, {Barclay}, {Hall},
  {Sagear}, {Turtelboom}, {Zhang}, {Tzanidakis}, {Mighell}, {Coughlin}, {Bell},
  {Berta-Thompson}, {Williams}, {Dotson}, \& {Barentsen}}]{lightkurve}
{Lightkurve Collaboration}, {Cardoso}, J.~V.~d.~M., {Hedges}, C., {et~al.}
  2018, {Lightkurve: Kepler and TESS time series analysis in Python},
  Astrophysics Source Code Library

\bibitem[{{Lind} {et~al.}(2011){Lind}, {Asplund}, {Barklem}, \&
  {Belyaev}}]{Lind11}
{Lind}, K., {Asplund}, M., {Barklem}, P.~S., \& {Belyaev}, A.~K. 2011, \aap,
  528, A103

\bibitem[{{Lind} {et~al.}(2012){Lind}, {Bergemann}, \& {Asplund}}]{Lind12}
{Lind}, K., {Bergemann}, M., \& {Asplund}, M. 2012, \mnras, 427, 50

\bibitem[{{Lindegren} {et~al.}(2018){Lindegren}, {Hern{\'a}ndez}, {Bombrun},
  {Klioner}, {Bastian}, {Ramos-Lerate}, {de Torres}, {Steidelm{\"u}ller},
  {Stephenson}, {Hobbs}, {Lammers}, {Biermann}, {Geyer}, {Hilger}, {Michalik},
  {Stampa}, {McMillan}, {Casta{\~n}eda}, {Clotet}, {Comoretto}, {Davidson},
  {Fabricius}, {Gracia}, {Hambly}, {Hutton}, {Mora}, {Portell}, {van Leeuwen},
  {Abbas}, {Abreu}, {Altmann}, {Andrei}, {Anglada}, {Balaguer-N{\'u}{\~n}ez},
  {Barache}, {Becciani}, {Bertone}, {Bianchi}, {Bouquillon}, {Bourda},
  {Br{\"u}semeister}, {Bucciarelli}, {Busonero}, {Buzzi}, {Cancelliere},
  {Carlucci}, {Charlot}, {Cheek}, {Crosta}, {Crowley}, {de Bruijne}, {de
  Felice}, {Drimmel}, {Esquej}, {Fienga}, {Fraile}, {Gai}, {Garralda},
  {Gonz{\'a}lez-Vidal}, {Guerra}, {Hauser}, {Hofmann}, {Holl}, {Jordan},
  {Lattanzi}, {Lenhardt}, {Liao}, {Licata}, {Lister}, {L{\"o}ffler},
  {Marchant}, {Martin-Fleitas}, {Messineo}, {Mignard}, {Morbidelli}, {Poggio},
  {Riva}, {Rowell}, {Salguero}, {Sarasso}, {Sciacca}, {Siddiqui}, {Smart},
  {Spagna}, {Steele}, {Taris}, {Torra}, {van Elteren}, {van Reeven}, \&
  {Vecchiato}}]{Lindegren18}
{Lindegren}, L., {Hern{\'a}ndez}, J., {Bombrun}, A., {et~al.} 2018, \aap, 616,
  A2

\bibitem[{{Ljung} {et~al.}(2006){Ljung}, {Nilsson}, {Asplund}, \&
  {Johansson}}]{LNAJ}
{Ljung}, G., {Nilsson}, H., {Asplund}, M., \& {Johansson}, S. 2006, \aap, 456,
  1181

\bibitem[{{Ludwig} {et~al.}(2010){Ludwig}, {Caffau}, {Steffen}, {Bonifacio}, \&
  {Sbordone}}]{Ludwig10}
{Ludwig}, H.~G., {Caffau}, E., {Steffen}, M., {Bonifacio}, P., \& {Sbordone},
  L. 2010, \aap, 509, A84

\bibitem[{{Lugaro} {et~al.}(2012){Lugaro}, {Karakas}, {Stancliffe}, \&
  {Rijs}}]{Lugaro12}
{Lugaro}, M., {Karakas}, A.~I., {Stancliffe}, R.~J., \& {Rijs}, C. 2012, \apj,
  747, 2

\bibitem[{{Mamajek} {et~al.}(2015){Mamajek}, {Torres}, {Prsa}, {Harmanec},
  {Asplund}, {Bennett}, {Capitaine}, {Christensen-Dalsgaard}, {Depagne},
  {Folkner}, {Haberreiter}, {Hekker}, {Hilton}, {Kostov}, {Kurtz}, {Laskar},
  {Mason}, {Milone}, {Montgomery}, {Richards}, {Schou}, \&
  {Stewart}}]{Mamajek15}
{Mamajek}, E.~E., {Torres}, G., {Prsa}, A., {et~al.} 2015, ArXiv e-prints
  [\eprint[arXiv]{1510.06262}]

\bibitem[{{Martin} {et~al.}(1988){Martin}, {Fuhr}, \& {Wiese}}]{MFW}
{Martin}, G., {Fuhr}, J., \& {Wiese}, W. 1988, J. Phys. Chem. Ref. Data Suppl.,
  17

\bibitem[{{Mashonkina} \& {Belyaev}(2019)}]{Mashonkina19}
{Mashonkina}, L.~I. \& {Belyaev}, A.~K. 2019, Astronomy Letters, 45, 341

\bibitem[{{Masseron} {et~al.}(2014){Masseron}, {Plez}, {Van Eck}, {Colin},
  {Daoutidis}, {Godefroid}, {Coheur}, {Bernath}, {Jorissen}, \&
  {Christlieb}}]{Masseron14}
{Masseron}, T., {Plez}, B., {Van Eck}, S., {et~al.} 2014, \aap, 571, A47

\bibitem[{{Mayor} {et~al.}(2003){Mayor}, {Pepe}, {Queloz}, {Bouchy},
  {Rupprecht}, {Lo Curto}, {Avila}, {Benz}, {Bertaux}, {Bonfils}, {Dall},
  {Dekker}, {Delabre}, {Eckert}, {Fleury}, {Gilliotte}, {Gojak}, {Guzman},
  {Kohler}, {Lizon}, {Longinotti}, {Lovis}, {Megevand}, {Pasquini}, {Reyes},
  {Sivan}, {Sosnowska}, {Soto}, {Udry}, {van Kesteren}, {Weber}, \&
  {Weilenmann}}]{Mayor03}
{Mayor}, M., {Pepe}, F., {Queloz}, D., {et~al.} 2003, The Messenger, 114, 20

\bibitem[{{McWilliam}(1997)}]{McWilliam97}
{McWilliam}, A. 1997, \araa, 35, 503

\bibitem[{{McWilliam}(1998)}]{McWilliam98}
{McWilliam}, A. 1998, \aj, 115, 1640

\bibitem[{{Meggers} {et~al.}(1975){Meggers}, {Corliss}, \& {Scribner}}]{MC}
{Meggers}, W.~F., {Corliss}, C.~H., \& {Scribner}, B.~F. 1975, {Tables of
  spectral-line intensities. Part I, II\_- arranged by elements.}

\bibitem[{{Meyer}(1994)}]{Meyer94}
{Meyer}, B.~S. 1994, \araa, 32, 153

\bibitem[{{Migdalek}(1978)}]{JMG}
{Migdalek}, J. 1978, \jqsrt, 20, 81

\bibitem[{{Milone} {et~al.}(2018){Milone}, {Marino}, {Renzini}, {D'Antona},
  {Anderson}, {Barbuy}, {Bedin}, {Bellini}, {Brown}, {Cassisi}, {Cordoni},
  {Lagioia}, {Nardiello}, {Ortolani}, {Piotto}, {Sarajedini}, {Tailo}, {van der
  Marel}, \& {Vesperini}}]{Milone18}
{Milone}, A.~P., {Marino}, A.~F., {Renzini}, A., {et~al.} 2018, \mnras, 481,
  5098

\bibitem[{{Mishenina} {et~al.}(2015){Mishenina}, {Gorbaneva}, {Pignatari},
  {Thielemann}, \& {Korotin}}]{Mishenina15}
{Mishenina}, T., {Gorbaneva}, T., {Pignatari}, M., {Thielemann}, F.-K., \&
  {Korotin}, S.~A. 2015, \mnras, 454, 1585

\bibitem[{{Mosser} {et~al.}(2012{\natexlab{a}}){Mosser}, {Goupil}, {Belkacem},
  {Marques}, {Beck}, {Bloemen}, {De Ridder}, {Barban}, {Deheuvels}, {Elsworth},
  {Hekker}, {Kallinger}, {Ouazzani}, {Pinsonneault}, {Samadi}, {Stello},
  {Garc{\'\i}a}, {Klaus}, {Li}, {Mathur}, \& {Morris}}]{Mosser12}
{Mosser}, B., {Goupil}, M.~J., {Belkacem}, K., {et~al.} 2012{\natexlab{a}},
  \aap, 548, A10

\bibitem[{{Mosser} {et~al.}(2012{\natexlab{b}}){Mosser}, {Goupil}, {Belkacem},
  {Michel}, {Stello}, {Marques}, {Elsworth}, {Barban}, {Beck}, {Bedding}, {De
  Ridder}, {Garc{\'\i}a}, {Hekker}, {Kallinger}, {Samadi}, {Stumpe}, {Barclay},
  \& {Burke}}]{Mosser12b}
{Mosser}, B., {Goupil}, M.~J., {Belkacem}, K., {et~al.} 2012{\natexlab{b}},
  \aap, 540, A143

\bibitem[{{M{\"o}sta} {et~al.}(2018){M{\"o}sta}, {Roberts}, {Halevi}, {Ott},
  {Lippuner}, {Haas}, \& {Schnetter}}]{Moesta18}
{M{\"o}sta}, P., {Roberts}, L.~F., {Halevi}, G., {et~al.} 2018, \apj, 864, 171

\bibitem[{{Mucciarelli} {et~al.}(2017){Mucciarelli}, {Merle}, \&
  {Bellazzini}}]{Mucciarelli17}
{Mucciarelli}, A., {Merle}, T., \& {Bellazzini}, M. 2017, \aap, 600, A104

\bibitem[{{Nilsson} {et~al.}(2002{\natexlab{a}}){Nilsson}, {Ivarsson},
  {Johansson}, \& {Lundberg}}]{NIJL}
{Nilsson}, H., {Ivarsson}, S., {Johansson}, S., \& {Lundberg}, H.
  2002{\natexlab{a}}, \aap, 381, 1090

\bibitem[{{Nilsson} {et~al.}(2002{\natexlab{b}}){Nilsson}, {Zhang}, {Lundberg},
  {Johansson}, \& {Nordstr{\"o}m}}]{NZL}
{Nilsson}, H., {Zhang}, Z.~G., {Lundberg}, H., {Johansson}, S., \&
  {Nordstr{\"o}m}, B. 2002{\natexlab{b}}, \aap, 382, 368

\bibitem[{{Nitz} {et~al.}(1998){Nitz}, {Wickliffe}, \& {Lawler}}]{NWL}
{Nitz}, D.~E., {Wickliffe}, M.~E., \& {Lawler}, J.~E. 1998, \apjs, 117, 313

\bibitem[{{Noll} {et~al.}(2012){Noll}, {Kausch}, {Barden}, {Jones}, {Szyszka},
  {Kimeswenger}, \& {Vinther}}]{Noll12}
{Noll}, S., {Kausch}, W., {Barden}, M., {et~al.} 2012, \aap, 543, A92

\bibitem[{{Nomoto} {et~al.}(2006){Nomoto}, {Tominaga}, {Umeda}, {Kobayashi}, \&
  {Maeda}}]{Nomoto06}
{Nomoto}, K., {Tominaga}, N., {Umeda}, H., {Kobayashi}, C., \& {Maeda}, K.
  2006, \nphysa, 777, 424

\bibitem[{{Nordlander} \& {Lind}(2017)}]{Nordlander17}
{Nordlander}, T. \& {Lind}, K. 2017, \aap, 607, A75

\bibitem[{{O'Brian} {et~al.}(1991){O'Brian}, {Wickliffe}, {Lawler}, {Whaling},
  \& {Brault}}]{BWL}
{O'Brian}, T.~R., {Wickliffe}, M.~E., {Lawler}, J.~E., {Whaling}, W., \&
  {Brault}, J.~W. 1991, Journal of the Optical Society of America B Optical
  Physics, 8, 1185

\bibitem[{{{\"O}nehag} {et~al.}(2009){{\"O}nehag}, {Gustafsson}, {Eriksson}, \&
  {Edvardsson}}]{Oenehag09}
{{\"O}nehag}, A., {Gustafsson}, B., {Eriksson}, K., \& {Edvardsson}, B. 2009,
  \aap, 498, 527

\bibitem[{{Parkinson} {et~al.}(1976){Parkinson}, {Reeves}, \& {Tomkins}}]{PRT}
{Parkinson}, W.~H., {Reeves}, E.~M., \& {Tomkins}, F.~S. 1976, Journal of
  Physics B Atomic Molecular Physics, 9, 157

\bibitem[{{Pehlivan Rhodin} {et~al.}(2017){Pehlivan Rhodin}, {Hartman},
  {Nilsson}, \& {J{\"o}nsson}}]{Pehlivan17}
{Pehlivan Rhodin}, A., {Hartman}, H., {Nilsson}, H., \& {J{\"o}nsson}, P. 2017,
  \aap, 598, A102

\bibitem[{{P{\'e}rez-Mesa} {et~al.}(2017){P{\'e}rez-Mesa}, {Zamora},
  {Garc{\'\i}a-Hern{\'a}ndez}, {Plez}, {Manchado}, {Karakas}, \&
  {Lugaro}}]{Perez-Mesa17}
{P{\'e}rez-Mesa}, V., {Zamora}, O., {Garc{\'\i}a-Hern{\'a}ndez}, D.~A.,
  {et~al.} 2017, \aap, 606, A20

\bibitem[{{Pian} {et~al.}(2017){Pian}, {D'Avanzo}, {Benetti}, {Branchesi},
  {Brocato}, {Campana}, {Cappellaro}, {Covino}, {D'Elia}, {Fynbo}, {Getman},
  {Ghirland a}, {Ghisellini}, {Grado}, {Greco}, {Hjorth}, {Kouveliotou},
  {Levan}, {Limatola}, {Malesani}, {Mazzali}, {Melandri}, {M{\o}ller},
  {Nicastro}, {Palazzi}, {Piranomonte}, {Rossi}, {Salafia}, {Selsing},
  {Stratta}, {Tanaka}, {Tanvir}, {Tomasella}, {Watson}, {Yang}, {Amati},
  {Antonelli}, {Ascenzi}, {Bernardini}, {Bo{\"e}r}, {Bufano}, {Bulgarelli},
  {Capaccioli}, {Casella}, {Castro-Tirado}, {Chassande-Mottin}, {Ciolfi},
  {Copperwheat}, {Dadina}, {De Cesare}, {di Paola}, {Fan}, {Gendre},
  {Giuffrida}, {Giunta}, {Hunt}, {Israel}, {Jin}, {Kasliwal}, {Klose}, {Lisi},
  {Longo}, {Maiorano}, {Mapelli}, {Masetti}, {Nava}, {Patricelli}, {Perley},
  {Pescalli}, {Piran}, {Possenti}, {Pulone}, {Razzano}, {Salvaterra},
  {Schipani}, {Spera}, {Stamerra}, {Stella}, {Tagliaferri}, {Testa}, {Troja},
  {Turatto}, {Vergani}, \& {Vergani}}]{Pian17}
{Pian}, E., {D'Avanzo}, P., {Benetti}, S., {et~al.} 2017, \nat, 551, 67

\bibitem[{{Pickering} {et~al.}(2001){Pickering}, {Thorne}, \& {Perez}}]{PTP}
{Pickering}, J.~C., {Thorne}, A.~P., \& {Perez}, R. 2001, \apjs, 132, 403

\bibitem[{{Piersanti} {et~al.}(2013){Piersanti}, {Cristallo}, \&
  {Straniero}}]{Piersanti13}
{Piersanti}, L., {Cristallo}, S., \& {Straniero}, O. 2013, \apj, 774, 98

\bibitem[{{Pinnington} {et~al.}(1993){Pinnington}, {Ji}, {Guo}, {Berends}, {van
  Hunen}, \& {Biemont}}]{PGBH}
{Pinnington}, E.~H., {Ji}, Q., {Guo}, B., {et~al.} 1993, Canadian Journal of
  Physics, 71, 470

\bibitem[{{Piskunov} {et~al.}(1995){Piskunov}, {Kupka}, {Ryabchikova}, {Weiss},
  \& {Jeffery}}]{Piskunov95}
{Piskunov}, N.~E., {Kupka}, F., {Ryabchikova}, T.~A., {Weiss}, W.~W., \&
  {Jeffery}, C.~S. 1995, \aaps, 112, 525

\bibitem[{{Pitts} \& {Newsom}(1986)}]{PN}
{Pitts}, R.~E. \& {Newsom}, G.~H. 1986, \jqsrt, 35, 383

\bibitem[{{Placco} {et~al.}(2016){Placco}, {Frebel}, {Beers}, {Yoon}, {Chiti},
  {Heger}, {Chan}, {Casey}, \& {Christlieb}}]{Placco16}
{Placco}, V.~M., {Frebel}, A., {Beers}, T.~C., {et~al.} 2016, \apj, 833, 21

\bibitem[{{Prantzos} {et~al.}(2019){Prantzos}, {Abia}, {Cristallo}, {Limongi},
  \& {Chieffi}}]{Prantzos19}
{Prantzos}, N., {Abia}, C., {Cristallo}, S., {Limongi}, M., \& {Chieffi}, A.
  2019, \mnras, 2748

\bibitem[{{Preston} {et~al.}(2019){Preston}, {Sneden}, {Chadid}, {Thompson}, \&
  {Shectman}}]{Preston19}
{Preston}, G.~W., {Sneden}, C., {Chadid}, M., {Thompson}, I.~B., \& {Shectman},
  S.~A. 2019, \aj, 157, 153

\bibitem[{{Pr{\v{s}}a} {et~al.}(2016){Pr{\v{s}}a}, {Harmanec}, {Torres},
  {Mamajek}, {Asplund}, {Capitaine}, {Christensen-Dalsgaard}, {Depagne},
  {Haberreiter}, {Hekker}, {Hilton}, {Kopp}, {Kostov}, {Kurtz}, {Laskar},
  {Mason}, {Milone}, {Montgomery}, {Richards}, {Schmutz}, {Schou}, \&
  {Stewart}}]{Prsa16}
{Pr{\v{s}}a}, A., {Harmanec}, P., {Torres}, G., {et~al.} 2016, \aj, 152, 41

\bibitem[{{Raassen} \& {Uylings}(1998)}]{RU}
{Raassen}, A.~J.~J. \& {Uylings}, P.~H.~M. 1998, \aap, 340, 300, (RU)

\bibitem[{{Ram{\'{\i}}rez} \& {Mel{\'e}ndez}(2005)}]{Ramirez05}
{Ram{\'{\i}}rez}, I. \& {Mel{\'e}ndez}, J. 2005, \apj, 626, 465

\bibitem[{{Ricker} {et~al.}(2015){Ricker}, {Winn}, {Vanderspek}, {Latham},
  {Bakos}, {Bean}, {Berta-Thompson}, {Brown}, {Buchhave}, {Butler}, {Butler},
  {Chaplin}, {Charbonneau}, {Christensen-Dalsgaard}, {Clampin}, {Deming},
  {Doty}, {De Lee}, {Dressing}, {Dunham}, {Endl}, {Fressin}, {Ge}, {Henning},
  {Holman}, {Howard}, {Ida}, {Jenkins}, {Jernigan}, {Johnson}, {Kaltenegger},
  {Kawai}, {Kjeldsen}, {Laughlin}, {Levine}, {Lin}, {Lissauer}, {MacQueen},
  {Marcy}, {McCullough}, {Morton}, {Narita}, {Paegert}, {Palle}, {Pepe},
  {Pepper}, {Quirrenbach}, {Rinehart}, {Sasselov}, {Sato}, {Seager},
  {Sozzetti}, {Stassun}, {Sullivan}, {Szentgyorgyi}, {Torres}, {Udry}, \&
  {Villasenor}}]{Ricker15}
{Ricker}, G.~R., {Winn}, J.~N., {Vanderspek}, R., {et~al.} 2015, Journal of
  Astronomical Telescopes, Instruments, and Systems, 1, 014003

\bibitem[{{Roederer}(2012)}]{Roederer12}
{Roederer}, I.~U. 2012, \apj, 756, 36

\bibitem[{{Roederer} {et~al.}(2018{\natexlab{a}}){Roederer}, {Hattori}, \&
  {Valluri}}]{Roederer18a}
{Roederer}, I.~U., {Hattori}, K., \& {Valluri}, M. 2018{\natexlab{a}}, \aj,
  156, 179

\bibitem[{{Roederer} {et~al.}(2016){Roederer}, {Karakas}, {Pignatari}, \&
  {Herwig}}]{Roederer16}
{Roederer}, I.~U., {Karakas}, A.~I., {Pignatari}, M., \& {Herwig}, F. 2016,
  \apj, 821, 37

\bibitem[{{Roederer} {et~al.}(2009){Roederer}, {Kratz}, {Frebel}, {Christlieb},
  {Pfeiffer}, {Cowan}, \& {Sneden}}]{Roederer09}
{Roederer}, I.~U., {Kratz}, K.-L., {Frebel}, A., {et~al.} 2009, \apj, 698, 1963

\bibitem[{{Roederer} {et~al.}(2014){Roederer}, {Preston}, {Thompson},
  {Shectman}, {Sneden}, {Burley}, \& {Kelson}}]{Roederer14}
{Roederer}, I.~U., {Preston}, G.~W., {Thompson}, I.~B., {et~al.} 2014, \aj,
  147, 136

\bibitem[{{Roederer} {et~al.}(2018{\natexlab{b}}){Roederer}, {Sakari},
  {Placco}, {Beers}, {Ezzeddine}, {Frebel}, \& {Hansen}}]{Roederer18}
{Roederer}, I.~U., {Sakari}, C.~M., {Placco}, V.~M., {et~al.}
  2018{\natexlab{b}}, \apj, 865, 129

\bibitem[{{Roederer} {et~al.}(2010){Roederer}, {Sneden}, {Lawler}, \&
  {Cowan}}]{Roederer10}
{Roederer}, I.~U., {Sneden}, C., {Lawler}, J.~E., \& {Cowan}, J.~J. 2010,
  \apjl, 714, L123

\bibitem[{{Ryabchikova} {et~al.}(2015){Ryabchikova}, {Piskunov}, {Kurucz},
  {Stempels}, {Heiter}, {Pakhomov}, \& {Barklem}}]{Ryabchikova15}
{Ryabchikova}, T., {Piskunov}, N., {Kurucz}, R.~L., {et~al.} 2015, \physscr,
  90, 054005

\bibitem[{{Ryabchikova} {et~al.}(1994){Ryabchikova}, {Hill}, {Landstreet},
  {Piskunov}, \& {Sigut}}]{RHL}
{Ryabchikova}, T.~A., {Hill}, G.~M., {Landstreet}, J.~D., {Piskunov}, N., \&
  {Sigut}, T.~A.~A. 1994, \mnras, 267, 697

\bibitem[{{Sakari} {et~al.}(2018){Sakari}, {Placco}, {Farrell}, {Roederer},
  {Wallerstein}, {Beers}, {Ezzeddine}, {Frebel}, {Hansen}, {Holmbeck},
  {Sneden}, {Cowan}, {Venn}, {Davis}, {Matijevi{\v{c}}}, {Wyse},
  {Bland-Hawthorn}, {Chiappini}, {Freeman}, {Gibson}, {Grebel}, {Helmi},
  {Kordopatis}, {Kunder}, {Navarro}, {Reid}, {Seabroke}, {Steinmetz}, \&
  {Watson}}]{Sakari18}
{Sakari}, C.~M., {Placco}, V.~M., {Farrell}, E.~M., {et~al.} 2018, \apj, 868,
  110

\bibitem[{{Schatz} {et~al.}(2002){Schatz}, {Toenjes}, {Pfeiffer}, {Beers},
  {Cowan}, {Hill}, \& {Kratz}}]{Schatz02}
{Schatz}, H., {Toenjes}, R., {Pfeiffer}, B., {et~al.} 2002, \apj, 579, 626

\bibitem[{{Schlafly} \& {Finkbeiner}(2011)}]{Schlafly11}
{Schlafly}, E.~F. \& {Finkbeiner}, D.~P. 2011, \apj, 737, 103

\bibitem[{{Schnabel} {et~al.}(2004){Schnabel}, {Schultz-Johanning}, \&
  {Kock}}]{Schnabel04}
{Schnabel}, R., {Schultz-Johanning}, M., \& {Kock}, M. 2004, \aap, 414, 1169

\bibitem[{{Shi} {et~al.}(2018){Shi}, {Yan}, {Zhou}, \& {Zhao}}]{Shi18}
{Shi}, J.~R., {Yan}, H.~L., {Zhou}, Z.~M., \& {Zhao}, G. 2018, \apj, 862, 71

\bibitem[{{Shirai} {et~al.}(2007){Shirai}, {Reader}, {Kramida}, \&
  {Sugar}}]{SRKS}
{Shirai}, T., {Reader}, J., {Kramida}, A.~E., \& {Sugar}, J. 2007, Journal of
  Physical and Chemical Reference Data, 36, 509

\bibitem[{{Siess} {et~al.}(2004){Siess}, {Goriely}, \& {Langer}}]{Siess04}
{Siess}, L., {Goriely}, S., \& {Langer}, N. 2004, \aap, 415, 1089

\bibitem[{{Sigut} \& {Landstreet}(1990)}]{SLd}
{Sigut}, T.~A.~A. \& {Landstreet}, J.~D. 1990, \mnras, 247, 611

\bibitem[{{Sitnova} {et~al.}(2013){Sitnova}, {Mashonkina}, \&
  {Ryabchikova}}]{Sitnova13}
{Sitnova}, T.~M., {Mashonkina}, L.~I., \& {Ryabchikova}, T.~A. 2013, Astronomy
  Letters, 39, 126

\bibitem[{{Sitnova} {et~al.}(2016){Sitnova}, {Mashonkina}, \&
  {Ryabchikova}}]{Sitnova16}
{Sitnova}, T.~M., {Mashonkina}, L.~I., \& {Ryabchikova}, T.~A. 2016, \mnras,
  461, 1000

\bibitem[{{Sivarani} {et~al.}(2004){Sivarani}, {Bonifacio}, {Molaro}, {Cayrel},
  {Spite}, {Spite}, {Plez}, {Andersen}, {Barbuy}, {Beers}, {Depagne}, {Hill},
  {Fran{\c c}ois}, {Nordstr{\"o}m}, \& {Primas}}]{Sivarani04}
{Sivarani}, T., {Bonifacio}, P., {Molaro}, P., {et~al.} 2004, \aap, 413, 1073

\bibitem[{{Skrutskie} {et~al.}(2006){Skrutskie}, {Cutri}, {Stiening},
  {Weinberg}, {Schneider}, {Carpenter}, {Beichman}, {Capps}, {Chester},
  {Elias}, {Huchra}, {Liebert}, {Lonsdale}, {Monet}, {Price}, {Seitzer},
  {Jarrett}, {Kirkpatrick}, {Gizis}, {Howard}, {Evans}, {Fowler}, {Fullmer},
  {Hurt}, {Light}, {Kopan}, {Marsh}, {McCallon}, {Tam}, {Van Dyk}, \&
  {Wheelock}}]{Skrutskie06}
{Skrutskie}, M.~F., {Cutri}, R.~M., {Stiening}, R., {et~al.} 2006, \aj, 131,
  1163

\bibitem[{{Smith}(1981)}]{Sm}
{Smith}, G. 1981, \aap, 103, 351

\bibitem[{{Smith}(1988)}]{S}
{Smith}, G. 1988, Journal of Physics B Atomic Molecular Physics, 21, 2827

\bibitem[{{Smith} \& {O'Neill}(1975)}]{SN}
{Smith}, G. \& {O'Neill}, J.~A. 1975, \aap, 38, 1

\bibitem[{{Smith} \& {Raggett}(1981)}]{SR}
{Smith}, G. \& {Raggett}, D. S.~J. 1981, Journal of Physics B Atomic Molecular
  Physics, 14, 4015

\bibitem[{{Smith} \& {Liszt}(1971)}]{SLa}
{Smith}, W.~H. \& {Liszt}, H.~S. 1971, Journal of the Optical Society of
  America (1917-1983), 61, 938

\bibitem[{{Sneden}(1973)}]{Sneden73}
{Sneden}, C. 1973, PhD thesis, Univ. of Texas at Austin

\bibitem[{{Sneden} {et~al.}(2008){Sneden}, {Cowan}, \& {Gallino}}]{Sneden08}
{Sneden}, C., {Cowan}, J.~J., \& {Gallino}, R. 2008, \araa, 46, 241

\bibitem[{{Sneden} {et~al.}(2003){Sneden}, {Cowan}, {Lawler}, {Ivans},
  {Burles}, {Beers}, {Primas}, {Hill}, {Truran}, {Fuller}, {Pfeiffer}, \&
  {Kratz}}]{Sneden03}
{Sneden}, C., {Cowan}, J.~J., {Lawler}, J.~E., {et~al.} 2003, \apj, 591, 936

\bibitem[{{Sneden} {et~al.}(2009){Sneden}, {Lawler}, {Cowan}, {Ivans}, \& {Den
  Hartog}}]{Sneden09}
{Sneden}, C., {Lawler}, J.~E., {Cowan}, J.~J., {Ivans}, I.~I., \& {Den Hartog},
  E.~A. 2009, \apjs, 182, 80

\bibitem[{{Sobeck} {et~al.}(2011){Sobeck}, {Kraft}, {Sneden}, {Preston},
  {Cowan}, {Smith}, {Thompson}, {Shectman}, \& {Burley}}]{Sobeck11}
{Sobeck}, J.~S., {Kraft}, R.~P., {Sneden}, C., {et~al.} 2011, \aj, 141, 175

\bibitem[{{Sobeck} {et~al.}(2007){Sobeck}, {Lawler}, \& {Sneden}}]{SLS}
{Sobeck}, J.~S., {Lawler}, J.~E., \& {Sneden}, C. 2007, \apj, 667, 1267

\bibitem[{{Soubiran} {et~al.}(2018){Soubiran}, {Jasniewicz}, {Chemin},
  {Zurbach}, {Brouillet}, {Panuzzo}, {Sartoretti}, {Katz}, {Le Campion},
  {Marchal}, {Hestroffer}, {Th{\'e}venin}, {Crifo}, {Udry}, {Cropper},
  {Seabroke}, {Viala}, {Benson}, {Blomme}, {Jean-Antoine}, {Huckle}, {Smith},
  {Baker}, {Damerdji}, {Dolding}, {Fr{\'e}mat}, {Gosset}, {Guerrier}, {Guy},
  {Haigron}, {Jan{\ss}en}, {Plum}, {Fabre}, {Lasne}, {Pailler}, {Panem},
  {Riclet}, {Royer}, {Tauran}, {Zwitter}, {Gueguen}, \& {Turon}}]{Soubiran18}
{Soubiran}, C., {Jasniewicz}, G., {Chemin}, L., {et~al.} 2018, \aap, 616, A7

\bibitem[{{Spite} {et~al.}(2006){Spite}, {Cayrel}, {Hill}, {Spite},
  {Fran{\c{c}}ois}, {Plez}, {Bonifacio}, {Molaro}, {Depagne}, \&
  {Andersen}}]{Spite06}
{Spite}, M., {Cayrel}, R., {Hill}, V., {et~al.} 2006, \aap, 455, 291

\bibitem[{{Spite} {et~al.}(2005){Spite}, {Cayrel}, {Plez}, {Hill}, {Spite},
  {Depagne}, {Fran{\c{c}}ois}, {Bonifacio}, {Barbuy}, \& {Beers}}]{Spite05}
{Spite}, M., {Cayrel}, R., {Plez}, B., {et~al.} 2005, \aap, 430, 655

\bibitem[{{Straniero} {et~al.}(1995){Straniero}, {Gallino}, {Busso}, {Chiefei},
  {Raiteri}, {Limongi}, \& {Salaris}}]{Straniero95}
{Straniero}, O., {Gallino}, R., {Busso}, M., {et~al.} 1995, \apjl, 440, L85

\bibitem[{{Straniero} {et~al.}(2006){Straniero}, {Gallino}, \&
  {Cristallo}}]{Straniero06}
{Straniero}, O., {Gallino}, R., \& {Cristallo}, S. 2006, \nphysa, 777, 311

\bibitem[{{Travaglio} {et~al.}(2004){Travaglio}, {Gallino}, {Arnone}, {Cowan},
  {Jordan}, \& {Sneden}}]{Travaglio04}
{Travaglio}, C., {Gallino}, R., {Arnone}, E., {et~al.} 2004, \apj, 601, 864

\bibitem[{{Velichko} {et~al.}(2010){Velichko}, {Mashonkina}, \&
  {Nilsson}}]{Velichko10}
{Velichko}, A.~B., {Mashonkina}, L.~I., \& {Nilsson}, H. 2010, Astronomy
  Letters, 36, 664

\bibitem[{{Wanajo}(2013)}]{Wanajo13}
{Wanajo}, S. 2013, \apjl, 770, L22

\bibitem[{{Warner}(1968{\natexlab{a}})}]{Wc}
{Warner}, B. 1968{\natexlab{a}}, \mnras, 139, 115

\bibitem[{{Warner}(1968{\natexlab{b}})}]{Wa}
{Warner}, B. 1968{\natexlab{b}}, \mnras, 140, 53

\bibitem[{{Watson} {et~al.}(2019){Watson}, {Hansen}, {Selsing}, {Koch},
  {Malesani}, {Andersen}, {Fynbo}, {Arcones}, {Bauswein}, {Covino}, {Grado},
  {Heintz}, {Hunt}, {Kouveliotou}, {Leloudas}, {Levan}, {Mazzali}, \&
  {Pian}}]{Watson19}
{Watson}, D., {Hansen}, C.~J., {Selsing}, J., {et~al.} 2019, \nat, 574, 497

\bibitem[{{Whaling} \& {Brault}(1988)}]{WBb}
{Whaling}, W. \& {Brault}, J.~W. 1988, \physscr, 38, 707

\bibitem[{{Wickliffe} \& {Lawler}(1997)}]{WL}
{Wickliffe}, M.~E. \& {Lawler}, J.~E. 1997, Journal of the Optical Society of
  America B Optical Physics, 14, 737

\bibitem[{{Wickliffe} {et~al.}(2000){Wickliffe}, {Lawler}, \& {Nave}}]{WLN}
{Wickliffe}, M.~E., {Lawler}, J.~E., \& {Nave}, G. 2000, \jqsrt, 66, 363

\bibitem[{{Wickliffe} {et~al.}(1994){Wickliffe}, {Salih}, \& {Lawler}}]{WSL}
{Wickliffe}, M.~E., {Salih}, S., \& {Lawler}, J.~E. 1994, \jqsrt, 51, 545

\bibitem[{{Wiese} {et~al.}(1966){Wiese}, {Smith}, \& {Glennon}}]{WSG}
{Wiese}, W.~L., {Smith}, M.~W., \& {Glennon}, B.~M. 1966, {Atomic transition
  probabilities. Vol.: Hydrogen through Neon. A critical data compilation}

\bibitem[{{Wiese} {et~al.}(1969){Wiese}, {Smith}, \& {Miles}}]{WSM}
{Wiese}, W.~L., {Smith}, M.~W., \& {Miles}, B.~M. 1969, {Atomic transition
  probabilities. Vol. 2: Sodium through Calcium. A critical data compilation}

\bibitem[{{Wood} {et~al.}(2014{\natexlab{a}}){Wood}, {Lawler}, {Den Hartog},
  {Sneden}, \& {Cowan}}]{WLDSC}
{Wood}, M.~P., {Lawler}, J.~E., {Den Hartog}, E.~A., {Sneden}, C., \& {Cowan},
  J.~J. 2014{\natexlab{a}}, \apjs, 214, 18

\bibitem[{{Wood} {et~al.}(2013){Wood}, {Lawler}, {Sneden}, \& {Cowan}}]{WLSC}
{Wood}, M.~P., {Lawler}, J.~E., {Sneden}, C., \& {Cowan}, J.~J. 2013, \apjs,
  208, 27

\bibitem[{{Wood} {et~al.}(2014{\natexlab{b}}){Wood}, {Lawler}, {Sneden}, \&
  {Cowan}}]{WLSCow}
{Wood}, M.~P., {Lawler}, J.~E., {Sneden}, C., \& {Cowan}, J.~J.
  2014{\natexlab{b}}, \apjs, 211, 20

\bibitem[{{Woosley} \& {Weaver}(1995)}]{Woosley95}
{Woosley}, S.~E. \& {Weaver}, T.~A. 1995, \apjs, 101, 181

\bibitem[{{Yan} {et~al.}(2015){Yan}, {Shi}, \& {Zhao}}]{Yan15}
{Yan}, H.~L., {Shi}, J.~R., \& {Zhao}, G. 2015, \apj, 802, 36

\bibitem[{{Zennaro} {et~al.}(2019){Zennaro}, {Milone}, {Marino}, {Cordoni},
  {Lagioia}, \& {Tailo}}]{Zennaro19}
{Zennaro}, M., {Milone}, A.~P., {Marino}, A.~F., {et~al.} 2019, \mnras, 487,
  3239

\end{thebibliography}

\begin{appendix}
 \section{Alternative methods for determining stellar parameters}\label{Sec: Alternative methods for determining stellar parameters}
 \subsection{Effective temperature}\label{Subec: A.1}
 \begin{figure*}
    \centering
    \resizebox{\hsize}{!}{\includegraphics{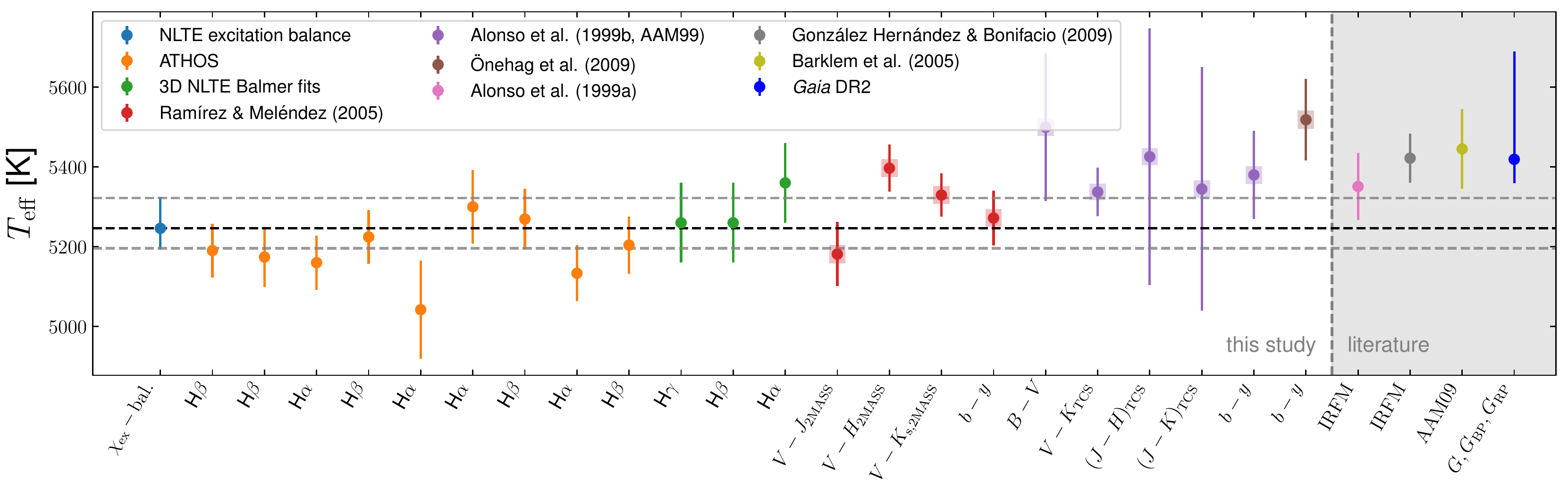}}
      \caption{Individual photometric and spectroscopic temperature measures for HD~20 obtained in this work. On the abscissa either the photometric color or the spectral region that was used to deduce $T_\mathrm{eff}$ is labeled. Different colors indicate the different scales and methods employed (see legend and main text for details). For each $T_\mathrm{eff}$ that was deduced from a color index, the negligible effect of no reddening and twice the applied reddening \citep[$E(B-V)=0.0149$~mag,][]{Schlafly11} is denoted by light-colored ranges behind the circles that are barely visible. The IRFM findings for HD~20 by \citet{Alonso99a} and \citet{Gonzalez09} are shown together with the literature values by \citet{Barklem05} and \textit{Gaia} DR2. The finally adopted temperature and its error ($5246^{+76}_{-50}$~K) are shown by black- and gray-dashed lines, respectively. 
              }
      \label{Fig:Teff_comparison}
\end{figure*}
In order to put our adopted $T_\mathrm{eff}$ in context to other methods, we have derived this parameter from several other spectroscopic and photometric techniques that are summarized and presented together with existing literature values in Fig. \ref{Fig:Teff_comparison}.
 \subsubsection{ATHOS - temperatures from Balmer lines}\label{Subsubsec: ATHOS - temperatures from Balmer lines}
ATHOS\footnote{\url{https://github.com/mihanke/athos}} \citep[A Tool for Homogenizing Stellar parameters,][]{Hanke18} is a stellar parameter pipeline designed to acquire high-accuracy and high-precision stellar parameters from optical spectra of FGK stars. To that end, it employs flux ratios (FRs) of empirically defined wavelength ranges to compute the stellar parameters $T_\mathrm{eff}$, [Fe/H],  and $\log{g}$ from dedicated analytical relations that have been trained on a large sample of benchmark stars. The strategy adopted is model-dependent only to the extent that a considerable fraction of the original parameters of the benchmark sample have been determined through modeling.

For $T_\mathrm{eff}$, the tool incorporates nine FRs involving the wings of two of the Balmer lines of neutral hydrogen, H$\alpha$ and H$\beta$. Each of the nine FRs poses an independent measure of temperature. ATHOS was applied to all spectra containing H$\alpha$ and H$\beta$, i.e. the UVES~580, MIKE, and HARPS spectra. In order to account for the substantial line broadening present in HD~20 (see Sect. \ref{Subsec: Rotational velocity}), we provided ATHOS with an effective resolution  
\begin{equation}
 R_\mathrm{eff}=\left(\left(\frac{1}{R_0}\right)^2 + \left(\frac{v\sin{i}}{c}\right)^2\right)^{-\frac{1}{2}}
\end{equation}
under the assumption that rotational broadening behaves approximately Gaussian\footnote{We emphasize that this step is not utterly important at this point, because the ATHOS implementation for $T_\mathrm{eff}$ is largely insensitive to rotational broadening \citep[see][]{Hanke18}. Line broadening, however, does affect ATHOS' [Fe/H] estimators (Sect. \ref{Subsec: ATHOS - [Fe/H] from flux ratios}).}. Here, $R_0$ denotes the instrumental resolving power of the input spectra. The mean temperature and its error for each of the nine relations are depicted in Fig. \ref{Fig:Teff_comparison}, whereas the weighted mean $T_\mathrm{eff}$ from all ATHOS results is $5194\pm25$~K, a temperature in good agreement with our adopted value. The latter low uncertainty is typical for the very high internal precision of ATHOS temperatures from high-S/N data. Nevertheless, it is important to bear in mind that the initial temperatures of ATHOS' benchmark sample suffered from finite accuracy. Thus we note an additional systematic error of 97~K \citep{Hanke18}.

\subsubsection{3D NLTE modeling of Balmer lines}\label{Subsubsec: 3D NLTE modeling of Balmer lines}
The classical spectroscopic approach of inferring $T_\mathrm{eff}$ from Balmer lines relies on their theoretical modeling and comparison of the profile wings to observed spectra \citep{Barklem02}. As a consequence, the approach is strongly model-dependent and prone to inaccuracies and/or unknowns in the attempts to reproduce real physical processes. To date, \citet{Amarsi18} presented the most complex and potentially most accurate calculations of Balmer line formation in late-type stars involving 3D hydrodynamic atmosphere models and NLTE radiative transfer. The authors showed that departures from ordinary 1D LTE line formation can be substantial and their negligence could introduce temperature inaccuracies on the order of 100~K.

\begin{figure*}
    \centering
    \resizebox{\hsize}{!}{\includegraphics{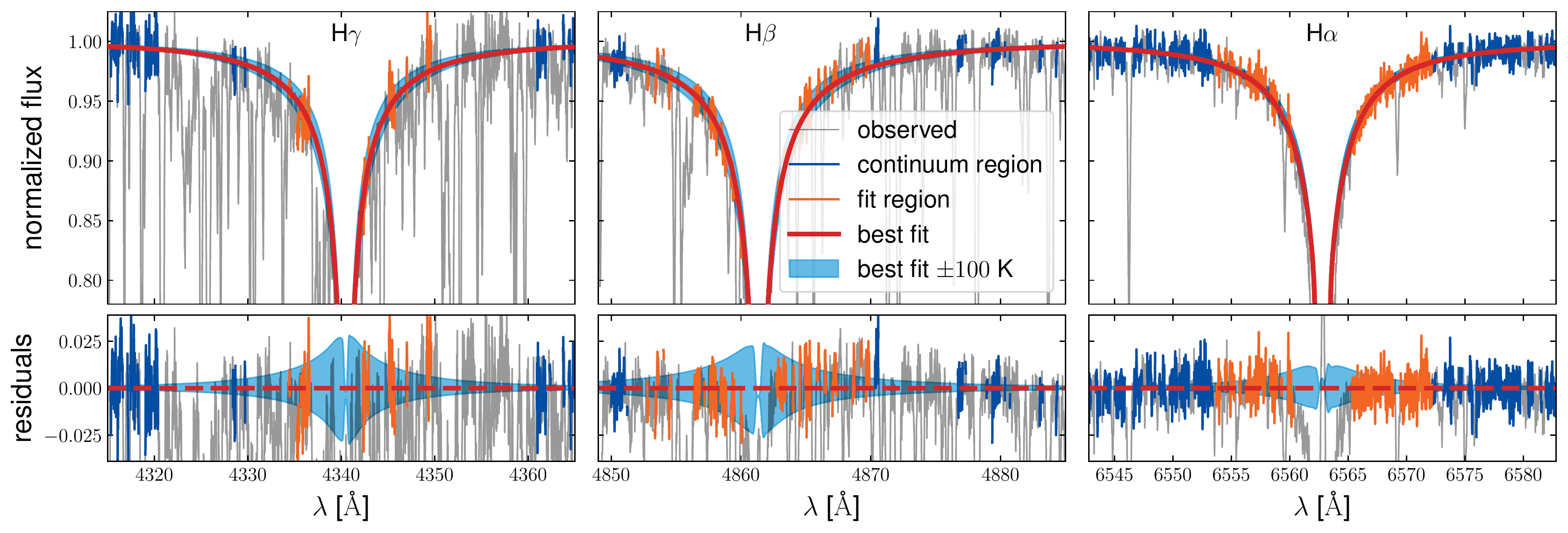}}
      \caption{$T_\mathrm{eff}$ fit results from fitting the wings of the Balmer lines H$\gamma$ (\textit{left}), H$\beta$ (\textit{middle}), and H$\alpha$ (\textit{right}). \textit{Upper panels}: Observed spectra (gray) with the best-fit 3D NLTE models (5260/5260/5360~K) by \citet{Amarsi18} and their error margins of 100~K depicted by red lines and blue areas, respectively. The wavelength regions used to obtain the continuum level are marked in blue, whereas orange lines highlight the parts of the spectrum that entered the $\chi^2$ minimization. \textit{Lower panels}: Residual spectrum.
              }
      \label{Fig:Amarsi_temperatures}
\end{figure*}
We took advantage of the extensive grid of 3D NLTE Balmer line models published by \citet{Amarsi18} and closely followed their fitting scheme to deduce $T_\mathrm{eff}$ for HD~20 from H$\gamma$, H$\beta$, and H$\alpha$ in the UVES spectra. In brief, for each profile, two 1D LTE spectra -- one including metal lines and one considering only the H-lines -- were modeled for the final parameters (Table \ref{Table: Target information}) and a line list including all transitions for the respective synthesis range found in VALD. We used these two artificial spectra to define ``clean'' wavelength regions free from substantial metal absorption by requesting the residual deviation to result in a change of less than 30~K in the derived temperature. Furthermore, for H$\alpha$, we employed SkyCalc \citep{Noll12, Jones13} to obtain a representative, synthetic telluric spectrum for the average observing conditions on Cerro Paranal and excluded all features above a threshold of 1\% in absorption. Any of the remaining wavelength ranges with fluxes above $98\%$ of the continuum flux were used to fit a linear continuum, while ranges of $\leq98\%$ of the continuum flux entered a $\chi^2$-minimization algorithm that interpolates between points of the Balmer model grid by employing cubic splines. For this purpose, all model parameters but $T_\mathrm{eff}$ were kept fixed at their recommended values (Table \ref{Table: Target information}). The resulting temperatures are 5260~K from H$\gamma$, 5260~K from H$\beta$, and 5360~K from H$\alpha$. Here, we caution against an over-interpretation of the deviation of the latter temperature, because it amounts to less than one combined error margin and H$\alpha$ is the least strong and least temperature-sensitive profile as can be seen in Fig. \ref{Fig:Amarsi_temperatures}. There, best-fit results are illustrated for all three profiles together with margins amounting to $\pm100$~K, which we adopt as error estimate for individual measures from this method. The straight average $T_\mathrm{eff}=5293\pm58$~K is in good agreement with our independently determined, adopted value (5246~K).

We would like to stress that -- apart from model uncertainties -- the accuracy of the outlined procedure is affected by non-linearities in the global continuum shape due to the blaze function, as has already been pointed out for UVES spectra by \citet{Amarsi18}. In fact, we see an asymmetric substructure in the residuals of H$\beta$ that cannot be explained by model deficiencies. For the same reason the Balmer profiles in the MIKE spectrum were not used as they show slightly stronger persistent distortions after performing the above simple normalization scheme. The HARPS spectra only cover H$\beta$ and H$\alpha$ (5190~K and 5300~K, respectively) with no apparent residual substructure after normalization. However, the noise level considerably exceeds the error margin of 100~K, which is why we excluded the HARPS spectrum from consideration, too. 

The treatment of normalization is one of the key advantages of the technique implemented in ATHOS over Balmer modeling: ATHOS does not rely on one global continuum for each Balmer profile, but rather computes its individual FRs from two wavelength regions that are spaced much less than the overall extent of the line. Indeed, this is based on the premise that between the two involved ranges the continuum stays constant. The narrow spacing, however, justifies the latter assumption. Moreover, typically, ATHOS provides four to five measures of temperature per Balmer line, such that any persistent effect induced by small-scale continuum variations can effectively be averaged out. This, on the other hand, would manifest itself in an increased relation-to-relation scatter, which is not observed for any of our HD~20 spectra.

\subsubsection{Color - [Fe/H] - $T_\mathrm{eff}$ calibrations}\label{Subsubsec: Color - [Fe/H] - Teff relations}
We have further used the available photometry to compute $T_\mathrm{eff}$ from several empirical relations in the literature. The first one was introduced by AAM99, who calibrated their analytical functions against a large sample of known [Fe/H] and $T_\mathrm{eff}$, which themselves were inferred from the infrared flux method \citep[IRFM, e.g.,][]{Blackwell77}. Since HD~20 was part of their sample, we mention here their IRFM-based temperature of $5351\pm84$~K \citep{Alonso99a}, which is slightly warmer compared to our adopted value. Unfortunately, most of the relations provided by AAM99 are not directly compatible with the photometry at hand, because AAM99 calibrated their relations for the infrared $JHK$ bands in the Telescopio Carlos S{\'a}nchez (TCS) system instead of the 2MASS system. For this reason, we made use of a two-step conversion; first from the 2MASS to the CIT (California Institute of Technology) system as described in the supplemental material for the 2MASS mission\footnote{\url{http://www.astro.caltech.edu/~jmc/2mass/v3/transformations/}}, and secondly to the TCS system adopting the transformations given by \citet{Alonso94}. Errors were propagated through all conversion steps, which poses the dominant source of error in the derived individual temperatures. We find a weighted average temperature of $5362\pm52$~K.

\citet{Barklem05} report $5445\pm100$~K using $BVR_\mathrm{C}I_\mathrm{C}JHK_\mathrm{s}$ photometry ($b-y$ was not considered) by \citet{Beers07} as well as the same color transformations and calibration relations. Despite having rejected that photometry (see Sect. \ref{Subsec: Photometric and astrometric data}), we attempted to reproduce their value from their photometry. To this end, for the $R$ and $I$ bands we applied the transformations given in \citet{Bessell83} to convert the magnitudes by \citet{Beers07} to the Johnson ones. Nonetheless, using exactly the same averaging scheme -- that is, dropping the strongest outliers in either direction, not considering $b-y$, and taking an unweighted mean\footnote{We note that a weighted average would result in a substantially lower $T_\mathrm{eff}$, since the value from $(V-K)_\mathrm{TCS}$ is much less uncertain than all the others.} -- we cannot reproduce their rather hot value, but find 5270~K in accordance with our adopted estimate. \citet{Barklem05} already noted generally warmer temperatures when comparing their sample to existing literature values and claimed the origin to be the usage of different reddening maps. Adopting their slightly higher extinction value of 0.017 only marginally increases our value by 5~K. We suspect two plausible reasons for the strong discrepancy, or a mixture thereof: If we neglected the erratum to AAM99 \citep{Alonso01} that cautions to invert the sign of the cross-term of colors and [Fe/H] in the calibrations, we would end up with temperatures that are on average higher by almost 200~K. Moreover, looking at \citet{Sivarani04}, who introduce the color transformations used by \citet{Barklem05}, we found that they transformed $V-K$ colors to the Johnson system, while the AAM99 requires this color in the TCS system. 

Another empirical calibration was introduced by \citet{Ramirez05} who revisited the IRFM temperature scale by AAM99 and provided updated relations (here, we are only considering the scales for giants) for the filter systems given in Table \ref{Table: Target information}. For the $B-V$ color we would in principle have the necessary photometry, but the colors lie outside of the validity range of the relations. The weighted mean $T_\mathrm{eff}$ from the remaining four colors involving the redder two 2MASS filters and Strömgren photometry is $5294\pm79$~K (rms), which is cooler (70~K) than the value obtained from AAM99 and hence more in line with the spectroscopic results. 

The last photometric scale we consider is for the Strömgren color $b-y$ and was invented by \citet{Oenehag09}. It is based on synthetic colors from MARCS model atmospheres. At $5518\pm102$~K, we find the derived temperature to be much hotter ($\sim250$~K) than our adopted value. 

\textit{Gaia} DR2 provides temperature estimates for millions of sources based on \textit{Gaia} colors alone, as described in \citet{Andrae18}. Although the authors note that due to several limitations their temperatures are impractical for studies of individual stars, for completeness, we mention their value of $5419^{+267}_{-57}$~K. Considering the small lower uncertainty, this again represents an unfeasibly high $T_\mathrm{eff}$.

\subsection{ATHOS - [Fe/H] from flux ratios}\label{Subsec: ATHOS - [Fe/H] from flux ratios}
ATHOS not only allows for the inference of $T_\mathrm{eff}$, but its FR-based method was also expanded to provide estimates for [Fe/H]. In total, there are 31 FRs involving \ion{Fe}{i} lines that -- together with the previously determined $T_\mathrm{eff}$ -- span hypersurfaces, which allow for [Fe/H] computations. The corresponding analytical relations where trained and fit on the same training sample as the temperature method. The metallicity labels were either extracted from the detailed studies of the \textit{Gaia} benchmark stars \citep{Jofre14, Hawkins16}, or stem from a homogeneous analysis of \ion{Fe}{ii} lines in LTE \citep[see][for details]{Hanke18}. For HD~20, we found [Fe/H$]=-1.62\pm0.06$~dex from the median and rms scatter of all 31 FRs, respectively. This finding is in excellent agreement with our adopted [Fe/H] of $-1.60$~dex and therefore poses and independent validation.

\subsection{The width of the H$\alpha$ core as mass indicator}\label{Subsubsec: Width of Halpha as mass indicator}
\begin{table}
\caption{Stellar masses and $\log{g}$ from the core of H$\alpha$.}
\label{Table: Balmer masses}
\centering
\resizebox{\columnwidth}{!}{%
\begin{tabular}{lcccc}
\hline\hline
Spectrum & $W_{\mathrm{H}_\alpha}$ & $\log_{10}{m/m_\odot}^a$ & $m/m_\odot$ & $\log{g}^b$\\
         & {\AA}                   & [dex]                   &  & [dex]\\
\hline\\[-5pt]
UVES 580 & 0.863 & $-0.282\pm0.095$ & $0.52_{-0.10}^{+0.12}$ & $2.17\pm0.10$\\[5pt]
MIKE & 0.863 & $-0.282\pm0.095$ & $0.52_{-0.11}^{+0.13}$ & $2.17\pm0.10$\\[5pt]
HARPS & 0.861 & $-0.277\pm0.094$ & $0.53_{-0.10}^{+0.14}$ & $2.18\pm0.09$\\[5pt]

\hline
\end{tabular}}
\tablefoot{
\tablefoottext{a}{Calculated from Eq. (3) in \citet{Bergemann16}.}
\tablefoottext{b}{Derived from Eq. \ref{Eq: Stellar structure}.}
}
\end{table}
While the wings of the Balmer line H$\alpha$ were used earlier to infer $T_\mathrm{eff}$, we will now address the usage of its line core to derive the stellar mass. \citet{Bergemann16} have shown that even in the face of current, state-of-the-art modeling techniques, it is not possible to reliably synthesize this part of the line. However, adopting an empirical approach, the authors discovered a connection between the H$\alpha$ core width and the stellar mass. The latter originated from CoRoT and Kepler asteroseismology. 

We pursued the strategy outlined in \citet{Bergemann16} and fit the blue profile wing ($6562.0\,\AA<\lambda<6562.8\,\AA$) via the function
\begin{equation}\label{Eq: Bergemann16}
 f(\lambda)=1-f_0 \exp{\left(-\left(\frac{\lambda_0 - \lambda}{W_{\mathrm{H}_\alpha}}\right)^3\right)},
\end{equation}
with free parameters $f_0$ and $W_{\mathrm{H}_\alpha}$, and the central position of the line core $\lambda_0=6562.819$~{\AA}. From the width $W_{\mathrm{H}_\alpha}$, we then computed the mass parameter $\log_{10}{m/m_\sun}$ using the relation given in \citet{Bergemann16} and subsequently the surface gravity through inversion of Eq. \ref{Eq: Stellar structure}. The involved solar reference values can be found in Table \ref{Table: Target information}. As for Eq. \ref{Eq: luminosity}, we computed the bolometric magnitude $M_\mathrm{bol}$ from the $V$-band photometry and $BC_V$ by AAM99. The measurements and results for individual spectra covering H$\alpha$ are presented in Table \ref{Table: Balmer masses}. The error in $\log{g}$ is largely governed by the uncertainty in the mass and for the gravity from this method we obtained $\log{g}=2.17\pm0.10$~dex in line with our measurements based on NLTE ionization equilibrium and about $2\sigma$ lower than our asteroseismic finding. A plausible reason for this discrepancy may be found in the circumstance that, strictly speaking, HD~20 is about 250~K warmer than the upper validity bound for $T_\mathrm{eff}$ in the calibration relation by \citet{Bergemann16}.
 
 \section{Abundance systematics}\label{Sec: Abundance systematics}
  \subsection{Instrument-induced versus other systematics}\label{Subsec: Spectrograph comparison}
  \begin{figure*}
    \centering
    \resizebox{\hsize}{!}{\includegraphics{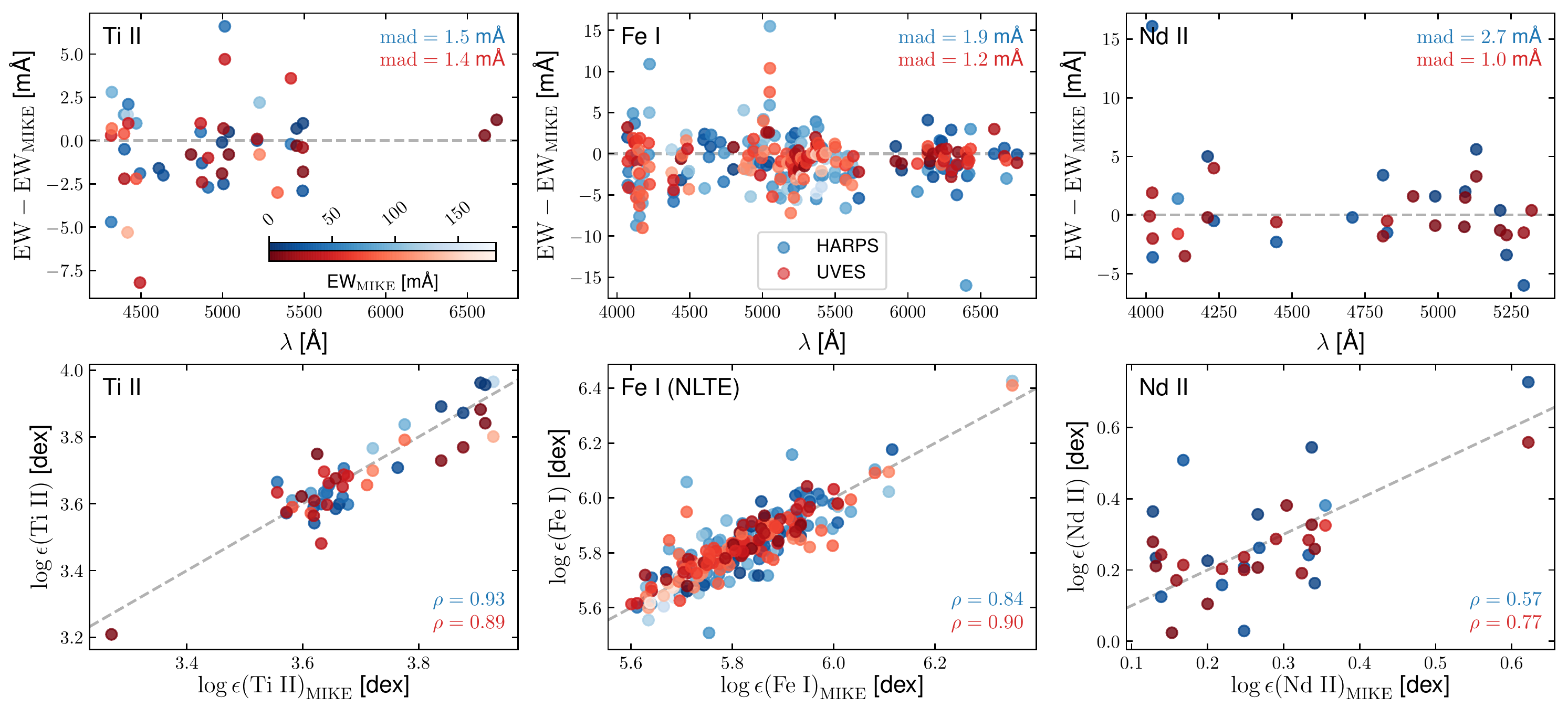}}
      \caption{Comparison of EWs (\textit{upper panels}) and deduced abundances (\textit{lower panels}) obtained from the same lines that were measured with three different instruments. Panels are horizontally separated by the three representative chemical species (from \textit{left} to \textit{right}: \ion{Ti}{ii}, \ion{Fe}{i}, and \ion{Nd}{ii}). \textit{Upper panels}: Residual EWs between HARPS (blue) and UVES (red) measurements with respect to the corresponding MIKE EWs as a function of wavelength. The lightness of the color stands for the measured EW in the MIKE spectrum as indicated by the color bars in the \textit{upper left panel}. \textit{Lower panels}: Abundances from the MIKE spectrum are shown on the abscissas, whereas HARPS and UVES findings are given along the ordinates. In each panel the perfect one-to-one correlation is represented by gray dashed lines and the correlation coefficients computed for the samples are presented on the \textit{lower right}.
              }
      \label{Fig:spectrgraph_comparison_corr}  
\end{figure*}
  As pointed out by the referee -- given our high-quality spectra gathered with three different instruments -- it is possible to investigate the presence of systematics originating from the choice of different resolutions and/or fiber-fed (in case of HARPS) versus slit spectrographs (MIKE and UVES). To this end, we performed tests using lines of the species \ion{Ti}{ii}, \ion{Fe}{i}, and \ion{Nd}{ii} that are distributed between 4000~{\AA} and 6800~{\AA}, which renders them accessible by all three instruments with only a few exceptions in the chip gaps. These three elements were chosen because they are on the one hand representatives for the main groups of $\alpha$, iron-peak, and neutron-capture elements and, on the other hand, allow for measurements of a sufficient number of lines (in this case more than 20) that permits meaningful number statistics.
   
   EWs for the sample of lines described above were measured in all three spectra using EWCODE. In the upper panels of Fig. \ref{Fig:spectrgraph_comparison_corr} we present the difference between measurements employing HARPS and UVES with respect to MIKE EWs. It is noteworthy that in principle the spread in this quantity is a convolution of both noise-induced errors from HARPS (UVES) and MIKE. However, in light of the substantially higher S/N of the MIKE spectrum at almost any wavelength, it appears safe to assume only a minor contribution due to noise in the MIKE spectrum. There are no obvious systematic trends or biases in the residuals, which leads us to the conclusion that for our analysis procedures of the star HD~20 the three spectrographs are entirely interchangeable without having to worry about introducing (additional) abundance systematics. The only notable difference is of a pure stochastic nature in the sense that HARPS EW residuals show larger spreads than UVES, which can be tied to the significantly lower S/N (see Fig. \ref{Fig:SN_allspec}).
   
   \begin{figure*}
    \centering
    \resizebox{0.85\hsize}{!}{\includegraphics{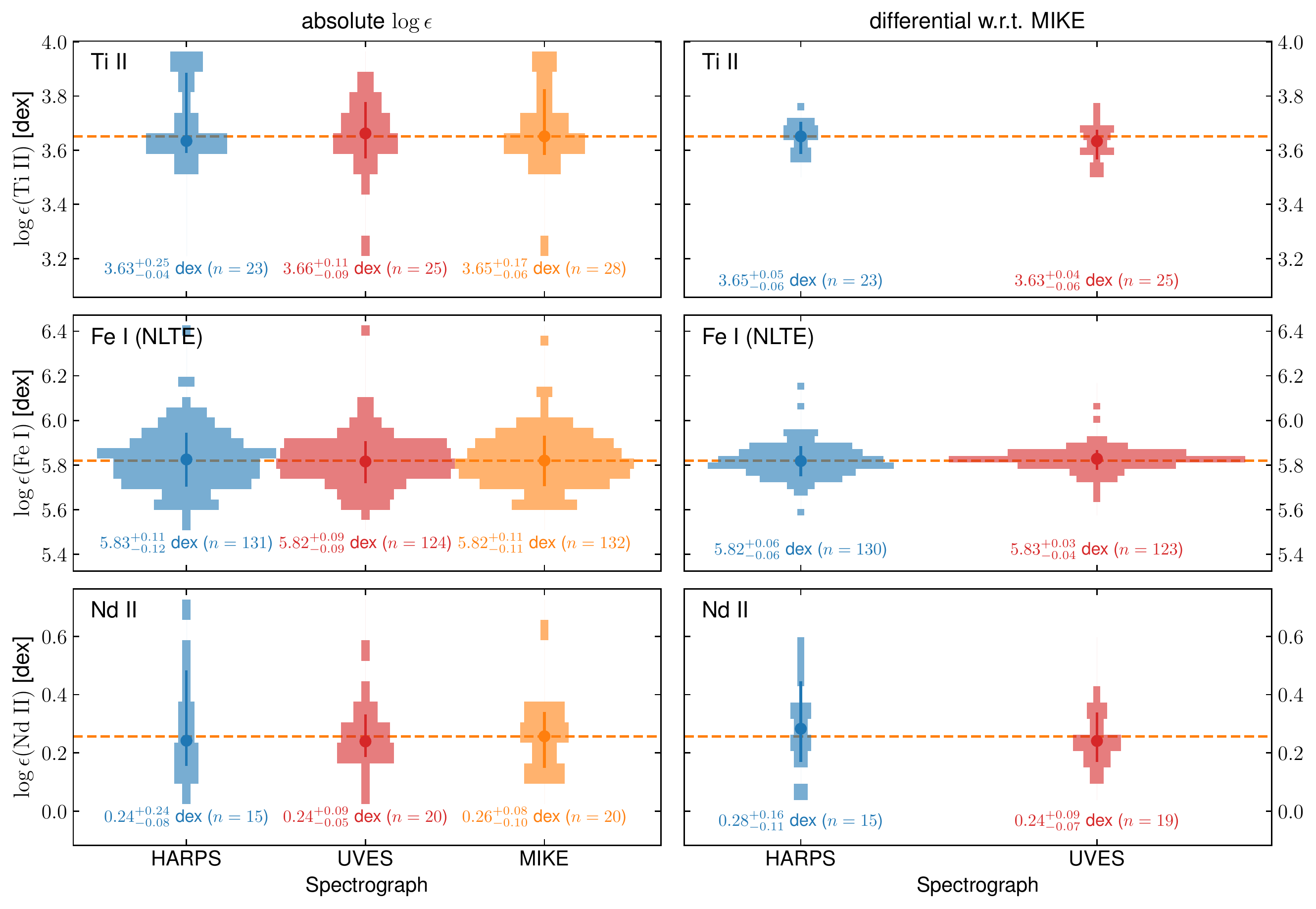}}
      \caption{Violin plots of absolute (\textit{left}) and line-by-line differential (\textit{right}) abundances for the same representative elements as in Fig. \ref{Fig:spectrgraph_comparison_corr}. Colors indicate spectrographs in the same way as in that figure with MIKE additionally being depicted in orange. Circles and vertical lines represent the median abundances, 15.9$^\mathrm{th}$, and 84.1$^\mathrm{th}$ percentiles, respectively. The latter are furthermore printed at the \textit{bottom} of each panel together with the number of involved lines, $n$.
              }
      \label{Fig:spectrgraph_comparison_hist}
\end{figure*}
    Once the EWs are propagated through the abundance analysis, it becomes obvious that noise is not the dominant source of error for the vast majority of lines when employing any of the tested instruments and their attributed S/N levels. This is illustrated in the lower panels of Fig. \ref{Fig:spectrgraph_comparison_corr}, where individual abundances from lines measured in the HARPS and UVES spectra are depicted as a function of their MIKE counterpart. NLTE corrections were applied to \ion{Fe}{i} and are expected to be negligible for the other two species. If spectrum noise were the sole reason for abundance errors the distributions would be completely uncorrelated and show ellipses that are aligned with the coordinate axes. Instead, we found strong correlations that imply governing systematic error components. We mention here possible origins for this observation to be uncertain oscillator strengths and/or shortcomings in the assumptions of one-dimensional and static atmospheres. 
    
    For Fig. \ref{Fig:spectrgraph_comparison_hist} we decoupled the systematic from the statistical component by performing line-by-line differential comparisons to the MIKE abundances. It is evident that the scatter in absolute abundances is hardly lower than 0.1~dex, while it is as low as 0.03~dex in the differential case for \ion{Fe}{i} and the UVES/MIKE combination. The spread in absolute abundances motivates the floor abundance error of 0.1~dex employed throughout this work in those cases ($n<4$) where the scatter could not be rigidly determined from the sample of lines themselves.
 
 \subsection{Impacts of model atmosphere errors}\label{Sec: Abundance systematics due to atmosphere}
\begin{figure}
    \centering
    \resizebox{0.76\hsize}{!}{\includegraphics{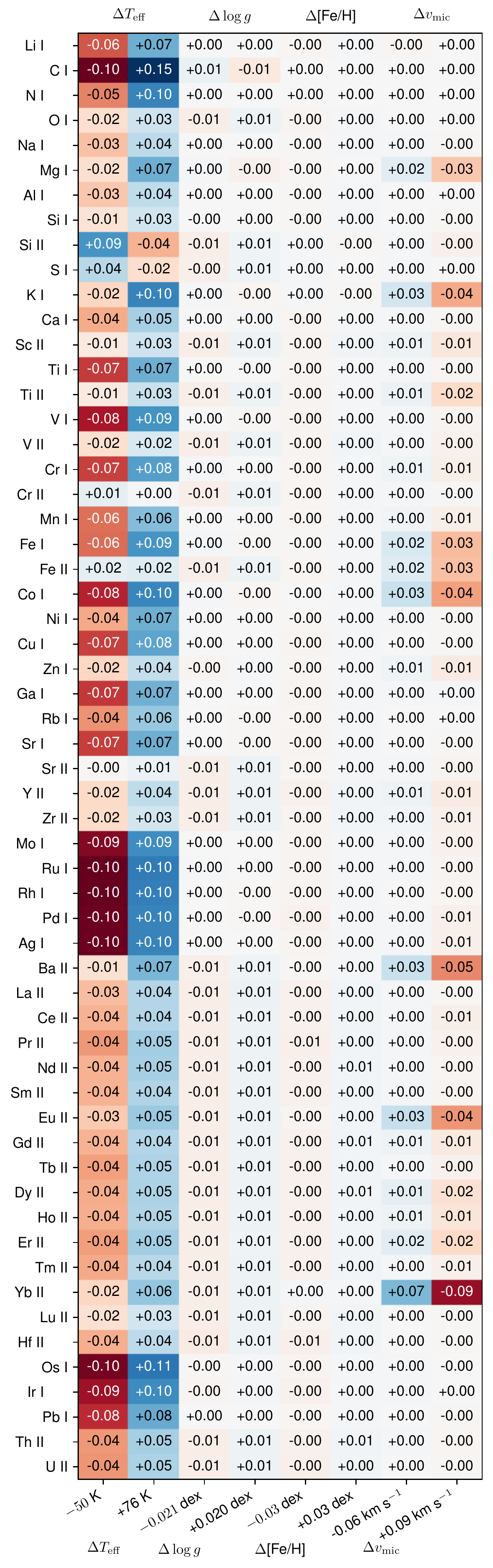}}
      \caption{Change in elemental abundances $\log{\epsilon}$ from individually varying the input model parameters by their error margins. Red and blue colors denote negative and positive residuals, respectively. The strength of the impact of an altered parameter (abscissa) on the elemental abundance (ordinate) is highlighted by the lightness of the color, where dark colors indicate strong departures. 
              }
      \label{Fig:systematics}
\end{figure}

 Here we present a detailed investigation of the propagation of errors on the key atmospheric parameters $T_\mathrm{eff}$, $\log{g}$, [Fe/H], and $v_\mathrm{mic}$ into the inferred individual stellar abundances (or upper limits) in LTE. To this end, eight model atmospheres were interpolated from the ATLAS grid, each denoting the departure of a stellar parameter from its optimal value by an amount dictated by our adopted errors (Table \ref{Table: Target information}). These altered atmospheres were used to redetermine the abundances from all transitions measured in this work based on their EW. In those cases where spectrum synthesis was used, a converted EW corresponding to the determined abundance was initially calculated through the MOOG driver \textit{ewfind} and the set of optimal atmospheric parameters. New average abundances were then derived using the median of all findings for one species. The resulting departures from the abundances listed in Table \ref{Table: Final abundances} can be found in Fig. \ref{Fig:systematics}.   

While the model metallicity can certainly be neglected as a factor of uncertainty, for the vast majority of elements, the model temperature appears to be the most critical parameter, in that changes induce the largest abundance deviations. Generally, the neutral species are more susceptible to $T_\mathrm{eff}$ than their ionized counterparts. Abundance deviations of the ionized species of the neutron-capture elements do not exceed the 0.05~dex level, therefore highlighting the robustness of the resulting pattern against model uncertainties.

 \begin{figure}
    \centering
    \resizebox{\hsize}{!}{\includegraphics{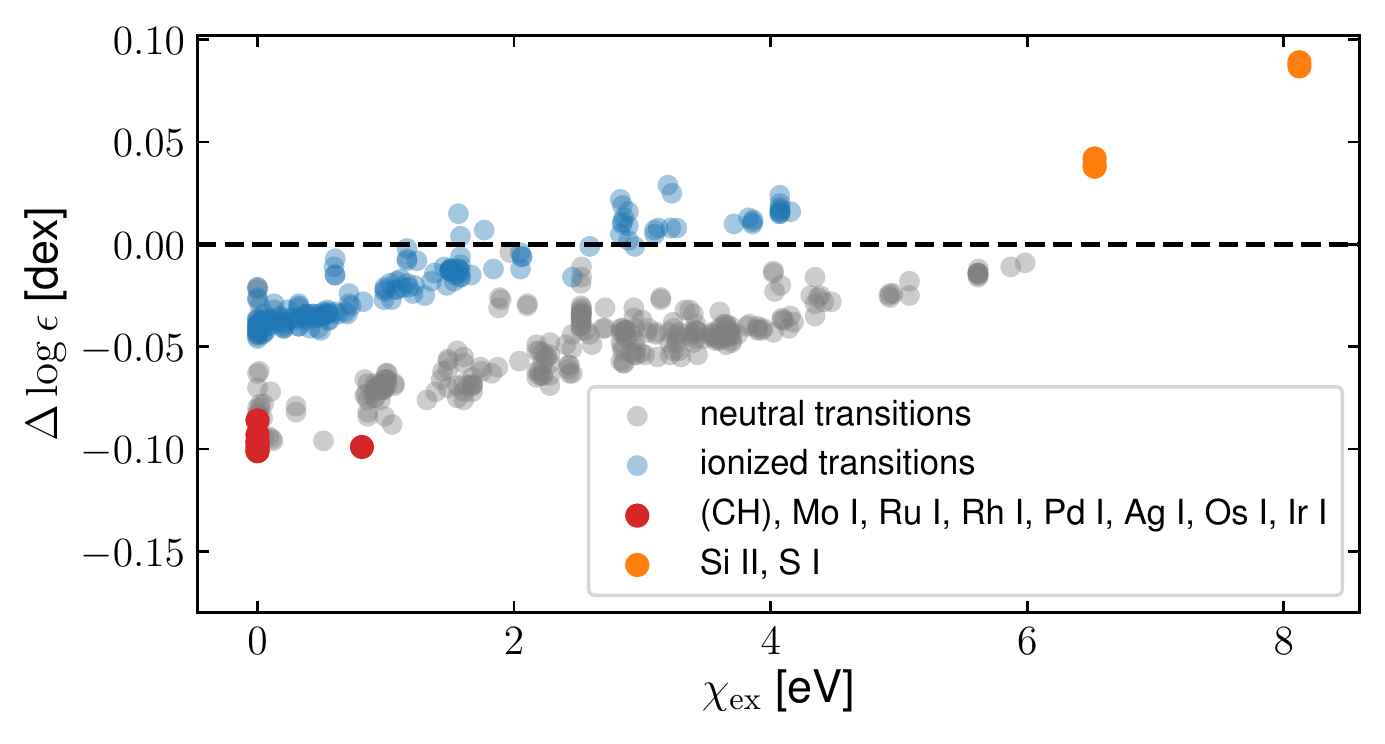}}
      \caption{Individual abundance changes from lowering the model $T_\mathrm{eff}$ by 50~K. Features from neutral species are shown in gray, whereas blue circles indicate ionized species. Highlighted in red and orange are the elements explicitly mentioned in the text. The manifold of CH lines used for synthesis and hence determination of the C abundance are not shown here. Their $\chi_\mathrm{ex}$ commonly resides around 0~eV.
              }
      \label{Fig:deviation_vs_ep}
\end{figure}
Interestingly, the overall trend of abundances correlating with temperature is reversed for \ion{Si}{ii} and \ion{S}{i}, where an anti-correlation is seen. We further note that considerable departures reaching or even exceeding the 0.10~dex level were found for C, Mo to Ag, \ion{Os}{i}, and \ion{Ir}{i}. Both effects can be linked to the lower energy level of the transitions as we show in Fig. \ref{Fig:deviation_vs_ep}. At the extreme end of temperature-related departures the lower level exclusively resides close to or at the ground level. In that regime, the number density is largely independent of temperature and the $T_\mathrm{eff}$ affects exclusively the H$^-$ continuous opacity with its strong temperature gradient. This leads to a strengthening of lines and  -- in turn -- lower abundances at fixed line strengths. With increasing $\chi_\mathrm{ex}$ the number density becomes susceptible to the $T_\mathrm{eff}$ change and increasingly counteracts the effect of the lower H$^-$ opacity. Hence, the abundance departures are reduced. For the high-$\chi_\mathrm{ex}$ lines, the impact of the change in number density exceeds the opacity effect, which leads to the inverse temperature dependence seen in Fig. \ref{Fig:deviation_vs_ep}.
 
Variations in the stellar surface gravity have their strongest effect on abundances of ionized species, though the overall magnitude remains low at $\sim\pm0.01$~dex. This can be understood in terms of gravity having a direct impact on the electron pressure which, in turn, determines the degree of ionization (Saha equation). Here, our \ion{O}{i} and \ion{S}{i} transitions behave as if they were ionized.

Deviations from changing $v_\mathrm{mic}$ exceed the 0.03~dex level in the mean abundances only for \ion{K}{i}, \ion{Co}{i}, \ion{Ba}{ii}, \ion{Eu}{ii}, and \ion{Yb}{ii}. The effect is limited to these species, as they show moderately strong lines with EWs of more than 80~m{\AA} and effects from microturbulence are limited to the higher line strength regime.  

\section{Additional tables}

\begin{table*}
\caption{Atomic transition parameters and abundances for individual lines.}
\label{Table: Line-by-line abundances}
\centering
\resizebox{2\columnwidth}{!}{%
\addtolength{\tabcolsep}{-3pt}
\begin{tabular}{llrrrrcrllllrrrrcrl}
\hline\hline\\[-7pt]
\multicolumn{1}{c}{$\lambda$} & \multicolumn{1}{c}{$X$} & $\chi_\mathrm{ex}$ & \multicolumn{1}{c}{$\log{gf}$} & \multicolumn{1}{c}{EW} & \multicolumn{2}{c}{$\log{\epsilon(X)}$} & $\Delta$ & Ref.$^b$ & & \multicolumn{1}{c}{$\lambda$} & \multicolumn{1}{c}{$X$} & $\chi_\mathrm{ex}$ & \multicolumn{1}{c}{$\log{gf}$} & \multicolumn{1}{c}{EW} & \multicolumn{2}{c}{$\log{\epsilon(X)}$} & $\Delta$ & Ref.$^b$\\[1pt]
\cline{6-7}\cline{16-17}\\[-5pt]
 & & & & & \multicolumn{1}{c}{LTE} & NLTE & & & & & & & & & \multicolumn{1}{c}{LTE} & NLTE & &\\
\multicolumn{1}{c}{$[${\AA}]}   &        &  [eV]  & \multicolumn{1}{c}{[dex]}  & \multicolumn{1}{c}{[m{\AA}]}  & \multicolumn{1}{c}{[dex]} & [dex] & \multicolumn{1}{c}{[dex]} & & & \multicolumn{1}{c}{$[${\AA}]}   &        &  [eV]  & \multicolumn{1}{c}{[dex]}  & \multicolumn{1}{c}{[m{\AA}]}  & \multicolumn{1}{c}{[dex]} & [dex] & \multicolumn{1}{c}{[dex]} & \\
\hline\\[-5pt]
6707.800$^{a}$ & \ion{Li}{i} & 0.000 & $0.174$ & $<0.3  $ & $<-0.34$ & \dots & \dots & 1 &  & 5667.151$^{a}$ & \ion{Sc}{ii} & 1.499 & $-1.309$ & $21.0 \pm0.8 $ & $1.80\pm0.02$ & \dots & \dots & 16,3\\
4300.000 & C (CH) & \dots & \dots & syn & $6.25\pm0.05$ & \dots & \dots & 2 &  & 5669.043$^{a}$ & \ion{Sc}{ii} & 1.499 & $-1.120$ & $23.5 \pm0.7 $ & $1.68\pm0.02$ & \dots & \dots & 3\\
3360.000 & N (NH) & \dots & \dots & syn & $6.21\pm0.10$ & \dots & \dots & 3 &  & 5684.202 & \ion{Sc}{ii} & 1.507 & $-1.074$ & $24.4 \pm0.6 $ & $1.66\pm0.02$ & \dots & \dots & 16\\
5577.339 & \ion{O}{i} & 1.967 & $-8.204$ & $2.8  \pm1.4 $ & $7.86\pm0.31$ & 7.86 & $0.00$ & 4 &  & 6245.639$^{a}$ & \ion{Sc}{ii} & 1.506 & $-0.980$ & $20.3 \pm0.6 $ & $1.44\pm0.02$ & \dots & \dots & 3\\
6363.776 & \ion{O}{i} & 0.020 & $-10.258$ & $2.7  \pm0.9 $ & $7.72\pm0.18$ & 7.72 & $0.00$ & 5 &  & 6309.920 & \ion{Sc}{ii} & 1.497 & $-1.618$ & $8.8  \pm1.0 $ & $1.64\pm0.06$ & \dots & \dots & 3\\
7771.944 & \ion{O}{i} & 9.146 & $0.369$ & $50.7 \pm0.7 $ & $8.27\pm0.01$ & 8.13 & $-0.14$ & 5 &  & 3717.391 & \ion{Ti}{i} & 0.000 & $-1.190$ & $23.5 \pm1.4 $ & $3.21\pm0.03$ & \dots & \dots & 17\\
7774.166 & \ion{O}{i} & 9.146 & $0.223$ & $40.8 \pm4.3 $ & $8.22\pm0.09$ & 8.08 & $-0.14$ & 5 &  & 3904.783 & \ion{Ti}{i} & 0.899 & $0.150$ & $44.2 \pm0.9 $ & $3.21\pm0.02$ & \dots & \dots & 17\\
7775.388 & \ion{O}{i} & 9.146 & $0.002$ & $29.4 \pm0.7 $ & $8.18\pm0.02$ & 8.06 & $-0.12$ & 5 &  & 4512.734 & \ion{Ti}{i} & 0.836 & $-0.400$ & $25.2 \pm1.4 $ & $3.27\pm0.03$ & 3.65 & $0.38$ & 17\\
5682.633 & \ion{Na}{i} & 2.102 & $-0.706$ & $10.1 \pm0.7 $ & $4.46\pm0.03$ & 4.39 & $-0.07$ & 5 &  & 4534.776 & \ion{Ti}{i} & 0.836 & $0.350$ & $63.7 \pm0.7 $ & $3.21\pm0.01$ & 3.61 & $0.40$ & 17\\
5688.205 & \ion{Na}{i} & 2.104 & $-0.452$ & $20.1 \pm0.7 $ & $4.55\pm0.02$ & 4.46 & $-0.09$ & 5 &  & 4617.269 & \ion{Ti}{i} & 1.749 & $0.440$ & $23.3 \pm0.7 $ & $3.36\pm0.02$ & 3.78 & $0.42$ & 17\\
4730.029 & \ion{Mg}{i} & 4.346 & $-2.379$ & $8.8  \pm0.8 $ & $6.13\pm0.04$ & 6.21 & $0.08$ & 6 &  & 4623.097 & \ion{Ti}{i} & 1.739 & $0.160$ & $12.9 \pm0.7 $ & $3.32\pm0.03$ & 3.78 & $0.46$ & 17\\
5528.405 & \ion{Mg}{i} & 4.346 & $-0.547$ & $120.6\pm0.8 $ & $6.25\pm0.01$ & 6.24 & $-0.01$ & 6 &  & 4758.118 & \ion{Ti}{i} & 2.249 & $0.510$ & $8.3  \pm0.7 $ & $3.29\pm0.04$ & 3.68 & $0.39$ & 17\\
5711.088 & \ion{Mg}{i} & 4.346 & $-1.842$ & $36.4 \pm0.7 $ & $6.32\pm0.01$ & 6.38 & $0.06$ & 6 &  & 4759.270 & \ion{Ti}{i} & 2.256 & $0.590$ & $13.7 \pm1.0 $ & $3.46\pm0.04$ & 3.85 & $0.39$ & 17\\
3961.520 & \ion{Al}{i} & 0.014 & $-0.323$ & syn & $3.58\pm0.20$ & 4.60 & $1.02$ & 7 &  & 4913.613 & \ion{Ti}{i} & 1.873 & $0.220$ & $13.1 \pm0.7 $ & $3.40\pm0.03$ & 3.81 & $0.41$ & 17\\
6696.023 & \ion{Al}{i} & 3.143 & $-1.347$ & syn & $4.54\pm0.20$ & 4.68 & $0.14$ & 8 &  & 4997.097 & \ion{Ti}{i} & 0.000 & $-2.070$ & $6.3  \pm0.7 $ & $3.31\pm0.05$ & 3.88 & $0.57$ & 17\\
6698.673 & \ion{Al}{i} & 3.143 & $-1.647$ & syn & $4.54\pm0.20$ & 4.69 & $0.15$ & 8 &  & 5016.161 & \ion{Ti}{i} & 0.848 & $-0.480$ & $24.1 \pm0.7 $ & $3.30\pm0.02$ & 3.67 & $0.37$ & 17\\
7835.309 & \ion{Al}{i} & 4.021 & $-0.649$ & syn & $4.44\pm0.40$ & 4.64 & $0.20$ & 3 &  & 5025.570 & \ion{Ti}{i} & 2.041 & $0.250$ & $9.7  \pm0.6 $ & $3.39\pm0.03$ & \dots & \dots & 18\\
7836.134 & \ion{Al}{i} & 4.021 & $-0.494$ & syn & $4.64\pm0.20$ & 4.79 & $0.15$ & 3 &  & 5039.957 & \ion{Ti}{i} & 0.021 & $-1.080$ & $38.7 \pm0.7 $ & $3.29\pm0.01$ & 3.82 & $0.53$ & 17\\
4947.607 & \ion{Si}{i} & 5.082 & $-1.760$ & $4.0  \pm0.8 $ & $5.88\pm0.10$ & \dots & \dots & 3 &  & 5087.058 & \ion{Ti}{i} & 1.430 & $-0.880$ & $4.0  \pm0.6 $ & $3.46\pm0.07$ & 3.87 & $0.41$ & 17\\
5665.555 & \ion{Si}{i} & 4.920 & $-2.040$ & $6.6  \pm0.5 $ & $6.20\pm0.04$ & 6.16 & $-0.04$ & 9 &  & 5147.478 & \ion{Ti}{i} & 0.000 & $-1.940$ & $9.6  \pm0.6 $ & $3.37\pm0.03$ & 3.94 & $0.57$ & 17\\
5666.677 & \ion{Si}{i} & 5.616 & $-1.797$ & $3.6  \pm0.7 $ & $6.40\pm0.10$ & \dots & \dots & 3 &  & 5192.969 & \ion{Ti}{i} & 0.021 & $-0.950$ & $53.7 \pm0.7 $ & $3.39\pm0.01$ & 3.93 & $0.54$ & 17\\
5684.484 & \ion{Si}{i} & 4.954 & $-1.650$ & $17.3 \pm0.7 $ & $6.30\pm0.02$ & 6.27 & $-0.03$ & 9 &  & 5210.384 & \ion{Ti}{i} & 0.048 & $-0.820$ & $52.3 \pm0.7 $ & $3.27\pm0.01$ & 3.79 & $0.52$ & 17\\
5690.425 & \ion{Si}{i} & 4.930 & $-1.870$ & $11.3 \pm0.7 $ & $6.29\pm0.03$ & 6.25 & $-0.04$ & 9 &  & 5866.451 & \ion{Ti}{i} & 1.067 & $-0.790$ & $11.3 \pm0.8 $ & $3.42\pm0.04$ & 3.93 & $0.51$ & 17\\
5701.104 & \ion{Si}{i} & 4.930 & $-2.050$ & $7.1  \pm0.6 $ & $6.25\pm0.04$ & 6.22 & $-0.03$ & 9 &  & 6126.216 & \ion{Ti}{i} & 1.067 & $-1.425$ & $3.6  \pm0.8 $ & $3.51\pm0.11$ & \dots & \dots & 19\\
5948.541 & \ion{Si}{i} & 5.082 & $-1.230$ & $26.0 \pm0.9 $ & $6.24\pm0.02$ & 6.23 & $-0.01$ & 9 &  & 6258.102 & \ion{Ti}{i} & 1.443 & $-0.390$ & $11.5 \pm0.6 $ & $3.42\pm0.03$ & \dots & \dots & 17\\
6125.021 & \ion{Si}{i} & 5.614 & $-1.465$ & $4.1  \pm0.7 $ & $6.12\pm0.08$ & \dots & \dots & 3 &  & 4316.794 & \ion{Ti}{ii} & 2.048 & $-1.620$ & $49.5 \pm1.0 $ & $3.72\pm0.02$ & \dots & \dots & 20\\
6142.483 & \ion{Si}{i} & 5.619 & $-1.296$ & $5.2  \pm0.8 $ & $6.06\pm0.08$ & \dots & \dots & 3 &  & 4320.950 & \ion{Ti}{ii} & 1.165 & $-1.880$ & $90.7 \pm2.1 $ & $3.79\pm0.04$ & \dots & \dots & 20\\
6145.016 & \ion{Si}{i} & 5.616 & $-1.311$ & $7.0  \pm0.7 $ & $6.21\pm0.05$ & \dots & \dots & 3 &  & 4395.839 & \ion{Ti}{ii} & 1.243 & $-1.930$ & $76.6 \pm0.9 $ & $3.62\pm0.02$ & \dots & \dots & 20\\
6155.134 & \ion{Si}{i} & 5.619 & $-0.755$ & $20.8 \pm0.9 $ & $6.20\pm0.02$ & \dots & \dots & 3 &  & 4398.271 & \ion{Ti}{ii} & 1.224 & $-2.780$ & $28.1 \pm1.0 $ & $3.60\pm0.02$ & \dots & \dots & 21\\
6237.319 & \ion{Si}{i} & 5.614 & $-0.975$ & $12.2 \pm0.7 $ & $6.14\pm0.03$ & \dots & \dots & 3 &  & 4417.714 & \ion{Ti}{ii} & 1.165 & $-1.190$ & $127.7\pm2.2 $ & $3.94\pm0.05$ & \dots & \dots & 22\\
6243.815 & \ion{Si}{i} & 5.616 & $-1.244$ & $6.2  \pm0.7 $ & $6.09\pm0.06$ & \dots & \dots & 3 &  & 4421.938 & \ion{Ti}{ii} & 2.061 & $-1.640$ & $49.1 \pm1.4 $ & $3.74\pm0.03$ & \dots & \dots & 20\\
6244.466 & \ion{Si}{i} & 5.616 & $-1.091$ & $7.0  \pm0.8 $ & $5.99\pm0.06$ & \dots & \dots & 3 &  & 4470.853 & \ion{Ti}{ii} & 1.165 & $-2.020$ & $73.7 \pm1.2 $ & $3.56\pm0.02$ & \dots & \dots & 22\\
6414.980 & \ion{Si}{i} & 5.871 & $-1.036$ & $7.8  \pm0.9 $ & $6.25\pm0.06$ & \dots & \dots & 3 &  & 4493.522 & \ion{Ti}{ii} & 1.080 & $-2.780$ & $30.4 \pm1.2 $ & $3.48\pm0.02$ & \dots & \dots & 20\\
6555.462 & \ion{Si}{i} & 5.984 & $-1.163$ & $7.4  \pm0.8 $ & $6.46\pm0.05$ & \dots & \dots & 3 &  & 4609.265 & \ion{Ti}{ii} & 1.180 & $-3.320$ & $12.2 \pm0.7 $ & $3.63\pm0.03$ & \dots & \dots & 20\\
6347.109 & \ion{Si}{ii} & 8.121 & $0.170$ & $16.1 \pm0.6 $ & $6.46\pm0.02$ & \dots & \dots & 10 &  & 4636.320 & \ion{Ti}{ii} & 1.165 & $-3.230$ & $20.1 \pm0.6 $ & $3.78\pm0.02$ & \dots & \dots & 21\\
6371.371 & \ion{Si}{ii} & 8.121 & $-0.040$ & $12.5 \pm0.7 $ & $6.52\pm0.03$ & \dots & \dots & 10 &  & 4794.814 & \ion{Ti}{ii} & 1.131 & $-4.190$ & $1.9  \pm0.4 $ & $3.59\pm0.10$ & \dots & \dots & 23\\
4694.113 & \ion{S}{i} & 6.525 & $-1.774$ & syn & $6.03\pm0.10$ & \dots & \dots & 3 &  & 4806.321 & \ion{Ti}{ii} & 1.084 & $-3.380$ & $5.5  \pm0.6 $ & $3.20\pm0.05$ & \dots & \dots & 21\\
4695.443 & \ion{S}{i} & 6.525 & $-1.919$ & syn & $5.98\pm0.20$ & \dots & \dots & 3 &  & 4865.610 & \ion{Ti}{ii} & 1.116 & $-2.700$ & $44.6 \pm0.8 $ & $3.66\pm0.01$ & \dots & \dots & 20\\
4696.252 & \ion{S}{i} & 6.525 & $-2.139$ & syn & $6.13\pm0.30$ & \dots & \dots & 3 &  & 4874.009 & \ion{Ti}{ii} & 3.095 & $-0.860$ & $24.1 \pm0.7 $ & $3.56\pm0.02$ & \dots & \dots & 20\\
6756.000 & \ion{S}{i} & \dots & \dots & syn & $6.03\pm0.10$ & \dots & \dots & 11 &  & 4911.194 & \ion{Ti}{ii} & 3.123 & $-0.640$ & $37.0 \pm0.7 $ & $3.63\pm0.01$ & \dots & \dots & 20\\
7664.899 & \ion{K}{i} & 0.000 & $0.149$ & $116.5\pm1.0 $ & $4.17\pm0.02$ & 3.62 & $-0.55$ & 3 &  & 4996.367 & \ion{Ti}{ii} & 1.582 & $-3.290$ & $9.1  \pm0.6 $ & $3.87\pm0.03$ & \dots & \dots & 21\\
7698.964 & \ion{K}{i} & 0.000 & $-0.154$ & $96.6 \pm1.1 $ & $4.14\pm0.02$ & 3.59 & $-0.55$ & 3 &  & 5005.167 & \ion{Ti}{ii} & 1.566 & $-2.730$ & $19.8 \pm0.6 $ & $3.68\pm0.02$ & \dots & \dots & 20\\
4094.925 & \ion{Ca}{i} & 2.523 & $-0.760$ & $23.2 \pm0.9 $ & $4.78\pm0.02$ & \dots & \dots & 3 &  & 5013.686 & \ion{Ti}{ii} & 1.582 & $-2.140$ & $47.4 \pm0.8 $ & $3.65\pm0.01$ & \dots & \dots & 20\\
4108.526 & \ion{Ca}{i} & 2.709 & $-0.740$ & $17.6 \pm8.4 $ & $4.80\pm0.32$ & \dots & \dots & 3 &  & 5037.809 & \ion{Ti}{ii} & 1.582 & $-3.586$ & $3.3  \pm0.5 $ & $3.70\pm0.07$ & \dots & \dots & 3\\
4578.551 & \ion{Ca}{i} & 2.521 & $-0.697$ & $29.2 \pm0.9 $ & $4.82\pm0.02$ & \dots & \dots & 12,13 &  & 5211.530 & \ion{Ti}{ii} & 2.590 & $-1.410$ & $24.7 \pm0.6 $ & $3.57\pm0.01$ & \dots & \dots & 20\\
4685.268 & \ion{Ca}{i} & 2.933 & $-0.879$ & $12.7 \pm0.9 $ & $4.98\pm0.04$ & \dots & \dots & 14 &  & 5226.538 & \ion{Ti}{ii} & 1.566 & $-1.260$ & $102.0\pm0.7 $ & $3.70\pm0.01$ & \dots & \dots & 22\\
5260.387 & \ion{Ca}{i} & 2.521 & $-1.719$ & $2.8  \pm0.5 $ & $4.67\pm0.09$ & \dots & \dots & 12,13 &  & 5336.786 & \ion{Ti}{ii} & 1.582 & $-1.600$ & $82.3 \pm0.9 $ & $3.67\pm0.02$ & \dots & \dots & 20\\
5261.704 & \ion{Ca}{i} & 2.521 & $-0.579$ & $39.6 \pm0.6 $ & $4.88\pm0.01$ & \dots & \dots & 12,13 &  & 5418.768 & \ion{Ti}{ii} & 1.582 & $-2.130$ & $52.5 \pm0.6 $ & $3.70\pm0.01$ & \dots & \dots & 20\\
5349.465 & \ion{Ca}{i} & 2.709 & $-0.310$ & $45.2 \pm0.9 $ & $4.90\pm0.02$ & \dots & \dots & 12,13 &  & 5454.099 & \ion{Ti}{ii} & 1.566 & $-3.600$ & $5.0  \pm0.5 $ & $3.86\pm0.05$ & \dots & \dots & 3\\
5512.980 & \ion{Ca}{i} & 2.933 & $-0.464$ & $27.2 \pm0.6 $ & $4.95\pm0.01$ & \dots & \dots & 14,13 &  & 5490.693 & \ion{Ti}{ii} & 1.566 & $-2.663$ & $22.3 \pm0.9 $ & $3.65\pm0.02$ & \dots & \dots & 3\\
5581.965 & \ion{Ca}{i} & 2.523 & $-0.555$ & $41.9 \pm0.6 $ & $4.89\pm0.01$ & \dots & \dots & 12,13 &  & 5492.862 & \ion{Ti}{ii} & 1.582 & $-3.310$ & $6.6  \pm0.6 $ & $3.72\pm0.04$ & \dots & \dots & 21\\
5588.749 & \ion{Ca}{i} & 2.526 & $0.358$ & $93.6 \pm0.8 $ & $4.88\pm0.02$ & \dots & \dots & 12,13 &  & 6606.950 & \ion{Ti}{ii} & 2.061 & $-2.790$ & $5.8  \pm0.8 $ & $3.60\pm0.07$ & \dots & \dots & 19\\
5590.114 & \ion{Ca}{i} & 2.521 & $-0.571$ & $42.5 \pm0.7 $ & $4.91\pm0.01$ & \dots & \dots & 12,13 &  & 6680.134 & \ion{Ti}{ii} & 3.095 & $-1.890$ & $4.9  \pm0.7 $ & $3.74\pm0.07$ & \dots & \dots & 20\\
5601.277 & \ion{Ca}{i} & 2.526 & $-0.523$ & $43.5 \pm0.6 $ & $4.89\pm0.01$ & \dots & \dots & 12,13 &  & 4111.779$^{a}$ & \ion{V}{i} & 0.300 & $0.400$ & $30.8 \pm1.1 $ & $2.01\pm0.02$ & \dots & \dots & 24\\
6102.723 & \ion{Ca}{i} & 1.879 & $-0.793$ & $77.2 \pm1.0 $ & $4.99\pm0.02$ & \dots & \dots & 15 &  & 4379.230$^{a}$ & \ion{V}{i} & 0.300 & $0.580$ & $42.1 \pm1.0 $ & $1.97\pm0.01$ & \dots & \dots & 24\\
6122.217 & \ion{Ca}{i} & 1.886 & $-0.316$ & $107.3\pm0.9 $ & $5.03\pm0.01$ & \dots & \dots & 15 &  & 4864.730$^{a}$ & \ion{V}{i} & 0.017 & $-0.960$ & $6.1  \pm0.8 $ & $2.23\pm0.06$ & \dots & \dots & 24\\
6161.297 & \ion{Ca}{i} & 2.523 & $-1.266$ & $13.3 \pm0.9 $ & $4.92\pm0.03$ & \dots & \dots & 12,13 &  & 4875.486$^{a}$ & \ion{V}{i} & 0.040 & $-0.790$ & $6.9  \pm0.7 $ & $2.13\pm0.05$ & \dots & \dots & 24\\
6162.173 & \ion{Ca}{i} & 1.899 & $-0.090$ & $121.0\pm1.0 $ & $5.03\pm0.02$ & \dots & \dots & 15 &  & 4036.764 & \ion{V}{ii} & 1.476 & $-1.570$ & $18.6 \pm1.0 $ & $2.37\pm0.03$ & \dots & \dots & 25\\
6166.439 & \ion{Ca}{i} & 2.521 & $-1.142$ & $17.6 \pm0.8 $ & $4.94\pm0.02$ & \dots & \dots & 12,13 &  & 4183.428 & \ion{V}{ii} & 2.050 & $-1.060$ & $25.3 \pm1.0 $ & $2.62\pm0.02$ & \dots & \dots & 25\\
6169.042 & \ion{Ca}{i} & 2.523 & $-0.797$ & $34.0 \pm0.9 $ & $4.97\pm0.02$ & \dots & \dots & 12,13 &  & 4616.124 & \ion{Cr}{i} & 0.983 & $-1.190$ & $39.9 \pm2.9 $ & $3.68\pm0.05$ & \dots & \dots & 26\\
6169.563 & \ion{Ca}{i} & 2.526 & $-0.478$ & $49.8 \pm0.9 $ & $4.93\pm0.01$ & \dots & \dots & 12,13 &  & 4626.174 & \ion{Cr}{i} & 0.968 & $-1.330$ & $32.1 \pm0.7 $ & $3.66\pm0.01$ & \dots & \dots & 26\\
6439.075 & \ion{Ca}{i} & 2.526 & $0.390$ & $108.2\pm0.8 $ & $5.08\pm0.02$ & \dots & \dots & 12,13 &  & 4651.284 & \ion{Cr}{i} & 0.983 & $-1.460$ & $26.4 \pm0.8 $ & $3.69\pm0.02$ & \dots & \dots & 26\\
6449.808 & \ion{Ca}{i} & 2.521 & $-0.502$ & $45.9 \pm0.8 $ & $4.88\pm0.01$ & \dots & \dots & 12,13 &  & 4652.157 & \ion{Cr}{i} & 1.004 & $-0.913$ & $48.7 \pm0.8 $ & $3.58\pm0.01$ & \dots & \dots & 3\\
6455.598 & \ion{Ca}{i} & 2.523 & $-1.340$ & $10.0 \pm3.3 $ & $4.85\pm0.19$ & \dots & \dots & 14,13 &  & 4756.112 & \ion{Cr}{i} & 3.104 & $0.090$ & $15.1 \pm0.9 $ & $4.10\pm0.03$ & \dots & \dots & 19\\
6471.662 & \ion{Ca}{i} & 2.526 & $-0.686$ & $38.5 \pm0.7 $ & $4.94\pm0.01$ & \dots & \dots & 12,13 &  & 4936.335 & \ion{Cr}{i} & 3.113 & $-0.250$ & $4.1  \pm0.6 $ & $3.82\pm0.07$ & \dots & \dots & 26\\
6493.781 & \ion{Ca}{i} & 2.521 & $-0.109$ & $76.7 \pm0.9 $ & $5.00\pm0.02$ & \dots & \dots & 12,13 &  & 5247.565 & \ion{Cr}{i} & 0.961 & $-1.590$ & $23.3 \pm0.7 $ & $3.69\pm0.02$ & \dots & \dots & 26\\
6499.650 & \ion{Ca}{i} & 2.523 & $-0.818$ & $31.9 \pm0.8 $ & $4.94\pm0.02$ & \dots & \dots & 12,13 &  & 5296.691 & \ion{Cr}{i} & 0.983 & $-1.360$ & $32.3 \pm0.6 $ & $3.67\pm0.01$ & \dots & \dots & 26\\
4320.732 & \ion{Sc}{ii} & 0.606 & $-0.252$ & $112.9\pm2.1 $ & $1.65\pm0.05$ & \dots & \dots & 16 &  & 5300.745 & \ion{Cr}{i} & 0.983 & $-2.016$ & $6.1  \pm0.6 $ & $3.47\pm0.05$ & \dots & \dots & 3\\
4415.557 & \ion{Sc}{ii} & 0.596 & $-0.668$ & $98.4 \pm1.6 $ & $1.70\pm0.03$ & \dots & \dots & 16 &  & 5329.138 & \ion{Cr}{i} & 2.913 & $-0.008$ & $10.9 \pm0.6 $ & $3.81\pm0.03$ & \dots & \dots & 3\\
5239.813 & \ion{Sc}{ii} & 1.455 & $-0.765$ & $42.6 \pm0.7 $ & $1.66\pm0.01$ & \dots & \dots & 16 &  & 5345.796 & \ion{Cr}{i} & 1.004 & $-0.896$ & $53.0 \pm0.7 $ & $3.58\pm0.01$ & \dots & \dots & 3\\
5318.346$^{a}$ & \ion{Sc}{ii} & 1.356 & $-1.793$ & $6.7  \pm0.7 $ & $1.61\pm0.05$ & \dots & \dots & 16,3 &  & 5348.314 & \ion{Cr}{i} & 1.004 & $-1.210$ & $35.9 \pm0.7 $ & $3.61\pm0.01$ & \dots & \dots & 26\\
5526.783$^{a}$ & \ion{Sc}{ii} & 1.767 & $0.134$ & $70.8 \pm0.7 $ & $1.54\pm0.01$ & \dots & \dots & 16,3 &  & 3484.147 & \ion{Cr}{ii} & 2.455 & $-2.139$ & $52.4 \pm2.1 $ & $4.06\pm0.04$ & \dots & \dots & 27\\
5641.002$^{a}$ & \ion{Sc}{ii} & 1.499 & $-1.353$ & $29.1 \pm1.2 $ & $2.03\pm0.02$ & \dots & \dots & 16,3 &  & 4588.199 & \ion{Cr}{ii} & 4.071 & $-0.627$ & $56.0 \pm0.9 $ & $4.09\pm0.02$ & \dots & \dots & 28\\

\hline
\end{tabular}
\addtolength{\tabcolsep}{3pt}}
\end{table*}

\begin{table*}\ContinuedFloat
\caption[]{continued.}
\centering
\resizebox{2\columnwidth}{!}{%
\addtolength{\tabcolsep}{-3pt}
\begin{tabular}{llrrrrcrllllrrrrcrl}
\hline\hline\\[-7pt]
\multicolumn{1}{c}{$\lambda$} & \multicolumn{1}{c}{$X$} & $\chi_\mathrm{ex}$ & \multicolumn{1}{c}{$\log{gf}$} & \multicolumn{1}{c}{EW} & \multicolumn{2}{c}{$\log{\epsilon(X)}$} & $\Delta$ & Ref.$^b$ & & \multicolumn{1}{c}{$\lambda$} & \multicolumn{1}{c}{$X$} & $\chi_\mathrm{ex}$ & \multicolumn{1}{c}{$\log{gf}$} & \multicolumn{1}{c}{EW} & \multicolumn{2}{c}{$\log{\epsilon(X)}$} & $\Delta$ & Ref.$^b$\\[1pt]
\cline{6-7}\cline{16-17}\\[-5pt]
 & & & & & \multicolumn{1}{c}{LTE} & NLTE & & & & & & & & & \multicolumn{1}{c}{LTE} & NLTE & &\\
\multicolumn{1}{c}{$[${\AA}]}   &        &  [eV]  & \multicolumn{1}{c}{[dex]}  & \multicolumn{1}{c}{[m{\AA}]}  & \multicolumn{1}{c}{[dex]} & [dex] & \multicolumn{1}{c}{[dex]} & & & \multicolumn{1}{c}{$[${\AA}]}   &        &  [eV]  & \multicolumn{1}{c}{[dex]}  & \multicolumn{1}{c}{[m{\AA}]}  & \multicolumn{1}{c}{[dex]} & [dex] & \multicolumn{1}{c}{[dex]} & \\
\hline\\[-5pt]
4592.049 & \ion{Cr}{ii} & 4.074 & $-1.221$ & $26.1 \pm1.2 $ & $4.12\pm0.03$ & \dots & \dots & 28 &  & 5083.338 & \ion{Fe}{i} & 0.958 & $-2.958$ & $82.7 \pm0.6 $ & $5.69\pm0.01$ & 5.89 & $0.19$ & 33\\
4634.070 & \ion{Cr}{ii} & 4.072 & $-0.990$ & $37.1 \pm1.0 $ & $4.11\pm0.02$ & \dots & \dots & 29 &  & 5110.413 & \ion{Fe}{i} & 0.000 & $-3.760$ & $103.1\pm0.9 $ & $5.84\pm0.02$ & 6.09 & $0.25$ & 33\\
4812.337 & \ion{Cr}{ii} & 3.864 & $-1.960$ & $13.4 \pm0.7 $ & $4.27\pm0.03$ & \dots & \dots & 29 &  & 5123.720 & \ion{Fe}{i} & 1.011 & $-3.068$ & $81.0 \pm0.8 $ & $5.83\pm0.02$ & 6.01 & $0.18$ & 33\\
4836.229 & \ion{Cr}{ii} & 3.858 & $-1.960$ & $14.6 \pm0.7 $ & $4.31\pm0.03$ & \dots & \dots & 29 &  & 5127.359 & \ion{Fe}{i} & 0.915 & $-3.307$ & $67.3 \pm0.7 $ & $5.71\pm0.01$ & 5.92 & $0.22$ & 33\\
4884.607 & \ion{Cr}{ii} & 3.858 & $-2.100$ & $6.3  \pm0.7 $ & $4.04\pm0.06$ & \dots & \dots & 29 &  & 5166.282 & \ion{Fe}{i} & 0.000 & $-4.195$ & $75.3 \pm0.7 $ & $5.71\pm0.01$ & 5.94 & $0.23$ & 33\\
5232.496 & \ion{Cr}{ii} & 4.071 & $-2.360$ & $7.7  \pm1.3 $ & $4.61\pm0.09$ & \dots & \dots & 27 &  & 5171.596 & \ion{Fe}{i} & 1.485 & $-1.793$ & $111.9\pm0.9 $ & $5.72\pm0.02$ & 5.89 & $0.17$ & 33\\
5237.328 & \ion{Cr}{ii} & 4.073 & $-1.350$ & $31.6 \pm0.6 $ & $4.35\pm0.01$ & \dots & \dots & 27 &  & 5194.941 & \ion{Fe}{i} & 1.557 & $-2.090$ & $92.8 \pm0.8 $ & $5.68\pm0.02$ & 5.87 & $0.19$ & 33\\
5246.768 & \ion{Cr}{ii} & 3.714 & $-2.466$ & $6.2  \pm0.7 $ & $4.24\pm0.06$ & \dots & \dots & 28 &  & 5198.711 & \ion{Fe}{i} & 2.223 & $-2.135$ & $51.9 \pm0.7 $ & $5.71\pm0.01$ & 5.90 & $0.19$ & 33\\
5305.853 & \ion{Cr}{ii} & 3.827 & $-2.160$ & $9.5  \pm0.6 $ & $4.25\pm0.03$ & \dots & \dots & 27 &  & 5202.335 & \ion{Fe}{i} & 2.176 & $-1.838$ & $82.3 \pm0.8 $ & $5.89\pm0.01$ & 6.09 & $0.20$ & 33\\
5308.408 & \ion{Cr}{ii} & 4.071 & $-2.058$ & $7.9  \pm0.7 $ & $4.32\pm0.04$ & \dots & \dots & 27 &  & 5215.180 & \ion{Fe}{i} & 3.266 & $-0.871$ & $54.0 \pm0.6 $ & $5.60\pm0.01$ & 5.76 & $0.16$ & 31\\
5313.563 & \ion{Cr}{ii} & 4.074 & $-1.779$ & $13.8 \pm0.5 $ & $4.31\pm0.02$ & \dots & \dots & 27 &  & 5216.274 & \ion{Fe}{i} & 1.608 & $-2.150$ & $86.2 \pm1.0 $ & $5.66\pm0.02$ & 5.85 & $0.18$ & 33\\
5334.869 & \ion{Cr}{ii} & 4.072 & $-1.826$ & $10.7 \pm0.6 $ & $4.23\pm0.03$ & \dots & \dots & 27 &  & 5217.389 & \ion{Fe}{i} & 3.211 & $-1.070$ & $48.5 \pm0.8 $ & $5.65\pm0.01$ & 5.82 & $0.17$ & 34\\
5508.606 & \ion{Cr}{ii} & 4.156 & $-2.110$ & $6.8  \pm0.8 $ & $4.38\pm0.06$ & \dots & \dots & 19 &  & 5225.526 & \ion{Fe}{i} & 0.110 & $-4.789$ & $27.1 \pm1.1 $ & $5.60\pm0.02$ & 5.60 & $0.00$ & 33\\
4754.043$^{a}$ & \ion{Mn}{i} & 2.280 & $-0.080$ & $40.4 \pm0.7 $ & $3.27\pm0.01$ & 3.68 & $0.41$ & 30 &  & 5232.940 & \ion{Fe}{i} & 2.940 & $-0.058$ & $116.3\pm1.0 $ & $5.58\pm0.02$ & 5.59 & $0.01$ & 32\\
4761.512$^{a}$ & \ion{Mn}{i} & 2.951 & $-0.274$ & $10.5 \pm0.9 $ & $3.48\pm0.04$ & 3.85 & $0.38$ & 30 &  & 5242.490 & \ion{Fe}{i} & 3.634 & $-0.967$ & $32.6 \pm0.7 $ & $5.72\pm0.01$ & 5.89 & $0.17$ & 32\\
4765.846$^{a}$ & \ion{Mn}{i} & 2.939 & $-0.086$ & $10.4 \pm1.0 $ & $3.27\pm0.05$ & 3.66 & $0.39$ & 30 &  & 5247.050 & \ion{Fe}{i} & 0.087 & $-4.946$ & $25.1 \pm0.9 $ & $5.68\pm0.02$ & 5.91 & $0.23$ & 33\\
4766.418$^{a}$ & \ion{Mn}{i} & 2.918 & $0.105$ & $19.9 \pm1.0 $ & $3.38\pm0.03$ & 3.77 & $0.39$ & 30 &  & 5250.209 & \ion{Fe}{i} & 0.121 & $-4.938$ & $24.3 \pm0.8 $ & $5.70\pm0.02$ & 5.80 & $0.10$ & 33\\
4072.502 & \ion{Fe}{i} & 3.430 & $-1.439$ & $24.6 \pm12.4$ & $5.87\pm0.37$ & 6.00 & $0.13$ & 31 &  & 5253.462 & \ion{Fe}{i} & 3.283 & $-1.573$ & $19.5 \pm0.6 $ & $5.65\pm0.02$ & 5.81 & $0.15$ & 31\\
4073.762 & \ion{Fe}{i} & 3.266 & $-0.902$ & $56.3 \pm1.0 $ & $5.76\pm0.02$ & 5.93 & $0.17$ & 32 &  & 5263.306 & \ion{Fe}{i} & 3.266 & $-0.879$ & $55.5 \pm0.8 $ & $5.64\pm0.01$ & 5.80 & $0.16$ & 34\\
4079.838 & \ion{Fe}{i} & 2.858 & $-1.360$ & $48.3 \pm1.1 $ & $5.65\pm0.02$ & 5.81 & $0.17$ & 32 &  & 5266.554 & \ion{Fe}{i} & 2.998 & $-0.386$ & $97.4 \pm0.8 $ & $5.61\pm0.02$ & 5.65 & $0.04$ & 32\\
4098.175 & \ion{Fe}{i} & 3.241 & $-0.879$ & $53.9 \pm0.9 $ & $5.67\pm0.02$ & 5.78 & $0.11$ & 32 &  & 5269.537 & \ion{Fe}{i} & 0.859 & $-1.321$ & $165.4\pm0.9 $ & $5.56\pm0.01$ & 5.60 & $0.04$ & 33\\
4109.801 & \ion{Fe}{i} & 2.845 & $-0.895$ & $73.1 \pm9.2 $ & $5.63\pm0.18$ & 5.79 & $0.16$ & 31 &  & 5281.789 & \ion{Fe}{i} & 3.038 & $-0.834$ & $72.3 \pm0.7 $ & $5.63\pm0.01$ & 5.75 & $0.12$ & 32\\
4120.206 & \ion{Fe}{i} & 2.990 & $-1.267$ & $53.6 \pm0.9 $ & $5.79\pm0.01$ & 5.95 & $0.16$ & 32 &  & 5283.621 & \ion{Fe}{i} & 3.241 & $-0.432$ & $80.6 \pm0.8 $ & $5.60\pm0.01$ & 5.74 & $0.14$ & 34\\
4121.802 & \ion{Fe}{i} & 2.832 & $-1.450$ & $51.9 \pm0.8 $ & $5.77\pm0.01$ & 5.93 & $0.17$ & 32 &  & 5288.525 & \ion{Fe}{i} & 3.695 & $-1.508$ & $9.4  \pm0.7 $ & $5.66\pm0.04$ & 5.82 & $0.16$ & 32\\
4132.899 & \ion{Fe}{i} & 2.845 & $-1.006$ & $63.0 \pm1.6 $ & $5.54\pm0.03$ & 5.68 & $0.15$ & 32 &  & 5302.300 & \ion{Fe}{i} & 3.283 & $-0.720$ & $62.0 \pm0.8 $ & $5.60\pm0.01$ & 5.75 & $0.15$ & 34\\
4136.998 & \ion{Fe}{i} & 3.415 & $-0.453$ & $60.8 \pm0.9 $ & $5.55\pm0.02$ & 5.73 & $0.18$ & 32 &  & 5307.360 & \ion{Fe}{i} & 1.608 & $-2.987$ & $42.4 \pm0.7 $ & $5.72\pm0.01$ & 5.72 & $0.00$ & 33\\
4139.927 & \ion{Fe}{i} & 0.990 & $-3.629$ & $37.4 \pm1.0 $ & $5.70\pm0.02$ & 5.88 & $0.18$ & 33 &  & 5332.899 & \ion{Fe}{i} & 1.557 & $-2.777$ & $51.9 \pm0.8 $ & $5.61\pm0.01$ & 5.61 & $0.00$ & 32\\
4143.414 & \ion{Fe}{i} & 3.047 & $-0.204$ & $103.4\pm1.1 $ & $5.85\pm0.03$ & 5.99 & $0.13$ & 32 &  & 5341.023 & \ion{Fe}{i} & 1.608 & $-1.953$ & $103.1\pm0.9 $ & $5.80\pm0.02$ & 5.81 & $0.00$ & 32\\
4143.868 & \ion{Fe}{i} & 1.557 & $-0.511$ & $151.5\pm1.1 $ & $5.52\pm0.02$ & 5.65 & $0.13$ & 32 &  & 5365.398 & \ion{Fe}{i} & 3.573 & $-1.020$ & $26.9 \pm0.7 $ & $5.58\pm0.02$ & 5.78 & $0.19$ & 32\\
4147.669 & \ion{Fe}{i} & 1.485 & $-2.104$ & $88.8 \pm0.9 $ & $5.70\pm0.02$ & 5.77 & $0.07$ & 33 &  & 5367.465 & \ion{Fe}{i} & 4.415 & $0.443$ & $60.0 \pm0.8 $ & $5.62\pm0.01$ & 5.84 & $0.22$ & 32\\
4153.899 & \ion{Fe}{i} & 3.397 & $-0.321$ & $72.8 \pm1.5 $ & $5.61\pm0.03$ & 5.79 & $0.17$ & 32 &  & 5369.961 & \ion{Fe}{i} & 4.371 & $0.536$ & $64.8 \pm0.9 $ & $5.56\pm0.02$ & 5.79 & $0.23$ & 32\\
4154.498 & \ion{Fe}{i} & 2.832 & $-0.688$ & $78.8 \pm1.2 $ & $5.52\pm0.03$ & 5.69 & $0.17$ & 32 &  & 5371.489 & \ion{Fe}{i} & 0.958 & $-1.645$ & $150.5\pm0.9 $ & $5.72\pm0.02$ & 5.72 & $-0.01$ & 33\\
4154.805 & \ion{Fe}{i} & 3.368 & $-0.400$ & $71.8 \pm1.2 $ & $5.64\pm0.02$ & 5.82 & $0.17$ & 32 &  & 5379.573 & \ion{Fe}{i} & 3.695 & $-1.514$ & $11.3 \pm0.6 $ & $5.75\pm0.03$ & 5.92 & $0.17$ & 32\\
4157.779 & \ion{Fe}{i} & 3.417 & $-0.403$ & $68.3 \pm1.0 $ & $5.64\pm0.02$ & 5.74 & $0.10$ & 32 &  & 5383.368 & \ion{Fe}{i} & 4.312 & $0.645$ & $73.7 \pm0.7 $ & $5.55\pm0.01$ & 5.80 & $0.25$ & 32\\
4174.912 & \ion{Fe}{i} & 0.915 & $-2.969$ & $74.4 \pm1.0 $ & $5.62\pm0.02$ & 5.80 & $0.18$ & 33 &  & 5393.167 & \ion{Fe}{i} & 3.241 & $-0.715$ & $69.0 \pm0.9 $ & $5.67\pm0.02$ & 5.67 & $0.00$ & 34\\
4175.636 & \ion{Fe}{i} & 2.845 & $-0.827$ & $79.4 \pm1.0 $ & $5.68\pm0.02$ & 5.85 & $0.17$ & 32 &  & 5397.127 & \ion{Fe}{i} & 0.915 & $-1.993$ & $132.4\pm0.8 $ & $5.68\pm0.02$ & 5.69 & $0.01$ & 33\\
4184.891 & \ion{Fe}{i} & 2.832 & $-0.869$ & $76.2 \pm0.8 $ & $5.65\pm0.02$ & 5.81 & $0.16$ & 32 &  & 5405.774 & \ion{Fe}{i} & 0.990 & $-1.844$ & $134.0\pm0.9 $ & $5.64\pm0.02$ & 5.63 & $-0.01$ & 33\\
4217.545 & \ion{Fe}{i} & 3.430 & $-0.484$ & $69.2 \pm1.1 $ & $5.74\pm0.02$ & 5.92 & $0.18$ & 32 &  & 5410.909 & \ion{Fe}{i} & 4.473 & $0.398$ & $53.2 \pm0.8 $ & $5.61\pm0.01$ & 5.83 & $0.22$ & 32\\
4222.213 & \ion{Fe}{i} & 2.450 & $-0.967$ & $93.0 \pm1.1 $ & $5.66\pm0.02$ & 5.75 & $0.10$ & 33 &  & 5415.198 & \ion{Fe}{i} & 4.386 & $0.642$ & $69.2 \pm0.7 $ & $5.55\pm0.01$ & 5.80 & $0.25$ & 32\\
4224.171 & \ion{Fe}{i} & 3.368 & $-0.506$ & $64.7 \pm1.2 $ & $5.61\pm0.02$ & 5.78 & $0.17$ & 32 &  & 5434.523 & \ion{Fe}{i} & 1.011 & $-2.122$ & $123.1\pm1.0 $ & $5.72\pm0.02$ & 5.74 & $0.02$ & 33\\
4375.929 & \ion{Fe}{i} & 0.000 & $-3.031$ & $118.7\pm1.5 $ & $5.66\pm0.04$ & 5.85 & $0.19$ & 33 &  & 5497.516 & \ion{Fe}{i} & 1.011 & $-2.849$ & $95.8 \pm0.7 $ & $5.86\pm0.01$ & 5.87 & $0.01$ & 33\\
4388.407 & \ion{Fe}{i} & 3.603 & $-0.682$ & $47.4 \pm1.0 $ & $5.71\pm0.02$ & 5.90 & $0.19$ & 32 &  & 5501.465 & \ion{Fe}{i} & 0.958 & $-3.047$ & $82.7 \pm0.9 $ & $5.74\pm0.02$ & 5.74 & $0.00$ & 32\\
4389.244 & \ion{Fe}{i} & 0.052 & $-4.583$ & $39.4 \pm1.0 $ & $5.63\pm0.02$ & 5.83 & $0.20$ & 33 &  & 5506.778 & \ion{Fe}{i} & 0.990 & $-2.797$ & $92.4 \pm0.8 $ & $5.71\pm0.01$ & 5.75 & $0.03$ & 33\\
4439.881 & \ion{Fe}{i} & 2.279 & $-3.002$ & $10.3 \pm1.1 $ & $5.73\pm0.06$ & 5.81 & $0.08$ & 33 &  & 5569.618 & \ion{Fe}{i} & 3.417 & $-0.486$ & $67.5 \pm0.7 $ & $5.59\pm0.01$ & 5.73 & $0.14$ & 31\\
4447.717 & \ion{Fe}{i} & 2.223 & $-1.342$ & $90.9 \pm1.1 $ & $5.70\pm0.02$ & 5.86 & $0.17$ & 33 &  & 5572.842 & \ion{Fe}{i} & 3.397 & $-0.275$ & $83.8 \pm0.9 $ & $5.65\pm0.02$ & 5.77 & $0.12$ & 34\\
4476.018 & \ion{Fe}{i} & 2.845 & $-0.819$ & $105.4\pm1.4 $ & $6.22\pm0.03$ & 6.38 & $0.15$ & 32 &  & 5586.755 & \ion{Fe}{i} & 3.368 & $-0.120$ & $91.5 \pm0.7 $ & $5.61\pm0.01$ & 5.71 & $0.10$ & 32\\
4484.220 & \ion{Fe}{i} & 3.603 & $-0.864$ & $41.5 \pm1.4 $ & $5.79\pm0.03$ & 5.95 & $0.17$ & 32 &  & 5615.644 & \ion{Fe}{i} & 3.332 & $0.050$ & $102.1\pm1.9 $ & $5.59\pm0.04$ & 5.72 & $0.13$ & 34\\
4494.563 & \ion{Fe}{i} & 2.198 & $-1.136$ & $103.5\pm1.5 $ & $5.71\pm0.03$ & 5.88 & $0.16$ & 33 &  & 5624.542 & \ion{Fe}{i} & 3.417 & $-0.755$ & $54.7 \pm0.9 $ & $5.65\pm0.02$ & 5.81 & $0.16$ & 34\\
4595.358 & \ion{Fe}{i} & 3.301 & $-1.758$ & $21.5 \pm0.9 $ & $5.94\pm0.02$ & 6.11 & $0.17$ & 32 &  & 5662.516 & \ion{Fe}{i} & 4.178 & $-0.573$ & $28.4 \pm0.7 $ & $5.80\pm0.02$ & 5.98 & $0.17$ & 32\\
4602.001 & \ion{Fe}{i} & 1.608 & $-3.154$ & $29.1 \pm0.8 $ & $5.69\pm0.02$ & 5.80 & $0.11$ & 33 &  & 5916.247 & \ion{Fe}{i} & 2.453 & $-2.994$ & $10.6 \pm0.8 $ & $5.84\pm0.04$ & 5.85 & $0.01$ & 33\\
4602.940 & \ion{Fe}{i} & 1.485 & $-2.209$ & $90.2 \pm0.9 $ & $5.75\pm0.02$ & 5.86 & $0.11$ & 32 &  & 5956.693 & \ion{Fe}{i} & 0.859 & $-4.605$ & $10.1 \pm0.7 $ & $5.69\pm0.03$ & 5.90 & $0.22$ & 33\\
4632.911 & \ion{Fe}{i} & 1.608 & $-2.913$ & $47.7 \pm0.7 $ & $5.79\pm0.01$ & 5.98 & $0.19$ & 33 &  & 6065.481 & \ion{Fe}{i} & 2.608 & $-1.530$ & $65.0 \pm0.8 $ & $5.70\pm0.01$ & 5.87 & $0.17$ & 33\\
4643.463 & \ion{Fe}{i} & 3.654 & $-1.147$ & $22.1 \pm0.8 $ & $5.72\pm0.02$ & 5.89 & $0.17$ & 32 &  & 6136.614 & \ion{Fe}{i} & 2.453 & $-1.400$ & $81.7 \pm0.9 $ & $5.69\pm0.02$ & 5.77 & $0.08$ & 33\\
4647.434 & \ion{Fe}{i} & 2.949 & $-1.351$ & $53.8 \pm0.7 $ & $5.78\pm0.01$ & 5.95 & $0.17$ & 32 &  & 6136.993 & \ion{Fe}{i} & 2.198 & $-2.950$ & $17.9 \pm0.8 $ & $5.77\pm0.02$ & 5.88 & $0.11$ & 33\\
4710.283 & \ion{Fe}{i} & 3.018 & $-1.612$ & $38.9 \pm0.9 $ & $5.85\pm0.02$ & 6.02 & $0.17$ & 32 &  & 6137.691 & \ion{Fe}{i} & 2.588 & $-1.403$ & $76.4 \pm0.8 $ & $5.74\pm0.01$ & 5.83 & $0.09$ & 33\\
4733.591 & \ion{Fe}{i} & 1.485 & $-2.988$ & $47.2 \pm0.9 $ & $5.71\pm0.01$ & 5.91 & $0.20$ & 32 &  & 6151.617 & \ion{Fe}{i} & 2.176 & $-3.299$ & $7.2  \pm0.8 $ & $5.66\pm0.05$ & 5.66 & $0.00$ & 33\\
4736.773 & \ion{Fe}{i} & 3.211 & $-0.752$ & $70.8 \pm0.8 $ & $5.75\pm0.02$ & 5.90 & $0.15$ & 32 &  & 6173.334 & \ion{Fe}{i} & 2.223 & $-2.880$ & $19.8 \pm0.7 $ & $5.78\pm0.02$ & 5.78 & $0.00$ & 33\\
4800.649 & \ion{Fe}{i} & 4.143 & $-1.029$ & $10.4 \pm1.2 $ & $5.72\pm0.06$ & 5.89 & $0.17$ & 32 &  & 6219.280 & \ion{Fe}{i} & 2.198 & $-2.433$ & $39.9 \pm0.8 $ & $5.72\pm0.01$ & 5.72 & $0.00$ & 33\\
4871.318 & \ion{Fe}{i} & 2.865 & $-0.363$ & $103.7\pm0.7 $ & $5.61\pm0.01$ & 5.68 & $0.07$ & 32 &  & 6230.722 & \ion{Fe}{i} & 2.559 & $-1.281$ & $84.2 \pm0.8 $ & $5.72\pm0.01$ & 5.86 & $0.13$ & 33\\
4890.755 & \ion{Fe}{i} & 2.876 & $-0.394$ & $107.7\pm0.7 $ & $5.73\pm0.01$ & 5.77 & $0.04$ & 32 &  & 6232.640 & \ion{Fe}{i} & 3.654 & $-1.223$ & $21.1 \pm0.7 $ & $5.71\pm0.02$ & 5.71 & $0.00$ & 31\\
4903.310 & \ion{Fe}{i} & 2.882 & $-0.926$ & $78.6 \pm0.8 $ & $5.70\pm0.01$ & 5.82 & $0.12$ & 32 &  & 6246.318 & \ion{Fe}{i} & 3.603 & $-0.733$ & $43.8 \pm0.8 $ & $5.62\pm0.01$ & 5.62 & $0.00$ & 34\\
4918.993 & \ion{Fe}{i} & 2.865 & $-0.342$ & $107.1\pm0.9 $ & $5.65\pm0.02$ & 5.71 & $0.07$ & 32 &  & 6252.555 & \ion{Fe}{i} & 2.404 & $-1.687$ & $70.9 \pm1.0 $ & $5.72\pm0.02$ & 5.90 & $0.18$ & 33\\
4938.814 & \ion{Fe}{i} & 2.876 & $-1.077$ & $68.9 \pm0.8 $ & $5.66\pm0.01$ & 5.78 & $0.12$ & 32 &  & 6265.132 & \ion{Fe}{i} & 2.176 & $-2.550$ & $36.5 \pm0.7 $ & $5.76\pm0.01$ & 5.95 & $0.20$ & 33\\
4939.686 & \ion{Fe}{i} & 0.859 & $-3.340$ & $68.0 \pm0.8 $ & $5.71\pm0.01$ & 5.82 & $0.11$ & 33 &  & 6297.792 & \ion{Fe}{i} & 2.223 & $-2.740$ & $21.0 \pm0.8 $ & $5.67\pm0.02$ & 5.86 & $0.20$ & 33\\
4985.253 & \ion{Fe}{i} & 3.929 & $-0.560$ & $38.5 \pm0.8 $ & $5.75\pm0.01$ & 5.90 & $0.15$ & 32 &  & 6301.499 & \ion{Fe}{i} & 3.654 & $-0.718$ & $43.2 \pm0.8 $ & $5.65\pm0.01$ & 5.81 & $0.16$ & 34\\
4994.129 & \ion{Fe}{i} & 0.915 & $-3.080$ & $78.8 \pm0.8 $ & $5.70\pm0.01$ & 5.87 & $0.17$ & 33 &  & 6336.823 & \ion{Fe}{i} & 3.686 & $-0.856$ & $36.5 \pm0.9 $ & $5.70\pm0.02$ & 5.86 & $0.16$ & 31\\
5006.118 & \ion{Fe}{i} & 2.833 & $-0.638$ & $96.0 \pm0.9 $ & $5.68\pm0.02$ & 5.80 & $0.12$ & 32 &  & 6344.147 & \ion{Fe}{i} & 2.433 & $-2.923$ & $13.4 \pm0.9 $ & $5.85\pm0.03$ & 5.85 & $0.00$ & 33\\
5012.068 & \ion{Fe}{i} & 0.859 & $-2.642$ & $109.5\pm1.2 $ & $5.85\pm0.03$ & 5.96 & $0.12$ & 33 &  & 6400.000 & \ion{Fe}{i} & 3.603 & $-0.290$ & $69.8 \pm0.9 $ & $5.60\pm0.01$ & 5.77 & $0.17$ & 34\\
5028.126 & \ion{Fe}{i} & 3.573 & $-1.123$ & $29.1 \pm0.8 $ & $5.75\pm0.02$ & 5.90 & $0.15$ & 32 &  & 6408.017 & \ion{Fe}{i} & 3.686 & $-1.018$ & $28.3 \pm0.9 $ & $5.70\pm0.02$ & 5.86 & $0.16$ & 34\\
5044.211 & \ion{Fe}{i} & 2.851 & $-2.038$ & $21.4 \pm0.8 $ & $5.72\pm0.02$ & 5.72 & $0.00$ & 32 &  & 6411.648 & \ion{Fe}{i} & 3.654 & $-0.595$ & $51.2 \pm0.9 $ & $5.65\pm0.02$ & 5.80 & $0.14$ & 34\\
5049.819 & \ion{Fe}{i} & 2.279 & $-1.355$ & $90.8 \pm0.8 $ & $5.71\pm0.02$ & 5.85 & $0.15$ & 32 &  & 6421.350 & \ion{Fe}{i} & 2.279 & $-2.027$ & $61.1 \pm0.9 $ & $5.75\pm0.01$ & 5.75 & $0.00$ & 33\\
5051.634 & \ion{Fe}{i} & 0.915 & $-2.795$ & $96.7 \pm0.8 $ & $5.78\pm0.02$ & 5.94 & $0.17$ & 33 &  & 6430.845 & \ion{Fe}{i} & 2.176 & $-2.006$ & $69.4 \pm0.8 $ & $5.75\pm0.01$ & 5.75 & $0.00$ & 33\\
5068.766 & \ion{Fe}{i} & 2.940 & $-1.042$ & $64.6 \pm0.9 $ & $5.61\pm0.01$ & 5.62 & $0.01$ & 32 &  & 6481.869 & \ion{Fe}{i} & 2.279 & $-2.984$ & $12.1 \pm0.9 $ & $5.68\pm0.04$ & 5.68 & $0.00$ & 33\\

\hline
\end{tabular}
\addtolength{\tabcolsep}{3pt}}
\end{table*}

\begin{table*}\ContinuedFloat
\caption[]{continued.}
\centering
\resizebox{2\columnwidth}{!}{%
\addtolength{\tabcolsep}{-3pt}
\begin{tabular}{llrrrrcrllllrrrrcrl}
\hline\hline\\[-7pt]
\multicolumn{1}{c}{$\lambda$} & \multicolumn{1}{c}{$X$} & $\chi_\mathrm{ex}$ & \multicolumn{1}{c}{$\log{gf}$} & \multicolumn{1}{c}{EW} & \multicolumn{2}{c}{$\log{\epsilon(X)}$} & $\Delta$ & Ref.$^b$ & & \multicolumn{1}{c}{$\lambda$} & \multicolumn{1}{c}{$X$} & $\chi_\mathrm{ex}$ & \multicolumn{1}{c}{$\log{gf}$} & \multicolumn{1}{c}{EW} & \multicolumn{2}{c}{$\log{\epsilon(X)}$} & $\Delta$ & Ref.$^b$\\[1pt]
\cline{6-7}\cline{16-17}\\[-5pt]
 & & & & & \multicolumn{1}{c}{LTE} & NLTE & & & & & & & & & \multicolumn{1}{c}{LTE} & NLTE & &\\
\multicolumn{1}{c}{$[${\AA}]}   &        &  [eV]  & \multicolumn{1}{c}{[dex]}  & \multicolumn{1}{c}{[m{\AA}]}  & \multicolumn{1}{c}{[dex]} & [dex] & \multicolumn{1}{c}{[dex]} & & & \multicolumn{1}{c}{$[${\AA}]}   &        &  [eV]  & \multicolumn{1}{c}{[dex]}  & \multicolumn{1}{c}{[m{\AA}]}  & \multicolumn{1}{c}{[dex]} & [dex] & \multicolumn{1}{c}{[dex]} & \\
\hline\\[-5pt]
6593.869 & \ion{Fe}{i} & 2.433 & $-2.422$ & $31.1 \pm0.9 $ & $5.79\pm0.02$ & 5.88 & $0.09$ & 33 &  & 4317.309 & \ion{Zr}{ii} & 0.713 & $-1.450$ & $18.4 \pm1.0 $ & $1.42\pm0.03$ & \dots & \dots & 43\\
6663.441 & \ion{Fe}{i} & 2.424 & $-2.479$ & $30.2 \pm0.8 $ & $5.82\pm0.02$ & 5.97 & $0.16$ & 33 &  & 4379.742 & \ion{Zr}{ii} & 1.532 & $-0.356$ & $20.4 \pm0.9 $ & $1.26\pm0.02$ & \dots & \dots & 44\\
6677.985 & \ion{Fe}{i} & 2.692 & $-1.418$ & $70.4 \pm0.8 $ & $5.74\pm0.01$ & 5.90 & $0.15$ & 32 &  & 5112.270 & \ion{Zr}{ii} & 1.665 & $-0.850$ & $6.4  \pm0.6 $ & $1.28\pm0.05$ & 1.46 & $0.18$ & 43\\
6750.151 & \ion{Fe}{i} & 2.424 & $-2.621$ & $22.0 \pm0.8 $ & $5.78\pm0.02$ & 5.78 & $0.00$ & 33 &  & 3864.103 & \ion{Mo}{i} & 0.000 & $-0.010$ & syn & $0.48\pm0.10$ & \dots & \dots & 45\\
4491.400 & \ion{Fe}{ii} & 2.856 & $-2.640$ & $71.1 \pm1.2 $ & $5.90\pm0.02$ & \dots & \dots & 35 &  & 3498.942 & \ion{Ru}{i} & 0.000 & $0.310$ & syn & $0.55\pm0.10$ & \dots & \dots & 46\\
4515.334 & \ion{Fe}{ii} & 2.844 & $-2.360$ & $82.3 \pm1.9 $ & $5.82\pm0.04$ & \dots & \dots & 35 &  & 3434.885 & \ion{Rh}{i} & 0.000 & $0.450$ & syn & $-0.19\pm0.40$ & \dots & \dots & 47\\
4555.888 & \ion{Fe}{ii} & 2.828 & $-2.250$ & $91.6 \pm1.1 $ & $5.88\pm0.02$ & \dots & \dots & 35 &  & 3404.579 & \ion{Pd}{i} & 0.814 & $0.320$ & syn & $-0.12\pm0.20$ & \dots & \dots & 48\\
4576.328 & \ion{Fe}{ii} & 2.844 & $-2.920$ & $60.9 \pm0.7 $ & $5.96\pm0.01$ & \dots & \dots & 35 &  & 3382.889$^{a}$ & \ion{Ag}{i} & 0.000 & $-0.377$ & syn & $-0.29\pm0.20$ & \dots & \dots & 49,50\\
4582.835 & \ion{Fe}{ii} & 2.844 & $-3.060$ & $47.9 \pm1.0 $ & $5.88\pm0.02$ & \dots & \dots & 35 &  & 5853.668$^{a}$ & \ion{Ba}{ii} & 0.604 & $-1.010$ & syn & $0.77\pm0.05$ & 0.67 & $-0.10$ & 51\\
4620.513 & \ion{Fe}{ii} & 2.828 & $-3.190$ & $41.9 \pm0.7 $ & $5.89\pm0.01$ & \dots & \dots & 35 &  & 6496.897$^{a}$ & \ion{Ba}{ii} & 0.604 & $-0.380$ & syn & $1.09\pm0.10$ & 0.82 & $-0.27$ & 51\\
4656.981 & \ion{Fe}{ii} & 2.891 & $-3.570$ & $23.6 \pm1.0 $ & $5.97\pm0.02$ & \dots & \dots & 35 &  & 4123.220$^{a}$ & \ion{La}{ii} & 0.321 & $0.130$ & $47.4 \pm0.9 $ & $-0.19\pm0.01$ & \dots & \dots & 52\\
4923.922 & \ion{Fe}{ii} & 2.891 & $-1.210$ & $147.8\pm0.9 $ & $5.91\pm0.01$ & \dots & \dots & 35 &  & 4322.500$^{a}$ & \ion{La}{ii} & 0.173 & $-0.930$ & $6.9  \pm0.7 $ & $-0.35\pm0.05$ & \dots & \dots & 52\\
5197.568 & \ion{Fe}{ii} & 3.230 & $-2.050$ & $80.4 \pm1.0 $ & $5.79\pm0.02$ & \dots & \dots & 35 &  & 5114.560$^{a}$ & \ion{La}{ii} & 0.235 & $-1.030$ & $10.1 \pm0.8 $ & $-0.09\pm0.04$ & \dots & \dots & 52\\
5275.997 & \ion{Fe}{ii} & 3.199 & $-1.900$ & $96.1 \pm1.0 $ & $5.90\pm0.02$ & \dots & \dots & 35 &  & 5122.990$^{a}$ & \ion{La}{ii} & 0.321 & $-0.910$ & $10.4 \pm0.8 $ & $-0.09\pm0.04$ & \dots & \dots & 52\\
5414.089 & \ion{Fe}{ii} & 3.221 & $-3.480$ & $10.9 \pm0.6 $ & $5.80\pm0.03$ & \dots & \dots & 35 &  & 6390.480$^{a}$ & \ion{La}{ii} & 0.321 & $-1.410$ & $4.5  \pm0.8 $ & $-0.05\pm0.09$ & \dots & \dots & 52\\
5525.117 & \ion{Fe}{ii} & 3.267 & $-3.940$ & $5.2  \pm0.5 $ & $5.95\pm0.02$ & \dots & \dots & 35 &  & 6774.268 & \ion{La}{ii} & 0.126 & $-1.820$ & $3.2  \pm0.6 $ & $-0.05\pm0.09$ & \dots & \dots & 53\\
6432.677 & \ion{Fe}{ii} & 2.891 & $-3.500$ & $27.2 \pm0.7 $ & $5.91\pm0.01$ & \dots & \dots & 35 &  & 4083.222 & \ion{Ce}{ii} & 0.701 & $0.270$ & $14.0 \pm1.1 $ & $0.16\pm0.04$ & \dots & \dots & 54\\
3412.336 & \ion{Co}{i} & 0.514 & $0.030$ & $96.4 \pm2.6 $ & $2.99\pm0.08$ & \dots & \dots & 33 &  & 4137.645 & \ion{Ce}{ii} & 0.516 & $0.400$ & $27.7 \pm1.2 $ & $0.19\pm0.03$ & \dots & \dots & 54\\
3518.348 & \ion{Co}{i} & 1.049 & $0.070$ & $89.4 \pm8.4 $ & $3.28\pm0.23$ & \dots & \dots & 33 &  & 4222.597 & \ion{Ce}{ii} & 0.122 & $-0.150$ & $22.9 \pm1.1 $ & $0.19\pm0.03$ & \dots & \dots & 54\\
4121.316$^{a}$ & \ion{Co}{i} & 0.922 & $-0.320$ & $90.8 \pm0.8 $ & $3.03\pm0.02$ & 3.61 & $0.58$ & 3 &  & 4364.653 & \ion{Ce}{ii} & 0.495 & $-0.170$ & $10.4 \pm0.8 $ & $0.19\pm0.04$ & \dots & \dots & 54\\
4813.476 & \ion{Co}{i} & 3.216 & $0.120$ & $6.3  \pm0.7 $ & $3.22\pm0.05$ & 3.57 & $0.35$ & 3 &  & 4382.165 & \ion{Ce}{ii} & 0.684 & $0.130$ & $9.5  \pm0.8 $ & $0.06\pm0.04$ & \dots & \dots & 54\\
4867.869 & \ion{Co}{i} & 3.117 & $0.302$ & $13.9 \pm0.7 $ & $3.30\pm0.03$ & 3.76 & $0.46$ & 3 &  & 4449.330 & \ion{Ce}{ii} & 0.609 & $0.040$ & syn & $0.19\pm0.10$ & \dots & \dots & 54\\
5342.701 & \ion{Co}{i} & 4.021 & $0.741$ & $5.4  \pm0.8 $ & $3.35\pm0.07$ & \dots & \dots & 3 &  & 4486.909 & \ion{Ce}{ii} & 0.295 & $-0.180$ & $16.3 \pm1.0 $ & $0.20\pm0.03$ & \dots & \dots & 54\\
4331.642 & \ion{Ni}{i} & 1.676 & $-2.270$ & $23.2 \pm1.0 $ & $4.47\pm0.02$ & \dots & \dots & 36 &  & 4562.359 & \ion{Ce}{ii} & 0.478 & $0.210$ & $22.5 \pm0.8 $ & $0.17\pm0.02$ & \dots & \dots & 54\\
4470.477 & \ion{Ni}{i} & 3.399 & $-0.300$ & $26.9 \pm1.1 $ & $4.41\pm0.03$ & \dots & \dots & 36 &  & 4628.161 & \ion{Ce}{ii} & 0.516 & $0.140$ & $18.5 \pm0.8 $ & $0.17\pm0.02$ & \dots & \dots & 54\\
4756.515 & \ion{Ni}{i} & 3.480 & $-0.270$ & $24.5 \pm0.8 $ & $4.40\pm0.02$ & \dots & \dots & 36 &  & 5187.458 & \ion{Ce}{ii} & 1.212 & $0.170$ & $5.2  \pm1.3 $ & $0.25\pm0.13$ & \dots & \dots & 54\\
4806.987 & \ion{Ni}{i} & 3.679 & $-0.640$ & $8.4  \pm0.6 $ & $4.43\pm0.03$ & \dots & \dots & 33 &  & 5274.229 & \ion{Ce}{ii} & 1.044 & $0.130$ & $8.4  \pm0.8 $ & $0.32\pm0.05$ & \dots & \dots & 54\\
4829.023 & \ion{Ni}{i} & 3.542 & $-0.330$ & $24.1 \pm1.2 $ & $4.51\pm0.03$ & \dots & \dots & 33 &  & 4405.829$^{a}$ & \ion{Pr}{ii} & 0.550 & $-0.060$ & $4.0  \pm0.9 $ & $-0.38\pm0.10$ & \dots & \dots & 55\\
4831.176 & \ion{Ni}{i} & 3.606 & $-0.320$ & $18.7 \pm0.7 $ & $4.43\pm0.02$ & \dots & \dots & 36 &  & 4408.810$^{a}$ & \ion{Pr}{ii} & 0.000 & $0.050$ & syn & $-0.35\pm0.10$ & \dots & \dots & 55\\
4866.271 & \ion{Ni}{i} & 3.539 & $-0.220$ & $21.8 \pm0.8 $ & $4.34\pm0.02$ & \dots & \dots & 36 &  & 4449.828$^{a}$ & \ion{Pr}{ii} & 0.204 & $-0.260$ & syn & $-0.31\pm0.20$ & \dots & \dots & 55\\
4873.442 & \ion{Ni}{i} & 3.699 & $-0.380$ & $11.8 \pm0.8 $ & $4.35\pm0.03$ & \dots & \dots & 36 &  & 3780.380 & \ion{Nd}{ii} & 0.471 & $-0.350$ & $6.9  \pm1.0 $ & $0.02\pm0.07$ & \dots & \dots & 56\\
4904.412 & \ion{Ni}{i} & 3.542 & $-0.170$ & $29.0 \pm0.8 $ & $4.45\pm0.02$ & \dots & \dots & 33 &  & 4012.700 & \ion{Nd}{ii} & 0.000 & $-0.600$ & $18.4 \pm2.3 $ & $0.20\pm0.07$ & \dots & \dots & 56\\
4913.973 & \ion{Ni}{i} & 3.743 & $-0.630$ & $7.3  \pm0.7 $ & $4.41\pm0.05$ & \dots & \dots & 33 &  & 4021.330 & \ion{Nd}{ii} & 0.321 & $-0.100$ & $21.8 \pm0.7 $ & $0.15\pm0.02$ & \dots & \dots & 56\\
4935.831 & \ion{Ni}{i} & 3.941 & $-0.350$ & $10.7 \pm0.7 $ & $4.52\pm0.03$ & \dots & \dots & 33 &  & 4023.000 & \ion{Nd}{ii} & 0.559 & $0.040$ & $21.6 \pm2.6 $ & $0.26\pm0.07$ & \dots & \dots & 56\\
4937.348 & \ion{Ni}{i} & 3.606 & $-0.390$ & $17.3 \pm0.8 $ & $4.45\pm0.03$ & \dots & \dots & 33 &  & 4109.450 & \ion{Nd}{ii} & 0.321 & $0.350$ & $51.5 \pm8.1 $ & $0.28\pm0.15$ & \dots & \dots & 56\\
4996.844 & \ion{Ni}{i} & 3.635 & $-0.980$ & $6.0  \pm0.8 $ & $4.56\pm0.07$ & \dots & \dots & 33 &  & 4133.350 & \ion{Nd}{ii} & 0.321 & $-0.490$ & $13.2 \pm1.1 $ & $0.26\pm0.04$ & \dots & \dots & 56\\
5017.576 & \ion{Ni}{i} & 3.539 & $-0.030$ & $29.1 \pm0.9 $ & $4.31\pm0.02$ & \dots & \dots & 36 &  & 4211.290 & \ion{Nd}{ii} & 0.205 & $-0.860$ & $9.6  \pm0.9 $ & $0.34\pm0.05$ & \dots & \dots & 56\\
5035.362 & \ion{Ni}{i} & 3.635 & $0.290$ & $40.1 \pm0.6 $ & $4.30\pm0.01$ & \dots & \dots & 36 &  & 4232.370 & \ion{Nd}{ii} & 0.064 & $-0.470$ & $21.4 \pm0.9 $ & $0.20\pm0.02$ & \dots & \dots & 56\\
5082.344 & \ion{Ni}{i} & 3.658 & $-0.540$ & $11.8 \pm0.7 $ & $4.46\pm0.03$ & \dots & \dots & 33 &  & 4446.380 & \ion{Nd}{ii} & 0.205 & $-0.350$ & $22.0 \pm1.0 $ & $0.23\pm0.03$ & \dots & \dots & 56\\
5084.096 & \ion{Ni}{i} & 3.679 & $0.030$ & $30.7 \pm0.7 $ & $4.43\pm0.01$ & \dots & \dots & 33 &  & 4706.540 & \ion{Nd}{ii} & 0.000 & $-0.710$ & $18.0 \pm0.9 $ & $0.22\pm0.03$ & \dots & \dots & 56\\
5099.930 & \ion{Ni}{i} & 3.679 & $-0.100$ & $21.3 \pm0.7 $ & $4.35\pm0.02$ & \dots & \dots & 33 &  & 4811.342 & \ion{Nd}{ii} & 0.064 & $-1.140$ & $13.2 \pm0.6 $ & $0.56\pm0.02$ & \dots & \dots & 57\\
5102.966 & \ion{Ni}{i} & 1.676 & $-2.870$ & $10.1 \pm0.8 $ & $4.58\pm0.04$ & \dots & \dots & 36 &  & 4825.480 & \ion{Nd}{ii} & 0.182 & $-0.420$ & $21.9 \pm1.0 $ & $0.23\pm0.02$ & \dots & \dots & 56\\
5115.392 & \ion{Ni}{i} & 3.834 & $-0.110$ & $20.3 \pm0.7 $ & $4.49\pm0.02$ & \dots & \dots & 33 &  & 4914.380 & \ion{Nd}{ii} & 0.380 & $-0.700$ & $8.6  \pm0.8 $ & $0.25\pm0.04$ & \dots & \dots & 56\\
5155.126 & \ion{Ni}{i} & 3.898 & $-0.650$ & $7.1  \pm0.7 $ & $4.58\pm0.05$ & \dots & \dots & 33 &  & 4989.950 & \ion{Nd}{ii} & 0.630 & $-0.500$ & $6.7  \pm0.6 $ & $0.21\pm0.04$ & \dots & \dots & 57\\
5155.764 & \ion{Ni}{i} & 3.898 & $0.074$ & $16.5 \pm0.7 $ & $4.26\pm0.02$ & \dots & \dots & 3 &  & 5089.832 & \ion{Nd}{ii} & 0.205 & $-1.160$ & $2.9  \pm0.4 $ & $0.01\pm0.07$ & \dots & \dots & 57\\
5176.560 & \ion{Ni}{i} & 3.898 & $-0.440$ & $6.4  \pm0.6 $ & $4.32\pm0.05$ & \dots & \dots & 33 &  & 5092.790 & \ion{Nd}{ii} & 0.380 & $-0.610$ & $9.8  \pm0.6 $ & $0.21\pm0.03$ & \dots & \dots & 56\\
5578.718 & \ion{Ni}{i} & 1.676 & $-2.830$ & $9.6  \pm0.5 $ & $4.49\pm0.03$ & \dots & \dots & 36 &  & 5130.590 & \ion{Nd}{ii} & 1.304 & $0.450$ & $10.6 \pm1.0 $ & $0.20\pm0.05$ & \dots & \dots & 56\\
5614.773 & \ion{Ni}{i} & 4.154 & $-0.573$ & $3.5  \pm0.6 $ & $4.43\pm0.08$ & \dots & \dots & 3 &  & 5212.360 & \ion{Nd}{ii} & 0.205 & $-0.960$ & $5.7  \pm0.6 $ & $0.11\pm0.05$ & \dots & \dots & 56\\
5682.199 & \ion{Ni}{i} & 4.105 & $-0.470$ & $6.6  \pm0.6 $ & $4.56\pm0.04$ & \dots & \dots & 33 &  & 5234.190 & \ion{Nd}{ii} & 0.550 & $-0.510$ & $8.7  \pm0.8 $ & $0.23\pm0.04$ & \dots & \dots & 56\\
6108.116 & \ion{Ni}{i} & 1.676 & $-2.600$ & $13.2 \pm0.8 $ & $4.40\pm0.03$ & \dots & \dots & 36 &  & 5293.160 & \ion{Nd}{ii} & 0.823 & $0.100$ & $15.0 \pm0.7 $ & $0.19\pm0.02$ & \dots & \dots & 56\\
6175.367 & \ion{Ni}{i} & 4.089 & $-0.530$ & $5.1  \pm0.7 $ & $4.47\pm0.07$ & \dots & \dots & 33 &  & 5319.810 & \ion{Nd}{ii} & 0.550 & $-0.140$ & $16.5 \pm0.7 $ & $0.17\pm0.02$ & \dots & \dots & 56\\
6176.811 & \ion{Ni}{i} & 4.088 & $-0.260$ & $8.8  \pm0.7 $ & $4.45\pm0.04$ & \dots & \dots & 36 &  & 5688.520 & \ion{Nd}{ii} & 0.986 & $-0.310$ & $6.2  \pm0.6 $ & $0.33\pm0.05$ & \dots & \dots & 56\\
6643.630 & \ion{Ni}{i} & 1.676 & $-2.220$ & $29.7 \pm0.7 $ & $4.42\pm0.02$ & \dots & \dots & 36 &  & 3896.970 & \ion{Sm}{ii} & 0.041 & $-0.670$ & $9.7  \pm0.6 $ & $-0.27\pm0.03$ & \dots & \dots & 58\\
6767.772 & \ion{Ni}{i} & 1.826 & $-2.140$ & $25.7 \pm0.9 $ & $4.42\pm0.02$ & \dots & \dots & 36 &  & 4220.660 & \ion{Sm}{ii} & 0.544 & $-0.440$ & $9.0  \pm0.8 $ & $-0.03\pm0.04$ & \dots & \dots & 58\\
6772.315 & \ion{Ni}{i} & 3.658 & $-0.980$ & $6.5  \pm1.0 $ & $4.56\pm0.08$ & \dots & \dots & 33 &  & 4318.927 & \ion{Sm}{ii} & 0.278 & $-0.250$ & $21.3 \pm1.1 $ & $-0.08\pm0.03$ & \dots & \dots & 58\\
5105.537 & \ion{Cu}{i} & 1.389 & $-1.542$ & $7.6  \pm0.8 $ & $1.71\pm0.05$ & \dots & \dots & 3 &  & 4421.130 & \ion{Sm}{ii} & 0.379 & $-0.490$ & $14.3 \pm1.4 $ & $0.05\pm0.05$ & \dots & \dots & 58\\
5218.198 & \ion{Cu}{i} & 3.817 & $0.364$ & $3.5  \pm0.5 $ & $2.04\pm0.07$ & \dots & \dots & 3 &  & 4434.320 & \ion{Sm}{ii} & 0.379 & $-0.070$ & $25.1 \pm1.1 $ & $-0.07\pm0.02$ & \dots & \dots & 58\\
5782.113$^{a}$ & \ion{Cu}{i} & 1.641 & $-1.720$ & $3.3  \pm0.6 $ & $1.76\pm0.08$ & \dots & \dots & 3 &  & 4519.630 & \ion{Sm}{ii} & 0.544 & $-0.350$ & $16.2 \pm1.7 $ & $0.15\pm0.06$ & \dots & \dots & 58\\
4722.153 & \ion{Zn}{i} & 4.030 & $-0.338$ & $32.3 \pm0.7 $ & $2.88\pm0.01$ & \dots & \dots & 37 &  & 4523.910 & \ion{Sm}{ii} & 0.434 & $-0.390$ & $11.1 \pm1.3 $ & $-0.12\pm0.06$ & \dots & \dots & 58\\
4810.528 & \ion{Zn}{i} & 4.078 & $-0.137$ & $40.5 \pm0.8 $ & $2.88\pm0.01$ & \dots & \dots & 37 &  & 4577.690 & \ion{Sm}{ii} & 0.248 & $-0.650$ & $11.2 \pm0.9 $ & $-0.07\pm0.04$ & \dots & \dots & 58\\
4172.042 & \ion{Ga}{i} & 0.102 & $-0.308$ & syn & $1.03\pm0.20$ & \dots & \dots & 38 &  & 4591.810 & \ion{Sm}{ii} & 0.185 & $-1.120$ & $5.0  \pm1.0 $ & $-0.05\pm0.10$ & \dots & \dots & 58\\
7800.259 & \ion{Rb}{i} & 0.000 & $0.137$ & syn & $<1.52$ & \dots & \dots & 39 &  & 4642.228 & \ion{Sm}{ii} & 0.379 & $-0.460$ & $13.1 \pm0.7 $ & $-0.05\pm0.03$ & \dots & \dots & 58\\
4607.331 & \ion{Sr}{i} & 0.000 & $0.283$ & $9.1  \pm1.2 $ & $1.00\pm0.07$ & 1.40 & $0.40$ & 40 &  & 4676.900 & \ion{Sm}{ii} & 0.041 & $-0.870$ & $11.1 \pm0.8 $ & $-0.09\pm0.04$ & \dots & \dots & 58\\
4161.792 & \ion{Sr}{ii} & 2.940 & $-0.502$ & $16.7 \pm6.7 $ & $1.50\pm0.26$ & \dots & \dots & 39 &  & 4704.400 & \ion{Sm}{ii} & 0.000 & $-0.860$ & $15.2 \pm0.8 $ & $0.01\pm0.03$ & \dots & \dots & 58\\
4398.010 & \ion{Y}{ii} & 0.130 & $-1.000$ & $47.2 \pm1.0 $ & $0.45\pm0.02$ & \dots & \dots & 41 &  & 4129.720$^{a}$ & \ion{Eu}{ii} & 0.000 & $0.220$ & syn & $-0.41\pm0.05$ & \dots & \dots & 59\\
5087.419 & \ion{Y}{ii} & 1.084 & $-0.170$ & $42.8 \pm0.6 $ & $0.53\pm0.01$ & \dots & \dots & 41 &  & 4205.040$^{a}$ & \ion{Eu}{ii} & 0.000 & $0.210$ & syn & $-0.35\pm0.05$ & \dots & \dots & 59\\
5119.112 & \ion{Y}{ii} & 0.992 & $-1.360$ & $6.5  \pm0.7 $ & $0.60\pm0.05$ & \dots & \dots & 41 &  & 6645.060$^{a}$ & \ion{Eu}{ii} & 1.379 & $0.120$ & $10.9 \pm0.8 $ & $-0.29\pm0.04$ & \dots & \dots & 59\\
5123.211 & \ion{Y}{ii} & 0.992 & $-0.830$ & $18.8 \pm0.9 $ & $0.58\pm0.03$ & \dots & \dots & 41 &  & 3481.802 & \ion{Gd}{ii} & 0.492 & $0.110$ & $10.0 \pm1.6 $ & $-0.29\pm0.08$ & \dots & \dots & 60\\
5200.410 & \ion{Y}{ii} & 0.992 & $-0.570$ & $25.1 \pm0.7 $ & $0.48\pm0.02$ & \dots & \dots & 41 &  & 3768.396 & \ion{Gd}{ii} & 0.079 & $0.210$ & syn & $0.00\pm0.10$ & \dots & \dots & 60\\
5205.722 & \ion{Y}{ii} & 1.032 & $-0.340$ & $37.6 \pm0.8 $ & $0.54\pm0.01$ & \dots & \dots & 41 &  & 4130.366 & \ion{Gd}{ii} & 0.731 & $0.139$ & syn & $0.12\pm0.05$ & \dots & \dots & 57\\
5402.774 & \ion{Y}{ii} & 1.839 & $-0.630$ & $6.9  \pm0.6 $ & $0.80\pm0.04$ & \dots & \dots & 42 &  & 4215.022 & \ion{Gd}{ii} & 0.427 & $-0.440$ & $10.3 \pm0.8 $ & $0.07\pm0.04$ & \dots & \dots & 60\\
3499.571 & \ion{Zr}{ii} & 0.409 & $-1.060$ & syn & $1.16\pm0.10$ & \dots & \dots & 43 &  & 4251.731 & \ion{Gd}{ii} & 0.382 & $-0.220$ & syn & $0.15\pm0.10$ & \dots & \dots & 60\\
4208.980 & \ion{Zr}{ii} & 0.713 & $-0.510$ & $53.5 \pm1.0 $ & $1.20\pm0.02$ & 1.35 & $0.15$ & 43 &  & 3848.730$^{a}$ & \ion{Tb}{ii} & 0.000 & $0.280$ & syn & $-0.72\pm0.15$ & \dots & \dots & 61\\

\hline
\end{tabular}
\addtolength{\tabcolsep}{3pt}}
\end{table*}

\begin{table*}\ContinuedFloat
\caption[]{continued.}
\centering
\resizebox{2\columnwidth}{!}{%
\addtolength{\tabcolsep}{-3pt}
\begin{tabular}{llrrrrcrllllrrrrcrl}
\hline\hline\\[-7pt]
\multicolumn{1}{c}{$\lambda$} & \multicolumn{1}{c}{$X$} & $\chi_\mathrm{ex}$ & \multicolumn{1}{c}{$\log{gf}$} & \multicolumn{1}{c}{EW} & \multicolumn{2}{c}{$\log{\epsilon(X)}$} & $\Delta$ & Ref.$^b$ & & \multicolumn{1}{c}{$\lambda$} & \multicolumn{1}{c}{$X$} & $\chi_\mathrm{ex}$ & \multicolumn{1}{c}{$\log{gf}$} & \multicolumn{1}{c}{EW} & \multicolumn{2}{c}{$\log{\epsilon(X)}$} & $\Delta$ & Ref.$^b$\\[1pt]
\cline{6-7}\cline{16-17}\\[-5pt]
 & & & & & \multicolumn{1}{c}{LTE} & NLTE & & & & & & & & & \multicolumn{1}{c}{LTE} & NLTE & &\\
\multicolumn{1}{c}{$[${\AA}]}   &        &  [eV]  & \multicolumn{1}{c}{[dex]}  & \multicolumn{1}{c}{[m{\AA}]}  & \multicolumn{1}{c}{[dex]} & [dex] & \multicolumn{1}{c}{[dex]} & & & \multicolumn{1}{c}{$[${\AA}]}   &        &  [eV]  & \multicolumn{1}{c}{[dex]}  & \multicolumn{1}{c}{[m{\AA}]}  & \multicolumn{1}{c}{[dex]} & [dex] & \multicolumn{1}{c}{[dex]} & \\
\hline\\[-5pt]
3874.170$^{a}$ & \ion{Tb}{ii} & 0.000 & $0.270$ & syn & $-0.77\pm0.15$ & \dots & \dots & 61 &  & 3795.760 & \ion{Tm}{ii} & 0.029 & $-0.230$ & syn & $-0.85\pm0.10$ & \dots & \dots & 65\\
3460.970 & \ion{Dy}{ii} & 0.000 & $-0.070$ & $35.2 \pm2.4 $ & $-0.12\pm0.05$ & \dots & \dots & 62 &  & 3848.020 & \ion{Tm}{ii} & 0.000 & $-0.140$ & syn & $-0.87\pm0.10$ & \dots & \dots & 65\\
3694.810 & \ion{Dy}{ii} & 0.103 & $-0.110$ & syn & $0.24\pm0.10$ & \dots & \dots & 62 &  & 3694.192$^{a}$ & \ion{Yb}{ii} & 0.000 & $-0.300$ & syn & $-0.06\pm0.20$ & \dots & \dots & 55\\
4103.310 & \ion{Dy}{ii} & 0.103 & $-0.380$ & $36.2 \pm1.0 $ & $0.17\pm0.02$ & \dots & \dots & 62 &  & 6221.890$^{a}$ & \ion{Lu}{ii} & 1.540 & $-0.760$ & syn & $-0.57\pm0.10$ & \dots & \dots & 54\\
4449.700 & \ion{Dy}{ii} & 0.000 & $-1.030$ & syn & $0.26\pm0.10$ & \dots & \dots & 62 &  & 3399.790 & \ion{Hf}{ii} & 0.000 & $-0.570$ & syn & $-0.18\pm0.10$ & \dots & \dots & 66\\
3796.730$^{a}$ & \ion{Ho}{ii} & 0.000 & $0.160$ & syn & $-0.54\pm0.10$ & \dots & \dots & 63 &  & 3918.090 & \ion{Hf}{ii} & 0.452 & $-1.140$ & syn & $-0.23\pm0.20$ & \dots & \dots & 66\\
4045.450$^{a}$ & \ion{Ho}{ii} & 0.000 & $-0.050$ & syn & $-0.44\pm0.10$ & \dots & \dots & 63 &  & 4093.150 & \ion{Hf}{ii} & 0.452 & $-1.150$ & syn & $-0.27\pm0.20$ & \dots & \dots & 66\\
3499.103 & \ion{Er}{ii} & 0.055 & $0.290$ & syn & $-0.09\pm0.10$ & \dots & \dots & 64 &  & 4260.849 & \ion{Os}{i} & 0.000 & $-1.430$ & syn & $0.39\pm0.10$ & \dots & \dots & 67\\
3692.649 & \ion{Er}{ii} & 0.055 & $0.280$ & syn & $-0.11\pm0.10$ & \dots & \dots & 64 &  & 4420.468 & \ion{Os}{i} & 0.000 & $-1.530$ & syn & $0.41\pm0.20$ & \dots & \dots & 53\\
3729.524 & \ion{Er}{ii} & 0.000 & $-0.590$ & $26.6 \pm1.2 $ & $0.01\pm0.03$ & \dots & \dots & 64 &  & 3800.124 & \ion{Ir}{i} & 0.000 & $-1.442$ & syn & $0.42\pm0.10$ & \dots & \dots & 67\\
3786.836 & \ion{Er}{ii} & 0.000 & $-0.520$ & syn & $0.13\pm0.10$ & \dots & \dots & 64 &  & 4057.807 & \ion{Pb}{i} & 1.320 & $-0.220$ & syn & $<0.37$ & \dots & \dots & 68\\
3700.256 & \ion{Tm}{ii} & 0.029 & $-0.380$ & syn & $-0.87\pm0.15$ & \dots & \dots & 65 &  & 4019.129 & \ion{Th}{ii} & 0.000 & $-0.228$ & syn & $-0.85\pm0.10$ & \dots & \dots & 69\\
3701.363 & \ion{Tm}{ii} & 0.000 & $-0.540$ & syn & $-0.87\pm0.15$ & \dots & \dots & 65 &  & 3859.571 & \ion{U}{ii} & 0.036 & $-0.067$ & syn & $<-1.21$ & \dots & \dots & 70\\

\hline
\end{tabular}
\addtolength{\tabcolsep}{3pt}}
\tablefoot{
\tablefoottext{a}{Additional HFS was considered.}
\tablefoottext{b}{References: \input{line_data_footnotes.dat}}
}
\end{table*}

\end{appendix}
\end{document}